\documentclass[9pt,sigconf]{acmart}

\usepackage{booktabs} 
\usepackage[normalem]{ulem}
\usepackage{hyperref}
\usepackage[export]{adjustbox}

\usepackage{cleveref}

\crefformat{section}{\S#2#1#3} 
\crefformat{subsection}{\S#2#1#3}
\crefformat{subsubsection}{\S#2#1#3}
\crefformat{appendix}{\S#2#1#3}

\usepackage{datetime}
\usepackage{url}
\usepackage[ruled, noend]{algorithm2e}
\usepackage{float}
\usepackage{color}
\usepackage{multirow}
\usepackage{soul}
\usepackage{array}
\usepackage{subcaption}
\usepackage{soul}
\usepackage{balance}
\usepackage{enumitem}
\usepackage{graphicx}
\usepackage{amsmath}
\usepackage{kotex}
\usepackage{mathtools}
\usepackage{physics}

\usepackage{stmaryrd}
\usepackage{trimclip}
\usepackage{xcolor}
\usepackage{fontawesome}
\usepackage{needspace}

\usepackage{amsfonts}

\usepackage{pifont}

\makeatletter
\DeclareRobustCommand{\shortto}{%
  \mathrel{\mathpalette\short@to\relax}%
}

\newcommand{\short@to}[2]{%
  \mkern2mu
  \clipbox{{.3\width} 0 0 0}{$\m@th#1\vphantom{+}{\shortrightarrow}$}%
  }
\makeatother

\newcolumntype{L}[1]{>{\raggedright\let\newline\\\arraybackslash\hspace{0pt}}m{#1}}

\newcommand{\eat}[1]{}

\usepackage{tikz}
\usetikzlibrary{calc}

\newtheorem{p-rule}{\bf Rule}

\begin{document}
\fancyhead{}

\let\oldnl\nl
\newcommand{\nonl}{\renewcommand{\nl}{\let\nl\oldnl}}

\DeclarePairedDelimiter{\ceil}{\lceil}{\rceil}


\newcommand{\FuncName}[1]{\textsc{{#1}}}
\newcommand{\QueryNum}{\abs{\mathcal{M}}}
\newcommand{\CondQueryNum}{c(q, g|M)}
\newcommand{\AGM}{AGM(q)}
\newcommand{\SampleSpace}{\Omega}
\newcommand{\database}{\mathcal{D}}
\newcommand{\Keys}{\mathcal{J}}
\newcommand{\tables}{\mathcal{T}}
\newcommand{\joins}{\mathcal{J}}
\newcommand{\ranges}{\mathcal{R}}
\newcommand{\filters}{\mathcal{F}}
\newcommand{\attributes}{\mathcal{A}}
\newcommand{\TablePair}{\mathcal{T}}
\newcommand{\Tables}{\mathcal{T}}
\newcommand{\Groups}{\mathcal{G}}
\newcommand{\Matches}{\mathcal{M}}
\newcommand{\Branch}{b \cdot w(i-1)}
\newcommand{\truecard}{card(Q)}
\newcommand{\estcard}{\widehat{card}(Q)}
\newcommand{\truesel}{P(Q)}
\newcommand{\estsel}{\widehat{P}(Q)}
\newcommand{\trueprob}{P(X)}
\newcommand{\estprob}{\widehat{P}(X)}
\newcommand{\requests}{\mathcal{R}}
\newcommand{\batch}{\mathcal{B}}

\newcommand{\bound}{I}
\newcommand{\mem}{m}
\newcommand{\decode}{d}
\newcommand{\evict}{e}
\newcommand{\has}{s}

\newcommand{\stat}{stat}
\newcommand{\eststat}{\widehat{stat}}

\newcommand{\I}{\mathbb{I}}
\newcommand{\J}{\mathbb{J}}
\newcommand{\JoinKeys}{\mathcal{X}}
\newcommand{\NonJoinKeys}{\mathcal{Y}}

\useunder{\uline}{\ul}{}
\newcommand{\veryshortarrow}[1][3pt]{\mathrel{%
   \hbox{\rule[\dimexpr\fontdimen22\textfont2-.2pt\relax]{#1}{.4pt}}%
   \mkern-4mu\hbox{\usefont{U}{lasy}{m}{n}\symbol{41}}}}

\newcommand{\DataGraph}{g}
\newcommand{\DataVertexSet}{V_G}
\newcommand{\DataEdgeSet}{E_G}
\newcommand{\QueryGraph}{q}
\newcommand{\QueryVertex}{u}
\newcommand{\DataVertex}{v}
\newcommand{\Degree}[1]{Deg({#1})}
\newcommand{\Neighbor}[1]{N({#1})}
\newcommand{\Edge}[2]{\left({#1},{#2}\right)}
\newcommand{\VSet}[1]{V({#1})}
\newcommand{\ESet}[1]{E({#1})}
\newcommand{\GESet}[1]{GE({#1})}
\newcommand{\ELabel}[1]{L({#1})}

\newif\ifFullVersion

\def\FullVersion{\let\ifFullVersion=\iftrue}
\def\ShortVersion{\let\ifFullVersion=\iffalse}
\FullVersion

\newcommand{\Break}{\STATE \algorithmicbreak}

\newcommand{\NEO}{\FuncName{NEO}\xspace}
\newcommand{\Gurobi}{\FuncName{Gurobi}\xspace}
\newcommand{\Orca}{\FuncName{Orca}\xspace}
\newcommand{\vLLM}{\FuncName{vLLM}\xspace}
\newcommand{\SARATHI}{\FuncName{Sarathi}\xspace}
\newcommand{\Sarathi}{\FuncName{Sarathi}\xspace}

\newcommand{\vLLMHY}{\FuncName{vLLM}$_{hy}$\xspace}
\newcommand{\SarathiPC}{\FuncName{Sarathi}$_{P=C}$\xspace}
\newcommand{\SarathiNOCP}{\FuncName{Sarathi}$_{nocp}$\xspace}
\newcommand{\SarathiNOHY}{\FuncName{Sarathi}$_{nohy}$\xspace}

\newcommand{\InstInfer}{\FuncName{InstInfer}\xspace}
\newcommand{\LLMViewer}{\FuncName{LLMViewer}\xspace}

\newcommand{\DistServe}{\FuncName{DistServe}\xspace}
\newcommand{\Dejavu}{\FuncName{Dejavu}\xspace}
\newcommand{\vTensor}{\FuncName{vTensor}\xspace}
\newcommand{\FlexInfer}{\FuncName{FlexInfer}\xspace}
\newcommand{\InfiniGen}{\FuncName{InfiniGen}\xspace}
\newcommand{\NanoFlow}{\FuncName{NanoFlow}\xspace}
\newcommand{\Vidur}{\FuncName{Vidur}\xspace}
\newcommand{\SCLS}{\FuncName{SCLS}\xspace}
\newcommand{\SSJF}{\FuncName{SSJF}\xspace}
\newcommand{\Llumnix}{\FuncName{Llumnix}\xspace}

\newcommand{\Scheduler}{\FuncName{Schedule}}
\newcommand{\InferenceEngine}{\FuncName{InferenceEngine}}
\newcommand{\GetNewRequests}{\FuncName{GetNewRequests}}
\newcommand{\Schedule}{\FuncName{Schedule}}
\newcommand{\GetNextBatch}{\FuncName{GetNextBatch}}
\newcommand{\Process}{\FuncName{Process}}
\newcommand{\vLLMSchedule}{\FuncName{Schedule_{vLLM}}}
\newcommand{\SarathiSchedule}{\FuncName{Schedule_{Sarathi}}}
\newcommand{\RankSchedule}{\FuncName{Schedule_{Rank}}}
\newcommand{\CanAllocate}{\FuncName{CanAllocate}}
\newcommand{\Promote}{\FuncName{Promote}}
\newcommand{\Demote}{\FuncName{Demote}}
\newcommand{\OrderRequests}{\FuncName{OrderRequests}}
\newcommand{\BreakOrEvictRequestsIfMIsInsufficient}{\FuncName{BreakOrEvictRequestsIfMIsInsufficient}}
\newcommand{\RemoveRequestFromOriginalQueue}{\FuncName{RemoveRequestFromOriginalQueue}}
\newcommand{\BreakIfHybridBatchingDisabled}{\FuncName{CheckAndBreakIfHybridBatchingDisabled}}

\newcommand{\vertex}{v}
\newcommand{\none}{-}
\newcommand{\QueryTree}{q'}
\newcommand{\convert}{convert}
\newcommand{\precondition}{precondition}
\SetKw{Continue}{continue}

\newcommand{\GMO}{mo_{g}}
\newcommand{\PartialOrders}{PO}

\newcommand{\SWITCH}[1]{\STATE \textbf{switch} (#1)}
\newcommand{\ENDSWITCH}{\STATE \textbf{end switch}}
\newcommand{\CASE}[1]{\STATE \textbf{case} #1\textbf{:} \begin{ALC@g}}
\newcommand{\ENDCASE}{\end{ALC@g}}
\newcommand{\CASELINE}[1]{\STATE \textbf{case} #1\textbf{:} }
\newcommand{\DEFAULT}{\STATE \textbf{default:} \begin{ALC@g}}
\newcommand{\ENDDEFAULT}{\end{ALC@g}}
\newcommand{\DEFAULTLINE}[1]{\STATE \textbf{default:} }

\def\QEDmark{\ensuremath{\square}}
\def\endproof{\hfill\QEDmark}

\newcommand{\bluecomment}[1]{}
\newcommand{\redcomment}[1]{\tcc{#1}}

\newcommand{\spellcheck}[1]{#1}
\newcommand{\redtext}[1]{\textcolor{red}{#1}}
\newcommand{\blue}[1]{\textcolor{blue}{#1}}
\newcommand{\DP}[1]{\textcolor{purple}{#1}}
\newcommand{\newredtext}[1]{#1}
\newcommand{\newbluetext}[1]{#1}

\newcommand{\revtext}[1]{#1}

\newcommand{\kmkim}[1]{\textcolor{blue}{#1}}

\newcommand{\profwshan}[1]{{#1}}
\newcommand{\profwshangreen}[1]{{#1}}
\newcommand{\profwshangr}[1]{#1}

\newcommand{\full}[1]{#1}

\ifFullVersion
\newcommand{\revision}[1]{#1}
\newcommand{\todelete}[1]{#1}
\newcommand{\release}[1]{#1}
\else
\newcommand{\revision}[1]{#1} 
\newcommand{\todelete}[1]{\textcolor{red}{#1}}
\newcommand{\release}[1]{#1}
\fi

\newcommand{\iseo}[1]{#1} 

\LinesNumbered
\SetAlgoCaptionSeparator{.}
\SetKwProg{Fn}{Function}{}{end}
\SetKwFor{uForEach}{foreach}{do}{}
\SetStartEndCondition{ (}{) }{)}
\SetNlSty{texttt}{}{:}
\SetArgSty{}

\def\ojoin{\setbox0=\hbox{$\bowtie$}%
  \rule[-.0ex]{.25em}{.4pt}\llap{\rule[\ht0]{.25em}{.4pt}}}
\def\leftouterjoin{\mathbin{\ojoin\mkern-5.8mu\bowtie}}
\def\rightouterjoin{\mathbin{\bowtie\mkern-5.8mu\ojoin}}
\def\fullouterjoin{\mathbin{\ojoin\mkern-5.8mu\bowtie\mkern-5.8mu\ojoin}}

\title{Trustworthy and Efficient LLMs Meet Databases}

\author{Kyoungmin Kim}
\email{kyoung-min.kim@epfl.ch}
\affiliation{%
  \institution{EPFL}
  \country{Switzerland}
}

\author{Anastasia Ailamaki}
\email{anastasia.ailamaki@epfl.ch}
\affiliation{%
  \institution{EPFL}
  \country{Switzerland}
}

\begin{abstract}
In the rapidly evolving AI era with large language models (LLMs) at the core, making LLMs more trustworthy and efficient, especially in output generation (inference), has gained significant attention. This is to reduce plausible but faulty LLM outputs (a.k.a hallucinations) and meet the highly increased inference demands.
This tutorial explores such efforts and makes them transparent to the database community. Understanding these efforts is essential in harnessing LLMs in database tasks and adapting database techniques to LLMs.
Furthermore, we delve into the synergy between LLMs and databases, highlighting new opportunities and challenges in their intersection.
This tutorial aims to share with database researchers and practitioners essential concepts and strategies around LLMs, reduce the unfamiliarity of LLMs, and inspire joining in the intersection between LLMs and databases.
\end{abstract}

\maketitle 

\setcounter{definition2}{0}
\setcounter{figure}{0}
\setcounter{section}{0}
\setcounter{page}{1}

\vspace*{-0.1cm}
\section{Introduction}\label{sec:introduction}





Large language models (LLMs) have recently transformed various fields with their ability to understand and generate human-like text. In the database domain, researchers are leveraging LLMs to tackle complex data management tasks \cite{SIGMOD2024tutorial, LLM4DB}.
LLMs can function not only as assistants for database administrators (DBAs) \cite{LLMasDBA, LLMDatabaseTuningManualDBBERT} but also as internal components of database systems, optimizing query plans \cite{LLMQueryOptLLMR2, LLMQueryOptUnreasonable} and 
translating natural languages to SQLs \cite{multipathchasesql2024}.




Beyond these applications, key concepts and advancements from the LLM community remain underexplored by database researchers. This tutorial aims to bridge that gap by focusing on enhancing the trustworthiness and efficiency of LLMs. Improving trustworthiness involves reducing hallucinations \cite{surveyofhallucination} to ensure LLMs generate accurate, factual responses, thereby increasing their reliability in database tasks requiring precise answers and reasoning. Enhancing efficiency focuses on decreasing inference latency and boosting throughput.


Inference efficiency is particularly important because, while training LLMs demands substantial resources and expertise, inference occurs daily across numerous users, leading to significant operational costs. For instance, OpenAI handles millions of requests, incurring substantial monthly expenses to run ChatGPT \cite{chatgptnumbers, chatgptcost}. Integrating LLMs with external data sources, such as vector databases and document retrieval systems in retrieval-augmented generation (RAG) \cite{RAG}, increases the number and complexity of LLM calls, especially with longer inputs. Recent trends in chain-of-thought and multi-path reasoning, exemplified by models like OpenAI's o1 \cite{openaio1}, further amplify inference demands, as generating final answers may require multiple LLM calls to enhance trustworthiness.


From a systems perspective, improving LLM inference efficiency parallels database management system (DBMS) development, presenting opportunities for database researchers to contribute to creating more efficient LLMs, promoting economic and environmental sustainability by reducing the CO2 footprint associated with extensive GPU usage.

After introducing the essential ideas in making LLMs more trustworthy and efficient, we will explain the intersection of LLMs and databases with new challenges and opportunities.



\vspace*{-0.1cm}
\subsection{Target Audience and Prerequisites}



Our tutorial is designed for conference attendees, focusing on three key areas to maximize engagement:


\noindent \textbf{Trustworthy LLMs} (Section \ref{sec:trust}): Aimed at individuals seeking to effectively utilize large language models (LLMs) in database tasks with minimal errors. Prerequisites include experience with LLMs like ChatGPT and the distinction between training and inference in machine learning. No in-depth knowledge of LLM internals is required. 


\noindent \textbf{Efficient LLMs} (Section \ref{sec:efficient}): Targeted at those interested in enhancing LLM inference efficiency or contributing to the development of fast LLM inference systems by applying database techniques. Prerequisites include basic database knowledge and an understanding of GPUs. Familiarity with Transformer architecture, attention mechanisms, and key-value (KV) caching is advantageous.





\noindent \textbf{LLMs Meet Databases} (Section \ref{sec:meet}): Intended for participants exploring new research opportunities at the intersection of databases and LLMs. A background in databases, including OLAP, relational algebra, cost-based query optimization, and approximate/adaptive query processing, will be helpful.




Our goal is to bridge the gap between essential LLM knowledge and the database community, enabling researchers already utilizing LLMs to uncover and develop unexplored ideas. Rather than merely listing state-of-the-art papers, we employ consistent visuals and focus on core concepts and insights, facilitating a deeper understanding and navigation of the evolving LLM landscape.





\subsection{Tutorial Length}

The intended length of this tutorial is 1.5 hour, with 40, 30, and 20 minutes each for Sections \ref{sec:trust}, \ref{sec:efficient}, and \ref{sec:meet}, respectively.






\section{Tutorial Outline}\label{sec:tutorial}


\begin{figure*}[htb] 
\centering
\includegraphics[width=1.8\columnwidth]{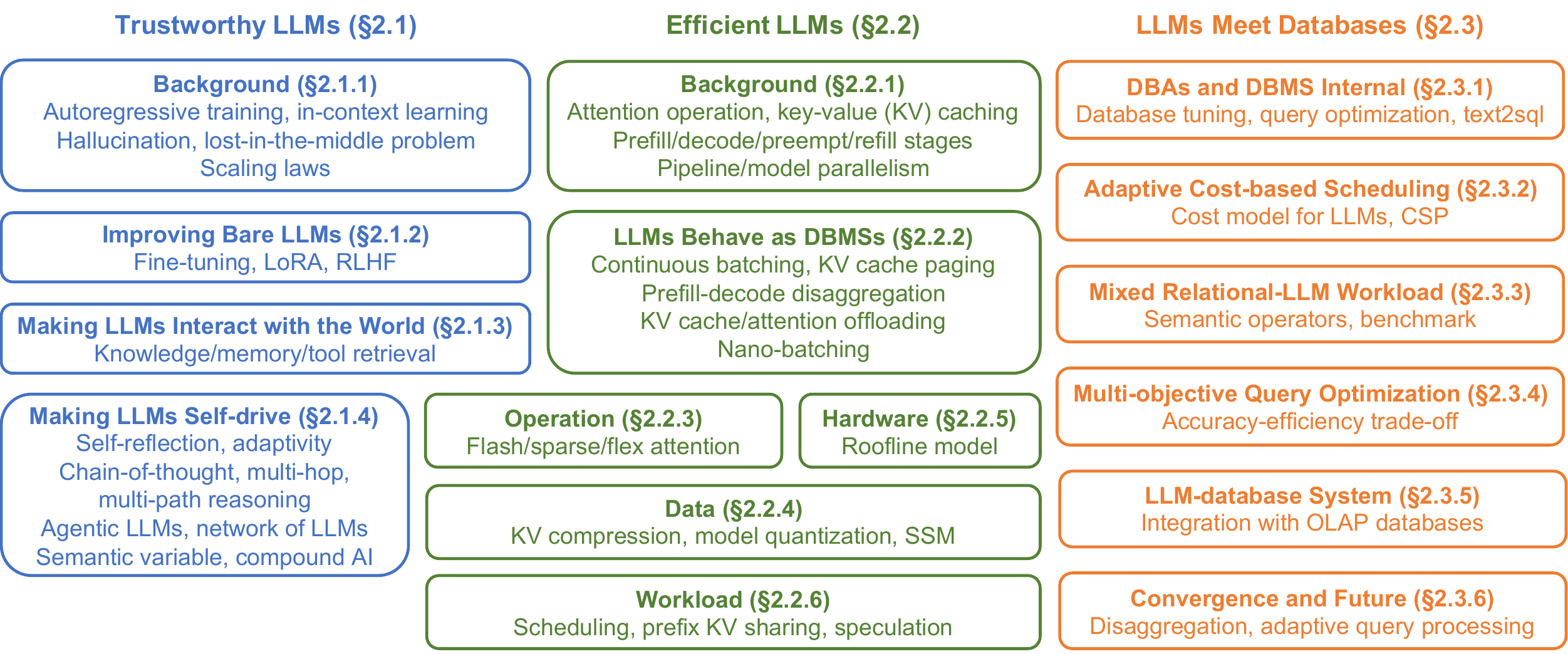}
\caption{Tutorial outline (each subsection with keywords).
}\label{fig:tutorial}
\vspace*{-0.3cm}
\end{figure*}


The tutorial is structured into three main sections, addressing critical aspects of LLMs and their interplay with database systems. Figure \ref{fig:tutorial} visualizes the outline with keywords for each subsection.
Section \ref{sec:trust} focuses on improving the trustworthiness of LLMs, exploring challenges such as hallucination and context limitations while presenting state-of-the-art solutions to improve the accuracy and reliability of generated outputs.
Section \ref{sec:efficient} emphasizes efficiency, covering optimization strategies for inference, data management, and hardware utilization.
Finally, Section \ref{sec:meet} highlights the convergence of LLMs and databases, exploring opportunities for integration, new workloads, and emerging system designs. 
Since the field is changing fast, we will regularly reflect new information until the tutorial date.

\subsection{Trustworthy LLMs}\label{sec:trust}

The first part of the tutorial explains the efforts to reduce hallucinations and make LLMs more trustworthy, using an analogy that LLMs resemble humans.
We explain background (Section \ref{sec:trust:background}), how LLMs can solely improve (Section \ref{sec:trust:model}), how LLMs can improve by interacting with the external world (Section \ref{sec:trust:env}), and how LLMs can automatically make such decisions and interact with other LLMs (Section \ref{sec:trust:autonomy}).

\subsubsection{Background}\label{sec:trust:background}


Large Language Models (LLMs) function as text-in, text-out systems, generating texts based on their training. Training an LLM is akin to nurturing a child: by exposing it to extensive text data, the model acquires world knowledge and reasoning abilities. This process involves predicting the most probable next token in a sequence, a type of self-supervised learning. For a sequence of tokens, the model learns to predict the latter tokens based on the preceding ones, enabling it to generate coherent text continuations.


Fine-tuning refines this process for specific tasks or domains, similar to how individuals specialize in particular professions. In contrast, in-context learning provides additional information or examples within the input without altering the model's parameters, akin to consulting external references during an open-book exam. Many prompting techniques \cite{reliablepromptgpt32023, promptreportsurvey2024, promptsurveychaintreegraph2024, surveyofpromptingnlp2024, systematicsurveyprompt2024, unleashingpromptsurvey2023, bettercotpromptsurvey2023, promptcanvas2024} including chain-of-thought prompting \cite{cot2022, zerocot2022} and its variants \cite{tot2023, got2024} may leverage in-context learning to enhance performance.



During inference, LLMs generate texts autoregressively, producing one token at a time. This process may involve deterministic methods like greedy or beam search, or probabilistic approaches such as nucleus sampling \cite{nucleussampling,closednucleussampling}, which helps avoid selecting low-probability tokens.




However, LLMs experience hallucinations \cite{hallucinationisinevitable, surveyofhallucination, surveyhallucinationmitigation2024, LLMsalwayshallucinate2024}, generating plausible-sounding but incorrect or fabricated information. This is an unavoidable aspect of LLMs \cite{hallucinationisinevitable, LLMsalwayshallucinate2024} which arises from limitations in capturing real-world knowledge, inherent approximations in training and inference, input noise, etc. 
Even slight input perturbations can significantly influence hallucinations \cite{inputperturbationhallucination2023, inputperturbationhallucination2024}, and the detection of hallucinations has been a major problem \cite{detecthallucinationentropy2024, lookbackdetecthallucinationattention2024, detectedithallucination2024, detecthallucinationgraph2024, detecthallucinationtokenprobability2024, realtimehallucinationdetect2024, unifieddetecthallucinationmultimodal2024}.




Additionally, the lost-in-the-middle problem \cite{lostinthemiddle,foundinthemiddle} indicates that LLMs may struggle to utilize information located in the middle of long contexts, often performing better when relevant information is at the beginning or end of the input, exhibiting a U-shaped performance curve. 
This phenomenon has been attributed to inherent attention biases within LLMs, where tokens at the start and end of the input receive higher attention, regardless of their relevance \cite{foundinthemiddle}.
This tendency can lead to increased hallucinations as context lengthens \cite{longcontexthallucinationbenchmark2024}.


Scaling laws \cite{scalinglaws2020, scalinglawschinchilla2022, reconcilingchinchilla2024} explain that error rates decrease as model size and training data increase, with optimal scaling requiring proportional growth in both \cite{scalinglawschinchilla2022}. However, this may not hold for smaller models \cite{reconcilingchinchilla2024}. 
Laws can also relate to temporal loss in the training curve \cite{temporalscalinglaws2024}, downstream tasks \cite{downstreamscalinglaws2024}, model quantization \cite{quantizedscalinglaws2024}, transfer learning \cite{scalinglawstransfer2021, scalinglawstransfer2024}, number of generated samples \cite{scalinglawsamplinglargelanguagemonkeys}, and inference time \cite{openaio1} with the advance of using long, complex reasoning paths.
Due to automatic prompting techniques \cite{DSPy, reprompt2024, repromptingcot2024, promptgen2024, autoprompt2020} and that larger models tend to be less sensitive to prompt variations \cite{largemodelslesssensitive2024}, 
we focus less on prompting techniques.


\noindent \textbf{Target.} The audience will distinguish pre-training, fine-tuning, and in-context learning phases of LLMs, and understand the inherent challenges in making LLMs trustworthy.

\subsubsection{Improving Bare LLMs}\label{sec:trust:model}


We briefly explain the approaches to improve the LLM itself to make it more trustworthy. Since LLM is a specific class of machine learning (ML) models, general ML approaches to enhance accuracy may work for LLMs. 
However, as such approaches have been extensively studied from the classic ML era, we target more LLM-specific approaches.

As it is infeasible to increase the model size indefinitely, and the models typically follow the Transformer architecture \cite{attention}, efforts have been put to increase or augment training data (where LLMs themselves can be used to generate data) \cite{piledata2021, texttotextdata2020, codedata2023, dataaugmentationusingllm2024, surveydatasynthesisaugmentation2024, dataaugmentforintentclassificationwithllm2022, surveydataaugment2024, llmfordataagumentation2024, surveydataaugment2023, nextgenerationdata2024, evaluatingLLMsasdatagenerators2024}, improve data quality (again, LLMs can be used to clean data) \cite{dataqualitycorpus2021, dataqualityfederated2024, datacleaningusingllm2024, llmtocleannoisydata2023, textbooksareallyouneed, licensedsourcecodedataset2023}, make inferences more robust \cite{contrastivedecoding2024, contrastingretrievalhead2024}, and apply better training and fine-tuning methods.

Specifically, fine-tuning covers a broad spectrum of work for example, parameter-efficient fine-tuning (PEFT) \cite{LoRA, adapterfusion2021, prefixtuning2021, prompttuning2021, adapterefficientfinetuning2019}, instruction tuning \cite{instructiontuning2021, instructiontuningcode, unifyingvisionlanguageinstruction2021, multitaskinstruction2022, GRITinstructiontuning2024, selfinstructtuning2023}, reinforcement learning from human feedback (RLHF) \cite{RLHF2017, RLHF2020, RLHF2022, safeRLHF2024, surveyRLHF2023, offlineRLHF2023, multiturnRLHF2024, openproblemlimitationRLHF2023, RLHFhowtoevaluate2024, ReSTRLHF2023}, and direct preference optimization (DPO) \cite{mitigatinghallucinationmultimodaldpo2024, DPO2023, RTODPOPPO2024, sDPOdontuseyourdataallatonce}.
RLHF leverages human feedback to train a reward model in reinforcement learning (RL), guiding the LLM through RL to produce desired outputs.
DPO simplifies the alignment process by directly optimizing the policy model without a separate reward model.
While these approaches rely on RL that continuously interacts with human or the world external to LLMs, such interactions are often limited to training and do not occur in inferences, which we explain in the following sections.


Other than the training methods, a new model architecture of differential Transformer \cite{differentialtransformer2024} reduces the distractions of the models to focus on unnecessary information in the long context, which works similarly to robust decoding strategies \cite{contrastingretrievalhead2024, contrastivedecoding2024}.

\noindent \textbf{Target.} The audience will learn about tuning LLMs to make them more trustworthy and aligned with user intentions.



\subsubsection{Making LLMs Interact with the World: Adding Eyes and Hands}\label{sec:trust:env}

LLMs alone can encounter knowledge, memory, and capability limitations \cite{RAGanything2023}. 
Their knowledge is confined to the static information encoded during training, leading to potential inaccuracies over time. Memory constraints arise from limited context windows, hindering the handling of extended conversations. Additionally, their text-based nature restricts interactions with the physical world.
To address these challenges, LLMs can retrieve knowledge, memory, and tools.

This section focuses on \emph{what} and \emph{how} to retrieve.
\emph{When} to retrieve is the key to autonomy and will be detailed in the next section.



Knowledge retrieval is represented by well-known retrieval-augmented generation (RAG) \cite{RAG, RAGsurvey2023, RAGsurvey2024, RAGorlongcontext2024, structgptgeneralstructureddata}. Based on the data type, it can fetch knowledge from knowledge graphs \cite{CRAG, graphRAGdualpathway2024, graphRAGgretriever2024, graphRAGhybrid2024, graphRAGroadmap2024, graphRAGsurvey2024, graphRAG2024, graphRAGplanongraph2024, graphRAGthinkongraph2023}, tables \cite{tablepromptinglearningtoreduce, tableRAGDL2023, tableRAGsummarization2024, tableRAGtransformer2023, TAG, tableRAGbenchmark2024, tableRAGchainofreasoning2022, tableRAGdenoisingtabletext2024, tableRAGfewshot2023, tableRAGhybridQAbenchmark2020, tableRAGmultigranularity2024, tableRAGsqlgenfinetune2024, tableRAGtext2sqlbenchmark2023}, images \cite{imageRAG2023, imageRAGopenQA2022, imageRAGtext2image2023}, not just documents. The data may be chunked/vectorized, stored in vector databases, then similar chunks are searched online.
While vector similarities are typically used, more advanced similarity scores are possible, e.g., using dual or cross encoders \cite{searchcrossencoder2022, largedualencoder2024}.


Memory retrieval attempts to overcome the limited context size of LLMs by storing previously seen tokens as key-value pairs \cite{memoryRAGaugmentmemory2023, memoryRAGmemorizingtransformer2022, memoryRAGtrilliontokens2022, memoryRAGmemlong2024, memoryretrievalmasseditingmemory2023} and fetching relevant pairs in upcoming requests, managing memory stores in hierarchical or partitioned way \cite{DSMR2024, partitionedmemoryretrieval2024} or even as a database \cite{ChatDBdatabasesassymbolicmemory}.
Fetching information from long input can also be done without maintaining a separate memory store, but by sparsifying the model layers \cite{SlidingWindowAttention, sparseattention2024}.
One can relate low-rank adapters and mixture-of-experts \cite{LoRA, moeglam2022, moelora2024, moeloramillion2024, moeswitchtransformer2022, loramixrora2024, moelightninghighthoughput2024, efficientMoEpipeline2024} with memory retrieval since lightweight model parameters are fine-tuned per specific task and domain, and dynamically fetched at online inferences.

Tool retrieval searches for the APIs to interact with external environments \cite{toolRAGcolt2024, toolRAGgorilla2023, toolRAGmetaknowledge2024, toolRAGtoolformer2023, iterativetoolretrieval2024, toolsandbox2024, toolRAGtoolLLM16000, toolRAGwhataretoolsanyway}.
One can connect LLMs with databases to call SQLs that can help answering user questions \cite{TAG}.
Constrained decoding \cite{xgrammarconstraineddecoding2024, constraineddecoding2023, constraineddecodingnoninvasive2024, constraineddecodingwithoutlogitaccess2024} allows output to follow specific structure which can increase correctness and efficiency.

The challenges in retrieval include the followings.
1) Heterogeneity: LLMs are text-based, but knowledge can be of any type. Even for text retrieval, heterogeneous lengths and intents between queries and documents can lead to suboptimal retrieval accuracy \cite{ARAGOG, HyDE}, and the vector-similarity search may be too simple to retrieve necessary information \cite{searchvectordatabasesurvey2024, searchsimilaritydiversity2024}.
2) Scalability: Not only that LLMs have limited context or data they can utilize per inference, but maintaining a large set of retrieval entities and retrieving a subset may incur overheads \cite{SPFresh, ANNS2024}.
While approximation can mitigate the search overhead and make the search negligible to LLM inference costs, it is limited to vector-similarity search, and generalization to more complex searches \cite{searchdualencoder2020, searchcolBERT2020, searchcrossencoder2022, RAGbridgingretrieverLLM2024} remains challenging.
3) Sparsity: This is also relevant to data sparsity and noise \cite{searchpowerofnoise2024}, where relevant data is sparse compared to large information pools.
4) Reliability: Retrieved knowledge may be imperfect \cite{astuteRAGimperfect}.

\noindent \textbf{Target.} The audience will understand how LLMs can interact with the world and exploit external knowledge to overcome the limits in using LLMs alone.

\subsubsection{Making LLMs Self-drive: Adding Brain}\label{sec:trust:autonomy}


Now we have more powerful models and interactions with the world. The last part is how we can make LLMs smart enough to maximize these capacities, adding autonomy.
Self-consistency and major voting enables a simple yet effective solution for increasing consistency \cite{selfconsistency2023}, however, it fails to generate accurate and diverse answers \cite{selfconsistencynotgoodmultihop2024, selfconsistencypara2024, selfconsistencyuniversal2023} and is yet passive.
More active approaches include self-reflection and adaptive retrieval \cite{selfreflectionhallucination2023, selfRAG, adaptiveRAG, selfreflectdual2024, adaptiveRAGshortformopendomainQA2024, reflectionbench2024, RAGiterativeselffeedback2024}, which adaptively retrieves information multiple times based on the generated output, model confidence, query complexity, or fine-tuned policies. 
This is particularly helpful for chain-of-thought/multi-hop reasoning and question answering \cite{CoTwithoutprompting2024, multihopQAhierarchicalrethink2024, multihopQAselfguidingzeroshotprompt2024, multihopQAselfpromptedcot2023, criticaltokensmatterDPO2024}. 


The next step is to use multiple reasoning paths instead of a single path.
This multi-path reasoning has been an effective approach for driving LLMs \cite{rstarmultipath2024, multipathllamaberrypairwiseoptimizationolympiad2024, multipathnashcot2024, multipathchasesql2024}.
While the exact mechanism remains closed, OpenAI's o1 model is assumed to plan subtasks, conduct these, and revise the results to decide whether to extend the current plan or generate different plans, forming a tree-like reasoning structure.
They suggest a new scaling law that LLM accuracy increases with inference time, not only with training time and data \cite{openaio1}.


Agentic LLM indicates that LLMs can act as agents, selecting actions based on observations \cite{agenticLLMagentictransformer2023, agenticLLMmultiagentLLMsurvey2024, agenticLLMreact2023, agenticLLMreflexion2023}.
Multiple agents exploit collaborative reasoning, parallel processing, diversity, and specialization akin to humans \cite{multiagentconsensus2023, multiagentLLMgame2023, multiagentLLMmagic2024, multiagentLLMstrategist2024, multiagentLLMsurvey2024, multiagentLLMverylargescale2024, multiagentscaling2024}.
Semantic variables \cite{Parrot} regard LLM input and output tokens as dependent variables to explicitly model control flows. 

A broader view includes compound AI \cite{compoundAI}
where AI and non-AI components interact with each other, including retrievals, control flows, agentic LLMs, and more.
An interesting example is automated research process \cite{AIscientist2024, AIscientistsofscientist2024}. 

\noindent \textbf{Target.} The audience will learn about approaches to make LLMs self-driving and build systems around LLMs for complex tasks.

\subsection{Efficient LLMs}\label{sec:efficient}

The second part of the tutorial demystifies the internals of LLM inference process and explains the efforts to make it more efficient, using an analogy that LLMs behave as DBMSs.
We explain background (Section \ref{sec:efficient:background}) and how LLM inference systems resemble DBMSs in improving their efficiency (Section \ref{sec:efficient:inference_system}).
We then explain further work for each dimension of operation (Section \ref{sec:efficient:operation}), data (Section \ref{sec:efficient:data}), hardware (Section \ref{sec:efficient:hardware}), and workload (Section \ref{sec:efficient:workload}). 


\subsubsection{Background}\label{sec:efficient:background}

The dominant Transformer architecture employs an attention mechanism \cite{attention} that calculates similarity scores between a token and its preceding tokens, effectively capturing inter-token relationships and managing extended contexts. This process has quadratic complexity, but key-value (KV) caching \cite{KVCache} optimizes it by storing and reusing these computations, reducing the complexity to linear during inference.
Non-attention operations mostly consist of matrix multiplications and activations.

Inference in Large Language Models (LLMs) involves two primary phases: prefill and decode. During the prefill, the model processes input tokens to generate the initial output token. The attention operates with quadratic complexity due to the absence of precomputed KVs, making it compute-intensive.
During the decode, the model generates subsequent tokens sequentially, each time using the last generated token as input. Here, the attention leverages KV caching, resulting in linear complexity relative to the number of processed tokens and reading their KVs, which makes this phase more memory-intensive.


In case of multiple requests, they face a race condition as in multi-tenant systems. If the GPU memory is insufficient to keep all requests' KVs, some running requests are preempted (evicted), releasing their KVs from the memory, and restarted (refilled) later \cite{vLLM}. Due to the low PCIe bandwidth, the released KVs are often recomputed when restarted, instead of offloading to other storage devices and loading back.
Multiple requests in either prefill or decode steps can be batched to amortize the cost of loading model weights from GPU memory.

Note that the model weights also occupy the GPU memory.
When the model size exceeds a single GPU capacity, techniques like model and pipeline parallelism \cite{megatronmodelparallelism2019, pipelineparallelismgpipe2019, pipelineparallelismpipedream2018} distribute model weights across multiple GPUs. This partitioning introduces data transfer overhead between GPUs. 


\noindent \textbf{Target.} The audience will understand the KV caching and different phases of LLM inference requests, and how they compete for the same GPU resource.

\subsubsection{LLM Inference Systems: LLMs Behave as DBMSs}\label{sec:efficient:inference_system}

LLM inference systems (e.g., \vLLM \cite{vLLM}) behave similarly to (in-memory) DBMS. KVs and model weights correspond to the data, which are maintained in GPU memory. Operators include matrix multiplications, activations, attentions, and data transfers. Compared to OLAP in databases, the operations are simpler yet much more time-sensitive, where the requests should be served in real-time.

Significant efforts have been made to increase the efficiency of LLM inference \cite{LLMinferenceefficiencysurvey2024}, largely based on operating and database systems.
\Orca \cite{Orca} forms a new batch of requests after each iteration (of prefills and decodes) whenever resources are available. Thus, a new request does not have to wait for all current running requests to finish, just like the pipelining in OS.
\vLLM \cite{vLLM} adopts paging and virtual memory to manage KVs, reducing memory fragmentation and enlarging the batch size.
Since prefills are typically more costly than decodes, making stalls for decodes when batched together, \Sarathi \cite{SARATHI, SARATHI_SERVE} chunks prefills to reduce pipeline bubbles.
Some other work \cite{Dejavu, Splitwise, DistServe} rather disaggregates the prefills and decodes into different GPUs, so the workload for each GPU is homogeneous. 
\vTensor \cite{vTensor} decouples the KV cache management and attention computation of \vLLM for better flexibility and performance. 
\NanoFlow \cite{NanoFlow} splits each batch into nano-batches for finer-grained pipelining, increasing the overlap of computation, memory operation, and data transfer between GPUs. It also hides CPU scheduling latency by asynchronous scheduling.
\InfiniGen \cite{InfiniGen} offloads the KVs to CPU memory to extend the KV cache and reloads the KVs from CPU layer-wise, but fetches a subset of KVs for efficiency, similarly to sparse attentions (Section \ref{sec:efficient:operation}).
\InstInfer \cite{InstInfer} offloads KVs and attention computations to flash drives, just like the storage-disaggregation and computation pushdown in databases \cite{FlexpushdownDB}.
\NEO \cite{NEO} selectively offloads attention computations and KVs from GPU to CPU, in order to maximize both GPU and CPU utilization.

\noindent \textbf{Target.} The audience will understand why LLMs behave similarly to DBMSs and how database techniques can improve LLM inference efficiency. In subsequent sections, the audience will learn about efforts and challenges in further improving efficiency in four dimensions: operation data, hardware, and workload.

\subsubsection{Operation: Attention}\label{sec:efficient:operation}

While matrix multiplications take the major portion in LLM latency in general \cite{SARATHI}, attentions can dominate the runtime for large inputs due to their quadratic complexity.
FlashAttention \cite{FlashAttention, FlashAttention_2, FlashAttention_3} has become a de facto standard as an efficient attention implementation, utilizing recent GPU technologies to boost the inference speed. The ideas include kernel operator fusion and GPU cache-aware KV transfer.
As in approximate query processing (AQP) in databases, sparse attentions \cite{SlidingWindowAttention, sparseattention2024, retrievalattention2024} do not compute the full attention scores for all preceding tokens but a subset as an approximation.
Some attentions rather optimize for long contexts \cite{flashdecodinglongcontext2023, parallelizemillioncontextlengthinferences}.
FlexAttention \cite{flexattention}
offers flexible and performant implementation of such attentions.


\subsubsection{Data: KV and Model Weights}\label{sec:efficient:data}

Reading KVs from GPU memory in decode-attentions is similar to sequential table scan. As KVs are maintained per each attention layer, reading KVs for a layer can overlap with other layers' operators \cite{Dejavu, InfiniGen}.
While offloading KVs and attention computation have been popular recently \cite{InfiniGen, InstInfer, NEO}, we need to be careful as it is challenging to predict the output lengths of LLM requests and thus their utilization patterns, and a KV for a single token may consume a few MBs.
KVs of long documents can be precomputed, compressed, and fetched for later retrievals \cite{CacheGenKVcompress}.
To reduce memory latency, one can opt for KV sharing across different attention heads \cite{KVsharecrosslayerattention2024, KVshareGQAgroupedqueryattention, KVshareWGQAgroupedqueryattention}, KV compression \cite{KVcompressKVcompress2024, KVcompressL22024, KVcompressminicache2024, CacheGenKVcompress, gearKVcachecompressionlossless2024}, model quantization \cite{modelquantizebenchmark2021, modelquantizeLLMCbenchmark2024, modelquantizeomniquant2024, modelquantizePTQQAT2024, modelquantizesmoothquant2023, modelquantizesurvey2021, modelquantizesurvey2024, modelquantizewhitepaper2021, quantizedscalinglaws2024}, or different model architectures than Transformer, such as State Space Models (SSMs) \cite{transformersaressms2024, SSM2022} that do not rely on attentions, thereby not generating KVs.
While hybrid architectures \cite{SSMblackmamba2024, SSMblockstatetransformer2023, SSMdensemamba2024, SSMjamba2024, SSMmamba2023, hymbaSSM2024, SSMsamba} can balance between the efficiency of SSMs and memorization capacity of Transformers, SSMs remain niche in the market \cite{stateofaireport}.
A recent work even shows that tokenizers can be removed from the models \cite{bytelatenttransformer2024}. 


\subsubsection{Hardware: Theory and Practice}\label{sec:efficient:hardware}

We briefly explain 1) the roofline model \cite{LLMViewer} and 2) some efforts to overcome the hardware limits \cite{FlexGen, vDNN, FlexNN, PowerInfer, TwinPilots, HeteGen} or leverage advanced hardware for LLM inference \cite{PIEpoolingcpu2024, cerebrasHWSWcodesign2023}. The roofline model is based on the computation speed (e.g., GPU FLOPS) and memory bandwidth, which acts as a theoretical hardware bound and determines whether an operator is compute-intensive or memory-intensive across different inputs.








\subsubsection{Workload: Scheduling, Prefix Sharing, Speculation}\label{sec:efficient:workload}


To handle multiple LLM requests, LLM inference systems implement request scheduler to send LLM requests to appropriate machines or GPUs to maximize throughput or minimize latency. 
Assuming independent requests, early schedulers either prioritize prefills \cite{vLLM} or decodes \cite{SARATHI_SERVE}, which tend to optimize latency or throughput, respectively. More complex schedulers consider fairness \cite{VTCvirtualtokencounter2024, FastServemultilevelfeedbackqueue2023} while compromising performance, or predict the output lengths of requests (not known in advance) and schedule shorter requests first \cite{OPrediction, OPrediction2, ORank}.

If different LLM requests can share a prefix in their inputs, the KVs of the prefix can be stored just once and reused for multiple requests \cite{SGLang, BlendServeprefixsharing2024}. This forms a trie structure with shared prefixes.
However, a single-token difference in inputs may invalidate the sharing of KVs of all subsequent tokens.
To increase the sharing opportunity, \cite{CacheBlend} uses the KVs of multiple token sequences to approximate the KVs of the concatenated sequence. The mechanism is similar to the speculation in OLAP \cite{speculation} and healing protocol in transactions \cite{healing} in databases.

This speculation and healing patterns also appear in speculative decoding \cite{SpeculativeDecoding} and model cascades \cite{frugalGPT2023, cascademultiobjective2024, casecadesharedprefixbatchdecode2024, palimpzest2024, CITERmodelcascadestokenlevelrouting2024}, accelerating the generation of tokens by leveraging smaller, faster models then validate the tokens using larger models, since the validation costs less than the generation.

\subsection{LLMs Meet Databases}\label{sec:meet}

The last part of the tutorial discusses the intersection between LLMs and databases, opportunities and challenges in how we can exploit LLMs for databases, how the development of databases can help LLMs, and how we can exploit new types of workloads and integrations of LLMs and databases.
We explain from more well-known to more untapped, deeper integrations in Sections \ref{sec:meet:dba_internal}-\ref{sec:meet:system} and provide more proactive visions in Section \ref{sec:meet:convergence}.

\subsubsection{LLMs for DBs: DBAs and DBMS Internal}\label{sec:meet:dba_internal}

We briefly explain how LLMs are utilized for well-known tasks of DBAs and DBMS internals such as database tuning \cite{LLMasDBA, LLMDatabaseTuningManualDBBERT}, text2sql \cite{multipathchasesql2024, spider2} and query optimization \cite{LLMQueryOptLLMR2, LLMQueryOptUnreasonable}.
As we mentioned in Section \ref{sec:introduction}, we will not cover every detail, as many of these efforts are covered in a previous tutorial \cite{SIGMOD2024tutorial} and its additional list of papers \cite{LLM4DB}.



\subsubsection{DBs for LLMs: Adaptive Cost-based Scheduling}\label{sec:meet:cost_based_schedule}

Unlike the sophisticated query optimizers in DBMSs, LLMs lack cost models and cost-based scheduling of LLM requests. \cite{Vidur} measures the batch times across various inputs (number of tokens to process and KV size to read).
\cite{LLMViewer} computes batch times based on the roofline models. These can be used to model batch times and formulate the problem of finding optimal schedules as a constrained satisfaction problem (CSP) \cite{Ours}.
While schedulers try to avoid preemptions to maximize performance, \cite{Ours} shows that harnessing preemptions can rather reduce overall latency compared to zero-preemptions.
As the exact hardware utilization of each request is not known in advance, the scheduling should be adaptive based on the observations, and it has not been explored much to schedule dependent requests connected via semantic variables or shared prefixes \cite{Ours}.

\subsubsection{DBs with LLMs: Mixed Relational-LLM Workload}\label{sec:meet:mixed_workload}



Not only solving existing tasks with LLMs, LLMs offer new functionalities when integrated into DBMSs. Semantic operators \cite{LOTUS} extend relational operators to batch-process the tabular data with LLMs (e.g., filters and joins using LLMs), which can be regarded as an AQP.
Workloads with LLMs provide a justification to use LLMs inside DBMSs (heavy LLM calls in plan optimization can be negligible compared to query execution with LLMs).
However, different pipelines (with semantic operators) lead to different accuracy and efficiency, thus defining the equivalence between two pipelines is non-trivial.
Furthermore, more complex pipelines or LLM calls do not always guarantee higher accuracy 
\cite{ARAGOG, scalingpropertiescompoundAI2024}, and searching similar entities with LLMs can be replaced with efficient vector-similarity searches \cite{palimpzest2024, relationalvector} as a type of model cascade.

\subsubsection{DBs with LLMs: Multi-objective Query Optimization and Benchmarks}\label{sec:meet:multi_objective_opt}



The challenge is therefore how we can automatically find good pipelines for mixed relational-LLM workloads under the multi-objective of accuracy and efficiency \cite{CAESURA, TAG, nsDB} as in compound AI systems \cite{ALTOcompoundAI, teolacompoundAI, asyncLLMfunctioncalling2024}. This calls for development of accurate cost models and accuracy-prediction models for LLMs and mixed relational-LLM workloads, in order to enable the holistic optimization of query plans consisting of both relational and non-relational operators.
The cost model itself can be learned via LLMs (or any ML models), possibly using RLHF or feedback from query execution without human intervention, where such an automatic training data generation is one of the advantages of solving database tasks compared to conventional ML tasks (e.g., natural language processing with human-labeled translation data) \cite{LearnedCE}.
Another model for predicting the output accuracy or detecting hallucination may be chosen from the scaling laws (using the general fact that larger models are more trustworthy) or separately trained.

To balance efficiency and accuracy, during the physical query optimization we should select proper models (ones used for execution) to avoid calling heavy LLMs unnecessarily. 
Depending on the complexity of the task, simple ML models with a small set of supervised data \cite{adaptiveRAG}, or larger deep generative models such as in tabular foundation models tailored to domain-specific data \cite{ASM, TableGPTfindtunedGPTfortabletasks2024, ingestablestabularfoundationmodel2023}, or LLMs with world knowledge and reasoning capacity \cite{LLMforDBtableLLMspecialist2024} can fit the task with different accuracy-efficiency trade-offs.
Small language models (SLMs) \cite{rstarmultipath2024, SLMincontextlearning2024, SLMPhi3, SLMsurvey2024, SLMsurvey2024.2, hymbaSSM2024, SLMSmolLM2024} are also a good choice.
Automatically finding the best prompt configuration \cite{DSPy, semanticbackpropagationagenticsystems2024} tailored to the mixed workloads and more (e.g., previously mentioned fine-tuning or multi-hop/multi-path reasoning with adaptivity during inference) might be desired.

Furthermore, unlike the TPC benchmarks for databases, another problem is that there is no comprehensive benchmark for relational-LLM workloads yet. \cite{TAG, optimizingrelationalLLMqueries} provide exploratory benchmarks without focusing on semantic joins.

\subsubsection{DBs with LLMs: Integrated System}\label{sec:meet:system}

Other than the cost models, we also need DBMSs with native LLM support to increase the optimization opportunities, 
alike systems for relational-vector workloads \cite{relationalvector, relationalvectorvbase2023}.
Current prototypes for relational-LLM workloads \cite{LOTUS, palimpzest2024, optimizingrelationalLLMqueries} separate table processing (e.g., pandas \cite{pandas}) and LLM inference engine (e.g., \vLLM \cite{vLLM}).


To maximize efficiency and scalability, we should focus on hardware utilization, data movement \cite{datamovementisallyouneed}, caching hot data, locating computations close to data (e.g., computation pushdown in storage-aggregation setting) \cite{serverlessLLMlocalityenhanced, skyservedatalocality2024}, asynchronous API calls \cite{asyncLLMfunctioncalling2024}, balancing loads, and multi-tenancy just like in DBMSs \cite{FlexpushdownDB, ClickHouse}.
One also has to decide whether to maintain a separate vector database for faster online vector retrievals, or use just-in-time vectorization for reducing storage overhead.
This also applies to precomputing KVs of data tokens \cite{CacheGenKVcompress} for faster LLM inferences or not, but with a higher caution as KVs are typically larger than vectors.



\subsubsection{Convergence and Future}\label{sec:meet:convergence}


We envision LLMs and databases to converge (e.g., neuro-symbolic systems \cite{nsDB, neurosymbolicAIlandscape2024, neuraldatabases2021, symphonydatalakes}), more than just applying the techniques from one domain to another.
A new LLM inference system optimized for DBMSs might be developed from an open-source cloud DBMS, utilizing recent implementations and optimizations for processing relational operators, such as storage-disaggregation and computation pushdown for scalable data and model management \cite{FlexpushdownDB}, GPU-based OLAP processing \cite{DogQC, Pyper, databaseonGPU} for the full use of GPUs for both relational operators and LLMs, hybrid operators with heterogeneous data transfer paths \cite{FlexpushdownDB, HetExchange}, adaptive query execution \cite{DatabricksAdaptiveQE} and more.
A unified query optimizer and data model for both relational data, KVs, and model weights, could offer opportunities for better data management and hardware utilization.
Finally, if we look into the near future, we could also harness the emerging CXL technology for memory disaggregation \cite{CXLdatabasekernels, CXLdissectingmemoryperformance, CXLmemoryusecasesindatabases} to manage model weights and KVs, and increased interest in pruning unnecessary data in OLAP \cite{PLAQUE, ThesholdQueries, diPs, Diamond} could lead to higher trustworthiness (due to reduced noise) and efficiency (due to less data to process) in the relational-LLM workloads, with connections to online aggregation \cite{DeepOnlineAgg2023} and incremental view maintenance \cite{DBSP}.








\noindent \textbf{Target.} The audience will understand the different depths of LLM-database integrations and be able to find interesting research topics from each of the integration, which are closely related to the current and near-future trends of databases.



%

\section{Related Tutorial}\label{sec:related}


Xupeng et al. \cite{SIGMOD2024tutorial} presented a tutorial at SIGMOD 2024 about the role of data management in the development (training, fine-tuning) and deployment (inference) of LLMs. It focused on how the knowledge is encoded into model parameters and extracted during the inference, and explained the concept of KV caching but not LLM inference systems.
Trummer \cite{VLDB2023tutorial} presented at VLDB 2023 about Transformer architecture, pre-training/fine-tuning/prompting in LLMs, and LLM applications in data management.
As pointed out by \cite{SIGMOD2024tutorial}, most of other tutorials presented at SIGMOD and VLDB about AI and databases focused on traditional machine learning and deep learning tasks not tailored to LLMs \cite{AItutorial1, AItutorial2, AItutorial3, AItutorial4, AItutorial5, AItutorial6, AItutorial7}, or specific LLM-related applications such as tabular data understanding \cite{tabulardataunderstandingtutorial} and queries with natural languages \cite{NLquerytutorial2021, NLquerytutorial2023}.
Dong et al. \cite{KDD2023tutorial} presented at SIGKDD 2023 about the role of LLMs in building intelligent AR/VR assistants.

In this tutorial, we will focus on more recent, general approaches to enhance the trustworthiness and efficiency of LLMs, which have not been addressed in previous tutorials. For trustworthiness, we will start with enhancing LLMs alone, LLMs with tools, and agentic LLMs and collaboration. For efficiency, we will explain how LLM inference systems resemble DBMSs. Then we will discuss how we can integrate LLMs and databases in depth. 
We expect that these are what researchers, who aim to use LLMs in their applications or optimize LLMs using database techniques, need to know about.
Instead of a common analogy that LLMs are knowledge bases as they generate plausible facts, we will use analogies that LLMs behave as DBMSs and improve as how humans solve challenging problems.

\bibliographystyle{ACM-Reference-Format}
\balance
\typeout{}
\bibliography{main}


\begin{thebibliography}{343}


\ifx \showCODEN    \undefined \def \showCODEN     #1{\unskip}     \fi
\ifx \showDOI      \undefined \def \showDOI       #1{#1}\fi
\ifx \showISBNx    \undefined \def \showISBNx     #1{\unskip}     \fi
\ifx \showISBNxiii \undefined \def \showISBNxiii  #1{\unskip}     \fi
\ifx \showISSN     \undefined \def \showISSN      #1{\unskip}     \fi
\ifx \showLCCN     \undefined \def \showLCCN      #1{\unskip}     \fi
\ifx \shownote     \undefined \def \shownote      #1{#1}          \fi
\ifx \showarticletitle \undefined \def \showarticletitle #1{#1}   \fi
\ifx \showURL      \undefined \def \showURL       {\relax}        \fi
\providecommand\bibfield[2]{#2}
\providecommand\bibinfo[2]{#2}
\providecommand\natexlab[1]{#1}
\providecommand\showeprint[2][]{arXiv:#2}

\bibitem[Abdin et~al\mbox{.}(2024)]%
        {SLMPhi3}
\bibfield{author}{\bibinfo{person}{Marah~I Abdin}, \bibinfo{person}{Sam~Ade Jacobs}, \bibinfo{person}{Ammar~Ahmad Awan}, \bibinfo{person}{Jyoti Aneja}, \bibinfo{person}{Ahmed Awadallah}, \bibinfo{person}{Hany Awadalla}, \bibinfo{person}{Nguyen Bach}, \bibinfo{person}{Amit Bahree}, \bibinfo{person}{Arash Bakhtiari}, \bibinfo{person}{Harkirat~S. Behl}, \bibinfo{person}{Alon Benhaim}, \bibinfo{person}{Misha Bilenko}, \bibinfo{person}{Johan Bjorck}, \bibinfo{person}{S{\'{e}}bastien Bubeck}, \bibinfo{person}{Martin Cai}, \bibinfo{person}{Caio C{\'{e}}sar~Teodoro Mendes}, \bibinfo{person}{Weizhu Chen}, \bibinfo{person}{Vishrav Chaudhary}, \bibinfo{person}{Parul Chopra}, \bibinfo{person}{Allie~Del Giorno}, \bibinfo{person}{Gustavo de Rosa}, \bibinfo{person}{Matthew Dixon}, \bibinfo{person}{Ronen Eldan}, \bibinfo{person}{Dan Iter}, \bibinfo{person}{Amit Garg}, \bibinfo{person}{Abhishek Goswami}, \bibinfo{person}{Suriya Gunasekar}, \bibinfo{person}{Emman Haider}, \bibinfo{person}{Junheng Hao},
  \bibinfo{person}{Russell~J. Hewett}, \bibinfo{person}{Jamie Huynh}, \bibinfo{person}{Mojan Javaheripi}, \bibinfo{person}{Xin Jin}, \bibinfo{person}{Piero Kauffmann}, \bibinfo{person}{Nikos Karampatziakis}, \bibinfo{person}{Dongwoo Kim}, \bibinfo{person}{Mahoud Khademi}, \bibinfo{person}{Lev Kurilenko}, \bibinfo{person}{James~R. Lee}, \bibinfo{person}{Yin~Tat Lee}, \bibinfo{person}{Yuanzhi Li}, \bibinfo{person}{Chen Liang}, \bibinfo{person}{Weishung Liu}, \bibinfo{person}{Eric Lin}, \bibinfo{person}{Zeqi Lin}, \bibinfo{person}{Piyush Madan}, \bibinfo{person}{Arindam Mitra}, \bibinfo{person}{Hardik Modi}, \bibinfo{person}{Anh Nguyen}, \bibinfo{person}{Brandon Norick}, \bibinfo{person}{Barun Patra}, \bibinfo{person}{Daniel Perez{-}Becker}, \bibinfo{person}{Thomas Portet}, \bibinfo{person}{Reid Pryzant}, \bibinfo{person}{Heyang Qin}, \bibinfo{person}{Marko Radmilac}, \bibinfo{person}{Corby Rosset}, \bibinfo{person}{Sambudha Roy}, \bibinfo{person}{Olatunji Ruwase}, \bibinfo{person}{Olli Saarikivi},
  \bibinfo{person}{Amin Saied}, \bibinfo{person}{Adil Salim}, \bibinfo{person}{Michael Santacroce}, \bibinfo{person}{Shital Shah}, \bibinfo{person}{Ning Shang}, \bibinfo{person}{Hiteshi Sharma}, \bibinfo{person}{Xia Song}, \bibinfo{person}{Masahiro Tanaka}, \bibinfo{person}{Xin Wang}, \bibinfo{person}{Rachel Ward}, \bibinfo{person}{Guanhua Wang}, \bibinfo{person}{Philipp Witte}, \bibinfo{person}{Michael Wyatt}, \bibinfo{person}{Can Xu}, \bibinfo{person}{Jiahang Xu}, \bibinfo{person}{Sonali Yadav}, \bibinfo{person}{Fan Yang}, \bibinfo{person}{Ziyi Yang}, \bibinfo{person}{Donghan Yu}, \bibinfo{person}{Chengruidong Zhang}, \bibinfo{person}{Cyril Zhang}, \bibinfo{person}{Jianwen Zhang}, \bibinfo{person}{Li~Lyna Zhang}, \bibinfo{person}{Yi Zhang}, \bibinfo{person}{Yue Zhang}, \bibinfo{person}{Yunan Zhang}, {and} \bibinfo{person}{Xiren Zhou}.} \bibinfo{year}{2024}\natexlab{}.
\newblock \showarticletitle{Phi-3 Technical Report: {A} Highly Capable Language Model Locally on Your Phone}.
\newblock \bibinfo{journal}{\emph{CoRR}}  \bibinfo{volume}{abs/2404.14219} (\bibinfo{year}{2024}).
\newblock
\urldef\tempurl%
\url{https://doi.org/10.48550/ARXIV.2404.14219}
\showDOI{\tempurl}
\showeprint[arXiv]{2404.14219}


\bibitem[Agrawal et~al\mbox{.}(2024a)]%
        {parallelizemillioncontextlengthinferences}
\bibfield{author}{\bibinfo{person}{Amey Agrawal}, \bibinfo{person}{Junda Chen}, \bibinfo{person}{{\'{I}}{\~{n}}igo Goiri}, \bibinfo{person}{Ramachandran Ramjee}, \bibinfo{person}{Chaojie Zhang}, \bibinfo{person}{Alexey Tumanov}, {and} \bibinfo{person}{Esha Choukse}.} \bibinfo{year}{2024}\natexlab{a}.
\newblock \showarticletitle{Mnemosyne: Parallelization Strategies for Efficiently Serving Multi-Million Context Length {LLM} Inference Requests Without Approximations}.
\newblock \bibinfo{journal}{\emph{CoRR}}  \bibinfo{volume}{abs/2409.17264} (\bibinfo{year}{2024}).
\newblock
\urldef\tempurl%
\url{https://doi.org/10.48550/ARXIV.2409.17264}
\showDOI{\tempurl}
\showeprint[arXiv]{2409.17264}


\bibitem[Agrawal et~al\mbox{.}(2024b)]%
        {Vidur}
\bibfield{author}{\bibinfo{person}{Amey Agrawal}, \bibinfo{person}{Nitin Kedia}, \bibinfo{person}{Jayashree Mohan}, \bibinfo{person}{Ashish Panwar}, \bibinfo{person}{Nipun Kwatra}, \bibinfo{person}{Bhargav~S. Gulavani}, \bibinfo{person}{Ramachandran Ramjee}, {and} \bibinfo{person}{Alexey Tumanov}.} \bibinfo{year}{2024}\natexlab{b}.
\newblock \showarticletitle{{VIDUR:} {A} Large-Scale Simulation Framework for {LLM} Inference}. In \bibinfo{booktitle}{\emph{Proceedings of the Seventh Annual Conference on Machine Learning and Systems, MLSys 2024, Santa Clara, CA, USA, May 13-16, 2024}}, \bibfield{editor}{\bibinfo{person}{Phillip~B. Gibbons}, \bibinfo{person}{Gennady Pekhimenko}, {and} \bibinfo{person}{Christopher~De Sa}} (Eds.). \bibinfo{publisher}{mlsys.org}.
\newblock
\urldef\tempurl%
\url{https://proceedings.mlsys.org/paper\_files/paper/2024/hash/b74a8de47d2b3c928360e0a011f48351-Abstract-Conference.html}
\showURL{%
\tempurl}


\bibitem[Agrawal et~al\mbox{.}(2024c)]%
        {SARATHI_SERVE}
\bibfield{author}{\bibinfo{person}{Amey Agrawal}, \bibinfo{person}{Nitin Kedia}, \bibinfo{person}{Ashish Panwar}, \bibinfo{person}{Jayashree Mohan}, \bibinfo{person}{Nipun Kwatra}, \bibinfo{person}{Bhargav~S. Gulavani}, \bibinfo{person}{Alexey Tumanov}, {and} \bibinfo{person}{Ramachandran Ramjee}.} \bibinfo{year}{2024}\natexlab{c}.
\newblock \showarticletitle{Taming Throughput-Latency Tradeoff in {LLM} Inference with Sarathi-Serve}. In \bibinfo{booktitle}{\emph{18th {USENIX} Symposium on Operating Systems Design and Implementation, {OSDI} 2024, Santa Clara, CA, USA, July 10-12, 2024}}, \bibfield{editor}{\bibinfo{person}{Ada Gavrilovska} {and} \bibinfo{person}{Douglas~B. Terry}} (Eds.). \bibinfo{publisher}{{USENIX} Association}, \bibinfo{pages}{117--134}.
\newblock
\urldef\tempurl%
\url{https://www.usenix.org/conference/osdi24/presentation/agrawal}
\showURL{%
\tempurl}


\bibitem[Agrawal et~al\mbox{.}(2023)]%
        {SARATHI}
\bibfield{author}{\bibinfo{person}{Amey Agrawal}, \bibinfo{person}{Ashish Panwar}, \bibinfo{person}{Jayashree Mohan}, \bibinfo{person}{Nipun Kwatra}, \bibinfo{person}{Bhargav~S. Gulavani}, {and} \bibinfo{person}{Ramachandran Ramjee}.} \bibinfo{year}{2023}\natexlab{}.
\newblock \showarticletitle{{SARATHI:} Efficient {LLM} Inference by Piggybacking Decodes with Chunked Prefills}.
\newblock \bibinfo{journal}{\emph{CoRR}}  \bibinfo{volume}{abs/2308.16369} (\bibinfo{year}{2023}).
\newblock
\urldef\tempurl%
\url{https://doi.org/10.48550/ARXIV.2308.16369}
\showDOI{\tempurl}
\showeprint[arXiv]{2308.16369}


\bibitem[Ahn et~al\mbox{.}(2024)]%
        {CXLmemoryusecasesindatabases}
\bibfield{author}{\bibinfo{person}{Minseon Ahn}, \bibinfo{person}{Thomas Willhalm}, \bibinfo{person}{Norman May}, \bibinfo{person}{Donghun Lee}, \bibinfo{person}{Suprasad~Mutalik Desai}, \bibinfo{person}{Daniel Booss}, \bibinfo{person}{Jungmin Kim}, \bibinfo{person}{Navneet Singh}, \bibinfo{person}{Daniel Ritter}, {and} \bibinfo{person}{Oliver Rebholz}.} \bibinfo{year}{2024}\natexlab{}.
\newblock \showarticletitle{An Examination of {CXL} Memory Use Cases for In-Memory Database Management Systems using {SAP} {HANA}}.
\newblock \bibinfo{journal}{\emph{Proc. {VLDB} Endow.}} \bibinfo{volume}{17}, \bibinfo{number}{12} (\bibinfo{year}{2024}), \bibinfo{pages}{3827--3840}.
\newblock
\urldef\tempurl%
\url{https://www.vldb.org/pvldb/vol17/p3827-ahn.pdf}
\showURL{%
\tempurl}


\bibitem[Ainslie et~al\mbox{.}(2023)]%
        {KVshareGQAgroupedqueryattention}
\bibfield{author}{\bibinfo{person}{Joshua Ainslie}, \bibinfo{person}{James Lee{-}Thorp}, \bibinfo{person}{Michiel de Jong}, \bibinfo{person}{Yury Zemlyanskiy}, \bibinfo{person}{Federico Lebr{\'{o}}n}, {and} \bibinfo{person}{Sumit Sanghai}.} \bibinfo{year}{2023}\natexlab{}.
\newblock \showarticletitle{{GQA:} Training Generalized Multi-Query Transformer Models from Multi-Head Checkpoints}. In \bibinfo{booktitle}{\emph{Proceedings of the 2023 Conference on Empirical Methods in Natural Language Processing, {EMNLP} 2023, Singapore, December 6-10, 2023}}, \bibfield{editor}{\bibinfo{person}{Houda Bouamor}, \bibinfo{person}{Juan Pino}, {and} \bibinfo{person}{Kalika Bali}} (Eds.). \bibinfo{publisher}{Association for Computational Linguistics}, \bibinfo{pages}{4895--4901}.
\newblock
\urldef\tempurl%
\url{https://doi.org/10.18653/V1/2023.EMNLP-MAIN.298}
\showDOI{\tempurl}


\bibitem[Akioyamen et~al\mbox{.}(2024)]%
        {LLMQueryOptUnreasonable}
\bibfield{author}{\bibinfo{person}{Peter Akioyamen}, \bibinfo{person}{Zixuan Yi}, {and} \bibinfo{person}{Ryan Marcus}.} \bibinfo{year}{2024}\natexlab{}.
\newblock \showarticletitle{The Unreasonable Effectiveness of LLMs for Query Optimization}.
\newblock \bibinfo{journal}{\emph{arXiv preprint arXiv:2411.02862}} (\bibinfo{year}{2024}).
\newblock


\bibitem[Allu et~al\mbox{.}(2024)]%
        {tableRAGsummarization2024}
\bibfield{author}{\bibinfo{person}{Uday Allu}, \bibinfo{person}{Biddwan Ahmed}, {and} \bibinfo{person}{Vishesh Tripathi}.} \bibinfo{year}{2024}\natexlab{}.
\newblock \showarticletitle{Beyond Extraction: Contextualising Tabular Data for Efficient Summarisation by Language Models}.
\newblock \bibinfo{journal}{\emph{CoRR}}  \bibinfo{volume}{abs/2401.02333} (\bibinfo{year}{2024}).
\newblock
\urldef\tempurl%
\url{https://doi.org/10.48550/ARXIV.2401.02333}
\showDOI{\tempurl}
\showeprint[arXiv]{2401.02333}


\bibitem[Anthony et~al\mbox{.}(2024)]%
        {SSMblackmamba2024}
\bibfield{author}{\bibinfo{person}{Quentin Anthony}, \bibinfo{person}{Yury Tokpanov}, \bibinfo{person}{Paolo Glorioso}, {and} \bibinfo{person}{Beren Millidge}.} \bibinfo{year}{2024}\natexlab{}.
\newblock \showarticletitle{BlackMamba: Mixture of Experts for State-Space Models}.
\newblock \bibinfo{journal}{\emph{CoRR}}  \bibinfo{volume}{abs/2402.01771} (\bibinfo{year}{2024}).
\newblock
\urldef\tempurl%
\url{https://doi.org/10.48550/ARXIV.2402.01771}
\showDOI{\tempurl}
\showeprint[arXiv]{2402.01771}


\bibitem[Asai et~al\mbox{.}(2024)]%
        {selfRAG}
\bibfield{author}{\bibinfo{person}{Akari Asai}, \bibinfo{person}{Zeqiu Wu}, \bibinfo{person}{Yizhong Wang}, \bibinfo{person}{Avirup Sil}, {and} \bibinfo{person}{Hannaneh Hajishirzi}.} \bibinfo{year}{2024}\natexlab{}.
\newblock \showarticletitle{Self-RAG: Learning to Retrieve, Generate, and Critique through Self-Reflection}. In \bibinfo{booktitle}{\emph{The Twelfth International Conference on Learning Representations, {ICLR} 2024, Vienna, Austria, May 7-11, 2024}}. \bibinfo{publisher}{OpenReview.net}.
\newblock
\urldef\tempurl%
\url{https://openreview.net/forum?id=hSyW5go0v8}
\showURL{%
\tempurl}


\bibitem[Badaro and Papotti(2022)]%
        {tabulardataunderstandingtutorial}
\bibfield{author}{\bibinfo{person}{Gilbert Badaro} {and} \bibinfo{person}{Paolo Papotti}.} \bibinfo{year}{2022}\natexlab{}.
\newblock \showarticletitle{Transformers for Tabular Data Representation: {A} Tutorial on Models and Applications}.
\newblock \bibinfo{journal}{\emph{Proc. {VLDB} Endow.}} \bibinfo{volume}{15}, \bibinfo{number}{12} (\bibinfo{year}{2022}), \bibinfo{pages}{3746--3749}.
\newblock
\urldef\tempurl%
\url{https://doi.org/10.14778/3554821.3554890}
\showDOI{\tempurl}


\bibitem[Banerjee et~al\mbox{.}(2024)]%
        {LLMsalwayshallucinate2024}
\bibfield{author}{\bibinfo{person}{Sourav Banerjee}, \bibinfo{person}{Ayushi Agarwal}, {and} \bibinfo{person}{Saloni Singla}.} \bibinfo{year}{2024}\natexlab{}.
\newblock \showarticletitle{LLMs Will Always Hallucinate, and We Need to Live With This}.
\newblock \bibinfo{journal}{\emph{CoRR}}  \bibinfo{volume}{abs/2409.05746} (\bibinfo{year}{2024}).
\newblock
\urldef\tempurl%
\url{https://doi.org/10.48550/ARXIV.2409.05746}
\showDOI{\tempurl}
\showeprint[arXiv]{2409.05746}


\bibitem[Barnett(2024)]%
        {scalinglawstransfer2024}
\bibfield{author}{\bibinfo{person}{Matthew Barnett}.} \bibinfo{year}{2024}\natexlab{}.
\newblock \showarticletitle{An Empirical Study of Scaling Laws for Transfer}.
\newblock \bibinfo{journal}{\emph{CoRR}}  \bibinfo{volume}{abs/2408.16947} (\bibinfo{year}{2024}).
\newblock
\urldef\tempurl%
\url{https://doi.org/10.48550/ARXIV.2408.16947}
\showDOI{\tempurl}
\showeprint[arXiv]{2408.16947}


\bibitem[Beltagy et~al\mbox{.}(2020)]%
        {SlidingWindowAttention}
\bibfield{author}{\bibinfo{person}{Iz Beltagy}, \bibinfo{person}{Matthew~E. Peters}, {and} \bibinfo{person}{Arman Cohan}.} \bibinfo{year}{2020}\natexlab{}.
\newblock \showarticletitle{Longformer: The Long-Document Transformer}.
\newblock \bibinfo{journal}{\emph{CoRR}}  \bibinfo{volume}{abs/2004.05150} (\bibinfo{year}{2020}).
\newblock
\showeprint[arXiv]{2004.05150}
\urldef\tempurl%
\url{https://arxiv.org/abs/2004.05150}
\showURL{%
\tempurl}


\bibitem[Ben~Allal et~al\mbox{.}(2024)]%
        {SLMSmolLM2024}
\bibfield{author}{\bibinfo{person}{Loubna Ben~Allal}, \bibinfo{person}{Anton Lozhkov}, {and} \bibinfo{person}{Elie Bakouch}.} \bibinfo{year}{2024}\natexlab{}.
\newblock \bibinfo{title}{SmolLM - blazingly fast and remarkably powerful}.
\newblock \bibinfo{howpublished}{\url{https://huggingface.co/blog/smollm}}.
\newblock
\newblock
\shownote{Accessed: December 15, 2024}.


\bibitem[Benaich and Hogarth(2024)]%
        {stateofaireport}
\bibfield{author}{\bibinfo{person}{Nathan Benaich} {and} \bibinfo{person}{Ian Hogarth}.} \bibinfo{year}{2024}\natexlab{}.
\newblock \bibinfo{title}{State of AI Report 2024}.
\newblock \bibinfo{howpublished}{\url{https://www.stateof.ai}}.
\newblock


\bibitem[Besta et~al\mbox{.}(2024a)]%
        {got2024}
\bibfield{author}{\bibinfo{person}{Maciej Besta}, \bibinfo{person}{Nils Blach}, \bibinfo{person}{Ales Kubicek}, \bibinfo{person}{Robert Gerstenberger}, \bibinfo{person}{Michal Podstawski}, \bibinfo{person}{Lukas Gianinazzi}, \bibinfo{person}{Joanna Gajda}, \bibinfo{person}{Tomasz Lehmann}, \bibinfo{person}{Hubert Niewiadomski}, \bibinfo{person}{Piotr Nyczyk}, {and} \bibinfo{person}{Torsten Hoefler}.} \bibinfo{year}{2024}\natexlab{a}.
\newblock \showarticletitle{Graph of Thoughts: Solving Elaborate Problems with Large Language Models}. In \bibinfo{booktitle}{\emph{Thirty-Eighth {AAAI} Conference on Artificial Intelligence, {AAAI} 2024, Thirty-Sixth Conference on Innovative Applications of Artificial Intelligence, {IAAI} 2024, Fourteenth Symposium on Educational Advances in Artificial Intelligence, {EAAI} 2014, February 20-27, 2024, Vancouver, Canada}}, \bibfield{editor}{\bibinfo{person}{Michael~J. Wooldridge}, \bibinfo{person}{Jennifer~G. Dy}, {and} \bibinfo{person}{Sriraam Natarajan}} (Eds.). \bibinfo{publisher}{{AAAI} Press}, \bibinfo{pages}{17682--17690}.
\newblock
\urldef\tempurl%
\url{https://doi.org/10.1609/AAAI.V38I16.29720}
\showDOI{\tempurl}


\bibitem[Besta et~al\mbox{.}(2024b)]%
        {promptsurveychaintreegraph2024}
\bibfield{author}{\bibinfo{person}{Maciej Besta}, \bibinfo{person}{Florim Memedi}, \bibinfo{person}{Zhenyu Zhang}, \bibinfo{person}{Robert Gerstenberger}, \bibinfo{person}{Nils Blach}, \bibinfo{person}{Piotr Nyczyk}, \bibinfo{person}{Marcin Copik}, \bibinfo{person}{Grzegorz Kwasniewski}, \bibinfo{person}{J{\"{u}}rgen M{\"{u}}ller}, \bibinfo{person}{Lukas Gianinazzi}, \bibinfo{person}{Ales Kubicek}, \bibinfo{person}{Hubert Niewiadomski}, \bibinfo{person}{Onur Mutlu}, {and} \bibinfo{person}{Torsten Hoefler}.} \bibinfo{year}{2024}\natexlab{b}.
\newblock \showarticletitle{Topologies of Reasoning: Demystifying Chains, Trees, and Graphs of Thoughts}.
\newblock \bibinfo{journal}{\emph{CoRR}}  \bibinfo{volume}{abs/2401.14295} (\bibinfo{year}{2024}).
\newblock
\urldef\tempurl%
\url{https://doi.org/10.48550/ARXIV.2401.14295}
\showDOI{\tempurl}
\showeprint[arXiv]{2401.14295}


\bibitem[Beurer{-}Kellner et~al\mbox{.}(2024)]%
        {constraineddecodingnoninvasive2024}
\bibfield{author}{\bibinfo{person}{Luca Beurer{-}Kellner}, \bibinfo{person}{Marc Fischer}, {and} \bibinfo{person}{Martin~T. Vechev}.} \bibinfo{year}{2024}\natexlab{}.
\newblock \showarticletitle{Guiding LLMs The Right Way: Fast, Non-Invasive Constrained Generation}. In \bibinfo{booktitle}{\emph{Forty-first International Conference on Machine Learning, {ICML} 2024, Vienna, Austria, July 21-27, 2024}}. \bibinfo{publisher}{OpenReview.net}.
\newblock
\urldef\tempurl%
\url{https://openreview.net/forum?id=pXaEYzrFae}
\showURL{%
\tempurl}


\bibitem[Birler et~al\mbox{.}(2024)]%
        {Diamond}
\bibfield{author}{\bibinfo{person}{Altan Birler}, \bibinfo{person}{Alfons Kemper}, {and} \bibinfo{person}{Thomas Neumann}.} \bibinfo{year}{2024}\natexlab{}.
\newblock \showarticletitle{Robust Join Processing with Diamond Hardened Joins}.
\newblock \bibinfo{journal}{\emph{Proc. {VLDB} Endow.}} \bibinfo{volume}{17}, \bibinfo{number}{11} (\bibinfo{year}{2024}), \bibinfo{pages}{3215--3228}.
\newblock
\urldef\tempurl%
\url{https://www.vldb.org/pvldb/vol17/p3215-birler.pdf}
\showURL{%
\tempurl}


\bibitem[Biswal et~al\mbox{.}(2024)]%
        {TAG}
\bibfield{author}{\bibinfo{person}{Asim Biswal}, \bibinfo{person}{Liana Patel}, \bibinfo{person}{Siddarth Jha}, \bibinfo{person}{Amog Kamsetty}, \bibinfo{person}{Shu Liu}, \bibinfo{person}{Joseph~E. Gonzalez}, \bibinfo{person}{Carlos Guestrin}, {and} \bibinfo{person}{Matei Zaharia}.} \bibinfo{year}{2024}\natexlab{}.
\newblock \showarticletitle{Text2SQL is Not Enough: Unifying {AI} and Databases with {TAG}}.
\newblock \bibinfo{journal}{\emph{CoRR}}  \bibinfo{volume}{abs/2408.14717} (\bibinfo{year}{2024}).
\newblock
\urldef\tempurl%
\url{https://doi.org/10.48550/ARXIV.2408.14717}
\showDOI{\tempurl}
\showeprint[arXiv]{2408.14717}


\bibitem[Bolding et~al\mbox{.}(2023)]%
        {llmtocleannoisydata2023}
\bibfield{author}{\bibinfo{person}{Quinten Bolding}, \bibinfo{person}{Baohao Liao}, \bibinfo{person}{Brandon~James Denis}, \bibinfo{person}{Jun Luo}, {and} \bibinfo{person}{Christof Monz}.} \bibinfo{year}{2023}\natexlab{}.
\newblock \showarticletitle{Ask Language Model to Clean Your Noisy Translation Data}. In \bibinfo{booktitle}{\emph{Findings of the Association for Computational Linguistics: {EMNLP} 2023, Singapore, December 6-10, 2023}}, \bibfield{editor}{\bibinfo{person}{Houda Bouamor}, \bibinfo{person}{Juan Pino}, {and} \bibinfo{person}{Kalika Bali}} (Eds.). \bibinfo{publisher}{Association for Computational Linguistics}, \bibinfo{pages}{3215--3236}.
\newblock
\urldef\tempurl%
\url{https://doi.org/10.18653/V1/2023.FINDINGS-EMNLP.212}
\showDOI{\tempurl}


\bibitem[Bonifati et~al\mbox{.}(2023)]%
        {ThesholdQueries}
\bibfield{author}{\bibinfo{person}{Angela Bonifati}, \bibinfo{person}{Stefania Dumbrava}, \bibinfo{person}{George Fletcher}, \bibinfo{person}{Jan Hidders}, \bibinfo{person}{Matthias Hofer}, \bibinfo{person}{Wim Martens}, \bibinfo{person}{Filip Murlak}, \bibinfo{person}{Joshua Shinavier}, \bibinfo{person}{Slawek Staworko}, {and} \bibinfo{person}{Dominik Tomaszuk}.} \bibinfo{year}{2023}\natexlab{}.
\newblock \showarticletitle{Threshold Queries}.
\newblock \bibinfo{journal}{\emph{{SIGMOD} Rec.}} \bibinfo{volume}{52}, \bibinfo{number}{1} (\bibinfo{year}{2023}), \bibinfo{pages}{64--73}.
\newblock
\urldef\tempurl%
\url{https://doi.org/10.1145/3604437.3604452}
\showDOI{\tempurl}


\bibitem[Borgeaud et~al\mbox{.}(2022)]%
        {memoryRAGtrilliontokens2022}
\bibfield{author}{\bibinfo{person}{Sebastian Borgeaud}, \bibinfo{person}{Arthur Mensch}, \bibinfo{person}{Jordan Hoffmann}, \bibinfo{person}{Trevor Cai}, \bibinfo{person}{Eliza Rutherford}, \bibinfo{person}{Katie Millican}, \bibinfo{person}{George van~den Driessche}, \bibinfo{person}{Jean{-}Baptiste Lespiau}, \bibinfo{person}{Bogdan Damoc}, \bibinfo{person}{Aidan Clark}, \bibinfo{person}{Diego de Las~Casas}, \bibinfo{person}{Aurelia Guy}, \bibinfo{person}{Jacob Menick}, \bibinfo{person}{Roman Ring}, \bibinfo{person}{Tom Hennigan}, \bibinfo{person}{Saffron Huang}, \bibinfo{person}{Loren Maggiore}, \bibinfo{person}{Chris Jones}, \bibinfo{person}{Albin Cassirer}, \bibinfo{person}{Andy Brock}, \bibinfo{person}{Michela Paganini}, \bibinfo{person}{Geoffrey Irving}, \bibinfo{person}{Oriol Vinyals}, \bibinfo{person}{Simon Osindero}, \bibinfo{person}{Karen Simonyan}, \bibinfo{person}{Jack~W. Rae}, \bibinfo{person}{Erich Elsen}, {and} \bibinfo{person}{Laurent Sifre}.} \bibinfo{year}{2022}\natexlab{}.
\newblock \showarticletitle{Improving Language Models by Retrieving from Trillions of Tokens}. In \bibinfo{booktitle}{\emph{International Conference on Machine Learning, {ICML} 2022, 17-23 July 2022, Baltimore, Maryland, {USA}}} \emph{(\bibinfo{series}{Proceedings of Machine Learning Research}, Vol.~\bibinfo{volume}{162})}, \bibfield{editor}{\bibinfo{person}{Kamalika Chaudhuri}, \bibinfo{person}{Stefanie Jegelka}, \bibinfo{person}{Le~Song}, \bibinfo{person}{Csaba Szepesv{\'{a}}ri}, \bibinfo{person}{Gang Niu}, {and} \bibinfo{person}{Sivan Sabato}} (Eds.). \bibinfo{publisher}{{PMLR}}, \bibinfo{pages}{2206--2240}.
\newblock
\urldef\tempurl%
\url{https://proceedings.mlr.press/v162/borgeaud22a.html}
\showURL{%
\tempurl}


\bibitem[Brandon et~al\mbox{.}(2024)]%
        {KVsharecrosslayerattention2024}
\bibfield{author}{\bibinfo{person}{William Brandon}, \bibinfo{person}{Mayank Mishra}, \bibinfo{person}{Aniruddha Nrusimha}, \bibinfo{person}{Rameswar Panda}, {and} \bibinfo{person}{Jonathan Ragan{-}Kelley}.} \bibinfo{year}{2024}\natexlab{}.
\newblock \showarticletitle{Reducing Transformer Key-Value Cache Size with Cross-Layer Attention}.
\newblock \bibinfo{journal}{\emph{CoRR}}  \bibinfo{volume}{abs/2405.12981} (\bibinfo{year}{2024}).
\newblock
\urldef\tempurl%
\url{https://doi.org/10.48550/ARXIV.2405.12981}
\showDOI{\tempurl}
\showeprint[arXiv]{2405.12981}


\bibitem[Brown et~al\mbox{.}(2024)]%
        {scalinglawsamplinglargelanguagemonkeys}
\bibfield{author}{\bibinfo{person}{Bradley C.~A. Brown}, \bibinfo{person}{Jordan Juravsky}, \bibinfo{person}{Ryan~Saul Ehrlich}, \bibinfo{person}{Ronald Clark}, \bibinfo{person}{Quoc~V. Le}, \bibinfo{person}{Christopher R{\'{e}}}, {and} \bibinfo{person}{Azalia Mirhoseini}.} \bibinfo{year}{2024}\natexlab{}.
\newblock \showarticletitle{Large Language Monkeys: Scaling Inference Compute with Repeated Sampling}.
\newblock \bibinfo{journal}{\emph{CoRR}}  \bibinfo{volume}{abs/2407.21787} (\bibinfo{year}{2024}).
\newblock
\urldef\tempurl%
\url{https://doi.org/10.48550/ARXIV.2407.21787}
\showDOI{\tempurl}
\showeprint[arXiv]{2407.21787}


\bibitem[Budiu et~al\mbox{.}(2023)]%
        {DBSP}
\bibfield{author}{\bibinfo{person}{Mihai Budiu}, \bibinfo{person}{Tej Chajed}, \bibinfo{person}{Frank McSherry}, \bibinfo{person}{Leonid Ryzhyk}, {and} \bibinfo{person}{Val Tannen}.} \bibinfo{year}{2023}\natexlab{}.
\newblock \showarticletitle{{DBSP:} Automatic Incremental View Maintenance for Rich Query Languages}.
\newblock \bibinfo{journal}{\emph{Proc. {VLDB} Endow.}} \bibinfo{volume}{16}, \bibinfo{number}{7} (\bibinfo{year}{2023}), \bibinfo{pages}{1601--1614}.
\newblock
\urldef\tempurl%
\url{https://doi.org/10.14778/3587136.3587137}
\showDOI{\tempurl}


\bibitem[Cao et~al\mbox{.}(2024)]%
        {moelightninghighthoughput2024}
\bibfield{author}{\bibinfo{person}{Shiyi Cao}, \bibinfo{person}{Shu Liu}, \bibinfo{person}{Tyler Griggs}, \bibinfo{person}{Peter Schafhalter}, \bibinfo{person}{Xiaoxuan Liu}, \bibinfo{person}{Ying Sheng}, \bibinfo{person}{Joseph~E Gonzalez}, \bibinfo{person}{Matei Zaharia}, {and} \bibinfo{person}{Ion Stoica}.} \bibinfo{year}{2024}\natexlab{}.
\newblock \showarticletitle{MoE-Lightning: High-Throughput MoE Inference on Memory-constrained GPUs}.
\newblock \bibinfo{journal}{\emph{arXiv preprint arXiv:2411.11217}} (\bibinfo{year}{2024}).
\newblock


\bibitem[Casper et~al\mbox{.}(2023)]%
        {openproblemlimitationRLHF2023}
\bibfield{author}{\bibinfo{person}{Stephen Casper}, \bibinfo{person}{Xander Davies}, \bibinfo{person}{Claudia Shi}, \bibinfo{person}{Thomas~Krendl Gilbert}, \bibinfo{person}{J{\'{e}}r{\'{e}}my Scheurer}, \bibinfo{person}{Javier Rando}, \bibinfo{person}{Rachel Freedman}, \bibinfo{person}{Tomasz Korbak}, \bibinfo{person}{David Lindner}, \bibinfo{person}{Pedro Freire}, \bibinfo{person}{Tony~Tong Wang}, \bibinfo{person}{Samuel Marks}, \bibinfo{person}{Charbel{-}Rapha{\"{e}}l S{\'{e}}gerie}, \bibinfo{person}{Micah Carroll}, \bibinfo{person}{Andi Peng}, \bibinfo{person}{Phillip J.~K. Christoffersen}, \bibinfo{person}{Mehul Damani}, \bibinfo{person}{Stewart Slocum}, \bibinfo{person}{Usman Anwar}, \bibinfo{person}{Anand Siththaranjan}, \bibinfo{person}{Max Nadeau}, \bibinfo{person}{Eric~J. Michaud}, \bibinfo{person}{Jacob Pfau}, \bibinfo{person}{Dmitrii Krasheninnikov}, \bibinfo{person}{Xin Chen}, \bibinfo{person}{Lauro Langosco}, \bibinfo{person}{Peter Hase}, \bibinfo{person}{Erdem Biyik}, \bibinfo{person}{Anca~D.
  Dragan}, \bibinfo{person}{David Krueger}, \bibinfo{person}{Dorsa Sadigh}, {and} \bibinfo{person}{Dylan Hadfield{-}Menell}.} \bibinfo{year}{2023}\natexlab{}.
\newblock \showarticletitle{Open Problems and Fundamental Limitations of Reinforcement Learning from Human Feedback}.
\newblock \bibinfo{journal}{\emph{Trans. Mach. Learn. Res.}}  \bibinfo{volume}{2023} (\bibinfo{year}{2023}).
\newblock
\urldef\tempurl%
\url{https://openreview.net/forum?id=bx24KpJ4Eb}
\showURL{%
\tempurl}


\bibitem[Chai et~al\mbox{.}(2023)]%
        {AItutorial5}
\bibfield{author}{\bibinfo{person}{Chengliang Chai}, \bibinfo{person}{Nan Tang}, \bibinfo{person}{Ju Fan}, {and} \bibinfo{person}{Yuyu Luo}.} \bibinfo{year}{2023}\natexlab{}.
\newblock \showarticletitle{Demystifying Artificial Intelligence for Data Preparation}. In \bibinfo{booktitle}{\emph{Companion of the 2023 International Conference on Management of Data, {SIGMOD/PODS} 2023, Seattle, WA, USA, June 18-23, 2023}}, \bibfield{editor}{\bibinfo{person}{Sudipto Das}, \bibinfo{person}{Ippokratis Pandis}, \bibinfo{person}{K.~Sel{\c{c}}uk Candan}, {and} \bibinfo{person}{Sihem Amer{-}Yahia}} (Eds.). \bibinfo{publisher}{{ACM}}, \bibinfo{pages}{13--20}.
\newblock
\urldef\tempurl%
\url{https://doi.org/10.1145/3555041.3589406}
\showDOI{\tempurl}


\bibitem[Chen et~al\mbox{.}(2024c)]%
        {selfreflectdual2024}
\bibfield{author}{\bibinfo{person}{Andong Chen}, \bibinfo{person}{Lianzhang Lou}, \bibinfo{person}{Kehai Chen}, \bibinfo{person}{Xuefeng Bai}, \bibinfo{person}{Yang Xiang}, \bibinfo{person}{Muyun Yang}, \bibinfo{person}{Tiejun Zhao}, {and} \bibinfo{person}{Min Zhang}.} \bibinfo{year}{2024}\natexlab{c}.
\newblock \showarticletitle{{DUAL-REFLECT:} Enhancing Large Language Models for Reflective Translation through Dual Learning Feedback Mechanisms}. In \bibinfo{booktitle}{\emph{Proceedings of the 62nd Annual Meeting of the Association for Computational Linguistics, {ACL} 2024 - Short Papers, Bangkok, Thailand, August 11-16, 2024}}, \bibfield{editor}{\bibinfo{person}{Lun{-}Wei Ku}, \bibinfo{person}{Andre Martins}, {and} \bibinfo{person}{Vivek Srikumar}} (Eds.). \bibinfo{publisher}{Association for Computational Linguistics}, \bibinfo{pages}{693--704}.
\newblock
\urldef\tempurl%
\url{https://aclanthology.org/2024.acl-short.64}
\showURL{%
\tempurl}


\bibitem[Chen et~al\mbox{.}(2024d)]%
        {selfconsistencynotgoodmultihop2024}
\bibfield{author}{\bibinfo{person}{Angelica Chen}, \bibinfo{person}{Jason Phang}, \bibinfo{person}{Alicia Parrish}, \bibinfo{person}{Vishakh Padmakumar}, \bibinfo{person}{Chen Zhao}, \bibinfo{person}{Samuel~R. Bowman}, {and} \bibinfo{person}{Kyunghyun Cho}.} \bibinfo{year}{2024}\natexlab{d}.
\newblock \showarticletitle{Two Failures of Self-Consistency in the Multi-Step Reasoning of LLMs}.
\newblock \bibinfo{journal}{\emph{Trans. Mach. Learn. Res.}}  \bibinfo{volume}{2024} (\bibinfo{year}{2024}).
\newblock
\urldef\tempurl%
\url{https://openreview.net/forum?id=5nBqY1y96B}
\showURL{%
\tempurl}


\bibitem[Chen et~al\mbox{.}(2023g)]%
        {unleashingpromptsurvey2023}
\bibfield{author}{\bibinfo{person}{Banghao Chen}, \bibinfo{person}{Zhaofeng Zhang}, \bibinfo{person}{Nicolas Langren{\'{e}}}, {and} \bibinfo{person}{Shengxin Zhu}.} \bibinfo{year}{2023}\natexlab{g}.
\newblock \showarticletitle{Unleashing the potential of prompt engineering in Large Language Models: a comprehensive review}.
\newblock \bibinfo{journal}{\emph{CoRR}}  \bibinfo{volume}{abs/2310.14735} (\bibinfo{year}{2023}).
\newblock
\urldef\tempurl%
\url{https://doi.org/10.48550/ARXIV.2310.14735}
\showDOI{\tempurl}
\showeprint[arXiv]{2310.14735}


\bibitem[Chen et~al\mbox{.}(2023d)]%
        {multiagentconsensus2023}
\bibfield{author}{\bibinfo{person}{Huaben Chen}, \bibinfo{person}{Wenkang Ji}, \bibinfo{person}{Lufeng Xu}, {and} \bibinfo{person}{Shiyu Zhao}.} \bibinfo{year}{2023}\natexlab{d}.
\newblock \showarticletitle{Multi-Agent Consensus Seeking via Large Language Models}.
\newblock \bibinfo{journal}{\emph{CoRR}}  \bibinfo{volume}{abs/2310.20151} (\bibinfo{year}{2023}).
\newblock
\urldef\tempurl%
\url{https://doi.org/10.48550/ARXIV.2310.20151}
\showDOI{\tempurl}
\showeprint[arXiv]{2310.20151}


\bibitem[Chen et~al\mbox{.}(2023e)]%
        {surveydataaugment2023}
\bibfield{author}{\bibinfo{person}{Jiaao Chen}, \bibinfo{person}{Derek Tam}, \bibinfo{person}{Colin Raffel}, \bibinfo{person}{Mohit Bansal}, {and} \bibinfo{person}{Diyi Yang}.} \bibinfo{year}{2023}\natexlab{e}.
\newblock \showarticletitle{An Empirical Survey of Data Augmentation for Limited Data Learning in {NLP}}.
\newblock \bibinfo{journal}{\emph{Trans. Assoc. Comput. Linguistics}}  \bibinfo{volume}{11} (\bibinfo{year}{2023}), \bibinfo{pages}{191--211}.
\newblock
\urldef\tempurl%
\url{https://doi.org/10.1162/TACL\_A\_00542}
\showDOI{\tempurl}


\bibitem[Chen et~al\mbox{.}(2024a)]%
        {scalingpropertiescompoundAI2024}
\bibfield{author}{\bibinfo{person}{Lingjiao Chen}, \bibinfo{person}{Jared~Quincy Davis}, \bibinfo{person}{Boris Hanin}, \bibinfo{person}{Peter Bailis}, \bibinfo{person}{Ion Stoica}, \bibinfo{person}{Matei Zaharia}, {and} \bibinfo{person}{James Zou}.} \bibinfo{year}{2024}\natexlab{a}.
\newblock \showarticletitle{Are more llm calls all you need? towards scaling laws of compound inference systems}.
\newblock \bibinfo{journal}{\emph{arXiv preprint arXiv:2403.02419}} (\bibinfo{year}{2024}).
\newblock


\bibitem[Chen et~al\mbox{.}(2024e)]%
        {graphRAGplanongraph2024}
\bibfield{author}{\bibinfo{person}{Liyi Chen}, \bibinfo{person}{Panrong Tong}, \bibinfo{person}{Zhongming Jin}, \bibinfo{person}{Ying Sun}, \bibinfo{person}{Jieping Ye}, {and} \bibinfo{person}{Hui Xiong}.} \bibinfo{year}{2024}\natexlab{e}.
\newblock \showarticletitle{Plan-on-Graph: Self-Correcting Adaptive Planning of Large Language Model on Knowledge Graphs}.
\newblock \bibinfo{journal}{\emph{CoRR}}  \bibinfo{volume}{abs/2410.23875} (\bibinfo{year}{2024}).
\newblock
\urldef\tempurl%
\url{https://doi.org/10.48550/ARXIV.2410.23875}
\showDOI{\tempurl}
\showeprint[arXiv]{2410.23875}


\bibitem[Chen et~al\mbox{.}(2023f)]%
        {frugalGPT2023}
\bibfield{author}{\bibinfo{person}{Lingjiao Chen}, \bibinfo{person}{Matei Zaharia}, {and} \bibinfo{person}{James Zou}.} \bibinfo{year}{2023}\natexlab{f}.
\newblock \showarticletitle{FrugalGPT: How to Use Large Language Models While Reducing Cost and Improving Performance}.
\newblock \bibinfo{journal}{\emph{CoRR}}  \bibinfo{volume}{abs/2305.05176} (\bibinfo{year}{2023}).
\newblock
\urldef\tempurl%
\url{https://doi.org/10.48550/ARXIV.2305.05176}
\showDOI{\tempurl}
\showeprint[arXiv]{2305.05176}


\bibitem[Chen(2023)]%
        {tableRAGfewshot2023}
\bibfield{author}{\bibinfo{person}{Wenhu Chen}.} \bibinfo{year}{2023}\natexlab{}.
\newblock \showarticletitle{Large Language Models are few(1)-shot Table Reasoners}. In \bibinfo{booktitle}{\emph{Findings of the Association for Computational Linguistics: {EACL} 2023, Dubrovnik, Croatia, May 2-6, 2023}}, \bibfield{editor}{\bibinfo{person}{Andreas Vlachos} {and} \bibinfo{person}{Isabelle Augenstein}} (Eds.). \bibinfo{publisher}{Association for Computational Linguistics}, \bibinfo{pages}{1090--1100}.
\newblock
\urldef\tempurl%
\url{https://doi.org/10.18653/V1/2023.FINDINGS-EACL.83}
\showDOI{\tempurl}


\bibitem[Chen et~al\mbox{.}(2022)]%
        {imageRAGopenQA2022}
\bibfield{author}{\bibinfo{person}{Wenhu Chen}, \bibinfo{person}{Hexiang Hu}, \bibinfo{person}{Xi Chen}, \bibinfo{person}{Pat Verga}, {and} \bibinfo{person}{William~W. Cohen}.} \bibinfo{year}{2022}\natexlab{}.
\newblock \showarticletitle{MuRAG: Multimodal Retrieval-Augmented Generator for Open Question Answering over Images and Text}. In \bibinfo{booktitle}{\emph{Proceedings of the 2022 Conference on Empirical Methods in Natural Language Processing, {EMNLP} 2022, Abu Dhabi, United Arab Emirates, December 7-11, 2022}}, \bibfield{editor}{\bibinfo{person}{Yoav Goldberg}, \bibinfo{person}{Zornitsa Kozareva}, {and} \bibinfo{person}{Yue Zhang}} (Eds.). \bibinfo{publisher}{Association for Computational Linguistics}, \bibinfo{pages}{5558--5570}.
\newblock
\urldef\tempurl%
\url{https://doi.org/10.18653/V1/2022.EMNLP-MAIN.375}
\showDOI{\tempurl}


\bibitem[Chen et~al\mbox{.}(2023c)]%
        {imageRAGtext2image2023}
\bibfield{author}{\bibinfo{person}{Wenhu Chen}, \bibinfo{person}{Hexiang Hu}, \bibinfo{person}{Chitwan Saharia}, {and} \bibinfo{person}{William~W. Cohen}.} \bibinfo{year}{2023}\natexlab{c}.
\newblock \showarticletitle{Re-Imagen: Retrieval-Augmented Text-to-Image Generator}. In \bibinfo{booktitle}{\emph{The Eleventh International Conference on Learning Representations, {ICLR} 2023, Kigali, Rwanda, May 1-5, 2023}}. \bibinfo{publisher}{OpenReview.net}.
\newblock
\urldef\tempurl%
\url{https://openreview.net/forum?id=XSEBx0iSjFQ}
\showURL{%
\tempurl}


\bibitem[Chen et~al\mbox{.}(2024b)]%
        {reprompt2024}
\bibfield{author}{\bibinfo{person}{Weizhe Chen}, \bibinfo{person}{Sven Koenig}, {and} \bibinfo{person}{Bistra Dilkina}.} \bibinfo{year}{2024}\natexlab{b}.
\newblock \showarticletitle{RePrompt: Planning by Automatic Prompt Engineering for Large Language Models Agents}.
\newblock \bibinfo{journal}{\emph{CoRR}}  \bibinfo{volume}{abs/2406.11132} (\bibinfo{year}{2024}).
\newblock
\urldef\tempurl%
\url{https://doi.org/10.48550/ARXIV.2406.11132}
\showDOI{\tempurl}
\showeprint[arXiv]{2406.11132}


\bibitem[Chen et~al\mbox{.}(2024f)]%
        {selfconsistencypara2024}
\bibfield{author}{\bibinfo{person}{Wenqing Chen}, \bibinfo{person}{Weicheng Wang}, \bibinfo{person}{Zhixuan Chu}, \bibinfo{person}{Kui Ren}, \bibinfo{person}{Zibin Zheng}, {and} \bibinfo{person}{Zhichao Lu}.} \bibinfo{year}{2024}\natexlab{f}.
\newblock \showarticletitle{Self-Para-Consistency: Improving Reasoning Tasks at Low Cost for Large Language Models}. In \bibinfo{booktitle}{\emph{Findings of the Association for Computational Linguistics, {ACL} 2024, Bangkok, Thailand and virtual meeting, August 11-16, 2024}}, \bibfield{editor}{\bibinfo{person}{Lun{-}Wei Ku}, \bibinfo{person}{Andre Martins}, {and} \bibinfo{person}{Vivek Srikumar}} (Eds.). \bibinfo{publisher}{Association for Computational Linguistics}, \bibinfo{pages}{14162--14167}.
\newblock
\urldef\tempurl%
\url{https://doi.org/10.18653/V1/2024.FINDINGS-ACL.842}
\showDOI{\tempurl}


\bibitem[Chen et~al\mbox{.}(2020)]%
        {tableRAGhybridQAbenchmark2020}
\bibfield{author}{\bibinfo{person}{Wenhu Chen}, \bibinfo{person}{Hanwen Zha}, \bibinfo{person}{Zhiyu Chen}, \bibinfo{person}{Wenhan Xiong}, \bibinfo{person}{Hong Wang}, {and} \bibinfo{person}{William~Yang Wang}.} \bibinfo{year}{2020}\natexlab{}.
\newblock \showarticletitle{HybridQA: {A} Dataset of Multi-Hop Question Answering over Tabular and Textual Data}. In \bibinfo{booktitle}{\emph{Findings of the Association for Computational Linguistics: {EMNLP} 2020, Online Event, 16-20 November 2020}} \emph{(\bibinfo{series}{Findings of {ACL}}, Vol.~\bibinfo{volume}{{EMNLP} 2020})}, \bibfield{editor}{\bibinfo{person}{Trevor Cohn}, \bibinfo{person}{Yulan He}, {and} \bibinfo{person}{Yang Liu}} (Eds.). \bibinfo{publisher}{Association for Computational Linguistics}, \bibinfo{pages}{1026--1036}.
\newblock
\urldef\tempurl%
\url{https://doi.org/10.18653/V1/2020.FINDINGS-EMNLP.91}
\showDOI{\tempurl}


\bibitem[Chen et~al\mbox{.}(2023a)]%
        {selfconsistencyuniversal2023}
\bibfield{author}{\bibinfo{person}{Xinyun Chen}, \bibinfo{person}{Renat Aksitov}, \bibinfo{person}{Uri Alon}, \bibinfo{person}{Jie Ren}, \bibinfo{person}{Kefan Xiao}, \bibinfo{person}{Pengcheng Yin}, \bibinfo{person}{Sushant Prakash}, \bibinfo{person}{Charles Sutton}, \bibinfo{person}{Xuezhi Wang}, {and} \bibinfo{person}{Denny Zhou}.} \bibinfo{year}{2023}\natexlab{a}.
\newblock \showarticletitle{Universal Self-Consistency for Large Language Model Generation}.
\newblock \bibinfo{journal}{\emph{CoRR}}  \bibinfo{volume}{abs/2311.17311} (\bibinfo{year}{2023}).
\newblock
\urldef\tempurl%
\url{https://doi.org/10.48550/ARXIV.2311.17311}
\showDOI{\tempurl}
\showeprint[arXiv]{2311.17311}


\bibitem[Chen et~al\mbox{.}(2024g)]%
        {unifieddetecthallucinationmultimodal2024}
\bibfield{author}{\bibinfo{person}{Xiang Chen}, \bibinfo{person}{Chenxi Wang}, \bibinfo{person}{Yida Xue}, \bibinfo{person}{Ningyu Zhang}, \bibinfo{person}{Xiaoyan Yang}, \bibinfo{person}{Qiang Li}, \bibinfo{person}{Yue Shen}, \bibinfo{person}{Lei Liang}, \bibinfo{person}{Jinjie Gu}, {and} \bibinfo{person}{Huajun Chen}.} \bibinfo{year}{2024}\natexlab{g}.
\newblock \showarticletitle{Unified Hallucination Detection for Multimodal Large Language Models}. In \bibinfo{booktitle}{\emph{Proceedings of the 62nd Annual Meeting of the Association for Computational Linguistics (Volume 1: Long Papers), {ACL} 2024, Bangkok, Thailand, August 11-16, 2024}}, \bibfield{editor}{\bibinfo{person}{Lun{-}Wei Ku}, \bibinfo{person}{Andre Martins}, {and} \bibinfo{person}{Vivek Srikumar}} (Eds.). \bibinfo{publisher}{Association for Computational Linguistics}, \bibinfo{pages}{3235--3252}.
\newblock
\urldef\tempurl%
\url{https://doi.org/10.18653/V1/2024.ACL-LONG.178}
\showDOI{\tempurl}


\bibitem[Chen et~al\mbox{.}(2023b)]%
        {symphonydatalakes}
\bibfield{author}{\bibinfo{person}{Zui Chen}, \bibinfo{person}{Zihui Gu}, \bibinfo{person}{Lei Cao}, \bibinfo{person}{Ju Fan}, \bibinfo{person}{Samuel Madden}, {and} \bibinfo{person}{Nan Tang}.} \bibinfo{year}{2023}\natexlab{b}.
\newblock \showarticletitle{Symphony: Towards Natural Language Query Answering over Multi-modal Data Lakes}. In \bibinfo{booktitle}{\emph{13th Conference on Innovative Data Systems Research, {CIDR} 2023, Amsterdam, The Netherlands, January 8-11, 2023}}. \bibinfo{publisher}{www.cidrdb.org}.
\newblock
\urldef\tempurl%
\url{https://www.cidrdb.org/cidr2023/papers/p51-chen.pdf}
\showURL{%
\tempurl}


\bibitem[Cheng et~al\mbox{.}(2024)]%
        {agenticLLMmultiagentLLMsurvey2024}
\bibfield{author}{\bibinfo{person}{Yuheng Cheng}, \bibinfo{person}{Ceyao Zhang}, \bibinfo{person}{Zhengwen Zhang}, \bibinfo{person}{Xiangrui Meng}, \bibinfo{person}{Sirui Hong}, \bibinfo{person}{Wenhao Li}, \bibinfo{person}{Zihao Wang}, \bibinfo{person}{Zekai Wang}, \bibinfo{person}{Feng Yin}, \bibinfo{person}{Junhua Zhao}, {and} \bibinfo{person}{Xiuqiang He}.} \bibinfo{year}{2024}\natexlab{}.
\newblock \showarticletitle{Exploring Large Language Model based Intelligent Agents: Definitions, Methods, and Prospects}.
\newblock \bibinfo{journal}{\emph{CoRR}}  \bibinfo{volume}{abs/2401.03428} (\bibinfo{year}{2024}).
\newblock
\urldef\tempurl%
\url{https://doi.org/10.48550/ARXIV.2401.03428}
\showDOI{\tempurl}
\showeprint[arXiv]{2401.03428}


\bibitem[Chinnakonduru and Mohapatra(2024)]%
        {KVshareWGQAgroupedqueryattention}
\bibfield{author}{\bibinfo{person}{Sai~Sena Chinnakonduru} {and} \bibinfo{person}{Astarag Mohapatra}.} \bibinfo{year}{2024}\natexlab{}.
\newblock \showarticletitle{Weighted Grouped Query Attention in Transformers}.
\newblock \bibinfo{journal}{\emph{CoRR}}  \bibinfo{volume}{abs/2407.10855} (\bibinfo{year}{2024}).
\newblock
\urldef\tempurl%
\url{https://doi.org/10.48550/ARXIV.2407.10855}
\showDOI{\tempurl}
\showeprint[arXiv]{2407.10855}


\bibitem[Cho et~al\mbox{.}(2021)]%
        {unifyingvisionlanguageinstruction2021}
\bibfield{author}{\bibinfo{person}{Jaemin Cho}, \bibinfo{person}{Jie Lei}, \bibinfo{person}{Hao Tan}, {and} \bibinfo{person}{Mohit Bansal}.} \bibinfo{year}{2021}\natexlab{}.
\newblock \showarticletitle{Unifying Vision-and-Language Tasks via Text Generation}. In \bibinfo{booktitle}{\emph{Proceedings of the 38th International Conference on Machine Learning, {ICML} 2021, 18-24 July 2021, Virtual Event}} \emph{(\bibinfo{series}{Proceedings of Machine Learning Research}, Vol.~\bibinfo{volume}{139})}, \bibfield{editor}{\bibinfo{person}{Marina Meila} {and} \bibinfo{person}{Tong Zhang}} (Eds.). \bibinfo{publisher}{{PMLR}}, \bibinfo{pages}{1931--1942}.
\newblock
\urldef\tempurl%
\url{http://proceedings.mlr.press/v139/cho21a.html}
\showURL{%
\tempurl}


\bibitem[Christiano et~al\mbox{.}(2017)]%
        {RLHF2017}
\bibfield{author}{\bibinfo{person}{Paul~F. Christiano}, \bibinfo{person}{Jan Leike}, \bibinfo{person}{Tom~B. Brown}, \bibinfo{person}{Miljan Martic}, \bibinfo{person}{Shane Legg}, {and} \bibinfo{person}{Dario Amodei}.} \bibinfo{year}{2017}\natexlab{}.
\newblock \showarticletitle{Deep Reinforcement Learning from Human Preferences}. In \bibinfo{booktitle}{\emph{Advances in Neural Information Processing Systems 30: Annual Conference on Neural Information Processing Systems 2017, December 4-9, 2017, Long Beach, CA, {USA}}}, \bibfield{editor}{\bibinfo{person}{Isabelle Guyon}, \bibinfo{person}{Ulrike von Luxburg}, \bibinfo{person}{Samy Bengio}, \bibinfo{person}{Hanna~M. Wallach}, \bibinfo{person}{Rob Fergus}, \bibinfo{person}{S.~V.~N. Vishwanathan}, {and} \bibinfo{person}{Roman Garnett}} (Eds.). \bibinfo{pages}{4299--4307}.
\newblock
\urldef\tempurl%
\url{https://proceedings.neurips.cc/paper/2017/hash/d5e2c0adad503c91f91df240d0cd4e49-Abstract.html}
\showURL{%
\tempurl}


\bibitem[Chrysogelos et~al\mbox{.}(2019)]%
        {HetExchange}
\bibfield{author}{\bibinfo{person}{Periklis Chrysogelos}, \bibinfo{person}{Manos Karpathiotakis}, \bibinfo{person}{Raja Appuswamy}, {and} \bibinfo{person}{Anastasia Ailamaki}.} \bibinfo{year}{2019}\natexlab{}.
\newblock \showarticletitle{HetExchange: Encapsulating heterogeneous {CPU-GPU} parallelism in {JIT} compiled engines}.
\newblock \bibinfo{journal}{\emph{Proc. {VLDB} Endow.}} \bibinfo{volume}{12}, \bibinfo{number}{5} (\bibinfo{year}{2019}), \bibinfo{pages}{544--556}.
\newblock
\urldef\tempurl%
\url{https://doi.org/10.14778/3303753.3303760}
\showDOI{\tempurl}


\bibitem[Chuang et~al\mbox{.}(2024)]%
        {lookbackdetecthallucinationattention2024}
\bibfield{author}{\bibinfo{person}{Yung{-}Sung Chuang}, \bibinfo{person}{Linlu Qiu}, \bibinfo{person}{Cheng{-}Yu Hsieh}, \bibinfo{person}{Ranjay Krishna}, \bibinfo{person}{Yoon Kim}, {and} \bibinfo{person}{James~R. Glass}.} \bibinfo{year}{2024}\natexlab{}.
\newblock \showarticletitle{Lookback Lens: Detecting and Mitigating Contextual Hallucinations in Large Language Models Using Only Attention Maps}. In \bibinfo{booktitle}{\emph{Proceedings of the 2024 Conference on Empirical Methods in Natural Language Processing, {EMNLP} 2024, Miami, FL, USA, November 12-16, 2024}}, \bibfield{editor}{\bibinfo{person}{Yaser Al{-}Onaizan}, \bibinfo{person}{Mohit Bansal}, {and} \bibinfo{person}{Yun{-}Nung Chen}} (Eds.). \bibinfo{publisher}{Association for Computational Linguistics}, \bibinfo{pages}{1419--1436}.
\newblock
\urldef\tempurl%
\url{https://aclanthology.org/2024.emnlp-main.84}
\showURL{%
\tempurl}


\bibitem[code4DB(2024)]%
        {LLM4DB}
\bibfield{author}{\bibinfo{person}{code4DB}.} \bibinfo{year}{2024}\natexlab{}.
\newblock \bibinfo{title}{LLM4DB: A Curated List of Resources on Large Language Models for Databases}.
\newblock \bibinfo{howpublished}{\url{https://github.com/code4DB/LLM4DB}}.
\newblock
\newblock
\shownote{Accessed: 2024-11-30}.


\bibitem[Cuconasu et~al\mbox{.}(2024)]%
        {searchpowerofnoise2024}
\bibfield{author}{\bibinfo{person}{Florin Cuconasu}, \bibinfo{person}{Giovanni Trappolini}, \bibinfo{person}{Federico Siciliano}, \bibinfo{person}{Simone Filice}, \bibinfo{person}{Cesare Campagnano}, \bibinfo{person}{Yoelle Maarek}, \bibinfo{person}{Nicola Tonellotto}, {and} \bibinfo{person}{Fabrizio Silvestri}.} \bibinfo{year}{2024}\natexlab{}.
\newblock \showarticletitle{The Power of Noise: Redefining Retrieval for {RAG} Systems}. In \bibinfo{booktitle}{\emph{Proceedings of the 47th International {ACM} {SIGIR} Conference on Research and Development in Information Retrieval, {SIGIR} 2024, Washington DC, USA, July 14-18, 2024}}, \bibfield{editor}{\bibinfo{person}{Grace~Hui Yang}, \bibinfo{person}{Hongning Wang}, \bibinfo{person}{Sam Han}, \bibinfo{person}{Claudia Hauff}, \bibinfo{person}{Guido Zuccon}, {and} \bibinfo{person}{Yi~Zhang}} (Eds.). \bibinfo{publisher}{{ACM}}, \bibinfo{pages}{719--729}.
\newblock
\urldef\tempurl%
\url{https://doi.org/10.1145/3626772.3657834}
\showDOI{\tempurl}


\bibitem[Dai et~al\mbox{.}(2024)]%
        {safeRLHF2024}
\bibfield{author}{\bibinfo{person}{Josef Dai}, \bibinfo{person}{Xuehai Pan}, \bibinfo{person}{Ruiyang Sun}, \bibinfo{person}{Jiaming Ji}, \bibinfo{person}{Xinbo Xu}, \bibinfo{person}{Mickel Liu}, \bibinfo{person}{Yizhou Wang}, {and} \bibinfo{person}{Yaodong Yang}.} \bibinfo{year}{2024}\natexlab{}.
\newblock \showarticletitle{Safe {RLHF:} Safe Reinforcement Learning from Human Feedback}. In \bibinfo{booktitle}{\emph{The Twelfth International Conference on Learning Representations, {ICLR} 2024, Vienna, Austria, May 7-11, 2024}}. \bibinfo{publisher}{OpenReview.net}.
\newblock
\urldef\tempurl%
\url{https://openreview.net/forum?id=TyFrPOKYXw}
\showURL{%
\tempurl}


\bibitem[Dai et~al\mbox{.}(2019)]%
        {KVCache}
\bibfield{author}{\bibinfo{person}{Zihang Dai}, \bibinfo{person}{Zhilin Yang}, \bibinfo{person}{Yiming Yang}, \bibinfo{person}{Jaime~G. Carbonell}, \bibinfo{person}{Quoc~Viet Le}, {and} \bibinfo{person}{Ruslan Salakhutdinov}.} \bibinfo{year}{2019}\natexlab{}.
\newblock \showarticletitle{Transformer-XL: Attentive Language Models beyond a Fixed-Length Context}. In \bibinfo{booktitle}{\emph{Proceedings of the 57th Conference of the Association for Computational Linguistics, {ACL} 2019, Florence, Italy, July 28- August 2, 2019, Volume 1: Long Papers}}, \bibfield{editor}{\bibinfo{person}{Anna Korhonen}, \bibinfo{person}{David~R. Traum}, {and} \bibinfo{person}{Llu{\'{\i}}s M{\`{a}}rquez}} (Eds.). \bibinfo{publisher}{Association for Computational Linguistics}, \bibinfo{pages}{2978--2988}.
\newblock
\urldef\tempurl%
\url{https://doi.org/10.18653/V1/P19-1285}
\showDOI{\tempurl}


\bibitem[Dao(2024)]%
        {FlashAttention_2}
\bibfield{author}{\bibinfo{person}{Tri Dao}.} \bibinfo{year}{2024}\natexlab{}.
\newblock \showarticletitle{FlashAttention-2: Faster Attention with Better Parallelism and Work Partitioning}. In \bibinfo{booktitle}{\emph{The Twelfth International Conference on Learning Representations, {ICLR} 2024, Vienna, Austria, May 7-11, 2024}}. \bibinfo{publisher}{OpenReview.net}.
\newblock
\urldef\tempurl%
\url{https://openreview.net/forum?id=mZn2Xyh9Ec}
\showURL{%
\tempurl}


\bibitem[Dao et~al\mbox{.}(2022)]%
        {FlashAttention}
\bibfield{author}{\bibinfo{person}{Tri Dao}, \bibinfo{person}{Daniel~Y. Fu}, \bibinfo{person}{Stefano Ermon}, \bibinfo{person}{Atri Rudra}, {and} \bibinfo{person}{Christopher R{\'{e}}}.} \bibinfo{year}{2022}\natexlab{}.
\newblock \showarticletitle{FlashAttention: Fast and Memory-Efficient Exact Attention with IO-Awareness}. In \bibinfo{booktitle}{\emph{Advances in Neural Information Processing Systems 35: Annual Conference on Neural Information Processing Systems 2022, NeurIPS 2022, New Orleans, LA, USA, November 28 - December 9, 2022}}, \bibfield{editor}{\bibinfo{person}{Sanmi Koyejo}, \bibinfo{person}{S.~Mohamed}, \bibinfo{person}{A.~Agarwal}, \bibinfo{person}{Danielle Belgrave}, \bibinfo{person}{K.~Cho}, {and} \bibinfo{person}{A.~Oh}} (Eds.).
\newblock
\urldef\tempurl%
\url{http://papers.nips.cc/paper\_files/paper/2022/hash/67d57c32e20fd0a7a302cb81d36e40d5-Abstract-Conference.html}
\showURL{%
\tempurl}


\bibitem[Dao and Gu(2024)]%
        {transformersaressms2024}
\bibfield{author}{\bibinfo{person}{Tri Dao} {and} \bibinfo{person}{Albert Gu}.} \bibinfo{year}{2024}\natexlab{}.
\newblock \showarticletitle{Transformers are SSMs: Generalized Models and Efficient Algorithms Through Structured State Space Duality}. In \bibinfo{booktitle}{\emph{Forty-first International Conference on Machine Learning, {ICML} 2024, Vienna, Austria, July 21-27, 2024}}. \bibinfo{publisher}{OpenReview.net}.
\newblock
\urldef\tempurl%
\url{https://openreview.net/forum?id=ztn8FCR1td}
\showURL{%
\tempurl}


\bibitem[Dao et~al\mbox{.}(2023)]%
        {flashdecodinglongcontext2023}
\bibfield{author}{\bibinfo{person}{Tri Dao}, \bibinfo{person}{Daniel Haziza}, \bibinfo{person}{Francisco Massa}, {and} \bibinfo{person}{Grigory Sizov}.} \bibinfo{year}{2023}\natexlab{}.
\newblock \bibinfo{title}{Flash-Decoding for long-context inference}.
\newblock \bibinfo{howpublished}{\url{https://pytorch.org/blog/flash-decoding/}}.
\newblock
\newblock
\shownote{Accessed: December 15, 2024}.


\bibitem[Devoto et~al\mbox{.}(2024)]%
        {KVcompressL22024}
\bibfield{author}{\bibinfo{person}{Alessio Devoto}, \bibinfo{person}{Yu Zhao}, \bibinfo{person}{Simone Scardapane}, {and} \bibinfo{person}{Pasquale Minervini}.} \bibinfo{year}{2024}\natexlab{}.
\newblock \showarticletitle{A Simple and Effective L{\_}2 Norm-Based Strategy for {KV} Cache Compression}. In \bibinfo{booktitle}{\emph{Proceedings of the 2024 Conference on Empirical Methods in Natural Language Processing, {EMNLP} 2024, Miami, FL, USA, November 12-16, 2024}}, \bibfield{editor}{\bibinfo{person}{Yaser Al{-}Onaizan}, \bibinfo{person}{Mohit Bansal}, {and} \bibinfo{person}{Yun{-}Nung Chen}} (Eds.). \bibinfo{publisher}{Association for Computational Linguistics}, \bibinfo{pages}{18476--18499}.
\newblock
\urldef\tempurl%
\url{https://aclanthology.org/2024.emnlp-main.1027}
\showURL{%
\tempurl}


\bibitem[Ding et~al\mbox{.}(2024a)]%
        {dataaugmentationusingllm2024}
\bibfield{author}{\bibinfo{person}{Bosheng Ding}, \bibinfo{person}{Chengwei Qin}, \bibinfo{person}{Ruochen Zhao}, \bibinfo{person}{Tianze Luo}, \bibinfo{person}{Xinze Li}, \bibinfo{person}{Guizhen Chen}, \bibinfo{person}{Wenhan Xia}, \bibinfo{person}{Junjie Hu}, \bibinfo{person}{Anh~Tuan Luu}, {and} \bibinfo{person}{Shafiq Joty}.} \bibinfo{year}{2024}\natexlab{a}.
\newblock \showarticletitle{Data Augmentation using LLMs: Data Perspectives, Learning Paradigms and Challenges}. In \bibinfo{booktitle}{\emph{Findings of the Association for Computational Linguistics, {ACL} 2024, Bangkok, Thailand and virtual meeting, August 11-16, 2024}}, \bibfield{editor}{\bibinfo{person}{Lun{-}Wei Ku}, \bibinfo{person}{Andre Martins}, {and} \bibinfo{person}{Vivek Srikumar}} (Eds.). \bibinfo{publisher}{Association for Computational Linguistics}, \bibinfo{pages}{1679--1705}.
\newblock
\urldef\tempurl%
\url{https://doi.org/10.18653/V1/2024.FINDINGS-ACL.97}
\showDOI{\tempurl}


\bibitem[Ding et~al\mbox{.}(2024b)]%
        {inputperturbationhallucination2024}
\bibfield{author}{\bibinfo{person}{Peng Ding}, \bibinfo{person}{Jingyu Wu}, \bibinfo{person}{Jun Kuang}, \bibinfo{person}{Dan Ma}, \bibinfo{person}{Xuezhi Cao}, \bibinfo{person}{Xunliang Cai}, \bibinfo{person}{Shi Chen}, \bibinfo{person}{Jiajun Chen}, {and} \bibinfo{person}{Shujian Huang}.} \bibinfo{year}{2024}\natexlab{b}.
\newblock \showarticletitle{Hallu-PI: Evaluating Hallucination in Multi-modal Large Language Models within Perturbed Inputs}. In \bibinfo{booktitle}{\emph{Proceedings of the 32nd {ACM} International Conference on Multimedia, {MM} 2024, Melbourne, VIC, Australia, 28 October 2024 - 1 November 2024}}, \bibfield{editor}{\bibinfo{person}{Jianfei Cai}, \bibinfo{person}{Mohan~S. Kankanhalli}, \bibinfo{person}{Balakrishnan Prabhakaran}, \bibinfo{person}{Susanne Boll}, \bibinfo{person}{Ramanathan Subramanian}, \bibinfo{person}{Liang Zheng}, \bibinfo{person}{Vivek~K. Singh}, \bibinfo{person}{Pablo C{\'{e}}sar}, \bibinfo{person}{Lexing Xie}, {and} \bibinfo{person}{Dong Xu}} (Eds.). \bibinfo{publisher}{{ACM}}, \bibinfo{pages}{10707--10715}.
\newblock
\urldef\tempurl%
\url{https://doi.org/10.1145/3664647.3681251}
\showDOI{\tempurl}


\bibitem[Dodge et~al\mbox{.}(2021)]%
        {dataqualitycorpus2021}
\bibfield{author}{\bibinfo{person}{Jesse Dodge}, \bibinfo{person}{Maarten Sap}, \bibinfo{person}{Ana Marasovic}, \bibinfo{person}{William Agnew}, \bibinfo{person}{Gabriel Ilharco}, \bibinfo{person}{Dirk Groeneveld}, \bibinfo{person}{Margaret Mitchell}, {and} \bibinfo{person}{Matt Gardner}.} \bibinfo{year}{2021}\natexlab{}.
\newblock \showarticletitle{Documenting Large Webtext Corpora: {A} Case Study on the Colossal Clean Crawled Corpus}. In \bibinfo{booktitle}{\emph{Proceedings of the 2021 Conference on Empirical Methods in Natural Language Processing, {EMNLP} 2021, Virtual Event / Punta Cana, Dominican Republic, 7-11 November, 2021}}, \bibfield{editor}{\bibinfo{person}{Marie{-}Francine Moens}, \bibinfo{person}{Xuanjing Huang}, \bibinfo{person}{Lucia Specia}, {and} \bibinfo{person}{Scott~Wen{-}tau Yih}} (Eds.). \bibinfo{publisher}{Association for Computational Linguistics}, \bibinfo{pages}{1286--1305}.
\newblock
\urldef\tempurl%
\url{https://doi.org/10.18653/V1/2021.EMNLP-MAIN.98}
\showDOI{\tempurl}


\bibitem[Dong et~al\mbox{.}(2024a)]%
        {hymbaSSM2024}
\bibfield{author}{\bibinfo{person}{Xin Dong}, \bibinfo{person}{Yonggan Fu}, \bibinfo{person}{Shizhe Diao}, \bibinfo{person}{Wonmin Byeon}, \bibinfo{person}{Zijia Chen}, \bibinfo{person}{Ameya~Sunil Mahabaleshwarkar}, \bibinfo{person}{Shih-Yang Liu}, \bibinfo{person}{Matthijs Van~Keirsbilck}, \bibinfo{person}{Min-Hung Chen}, \bibinfo{person}{Yoshi Suhara}, {et~al\mbox{.}}} \bibinfo{year}{2024}\natexlab{a}.
\newblock \showarticletitle{Hymba: A Hybrid-head Architecture for Small Language Models}.
\newblock \bibinfo{journal}{\emph{arXiv preprint arXiv:2411.13676}} (\bibinfo{year}{2024}).
\newblock


\bibitem[Dong et~al\mbox{.}(2023)]%
        {KDD2023tutorial}
\bibfield{author}{\bibinfo{person}{Xin~Luna Dong}, \bibinfo{person}{Seungwhan Moon}, \bibinfo{person}{Yifan~Ethan Xu}, \bibinfo{person}{Kshitiz Malik}, {and} \bibinfo{person}{Zhou Yu}.} \bibinfo{year}{2023}\natexlab{}.
\newblock \showarticletitle{Towards Next-Generation Intelligent Assistants Leveraging {LLM} Techniques}. In \bibinfo{booktitle}{\emph{Proceedings of the 29th {ACM} {SIGKDD} Conference on Knowledge Discovery and Data Mining, {KDD} 2023, Long Beach, CA, USA, August 6-10, 2023}}, \bibfield{editor}{\bibinfo{person}{Ambuj~K. Singh}, \bibinfo{person}{Yizhou Sun}, \bibinfo{person}{Leman Akoglu}, \bibinfo{person}{Dimitrios Gunopulos}, \bibinfo{person}{Xifeng Yan}, \bibinfo{person}{Ravi Kumar}, \bibinfo{person}{Fatma Ozcan}, {and} \bibinfo{person}{Jieping Ye}} (Eds.). \bibinfo{publisher}{{ACM}}, \bibinfo{pages}{5792--5793}.
\newblock
\urldef\tempurl%
\url{https://doi.org/10.1145/3580305.3599572}
\showDOI{\tempurl}


\bibitem[Dong et~al\mbox{.}(2024b)]%
        {xgrammarconstraineddecoding2024}
\bibfield{author}{\bibinfo{person}{Yixin Dong}, \bibinfo{person}{Charlie~F Ruan}, \bibinfo{person}{Yaxing Cai}, \bibinfo{person}{Ruihang Lai}, \bibinfo{person}{Ziyi Xu}, \bibinfo{person}{Yilong Zhao}, {and} \bibinfo{person}{Tianqi Chen}.} \bibinfo{year}{2024}\natexlab{b}.
\newblock \showarticletitle{XGrammar: Flexible and Efficient Structured Generation Engine for Large Language Models}.
\newblock \bibinfo{journal}{\emph{arXiv preprint arXiv:2411.15100}} (\bibinfo{year}{2024}).
\newblock


\bibitem[Drutsa et~al\mbox{.}(2020)]%
        {AItutorial6}
\bibfield{author}{\bibinfo{person}{Alexey Drutsa}, \bibinfo{person}{Valentina Fedorova}, \bibinfo{person}{Dmitry Ustalov}, \bibinfo{person}{Olga Megorskaya}, \bibinfo{person}{Evfrosiniya Zerminova}, {and} \bibinfo{person}{Daria Baidakova}.} \bibinfo{year}{2020}\natexlab{}.
\newblock \showarticletitle{Crowdsourcing Practice for Efficient Data Labeling: Aggregation, Incremental Relabeling, and Pricing}. In \bibinfo{booktitle}{\emph{Proceedings of the 2020 International Conference on Management of Data, {SIGMOD} Conference 2020, online conference [Portland, OR, USA], June 14-19, 2020}}, \bibfield{editor}{\bibinfo{person}{David Maier}, \bibinfo{person}{Rachel Pottinger}, \bibinfo{person}{AnHai Doan}, \bibinfo{person}{Wang{-}Chiew Tan}, \bibinfo{person}{Abdussalam Alawini}, {and} \bibinfo{person}{Hung~Q. Ngo}} (Eds.). \bibinfo{publisher}{{ACM}}, \bibinfo{pages}{2623--2627}.
\newblock
\urldef\tempurl%
\url{https://doi.org/10.1145/3318464.3383127}
\showDOI{\tempurl}


\bibitem[Du et~al\mbox{.}(2022)]%
        {moeglam2022}
\bibfield{author}{\bibinfo{person}{Nan Du}, \bibinfo{person}{Yanping Huang}, \bibinfo{person}{Andrew~M. Dai}, \bibinfo{person}{Simon Tong}, \bibinfo{person}{Dmitry Lepikhin}, \bibinfo{person}{Yuanzhong Xu}, \bibinfo{person}{Maxim Krikun}, \bibinfo{person}{Yanqi Zhou}, \bibinfo{person}{Adams~Wei Yu}, \bibinfo{person}{Orhan Firat}, \bibinfo{person}{Barret Zoph}, \bibinfo{person}{Liam Fedus}, \bibinfo{person}{Maarten~P. Bosma}, \bibinfo{person}{Zongwei Zhou}, \bibinfo{person}{Tao Wang}, \bibinfo{person}{Yu~Emma Wang}, \bibinfo{person}{Kellie Webster}, \bibinfo{person}{Marie Pellat}, \bibinfo{person}{Kevin Robinson}, \bibinfo{person}{Kathleen~S. Meier{-}Hellstern}, \bibinfo{person}{Toju Duke}, \bibinfo{person}{Lucas Dixon}, \bibinfo{person}{Kun Zhang}, \bibinfo{person}{Quoc~V. Le}, \bibinfo{person}{Yonghui Wu}, \bibinfo{person}{Zhifeng Chen}, {and} \bibinfo{person}{Claire Cui}.} \bibinfo{year}{2022}\natexlab{}.
\newblock \showarticletitle{GLaM: Efficient Scaling of Language Models with Mixture-of-Experts}. In \bibinfo{booktitle}{\emph{International Conference on Machine Learning, {ICML} 2022, 17-23 July 2022, Baltimore, Maryland, {USA}}} \emph{(\bibinfo{series}{Proceedings of Machine Learning Research}, Vol.~\bibinfo{volume}{162})}, \bibfield{editor}{\bibinfo{person}{Kamalika Chaudhuri}, \bibinfo{person}{Stefanie Jegelka}, \bibinfo{person}{Le~Song}, \bibinfo{person}{Csaba Szepesv{\'{a}}ri}, \bibinfo{person}{Gang Niu}, {and} \bibinfo{person}{Sivan Sabato}} (Eds.). \bibinfo{publisher}{{PMLR}}, \bibinfo{pages}{5547--5569}.
\newblock
\urldef\tempurl%
\url{https://proceedings.mlr.press/v162/du22c.html}
\showURL{%
\tempurl}


\bibitem[Du et~al\mbox{.}(2024)]%
        {dataqualityfederated2024}
\bibfield{author}{\bibinfo{person}{Yaxin Du}, \bibinfo{person}{Rui Ye}, \bibinfo{person}{Yuchi Fengting}, \bibinfo{person}{Wanru Zhao}, \bibinfo{person}{Jingjing Qu}, \bibinfo{person}{Yanfeng Wang}, {and} \bibinfo{person}{Siheng Chen}.} \bibinfo{year}{2024}\natexlab{}.
\newblock \showarticletitle{Data Quality Control in Federated Instruction-tuning of Large Language Models}.
\newblock \bibinfo{journal}{\emph{CoRR}}  \bibinfo{volume}{abs/2410.11540} (\bibinfo{year}{2024}).
\newblock
\urldef\tempurl%
\url{https://doi.org/10.48550/ARXIV.2410.11540}
\showDOI{\tempurl}
\showeprint[arXiv]{2410.11540}


\bibitem[Duarte(2024)]%
        {chatgptnumbers}
\bibfield{author}{\bibinfo{person}{Fabio Duarte}.} \bibinfo{year}{2024}\natexlab{}.
\newblock \bibinfo{title}{Number of ChatGPT Users (Dec 2024)}.
\newblock \bibinfo{howpublished}{\url{https://explodingtopics.com/blog/chatgpt-users}}.
\newblock
\newblock
\shownote{Accessed: 2024-12-15}.


\bibitem[Eibich et~al\mbox{.}(2024)]%
        {ARAGOG}
\bibfield{author}{\bibinfo{person}{Matous Eibich}, \bibinfo{person}{Shivay Nagpal}, {and} \bibinfo{person}{Alexander Fred{-}Ojala}.} \bibinfo{year}{2024}\natexlab{}.
\newblock \showarticletitle{{ARAGOG:} Advanced {RAG} Output Grading}.
\newblock \bibinfo{journal}{\emph{CoRR}}  \bibinfo{volume}{abs/2404.01037} (\bibinfo{year}{2024}).
\newblock
\urldef\tempurl%
\url{https://doi.org/10.48550/ARXIV.2404.01037}
\showDOI{\tempurl}
\showeprint[arXiv]{2404.01037}


\bibitem[Fagbohun et~al\mbox{.}(2024)]%
        {largemodelslesssensitive2024}
\bibfield{author}{\bibinfo{person}{Oluwole Fagbohun}, \bibinfo{person}{Rachel~M. Harrison}, {and} \bibinfo{person}{Anton Dereventsov}.} \bibinfo{year}{2024}\natexlab{}.
\newblock \showarticletitle{An Empirical Categorization of Prompting Techniques for Large Language Models: {A} Practitioner's Guide}.
\newblock \bibinfo{journal}{\emph{CoRR}}  \bibinfo{volume}{abs/2402.14837} (\bibinfo{year}{2024}).
\newblock
\urldef\tempurl%
\url{https://doi.org/10.48550/ARXIV.2402.14837}
\showDOI{\tempurl}
\showeprint[arXiv]{2402.14837}


\bibitem[Fan et~al\mbox{.}(2024)]%
        {RAGsurvey2024}
\bibfield{author}{\bibinfo{person}{Wenqi Fan}, \bibinfo{person}{Yujuan Ding}, \bibinfo{person}{Liangbo Ning}, \bibinfo{person}{Shijie Wang}, \bibinfo{person}{Hengyun Li}, \bibinfo{person}{Dawei Yin}, \bibinfo{person}{Tat{-}Seng Chua}, {and} \bibinfo{person}{Qing Li}.} \bibinfo{year}{2024}\natexlab{}.
\newblock \showarticletitle{A Survey on {RAG} Meeting LLMs: Towards Retrieval-Augmented Large Language Models}. In \bibinfo{booktitle}{\emph{Proceedings of the 30th {ACM} {SIGKDD} Conference on Knowledge Discovery and Data Mining, {KDD} 2024, Barcelona, Spain, August 25-29, 2024}}, \bibfield{editor}{\bibinfo{person}{Ricardo Baeza{-}Yates} {and} \bibinfo{person}{Francesco Bonchi}} (Eds.). \bibinfo{publisher}{{ACM}}, \bibinfo{pages}{6491--6501}.
\newblock
\urldef\tempurl%
\url{https://doi.org/10.1145/3637528.3671470}
\showDOI{\tempurl}


\bibitem[Farquhar et~al\mbox{.}(2024)]%
        {detecthallucinationentropy2024}
\bibfield{author}{\bibinfo{person}{Sebastian Farquhar}, \bibinfo{person}{Jannik Kossen}, \bibinfo{person}{Lorenz Kuhn}, {and} \bibinfo{person}{Yarin Gal}.} \bibinfo{year}{2024}\natexlab{}.
\newblock \showarticletitle{Detecting hallucinations in large language models using semantic entropy}.
\newblock \bibinfo{journal}{\emph{Nat.}} \bibinfo{volume}{630}, \bibinfo{number}{8017} (\bibinfo{year}{2024}), \bibinfo{pages}{625--630}.
\newblock
\urldef\tempurl%
\url{https://doi.org/10.1038/S41586-024-07421-0}
\showDOI{\tempurl}


\bibitem[Fedus et~al\mbox{.}(2022)]%
        {moeswitchtransformer2022}
\bibfield{author}{\bibinfo{person}{William Fedus}, \bibinfo{person}{Barret Zoph}, {and} \bibinfo{person}{Noam Shazeer}.} \bibinfo{year}{2022}\natexlab{}.
\newblock \showarticletitle{Switch Transformers: Scaling to Trillion Parameter Models with Simple and Efficient Sparsity}.
\newblock \bibinfo{journal}{\emph{J. Mach. Learn. Res.}}  \bibinfo{volume}{23} (\bibinfo{year}{2022}), \bibinfo{pages}{120:1--120:39}.
\newblock
\urldef\tempurl%
\url{https://jmlr.org/papers/v23/21-0998.html}
\showURL{%
\tempurl}


\bibitem[Feldstein et~al\mbox{.}(2024)]%
        {neurosymbolicAIlandscape2024}
\bibfield{author}{\bibinfo{person}{Jonathan Feldstein}, \bibinfo{person}{Paulius Dilkas}, \bibinfo{person}{Vaishak Belle}, {and} \bibinfo{person}{Efthymia Tsamoura}.} \bibinfo{year}{2024}\natexlab{}.
\newblock \showarticletitle{Mapping the Neuro-Symbolic {AI} Landscape by Architectures: {A} Handbook on Augmenting Deep Learning Through Symbolic Reasoning}.
\newblock \bibinfo{journal}{\emph{CoRR}}  \bibinfo{volume}{abs/2410.22077} (\bibinfo{year}{2024}).
\newblock
\urldef\tempurl%
\url{https://doi.org/10.48550/ARXIV.2410.22077}
\showDOI{\tempurl}
\showeprint[arXiv]{2410.22077}


\bibitem[Finlayson et~al\mbox{.}(2024)]%
        {closednucleussampling}
\bibfield{author}{\bibinfo{person}{Matthew Finlayson}, \bibinfo{person}{John Hewitt}, \bibinfo{person}{Alexander Koller}, \bibinfo{person}{Swabha Swayamdipta}, {and} \bibinfo{person}{Ashish Sabharwal}.} \bibinfo{year}{2024}\natexlab{}.
\newblock \showarticletitle{Closing the Curious Case of Neural Text Degeneration}. In \bibinfo{booktitle}{\emph{The Twelfth International Conference on Learning Representations, {ICLR} 2024, Vienna, Austria, May 7-11, 2024}}. \bibinfo{publisher}{OpenReview.net}.
\newblock
\urldef\tempurl%
\url{https://openreview.net/forum?id=dONpC9GL1o}
\showURL{%
\tempurl}


\bibitem[Frick et~al\mbox{.}(2024)]%
        {RLHFhowtoevaluate2024}
\bibfield{author}{\bibinfo{person}{Evan Frick}, \bibinfo{person}{Tianle Li}, \bibinfo{person}{Connor Chen}, \bibinfo{person}{Wei{-}Lin Chiang}, \bibinfo{person}{Anastasios~N. Angelopoulos}, \bibinfo{person}{Jiantao Jiao}, \bibinfo{person}{Banghua Zhu}, \bibinfo{person}{Joseph~E. Gonzalez}, {and} \bibinfo{person}{Ion Stoica}.} \bibinfo{year}{2024}\natexlab{}.
\newblock \showarticletitle{How to Evaluate Reward Models for {RLHF}}.
\newblock \bibinfo{journal}{\emph{CoRR}}  \bibinfo{volume}{abs/2410.14872} (\bibinfo{year}{2024}).
\newblock
\urldef\tempurl%
\url{https://doi.org/10.48550/ARXIV.2410.14872}
\showDOI{\tempurl}
\showeprint[arXiv]{2410.14872}


\bibitem[Fried et~al\mbox{.}(2023)]%
        {codedata2023}
\bibfield{author}{\bibinfo{person}{Daniel Fried}, \bibinfo{person}{Armen Aghajanyan}, \bibinfo{person}{Jessy Lin}, \bibinfo{person}{Sida Wang}, \bibinfo{person}{Eric Wallace}, \bibinfo{person}{Freda Shi}, \bibinfo{person}{Ruiqi Zhong}, \bibinfo{person}{Scott Yih}, \bibinfo{person}{Luke Zettlemoyer}, {and} \bibinfo{person}{Mike Lewis}.} \bibinfo{year}{2023}\natexlab{}.
\newblock \showarticletitle{InCoder: {A} Generative Model for Code Infilling and Synthesis}. In \bibinfo{booktitle}{\emph{The Eleventh International Conference on Learning Representations, {ICLR} 2023, Kigali, Rwanda, May 1-5, 2023}}. \bibinfo{publisher}{OpenReview.net}.
\newblock
\urldef\tempurl%
\url{https://openreview.net/forum?id=hQwb-lbM6EL}
\showURL{%
\tempurl}


\bibitem[Fu et~al\mbox{.}(2024a)]%
        {mitigatinghallucinationmultimodaldpo2024}
\bibfield{author}{\bibinfo{person}{Yuhan Fu}, \bibinfo{person}{Ruobing Xie}, \bibinfo{person}{Xingwu Sun}, \bibinfo{person}{Zhanhui Kang}, {and} \bibinfo{person}{Xirong Li}.} \bibinfo{year}{2024}\natexlab{a}.
\newblock \showarticletitle{Mitigating Hallucination in Multimodal Large Language Model via Hallucination-targeted Direct Preference Optimization}.
\newblock \bibinfo{journal}{\emph{arXiv preprint arXiv:2411.10436}} (\bibinfo{year}{2024}).
\newblock


\bibitem[Fu et~al\mbox{.}(2024b)]%
        {serverlessLLMlocalityenhanced}
\bibfield{author}{\bibinfo{person}{Yao Fu}, \bibinfo{person}{Leyang Xue}, \bibinfo{person}{Yeqi Huang}, \bibinfo{person}{Andrei{-}Octavian Brabete}, \bibinfo{person}{Dmitrii Ustiugov}, \bibinfo{person}{Yuvraj Patel}, {and} \bibinfo{person}{Luo Mai}.} \bibinfo{year}{2024}\natexlab{b}.
\newblock \showarticletitle{ServerlessLLM: Locality-Enhanced Serverless Inference for Large Language Models}.
\newblock \bibinfo{journal}{\emph{CoRR}}  \bibinfo{volume}{abs/2401.14351} (\bibinfo{year}{2024}).
\newblock
\urldef\tempurl%
\url{https://doi.org/10.48550/ARXIV.2401.14351}
\showDOI{\tempurl}
\showeprint[arXiv]{2401.14351}


\bibitem[Fu et~al\mbox{.}(2024c)]%
        {ORank}
\bibfield{author}{\bibinfo{person}{Yichao Fu}, \bibinfo{person}{Siqi Zhu}, \bibinfo{person}{Runlong Su}, \bibinfo{person}{Aurick Qiao}, \bibinfo{person}{Ion Stoica}, {and} \bibinfo{person}{Hao Zhang}.} \bibinfo{year}{2024}\natexlab{c}.
\newblock \showarticletitle{Efficient {LLM} Scheduling by Learning to Rank}.
\newblock \bibinfo{journal}{\emph{CoRR}}  \bibinfo{volume}{abs/2408.15792} (\bibinfo{year}{2024}).
\newblock
\urldef\tempurl%
\url{https://doi.org/10.48550/ARXIV.2408.15792}
\showDOI{\tempurl}
\showeprint[arXiv]{2408.15792}


\bibitem[Funke and Teubner(2020)]%
        {DogQC}
\bibfield{author}{\bibinfo{person}{Henning Funke} {and} \bibinfo{person}{Jens Teubner}.} \bibinfo{year}{2020}\natexlab{}.
\newblock \showarticletitle{Data-parallel query processing on non-uniform data}.
\newblock \bibinfo{journal}{\emph{Proceedings of the VLDB Endowment}} \bibinfo{volume}{13}, \bibinfo{number}{6} (\bibinfo{year}{2020}), \bibinfo{pages}{884--897}.
\newblock


\bibitem[Gao and Zhang(2024)]%
        {searchsimilaritydiversity2024}
\bibfield{author}{\bibinfo{person}{Hang Gao} {and} \bibinfo{person}{Yongfeng Zhang}.} \bibinfo{year}{2024}\natexlab{}.
\newblock \showarticletitle{{VRSD:} Rethinking Similarity and Diversity for Retrieval in Large Language Models}.
\newblock \bibinfo{journal}{\emph{CoRR}}  \bibinfo{volume}{abs/2407.04573} (\bibinfo{year}{2024}).
\newblock
\urldef\tempurl%
\url{https://doi.org/10.48550/ARXIV.2407.04573}
\showDOI{\tempurl}
\showeprint[arXiv]{2407.04573}


\bibitem[Gao et~al\mbox{.}(2021)]%
        {piledata2021}
\bibfield{author}{\bibinfo{person}{Leo Gao}, \bibinfo{person}{Stella Biderman}, \bibinfo{person}{Sid Black}, \bibinfo{person}{Laurence Golding}, \bibinfo{person}{Travis Hoppe}, \bibinfo{person}{Charles Foster}, \bibinfo{person}{Jason Phang}, \bibinfo{person}{Horace He}, \bibinfo{person}{Anish Thite}, \bibinfo{person}{Noa Nabeshima}, \bibinfo{person}{Shawn Presser}, {and} \bibinfo{person}{Connor Leahy}.} \bibinfo{year}{2021}\natexlab{}.
\newblock \showarticletitle{The Pile: An 800GB Dataset of Diverse Text for Language Modeling}.
\newblock \bibinfo{journal}{\emph{CoRR}}  \bibinfo{volume}{abs/2101.00027} (\bibinfo{year}{2021}).
\newblock
\showeprint[arXiv]{2101.00027}
\urldef\tempurl%
\url{https://arxiv.org/abs/2101.00027}
\showURL{%
\tempurl}


\bibitem[Gao et~al\mbox{.}(2023a)]%
        {HyDE}
\bibfield{author}{\bibinfo{person}{Luyu Gao}, \bibinfo{person}{Xueguang Ma}, \bibinfo{person}{Jimmy Lin}, {and} \bibinfo{person}{Jamie Callan}.} \bibinfo{year}{2023}\natexlab{a}.
\newblock \showarticletitle{Precise Zero-Shot Dense Retrieval without Relevance Labels}. In \bibinfo{booktitle}{\emph{Proceedings of the 61st Annual Meeting of the Association for Computational Linguistics (Volume 1: Long Papers), {ACL} 2023, Toronto, Canada, July 9-14, 2023}}, \bibfield{editor}{\bibinfo{person}{Anna Rogers}, \bibinfo{person}{Jordan~L. Boyd{-}Graber}, {and} \bibinfo{person}{Naoaki Okazaki}} (Eds.). \bibinfo{publisher}{Association for Computational Linguistics}, \bibinfo{pages}{1762--1777}.
\newblock
\urldef\tempurl%
\url{https://doi.org/10.18653/V1/2023.ACL-LONG.99}
\showDOI{\tempurl}


\bibitem[Gao et~al\mbox{.}(2023b)]%
        {RAGsurvey2023}
\bibfield{author}{\bibinfo{person}{Yunfan Gao}, \bibinfo{person}{Yun Xiong}, \bibinfo{person}{Xinyu Gao}, \bibinfo{person}{Kangxiang Jia}, \bibinfo{person}{Jinliu Pan}, \bibinfo{person}{Yuxi Bi}, \bibinfo{person}{Yi Dai}, \bibinfo{person}{Jiawei Sun}, \bibinfo{person}{Qianyu Guo}, \bibinfo{person}{Meng Wang}, {and} \bibinfo{person}{Haofen Wang}.} \bibinfo{year}{2023}\natexlab{b}.
\newblock \showarticletitle{Retrieval-Augmented Generation for Large Language Models: {A} Survey}.
\newblock \bibinfo{journal}{\emph{CoRR}}  \bibinfo{volume}{abs/2312.10997} (\bibinfo{year}{2023}).
\newblock
\urldef\tempurl%
\url{https://doi.org/10.48550/ARXIV.2312.10997}
\showDOI{\tempurl}
\showeprint[arXiv]{2312.10997}


\bibitem[Gema et~al\mbox{.}(2024)]%
        {contrastingretrievalhead2024}
\bibfield{author}{\bibinfo{person}{Aryo~Pradipta Gema}, \bibinfo{person}{Chen Jin}, \bibinfo{person}{Ahmed Abdulaal}, \bibinfo{person}{Tom Diethe}, \bibinfo{person}{Philip Teare}, \bibinfo{person}{Beatrice Alex}, \bibinfo{person}{Pasquale Minervini}, {and} \bibinfo{person}{Amrutha Saseendran}.} \bibinfo{year}{2024}\natexlab{}.
\newblock \showarticletitle{DeCoRe: Decoding by Contrasting Retrieval Heads to Mitigate Hallucinations}.
\newblock \bibinfo{journal}{\emph{arXiv preprint arXiv:2410.18860}} (\bibinfo{year}{2024}).
\newblock


\bibitem[Geng et~al\mbox{.}(2024)]%
        {constraineddecodingwithoutlogitaccess2024}
\bibfield{author}{\bibinfo{person}{Saibo Geng}, \bibinfo{person}{Berkay D{\"{o}}ner}, \bibinfo{person}{Chris Wendler}, \bibinfo{person}{Martin Josifoski}, {and} \bibinfo{person}{Robert West}.} \bibinfo{year}{2024}\natexlab{}.
\newblock \showarticletitle{Sketch-Guided Constrained Decoding for Boosting Blackbox Large Language Models without Logit Access}. In \bibinfo{booktitle}{\emph{Proceedings of the 62nd Annual Meeting of the Association for Computational Linguistics, {ACL} 2024 - Short Papers, Bangkok, Thailand, August 11-16, 2024}}, \bibfield{editor}{\bibinfo{person}{Lun{-}Wei Ku}, \bibinfo{person}{Andre Martins}, {and} \bibinfo{person}{Vivek Srikumar}} (Eds.). \bibinfo{publisher}{Association for Computational Linguistics}, \bibinfo{pages}{234--245}.
\newblock
\urldef\tempurl%
\url{https://aclanthology.org/2024.acl-short.23}
\showURL{%
\tempurl}


\bibitem[Geng et~al\mbox{.}(2023)]%
        {constraineddecoding2023}
\bibfield{author}{\bibinfo{person}{Saibo Geng}, \bibinfo{person}{Martin Josifoski}, \bibinfo{person}{Maxime Peyrard}, {and} \bibinfo{person}{Robert West}.} \bibinfo{year}{2023}\natexlab{}.
\newblock \showarticletitle{Grammar-Constrained Decoding for Structured {NLP} Tasks without Finetuning}. In \bibinfo{booktitle}{\emph{Proceedings of the 2023 Conference on Empirical Methods in Natural Language Processing, {EMNLP} 2023, Singapore, December 6-10, 2023}}, \bibfield{editor}{\bibinfo{person}{Houda Bouamor}, \bibinfo{person}{Juan Pino}, {and} \bibinfo{person}{Kalika Bali}} (Eds.). \bibinfo{publisher}{Association for Computational Linguistics}, \bibinfo{pages}{10932--10952}.
\newblock
\urldef\tempurl%
\url{https://doi.org/10.18653/V1/2023.EMNLP-MAIN.674}
\showDOI{\tempurl}


\bibitem[Gholami et~al\mbox{.}(2021)]%
        {modelquantizesurvey2021}
\bibfield{author}{\bibinfo{person}{Amir Gholami}, \bibinfo{person}{Sehoon Kim}, \bibinfo{person}{Zhen Dong}, \bibinfo{person}{Zhewei Yao}, \bibinfo{person}{Michael~W. Mahoney}, {and} \bibinfo{person}{Kurt Keutzer}.} \bibinfo{year}{2021}\natexlab{}.
\newblock \showarticletitle{A Survey of Quantization Methods for Efficient Neural Network Inference}.
\newblock \bibinfo{journal}{\emph{CoRR}}  \bibinfo{volume}{abs/2103.13630} (\bibinfo{year}{2021}).
\newblock
\showeprint[arXiv]{2103.13630}
\urldef\tempurl%
\url{https://arxiv.org/abs/2103.13630}
\showURL{%
\tempurl}


\bibitem[Gim et~al\mbox{.}(2024)]%
        {asyncLLMfunctioncalling2024}
\bibfield{author}{\bibinfo{person}{In Gim}, \bibinfo{person}{Seung-seob Lee}, {and} \bibinfo{person}{Lin Zhong}.} \bibinfo{year}{2024}\natexlab{}.
\newblock \showarticletitle{Asynchronous LLM Function Calling}.
\newblock \bibinfo{journal}{\emph{arXiv preprint arXiv:2412.07017}} (\bibinfo{year}{2024}).
\newblock


\bibitem[Glass et~al\mbox{.}(2023)]%
        {tableRAGtransformer2023}
\bibfield{author}{\bibinfo{person}{Michael~R. Glass}, \bibinfo{person}{Xueqing Wu}, \bibinfo{person}{Ankita~Rajaram Naik}, \bibinfo{person}{Gaetano Rossiello}, {and} \bibinfo{person}{Alfio Gliozzo}.} \bibinfo{year}{2023}\natexlab{}.
\newblock \showarticletitle{Retrieval-Based Transformer for Table Augmentation}. In \bibinfo{booktitle}{\emph{Findings of the Association for Computational Linguistics: {ACL} 2023, Toronto, Canada, July 9-14, 2023}}, \bibfield{editor}{\bibinfo{person}{Anna Rogers}, \bibinfo{person}{Jordan~L. Boyd{-}Graber}, {and} \bibinfo{person}{Naoaki Okazaki}} (Eds.). \bibinfo{publisher}{Association for Computational Linguistics}, \bibinfo{pages}{5635--5648}.
\newblock
\urldef\tempurl%
\url{https://doi.org/10.18653/V1/2023.FINDINGS-ACL.348}
\showDOI{\tempurl}


\bibitem[Gong et~al\mbox{.}(2024)]%
        {modelquantizeLLMCbenchmark2024}
\bibfield{author}{\bibinfo{person}{Ruihao Gong}, \bibinfo{person}{Yang Yong}, \bibinfo{person}{Shiqiao Gu}, \bibinfo{person}{Yushi Huang}, \bibinfo{person}{Chengtao Lv}, \bibinfo{person}{Yunchen Zhang}, \bibinfo{person}{Dacheng Tao}, {and} \bibinfo{person}{Xianglong Liu}.} \bibinfo{year}{2024}\natexlab{}.
\newblock \showarticletitle{{LLMC:} Benchmarking Large Language Model Quantization with a Versatile Compression Toolkit}. In \bibinfo{booktitle}{\emph{Proceedings of the 2024 Conference on Empirical Methods in Natural Language Processing: {EMNLP} 2024 - Industry Track, Miami, Florida, USA, November 12-16, 2024}}, \bibfield{editor}{\bibinfo{person}{Franck Dernoncourt}, \bibinfo{person}{Daniel Preotiuc{-}Pietro}, {and} \bibinfo{person}{Anastasia Shimorina}} (Eds.). \bibinfo{publisher}{Association for Computational Linguistics}, \bibinfo{pages}{132--152}.
\newblock
\urldef\tempurl%
\url{https://aclanthology.org/2024.emnlp-industry.12}
\showURL{%
\tempurl}


\bibitem[Gorishniy et~al\mbox{.}(2023)]%
        {tableRAGDL2023}
\bibfield{author}{\bibinfo{person}{Yury Gorishniy}, \bibinfo{person}{Ivan Rubachev}, \bibinfo{person}{Nikolay Kartashev}, \bibinfo{person}{Daniil Shlenskii}, \bibinfo{person}{Akim Kotelnikov}, {and} \bibinfo{person}{Artem Babenko}.} \bibinfo{year}{2023}\natexlab{}.
\newblock \showarticletitle{TabR: Unlocking the Power of Retrieval-Augmented Tabular Deep Learning}.
\newblock \bibinfo{journal}{\emph{CoRR}}  \bibinfo{volume}{abs/2307.14338} (\bibinfo{year}{2023}).
\newblock
\urldef\tempurl%
\url{https://doi.org/10.48550/ARXIV.2307.14338}
\showDOI{\tempurl}
\showeprint[arXiv]{2307.14338}


\bibitem[Gu and Dao(2023)]%
        {SSMmamba2023}
\bibfield{author}{\bibinfo{person}{Albert Gu} {and} \bibinfo{person}{Tri Dao}.} \bibinfo{year}{2023}\natexlab{}.
\newblock \showarticletitle{Mamba: Linear-Time Sequence Modeling with Selective State Spaces}.
\newblock \bibinfo{journal}{\emph{CoRR}}  \bibinfo{volume}{abs/2312.00752} (\bibinfo{year}{2023}).
\newblock
\urldef\tempurl%
\url{https://doi.org/10.48550/ARXIV.2312.00752}
\showDOI{\tempurl}
\showeprint[arXiv]{2312.00752}


\bibitem[Gu et~al\mbox{.}(2022)]%
        {SSM2022}
\bibfield{author}{\bibinfo{person}{Albert Gu}, \bibinfo{person}{Karan Goel}, {and} \bibinfo{person}{Christopher R{\'{e}}}.} \bibinfo{year}{2022}\natexlab{}.
\newblock \showarticletitle{Efficiently Modeling Long Sequences with Structured State Spaces}. In \bibinfo{booktitle}{\emph{The Tenth International Conference on Learning Representations, {ICLR} 2022, Virtual Event, April 25-29, 2022}}. \bibinfo{publisher}{OpenReview.net}.
\newblock
\urldef\tempurl%
\url{https://openreview.net/forum?id=uYLFoz1vlAC}
\showURL{%
\tempurl}


\bibitem[Guerreiro et~al\mbox{.}(2023)]%
        {inputperturbationhallucination2023}
\bibfield{author}{\bibinfo{person}{Nuno~Miguel Guerreiro}, \bibinfo{person}{Duarte~M. Alves}, \bibinfo{person}{Jonas Waldendorf}, \bibinfo{person}{Barry Haddow}, \bibinfo{person}{Alexandra Birch}, \bibinfo{person}{Pierre Colombo}, {and} \bibinfo{person}{Andr{\'{e}} F.~T. Martins}.} \bibinfo{year}{2023}\natexlab{}.
\newblock \showarticletitle{Hallucinations in Large Multilingual Translation Models}.
\newblock \bibinfo{journal}{\emph{Trans. Assoc. Comput. Linguistics}}  \bibinfo{volume}{11} (\bibinfo{year}{2023}), \bibinfo{pages}{1500--1517}.
\newblock
\urldef\tempurl%
\url{https://doi.org/10.1162/TACL\_A\_00615}
\showDOI{\tempurl}


\bibitem[G{\"{u}}l{\c{c}}ehre et~al\mbox{.}(2023)]%
        {ReSTRLHF2023}
\bibfield{author}{\bibinfo{person}{{\c{C}}aglar G{\"{u}}l{\c{c}}ehre}, \bibinfo{person}{Tom~Le Paine}, \bibinfo{person}{Srivatsan Srinivasan}, \bibinfo{person}{Ksenia Konyushkova}, \bibinfo{person}{Lotte Weerts}, \bibinfo{person}{Abhishek Sharma}, \bibinfo{person}{Aditya Siddhant}, \bibinfo{person}{Alex Ahern}, \bibinfo{person}{Miaosen Wang}, \bibinfo{person}{Chenjie Gu}, \bibinfo{person}{Wolfgang Macherey}, \bibinfo{person}{Arnaud Doucet}, \bibinfo{person}{Orhan Firat}, {and} \bibinfo{person}{Nando de Freitas}.} \bibinfo{year}{2023}\natexlab{}.
\newblock \showarticletitle{Reinforced Self-Training (ReST) for Language Modeling}.
\newblock \bibinfo{journal}{\emph{CoRR}}  \bibinfo{volume}{abs/2308.08998} (\bibinfo{year}{2023}).
\newblock
\urldef\tempurl%
\url{https://doi.org/10.48550/ARXIV.2308.08998}
\showDOI{\tempurl}
\showeprint[arXiv]{2308.08998}


\bibitem[Gunasekar et~al\mbox{.}(2023)]%
        {textbooksareallyouneed}
\bibfield{author}{\bibinfo{person}{Suriya Gunasekar}, \bibinfo{person}{Yi Zhang}, \bibinfo{person}{Jyoti Aneja}, \bibinfo{person}{Caio C{\'{e}}sar~Teodoro Mendes}, \bibinfo{person}{Allie~Del Giorno}, \bibinfo{person}{Sivakanth Gopi}, \bibinfo{person}{Mojan Javaheripi}, \bibinfo{person}{Piero Kauffmann}, \bibinfo{person}{Gustavo de Rosa}, \bibinfo{person}{Olli Saarikivi}, \bibinfo{person}{Adil Salim}, \bibinfo{person}{Shital Shah}, \bibinfo{person}{Harkirat~Singh Behl}, \bibinfo{person}{Xin Wang}, \bibinfo{person}{S{\'{e}}bastien Bubeck}, \bibinfo{person}{Ronen Eldan}, \bibinfo{person}{Adam~Tauman Kalai}, \bibinfo{person}{Yin~Tat Lee}, {and} \bibinfo{person}{Yuanzhi Li}.} \bibinfo{year}{2023}\natexlab{}.
\newblock \showarticletitle{Textbooks Are All You Need}.
\newblock \bibinfo{journal}{\emph{CoRR}}  \bibinfo{volume}{abs/2306.11644} (\bibinfo{year}{2023}).
\newblock
\urldef\tempurl%
\url{https://doi.org/10.48550/ARXIV.2306.11644}
\showDOI{\tempurl}
\showeprint[arXiv]{2306.11644}


\bibitem[Guo et~al\mbox{.}(2024)]%
        {multiagentLLMsurvey2024}
\bibfield{author}{\bibinfo{person}{Taicheng Guo}, \bibinfo{person}{Xiuying Chen}, \bibinfo{person}{Yaqi Wang}, \bibinfo{person}{Ruidi Chang}, \bibinfo{person}{Shichao Pei}, \bibinfo{person}{Nitesh~V. Chawla}, \bibinfo{person}{Olaf Wiest}, {and} \bibinfo{person}{Xiangliang Zhang}.} \bibinfo{year}{2024}\natexlab{}.
\newblock \showarticletitle{Large Language Model Based Multi-agents: {A} Survey of Progress and Challenges}. In \bibinfo{booktitle}{\emph{Proceedings of the Thirty-Third International Joint Conference on Artificial Intelligence, {IJCAI} 2024, Jeju, South Korea, August 3-9, 2024}}. \bibinfo{publisher}{ijcai.org}, \bibinfo{pages}{8048--8057}.
\newblock
\urldef\tempurl%
\url{https://www.ijcai.org/proceedings/2024/890}
\showURL{%
\tempurl}


\bibitem[Harlap et~al\mbox{.}(2018)]%
        {pipelineparallelismpipedream2018}
\bibfield{author}{\bibinfo{person}{Aaron Harlap}, \bibinfo{person}{Deepak Narayanan}, \bibinfo{person}{Amar Phanishayee}, \bibinfo{person}{Vivek Seshadri}, \bibinfo{person}{Nikhil~R. Devanur}, \bibinfo{person}{Gregory~R. Ganger}, {and} \bibinfo{person}{Phillip~B. Gibbons}.} \bibinfo{year}{2018}\natexlab{}.
\newblock \showarticletitle{PipeDream: Fast and Efficient Pipeline Parallel {DNN} Training}.
\newblock \bibinfo{journal}{\emph{CoRR}}  \bibinfo{volume}{abs/1806.03377} (\bibinfo{year}{2018}).
\newblock
\showeprint[arXiv]{1806.03377}
\urldef\tempurl%
\url{http://arxiv.org/abs/1806.03377}
\showURL{%
\tempurl}


\bibitem[Hasan(2024)]%
        {modelquantizePTQQAT2024}
\bibfield{author}{\bibinfo{person}{Jahid Hasan}.} \bibinfo{year}{2024}\natexlab{}.
\newblock \showarticletitle{Optimizing Large Language Models through Quantization: A Comparative Analysis of PTQ and QAT Techniques}.
\newblock \bibinfo{journal}{\emph{arXiv preprint arXiv:2411.06084}} (\bibinfo{year}{2024}).
\newblock


\bibitem[He et~al\mbox{.}(2024a)]%
        {flexattention}
\bibfield{author}{\bibinfo{person}{Horace He}, \bibinfo{person}{Driss Guessous}, \bibinfo{person}{Yanbo Liang}, {and} \bibinfo{person}{Joy Dong}.} \bibinfo{year}{2024}\natexlab{a}.
\newblock \bibinfo{title}{FlexAttention: The Flexibility of PyTorch with the Performance of FlashAttention}.
\newblock \bibinfo{howpublished}{\url{https://pytorch.org/blog/flexattention/}}.
\newblock


\bibitem[He et~al\mbox{.}(2024b)]%
        {SSMdensemamba2024}
\bibfield{author}{\bibinfo{person}{Wei He}, \bibinfo{person}{Kai Han}, \bibinfo{person}{Yehui Tang}, \bibinfo{person}{Chengcheng Wang}, \bibinfo{person}{Yujie Yang}, \bibinfo{person}{Tianyu Guo}, {and} \bibinfo{person}{Yunhe Wang}.} \bibinfo{year}{2024}\natexlab{b}.
\newblock \showarticletitle{DenseMamba: State Space Models with Dense Hidden Connection for Efficient Large Language Models}.
\newblock \bibinfo{journal}{\emph{CoRR}}  \bibinfo{volume}{abs/2403.00818} (\bibinfo{year}{2024}).
\newblock
\urldef\tempurl%
\url{https://doi.org/10.48550/ARXIV.2403.00818}
\showDOI{\tempurl}
\showeprint[arXiv]{2403.00818}


\bibitem[He et~al\mbox{.}(2024c)]%
        {graphRAGgretriever2024}
\bibfield{author}{\bibinfo{person}{Xiaoxin He}, \bibinfo{person}{Yijun Tian}, \bibinfo{person}{Yifei Sun}, \bibinfo{person}{Nitesh~V. Chawla}, \bibinfo{person}{Thomas Laurent}, \bibinfo{person}{Yann LeCun}, \bibinfo{person}{Xavier Bresson}, {and} \bibinfo{person}{Bryan Hooi}.} \bibinfo{year}{2024}\natexlab{c}.
\newblock \showarticletitle{G-Retriever: Retrieval-Augmented Generation for Textual Graph Understanding and Question Answering}.
\newblock \bibinfo{journal}{\emph{CoRR}}  \bibinfo{volume}{abs/2402.07630} (\bibinfo{year}{2024}).
\newblock
\urldef\tempurl%
\url{https://doi.org/10.48550/ARXIV.2402.07630}
\showDOI{\tempurl}
\showeprint[arXiv]{2402.07630}


\bibitem[He(2024)]%
        {moeloramillion2024}
\bibfield{author}{\bibinfo{person}{Xu~Owen He}.} \bibinfo{year}{2024}\natexlab{}.
\newblock \showarticletitle{Mixture of {A} Million Experts}.
\newblock \bibinfo{journal}{\emph{CoRR}}  \bibinfo{volume}{abs/2407.04153} (\bibinfo{year}{2024}).
\newblock
\urldef\tempurl%
\url{https://doi.org/10.48550/ARXIV.2407.04153}
\showDOI{\tempurl}
\showeprint[arXiv]{2407.04153}


\bibitem[Hernandez et~al\mbox{.}(2021)]%
        {scalinglawstransfer2021}
\bibfield{author}{\bibinfo{person}{Danny Hernandez}, \bibinfo{person}{Jared Kaplan}, \bibinfo{person}{Tom Henighan}, {and} \bibinfo{person}{Sam McCandlish}.} \bibinfo{year}{2021}\natexlab{}.
\newblock \showarticletitle{Scaling Laws for Transfer}.
\newblock \bibinfo{journal}{\emph{CoRR}}  \bibinfo{volume}{abs/2102.01293} (\bibinfo{year}{2021}).
\newblock
\showeprint[arXiv]{2102.01293}
\urldef\tempurl%
\url{https://arxiv.org/abs/2102.01293}
\showURL{%
\tempurl}


\bibitem[Hewing and Leinhos(2024)]%
        {promptcanvas2024}
\bibfield{author}{\bibinfo{person}{Michael Hewing} {and} \bibinfo{person}{Vincent Leinhos}.} \bibinfo{year}{2024}\natexlab{}.
\newblock \showarticletitle{The Prompt Canvas: A Literature-Based Practitioner Guide for Creating Effective Prompts in Large Language Models}.
\newblock \bibinfo{journal}{\emph{arXiv preprint arXiv:2412.05127}} (\bibinfo{year}{2024}).
\newblock


\bibitem[Hoffmann et~al\mbox{.}(2022)]%
        {scalinglawschinchilla2022}
\bibfield{author}{\bibinfo{person}{Jordan Hoffmann}, \bibinfo{person}{Sebastian Borgeaud}, \bibinfo{person}{Arthur Mensch}, \bibinfo{person}{Elena Buchatskaya}, \bibinfo{person}{Trevor Cai}, \bibinfo{person}{Eliza Rutherford}, \bibinfo{person}{Diego de Las~Casas}, \bibinfo{person}{Lisa~Anne Hendricks}, \bibinfo{person}{Johannes Welbl}, \bibinfo{person}{Aidan Clark}, \bibinfo{person}{Tom Hennigan}, \bibinfo{person}{Eric Noland}, \bibinfo{person}{Katie Millican}, \bibinfo{person}{George van~den Driessche}, \bibinfo{person}{Bogdan Damoc}, \bibinfo{person}{Aurelia Guy}, \bibinfo{person}{Simon Osindero}, \bibinfo{person}{Karen Simonyan}, \bibinfo{person}{Erich Elsen}, \bibinfo{person}{Jack~W. Rae}, \bibinfo{person}{Oriol Vinyals}, {and} \bibinfo{person}{Laurent Sifre}.} \bibinfo{year}{2022}\natexlab{}.
\newblock \showarticletitle{Training Compute-Optimal Large Language Models}.
\newblock \bibinfo{journal}{\emph{CoRR}}  \bibinfo{volume}{abs/2203.15556} (\bibinfo{year}{2022}).
\newblock
\urldef\tempurl%
\url{https://doi.org/10.48550/ARXIV.2203.15556}
\showDOI{\tempurl}
\showeprint[arXiv]{2203.15556}


\bibitem[Holtzman et~al\mbox{.}(2020)]%
        {nucleussampling}
\bibfield{author}{\bibinfo{person}{Ari Holtzman}, \bibinfo{person}{Jan Buys}, \bibinfo{person}{Li Du}, \bibinfo{person}{Maxwell Forbes}, {and} \bibinfo{person}{Yejin Choi}.} \bibinfo{year}{2020}\natexlab{}.
\newblock \showarticletitle{The Curious Case of Neural Text Degeneration}. In \bibinfo{booktitle}{\emph{8th International Conference on Learning Representations, {ICLR} 2020, Addis Ababa, Ethiopia, April 26-30, 2020}}. \bibinfo{publisher}{OpenReview.net}.
\newblock
\urldef\tempurl%
\url{https://openreview.net/forum?id=rygGQyrFvH}
\showURL{%
\tempurl}


\bibitem[Hong et~al\mbox{.}(2024)]%
        {tableRAGsqlgenfinetune2024}
\bibfield{author}{\bibinfo{person}{Zijin Hong}, \bibinfo{person}{Zheng Yuan}, \bibinfo{person}{Hao Chen}, \bibinfo{person}{Qinggang Zhang}, \bibinfo{person}{Feiran Huang}, {and} \bibinfo{person}{Xiao Huang}.} \bibinfo{year}{2024}\natexlab{}.
\newblock \showarticletitle{Knowledge-to-SQL: Enhancing {SQL} Generation with Data Expert {LLM}}. In \bibinfo{booktitle}{\emph{Findings of the Association for Computational Linguistics, {ACL} 2024, Bangkok, Thailand and virtual meeting, August 11-16, 2024}}, \bibfield{editor}{\bibinfo{person}{Lun{-}Wei Ku}, \bibinfo{person}{Andre Martins}, {and} \bibinfo{person}{Vivek Srikumar}} (Eds.). \bibinfo{publisher}{Association for Computational Linguistics}, \bibinfo{pages}{10997--11008}.
\newblock
\urldef\tempurl%
\url{https://doi.org/10.18653/V1/2024.FINDINGS-ACL.653}
\showDOI{\tempurl}


\bibitem[Houlsby et~al\mbox{.}(2019)]%
        {adapterefficientfinetuning2019}
\bibfield{author}{\bibinfo{person}{Neil Houlsby}, \bibinfo{person}{Andrei Giurgiu}, \bibinfo{person}{Stanislaw Jastrzebski}, \bibinfo{person}{Bruna Morrone}, \bibinfo{person}{Quentin de Laroussilhe}, \bibinfo{person}{Andrea Gesmundo}, \bibinfo{person}{Mona Attariyan}, {and} \bibinfo{person}{Sylvain Gelly}.} \bibinfo{year}{2019}\natexlab{}.
\newblock \showarticletitle{Parameter-Efficient Transfer Learning for {NLP}}. In \bibinfo{booktitle}{\emph{Proceedings of the 36th International Conference on Machine Learning, {ICML} 2019, 9-15 June 2019, Long Beach, California, {USA}}} \emph{(\bibinfo{series}{Proceedings of Machine Learning Research}, Vol.~\bibinfo{volume}{97})}, \bibfield{editor}{\bibinfo{person}{Kamalika Chaudhuri} {and} \bibinfo{person}{Ruslan Salakhutdinov}} (Eds.). \bibinfo{publisher}{{PMLR}}, \bibinfo{pages}{2790--2799}.
\newblock
\urldef\tempurl%
\url{http://proceedings.mlr.press/v97/houlsby19a.html}
\showURL{%
\tempurl}


\bibitem[Hsieh et~al\mbox{.}(2024)]%
        {foundinthemiddle}
\bibfield{author}{\bibinfo{person}{Cheng{-}Yu Hsieh}, \bibinfo{person}{Yung{-}Sung Chuang}, \bibinfo{person}{Chun{-}Liang Li}, \bibinfo{person}{Zifeng Wang}, \bibinfo{person}{Long~T. Le}, \bibinfo{person}{Abhishek Kumar}, \bibinfo{person}{James~R. Glass}, \bibinfo{person}{Alexander Ratner}, \bibinfo{person}{Chen{-}Yu Lee}, \bibinfo{person}{Ranjay Krishna}, {and} \bibinfo{person}{Tomas Pfister}.} \bibinfo{year}{2024}\natexlab{}.
\newblock \showarticletitle{Found in the middle: Calibrating Positional Attention Bias Improves Long Context Utilization}. In \bibinfo{booktitle}{\emph{Findings of the Association for Computational Linguistics, {ACL} 2024, Bangkok, Thailand and virtual meeting, August 11-16, 2024}}, \bibfield{editor}{\bibinfo{person}{Lun{-}Wei Ku}, \bibinfo{person}{Andre Martins}, {and} \bibinfo{person}{Vivek Srikumar}} (Eds.). \bibinfo{publisher}{Association for Computational Linguistics}, \bibinfo{pages}{14982--14995}.
\newblock
\urldef\tempurl%
\url{https://doi.org/10.18653/V1/2024.FINDINGS-ACL.890}
\showDOI{\tempurl}


\bibitem[Hu et~al\mbox{.}(2023)]%
        {ChatDBdatabasesassymbolicmemory}
\bibfield{author}{\bibinfo{person}{Chenxu Hu}, \bibinfo{person}{Jie Fu}, \bibinfo{person}{Chenzhuang Du}, \bibinfo{person}{Simian Luo}, \bibinfo{person}{Junbo Zhao}, {and} \bibinfo{person}{Hang Zhao}.} \bibinfo{year}{2023}\natexlab{}.
\newblock \showarticletitle{ChatDB: Augmenting LLMs with Databases as Their Symbolic Memory}.
\newblock \bibinfo{journal}{\emph{CoRR}}  \bibinfo{volume}{abs/2306.03901} (\bibinfo{year}{2023}).
\newblock
\urldef\tempurl%
\url{https://doi.org/10.48550/ARXIV.2306.03901}
\showDOI{\tempurl}
\showeprint[arXiv]{2306.03901}


\bibitem[Hu et~al\mbox{.}(2022)]%
        {LoRA}
\bibfield{author}{\bibinfo{person}{Edward~J. Hu}, \bibinfo{person}{Yelong Shen}, \bibinfo{person}{Phillip Wallis}, \bibinfo{person}{Zeyuan Allen{-}Zhu}, \bibinfo{person}{Yuanzhi Li}, \bibinfo{person}{Shean Wang}, \bibinfo{person}{Lu Wang}, {and} \bibinfo{person}{Weizhu Chen}.} \bibinfo{year}{2022}\natexlab{}.
\newblock \showarticletitle{LoRA: Low-Rank Adaptation of Large Language Models}. In \bibinfo{booktitle}{\emph{The Tenth International Conference on Learning Representations, {ICLR} 2022, Virtual Event, April 25-29, 2022}}. \bibinfo{publisher}{OpenReview.net}.
\newblock
\urldef\tempurl%
\url{https://openreview.net/forum?id=nZeVKeeFYf9}
\showURL{%
\tempurl}


\bibitem[Huang et~al\mbox{.}(2019)]%
        {pipelineparallelismgpipe2019}
\bibfield{author}{\bibinfo{person}{Yanping Huang}, \bibinfo{person}{Youlong Cheng}, \bibinfo{person}{Ankur Bapna}, \bibinfo{person}{Orhan Firat}, \bibinfo{person}{Dehao Chen}, \bibinfo{person}{Mia~Xu Chen}, \bibinfo{person}{HyoukJoong Lee}, \bibinfo{person}{Jiquan Ngiam}, \bibinfo{person}{Quoc~V. Le}, \bibinfo{person}{Yonghui Wu}, {and} \bibinfo{person}{Zhifeng Chen}.} \bibinfo{year}{2019}\natexlab{}.
\newblock \showarticletitle{GPipe: Efficient Training of Giant Neural Networks using Pipeline Parallelism}. In \bibinfo{booktitle}{\emph{Advances in Neural Information Processing Systems 32: Annual Conference on Neural Information Processing Systems 2019, NeurIPS 2019, December 8-14, 2019, Vancouver, BC, Canada}}, \bibfield{editor}{\bibinfo{person}{Hanna~M. Wallach}, \bibinfo{person}{Hugo Larochelle}, \bibinfo{person}{Alina Beygelzimer}, \bibinfo{person}{Florence d'Alch{\'{e}}{-}Buc}, \bibinfo{person}{Emily~B. Fox}, {and} \bibinfo{person}{Roman Garnett}} (Eds.). \bibinfo{pages}{103--112}.
\newblock
\urldef\tempurl%
\url{https://proceedings.neurips.cc/paper/2019/hash/093f65e080a295f8076b1c5722a46aa2-Abstract.html}
\showURL{%
\tempurl}


\bibitem[Isik et~al\mbox{.}(2024)]%
        {downstreamscalinglaws2024}
\bibfield{author}{\bibinfo{person}{Berivan Isik}, \bibinfo{person}{Natalia Ponomareva}, \bibinfo{person}{Hussein Hazimeh}, \bibinfo{person}{Dimitris Paparas}, \bibinfo{person}{Sergei Vassilvitskii}, {and} \bibinfo{person}{Sanmi Koyejo}.} \bibinfo{year}{2024}\natexlab{}.
\newblock \showarticletitle{Scaling Laws for Downstream Task Performance of Large Language Models}.
\newblock \bibinfo{journal}{\emph{CoRR}}  \bibinfo{volume}{abs/2402.04177} (\bibinfo{year}{2024}).
\newblock
\urldef\tempurl%
\url{https://doi.org/10.48550/ARXIV.2402.04177}
\showDOI{\tempurl}
\showeprint[arXiv]{2402.04177}


\bibitem[Ivanov et~al\mbox{.}(2021)]%
        {datamovementisallyouneed}
\bibfield{author}{\bibinfo{person}{Andrei Ivanov}, \bibinfo{person}{Nikoli Dryden}, \bibinfo{person}{Tal Ben{-}Nun}, \bibinfo{person}{Shigang Li}, {and} \bibinfo{person}{Torsten Hoefler}.} \bibinfo{year}{2021}\natexlab{}.
\newblock \showarticletitle{Data Movement Is All You Need: {A} Case Study on Optimizing Transformers}. In \bibinfo{booktitle}{\emph{Proceedings of the Fourth Conference on Machine Learning and Systems, MLSys 2021, virtual, April 5-9, 2021}}, \bibfield{editor}{\bibinfo{person}{Alex Smola}, \bibinfo{person}{Alex Dimakis}, {and} \bibinfo{person}{Ion Stoica}} (Eds.). \bibinfo{publisher}{mlsys.org}.
\newblock
\urldef\tempurl%
\url{https://proceedings.mlsys.org/paper\_files/paper/2021/hash/bc86e95606a6392f51f95a8de106728d-Abstract.html}
\showURL{%
\tempurl}


\bibitem[Jeong et~al\mbox{.}(2024)]%
        {adaptiveRAG}
\bibfield{author}{\bibinfo{person}{Soyeong Jeong}, \bibinfo{person}{Jinheon Baek}, \bibinfo{person}{Sukmin Cho}, \bibinfo{person}{Sung~Ju Hwang}, {and} \bibinfo{person}{Jong Park}.} \bibinfo{year}{2024}\natexlab{}.
\newblock \showarticletitle{Adaptive-RAG: Learning to Adapt Retrieval-Augmented Large Language Models through Question Complexity}. In \bibinfo{booktitle}{\emph{Proceedings of the 2024 Conference of the North American Chapter of the Association for Computational Linguistics: Human Language Technologies (Volume 1: Long Papers), {NAACL} 2024, Mexico City, Mexico, June 16-21, 2024}}, \bibfield{editor}{\bibinfo{person}{Kevin Duh}, \bibinfo{person}{Helena G{\'{o}}mez{-}Adorno}, {and} \bibinfo{person}{Steven Bethard}} (Eds.). \bibinfo{publisher}{Association for Computational Linguistics}, \bibinfo{pages}{7036--7050}.
\newblock
\urldef\tempurl%
\url{https://doi.org/10.18653/V1/2024.NAACL-LONG.389}
\showDOI{\tempurl}


\bibitem[Ji et~al\mbox{.}(2023a)]%
        {surveyofhallucination}
\bibfield{author}{\bibinfo{person}{Ziwei Ji}, \bibinfo{person}{Nayeon Lee}, \bibinfo{person}{Rita Frieske}, \bibinfo{person}{Tiezheng Yu}, \bibinfo{person}{Dan Su}, \bibinfo{person}{Yan Xu}, \bibinfo{person}{Etsuko Ishii}, \bibinfo{person}{Yejin Bang}, \bibinfo{person}{Andrea Madotto}, {and} \bibinfo{person}{Pascale Fung}.} \bibinfo{year}{2023}\natexlab{a}.
\newblock \showarticletitle{Survey of Hallucination in Natural Language Generation}.
\newblock \bibinfo{journal}{\emph{{ACM} Comput. Surv.}} \bibinfo{volume}{55}, \bibinfo{number}{12} (\bibinfo{year}{2023}), \bibinfo{pages}{248:1--248:38}.
\newblock
\urldef\tempurl%
\url{https://doi.org/10.1145/3571730}
\showDOI{\tempurl}


\bibitem[Ji et~al\mbox{.}(2023b)]%
        {selfreflectionhallucination2023}
\bibfield{author}{\bibinfo{person}{Ziwei Ji}, \bibinfo{person}{Tiezheng Yu}, \bibinfo{person}{Yan Xu}, \bibinfo{person}{Nayeon Lee}, \bibinfo{person}{Etsuko Ishii}, {and} \bibinfo{person}{Pascale Fung}.} \bibinfo{year}{2023}\natexlab{b}.
\newblock \showarticletitle{Towards Mitigating Hallucination in Large Language Models via Self-Reflection}.
\newblock \bibinfo{journal}{\emph{CoRR}}  \bibinfo{volume}{abs/2310.06271} (\bibinfo{year}{2023}).
\newblock
\urldef\tempurl%
\url{https://doi.org/10.48550/ARXIV.2310.06271}
\showDOI{\tempurl}
\showeprint[arXiv]{2310.06271}


\bibitem[Jiang et~al\mbox{.}(2023)]%
        {structgptgeneralstructureddata}
\bibfield{author}{\bibinfo{person}{Jinhao Jiang}, \bibinfo{person}{Kun Zhou}, \bibinfo{person}{Zican Dong}, \bibinfo{person}{Keming Ye}, \bibinfo{person}{Xin Zhao}, {and} \bibinfo{person}{Ji{-}Rong Wen}.} \bibinfo{year}{2023}\natexlab{}.
\newblock \showarticletitle{StructGPT: {A} General Framework for Large Language Model to Reason over Structured Data}. In \bibinfo{booktitle}{\emph{Proceedings of the 2023 Conference on Empirical Methods in Natural Language Processing, {EMNLP} 2023, Singapore, December 6-10, 2023}}, \bibfield{editor}{\bibinfo{person}{Houda Bouamor}, \bibinfo{person}{Juan Pino}, {and} \bibinfo{person}{Kalika Bali}} (Eds.). \bibinfo{publisher}{Association for Computational Linguistics}, \bibinfo{pages}{9237--9251}.
\newblock
\urldef\tempurl%
\url{https://doi.org/10.18653/V1/2023.EMNLP-MAIN.574}
\showDOI{\tempurl}


\bibitem[Jiang et~al\mbox{.}(2024)]%
        {NEO}
\bibfield{author}{\bibinfo{person}{Xuanlin Jiang}, \bibinfo{person}{Yang Zhou}, \bibinfo{person}{Shiyi Cao}, \bibinfo{person}{Ion Stoica}, {and} \bibinfo{person}{Minlan Yu}.} \bibinfo{year}{2024}\natexlab{}.
\newblock \showarticletitle{NEO: Saving GPU Memory Crisis with CPU Offloading for Online LLM Inference}.
\newblock \bibinfo{journal}{\emph{arXiv preprint arXiv:2411.01142}} (\bibinfo{year}{2024}).
\newblock


\bibitem[Kang et~al\mbox{.}(2024a)]%
        {tableRAGdenoisingtabletext2024}
\bibfield{author}{\bibinfo{person}{Deokhyung Kang}, \bibinfo{person}{Baikjin Jung}, \bibinfo{person}{Yunsu Kim}, {and} \bibinfo{person}{Gary~Geunbae Lee}.} \bibinfo{year}{2024}\natexlab{a}.
\newblock \showarticletitle{Denoising Table-Text Retrieval for Open-Domain Question Answering}. In \bibinfo{booktitle}{\emph{Proceedings of the 2024 Joint International Conference on Computational Linguistics, Language Resources and Evaluation, {LREC/COLING} 2024, 20-25 May, 2024, Torino, Italy}}, \bibfield{editor}{\bibinfo{person}{Nicoletta Calzolari}, \bibinfo{person}{Min{-}Yen Kan}, \bibinfo{person}{V{\'{e}}ronique Hoste}, \bibinfo{person}{Alessandro Lenci}, \bibinfo{person}{Sakriani Sakti}, {and} \bibinfo{person}{Nianwen Xue}} (Eds.). \bibinfo{publisher}{{ELRA} and {ICCL}}, \bibinfo{pages}{4634--4640}.
\newblock
\urldef\tempurl%
\url{https://aclanthology.org/2024.lrec-main.414}
\showURL{%
\tempurl}


\bibitem[Kang et~al\mbox{.}(2024b)]%
        {gearKVcachecompressionlossless2024}
\bibfield{author}{\bibinfo{person}{Hao Kang}, \bibinfo{person}{Qingru Zhang}, \bibinfo{person}{Souvik Kundu}, \bibinfo{person}{Geonhwa Jeong}, \bibinfo{person}{Zaoxing Liu}, \bibinfo{person}{Tushar Krishna}, {and} \bibinfo{person}{Tuo Zhao}.} \bibinfo{year}{2024}\natexlab{b}.
\newblock \showarticletitle{Gear: An efficient kv cache compression recipefor near-lossless generative inference of llm}.
\newblock \bibinfo{journal}{\emph{arXiv preprint arXiv:2403.05527}} (\bibinfo{year}{2024}).
\newblock


\bibitem[Kaplan et~al\mbox{.}(2020)]%
        {scalinglaws2020}
\bibfield{author}{\bibinfo{person}{Jared Kaplan}, \bibinfo{person}{Sam McCandlish}, \bibinfo{person}{Tom Henighan}, \bibinfo{person}{Tom~B. Brown}, \bibinfo{person}{Benjamin Chess}, \bibinfo{person}{Rewon Child}, \bibinfo{person}{Scott Gray}, \bibinfo{person}{Alec Radford}, \bibinfo{person}{Jeffrey Wu}, {and} \bibinfo{person}{Dario Amodei}.} \bibinfo{year}{2020}\natexlab{}.
\newblock \showarticletitle{Scaling Laws for Neural Language Models}.
\newblock \bibinfo{journal}{\emph{CoRR}}  \bibinfo{volume}{abs/2001.08361} (\bibinfo{year}{2020}).
\newblock
\showeprint[arXiv]{2001.08361}
\urldef\tempurl%
\url{https://arxiv.org/abs/2001.08361}
\showURL{%
\tempurl}


\bibitem[Karpukhin et~al\mbox{.}(2020)]%
        {searchdualencoder2020}
\bibfield{author}{\bibinfo{person}{Vladimir Karpukhin}, \bibinfo{person}{Barlas Oguz}, \bibinfo{person}{Sewon Min}, \bibinfo{person}{Patrick S.~H. Lewis}, \bibinfo{person}{Ledell Wu}, \bibinfo{person}{Sergey Edunov}, \bibinfo{person}{Danqi Chen}, {and} \bibinfo{person}{Wen{-}tau Yih}.} \bibinfo{year}{2020}\natexlab{}.
\newblock \showarticletitle{Dense Passage Retrieval for Open-Domain Question Answering}. In \bibinfo{booktitle}{\emph{Proceedings of the 2020 Conference on Empirical Methods in Natural Language Processing, {EMNLP} 2020, Online, November 16-20, 2020}}, \bibfield{editor}{\bibinfo{person}{Bonnie Webber}, \bibinfo{person}{Trevor Cohn}, \bibinfo{person}{Yulan He}, {and} \bibinfo{person}{Yang Liu}} (Eds.). \bibinfo{publisher}{Association for Computational Linguistics}, \bibinfo{pages}{6769--6781}.
\newblock
\urldef\tempurl%
\url{https://doi.org/10.18653/V1/2020.EMNLP-MAIN.550}
\showDOI{\tempurl}


\bibitem[Katsogiannis{-}Meimarakis and Koutrika(2021)]%
        {NLquerytutorial2021}
\bibfield{author}{\bibinfo{person}{George Katsogiannis{-}Meimarakis} {and} \bibinfo{person}{Georgia Koutrika}.} \bibinfo{year}{2021}\natexlab{}.
\newblock \showarticletitle{A Deep Dive into Deep Learning Approaches for Text-to-SQL Systems}. In \bibinfo{booktitle}{\emph{{SIGMOD} '21: International Conference on Management of Data, Virtual Event, China, June 20-25, 2021}}, \bibfield{editor}{\bibinfo{person}{Guoliang Li}, \bibinfo{person}{Zhanhuai Li}, \bibinfo{person}{Stratos Idreos}, {and} \bibinfo{person}{Divesh Srivastava}} (Eds.). \bibinfo{publisher}{{ACM}}, \bibinfo{pages}{2846--2851}.
\newblock
\urldef\tempurl%
\url{https://doi.org/10.1145/3448016.3457543}
\showDOI{\tempurl}


\bibitem[Katsogiannis{-}Meimarakis et~al\mbox{.}(2023)]%
        {NLquerytutorial2023}
\bibfield{author}{\bibinfo{person}{George Katsogiannis{-}Meimarakis}, \bibinfo{person}{Mike Xydas}, {and} \bibinfo{person}{Georgia Koutrika}.} \bibinfo{year}{2023}\natexlab{}.
\newblock \showarticletitle{Natural Language Interfaces for Databases with Deep Learning}.
\newblock \bibinfo{journal}{\emph{Proc. {VLDB} Endow.}} \bibinfo{volume}{16}, \bibinfo{number}{12} (\bibinfo{year}{2023}), \bibinfo{pages}{3878--3881}.
\newblock
\urldef\tempurl%
\url{https://doi.org/10.14778/3611540.3611575}
\showDOI{\tempurl}


\bibitem[Kaufmann et~al\mbox{.}(2023)]%
        {surveyRLHF2023}
\bibfield{author}{\bibinfo{person}{Timo Kaufmann}, \bibinfo{person}{Paul Weng}, \bibinfo{person}{Viktor Bengs}, {and} \bibinfo{person}{Eyke H{\"{u}}llermeier}.} \bibinfo{year}{2023}\natexlab{}.
\newblock \showarticletitle{A Survey of Reinforcement Learning from Human Feedback}.
\newblock \bibinfo{journal}{\emph{CoRR}}  \bibinfo{volume}{abs/2312.14925} (\bibinfo{year}{2023}).
\newblock
\urldef\tempurl%
\url{https://doi.org/10.48550/ARXIV.2312.14925}
\showDOI{\tempurl}
\showeprint[arXiv]{2312.14925}


\bibitem[Ke et~al\mbox{.}(2024)]%
        {RAGbridgingretrieverLLM2024}
\bibfield{author}{\bibinfo{person}{Zixuan Ke}, \bibinfo{person}{Weize Kong}, \bibinfo{person}{Cheng Li}, \bibinfo{person}{Mingyang Zhang}, \bibinfo{person}{Qiaozhu Mei}, {and} \bibinfo{person}{Michael Bendersky}.} \bibinfo{year}{2024}\natexlab{}.
\newblock \showarticletitle{Bridging the Preference Gap between Retrievers and LLMs}. In \bibinfo{booktitle}{\emph{Proceedings of the 62nd Annual Meeting of the Association for Computational Linguistics (Volume 1: Long Papers), {ACL} 2024, Bangkok, Thailand, August 11-16, 2024}}, \bibfield{editor}{\bibinfo{person}{Lun{-}Wei Ku}, \bibinfo{person}{Andre Martins}, {and} \bibinfo{person}{Vivek Srikumar}} (Eds.). \bibinfo{publisher}{Association for Computational Linguistics}, \bibinfo{pages}{10438--10451}.
\newblock
\urldef\tempurl%
\url{https://doi.org/10.18653/V1/2024.ACL-LONG.562}
\showDOI{\tempurl}


\bibitem[Khattab et~al\mbox{.}(2024)]%
        {DSPy}
\bibfield{author}{\bibinfo{person}{Omar Khattab}, \bibinfo{person}{Arnav Singhvi}, \bibinfo{person}{Paridhi Maheshwari}, \bibinfo{person}{Zhiyuan Zhang}, \bibinfo{person}{Keshav Santhanam}, \bibinfo{person}{Sri Vardhamanan}, \bibinfo{person}{Saiful Haq}, \bibinfo{person}{Ashutosh Sharma}, \bibinfo{person}{Thomas~T. Joshi}, \bibinfo{person}{Hanna Moazam}, \bibinfo{person}{Heather Miller}, \bibinfo{person}{Matei Zaharia}, {and} \bibinfo{person}{Christopher Potts}.} \bibinfo{year}{2024}\natexlab{}.
\newblock \showarticletitle{DSPy: Compiling Declarative Language Model Calls into State-of-the-Art Pipelines}. In \bibinfo{booktitle}{\emph{The Twelfth International Conference on Learning Representations, {ICLR} 2024, Vienna, Austria, May 7-11, 2024}}. \bibinfo{publisher}{OpenReview.net}.
\newblock
\urldef\tempurl%
\url{https://openreview.net/forum?id=sY5N0zY5Od}
\showURL{%
\tempurl}


\bibitem[Khattab and Zaharia(2020)]%
        {searchcolBERT2020}
\bibfield{author}{\bibinfo{person}{Omar Khattab} {and} \bibinfo{person}{Matei Zaharia}.} \bibinfo{year}{2020}\natexlab{}.
\newblock \showarticletitle{ColBERT: Efficient and Effective Passage Search via Contextualized Late Interaction over {BERT}}. In \bibinfo{booktitle}{\emph{Proceedings of the 43rd International {ACM} {SIGIR} conference on research and development in Information Retrieval, {SIGIR} 2020, Virtual Event, China, July 25-30, 2020}}, \bibfield{editor}{\bibinfo{person}{Jimmy~X. Huang}, \bibinfo{person}{Yi~Chang}, \bibinfo{person}{Xueqi Cheng}, \bibinfo{person}{Jaap Kamps}, \bibinfo{person}{Vanessa Murdock}, \bibinfo{person}{Ji{-}Rong Wen}, {and} \bibinfo{person}{Yiqun Liu}} (Eds.). \bibinfo{publisher}{{ACM}}, \bibinfo{pages}{39--48}.
\newblock
\urldef\tempurl%
\url{https://doi.org/10.1145/3397271.3401075}
\showDOI{\tempurl}


\bibitem[Kim et~al\mbox{.}(2024b)]%
        {sDPOdontuseyourdataallatonce}
\bibfield{author}{\bibinfo{person}{Dahyun Kim}, \bibinfo{person}{Yungi Kim}, \bibinfo{person}{Wonho Song}, \bibinfo{person}{Hyeonwoo Kim}, \bibinfo{person}{Yunsu Kim}, \bibinfo{person}{Sanghoon Kim}, {and} \bibinfo{person}{Chanjun Park}.} \bibinfo{year}{2024}\natexlab{b}.
\newblock \showarticletitle{sDPO: Don't Use Your Data All at Once}.
\newblock \bibinfo{journal}{\emph{CoRR}}  \bibinfo{volume}{abs/2403.19270} (\bibinfo{year}{2024}).
\newblock
\urldef\tempurl%
\url{https://doi.org/10.48550/ARXIV.2403.19270}
\showDOI{\tempurl}
\showeprint[arXiv]{2403.19270}


\bibitem[Kim et~al\mbox{.}(2024a)]%
        {Ours}
\bibfield{author}{\bibinfo{person}{Kyoungmin Kim}, \bibinfo{person}{Kijae Hong}, \bibinfo{person}{Caglar Gulcehre}, {and} \bibinfo{person}{Anastasia Ailamaki}.} \bibinfo{year}{2024}\natexlab{a}.
\newblock \showarticletitle{The Effect of Scheduling and Preemption on the Efficiency of LLM Inference Serving}.
\newblock \bibinfo{journal}{\emph{arXiv preprint arXiv:2411.07447}} (\bibinfo{year}{2024}).
\newblock


\bibitem[Kim et~al\mbox{.}(2022)]%
        {LearnedCE}
\bibfield{author}{\bibinfo{person}{Kyoungmin Kim}, \bibinfo{person}{Jisung Jung}, \bibinfo{person}{In Seo}, \bibinfo{person}{Wook{-}Shin Han}, \bibinfo{person}{Kangwoo Choi}, {and} \bibinfo{person}{Jaehyok Chong}.} \bibinfo{year}{2022}\natexlab{}.
\newblock \showarticletitle{Learned Cardinality Estimation: An In-depth Study}. In \bibinfo{booktitle}{\emph{{SIGMOD} '22: International Conference on Management of Data, Philadelphia, PA, USA, June 12 - 17, 2022}}, \bibfield{editor}{\bibinfo{person}{Zachary~G. Ives}, \bibinfo{person}{Angela Bonifati}, {and} \bibinfo{person}{Amr~El Abbadi}} (Eds.). \bibinfo{publisher}{{ACM}}, \bibinfo{pages}{1214--1227}.
\newblock
\urldef\tempurl%
\url{https://doi.org/10.1145/3514221.3526154}
\showDOI{\tempurl}


\bibitem[Kim et~al\mbox{.}(2024c)]%
        {ASM}
\bibfield{author}{\bibinfo{person}{Kyoungmin Kim}, \bibinfo{person}{Sangoh Lee}, \bibinfo{person}{Injung Kim}, {and} \bibinfo{person}{Wook{-}Shin Han}.} \bibinfo{year}{2024}\natexlab{c}.
\newblock \showarticletitle{{ASM:} Harmonizing Autoregressive Model, Sampling, and Multi-dimensional Statistics Merging for Cardinality Estimation}.
\newblock \bibinfo{journal}{\emph{Proc. {ACM} Manag. Data}} \bibinfo{volume}{2}, \bibinfo{number}{1} (\bibinfo{year}{2024}), \bibinfo{pages}{45:1--45:27}.
\newblock
\urldef\tempurl%
\url{https://doi.org/10.1145/3639300}
\showDOI{\tempurl}


\bibitem[Kim et~al\mbox{.}(2024d)]%
        {evaluatingLLMsasdatagenerators2024}
\bibfield{author}{\bibinfo{person}{Seungone Kim}, \bibinfo{person}{Juyoung Suk}, \bibinfo{person}{Xiang Yue}, \bibinfo{person}{Vijay Viswanathan}, \bibinfo{person}{Seongyun Lee}, \bibinfo{person}{Yizhong Wang}, \bibinfo{person}{Kiril Gashteovski}, \bibinfo{person}{Carolin Lawrence}, \bibinfo{person}{Sean Welleck}, {and} \bibinfo{person}{Graham Neubig}.} \bibinfo{year}{2024}\natexlab{d}.
\newblock \showarticletitle{Evaluating Language Models as Synthetic Data Generators}.
\newblock \bibinfo{journal}{\emph{arXiv preprint arXiv:2412.03679}} (\bibinfo{year}{2024}).
\newblock


\bibitem[Kocetkov et~al\mbox{.}(2023)]%
        {licensedsourcecodedataset2023}
\bibfield{author}{\bibinfo{person}{Denis Kocetkov}, \bibinfo{person}{Raymond Li}, \bibinfo{person}{Loubna~Ben Allal}, \bibinfo{person}{Jia Li}, \bibinfo{person}{Chenghao Mou}, \bibinfo{person}{Yacine Jernite}, \bibinfo{person}{Margaret Mitchell}, \bibinfo{person}{Carlos~Mu{\~{n}}oz Ferrandis}, \bibinfo{person}{Sean Hughes}, \bibinfo{person}{Thomas Wolf}, \bibinfo{person}{Dzmitry Bahdanau}, \bibinfo{person}{Leandro von Werra}, {and} \bibinfo{person}{Harm de Vries}.} \bibinfo{year}{2023}\natexlab{}.
\newblock \showarticletitle{The Stack: 3 {TB} of permissively licensed source code}.
\newblock \bibinfo{journal}{\emph{Trans. Mach. Learn. Res.}}  \bibinfo{volume}{2023} (\bibinfo{year}{2023}).
\newblock
\urldef\tempurl%
\url{https://openreview.net/forum?id=pxpbTdUEpD}
\showURL{%
\tempurl}


\bibitem[Kojima et~al\mbox{.}(2022)]%
        {zerocot2022}
\bibfield{author}{\bibinfo{person}{Takeshi Kojima}, \bibinfo{person}{Shixiang~Shane Gu}, \bibinfo{person}{Machel Reid}, \bibinfo{person}{Yutaka Matsuo}, {and} \bibinfo{person}{Yusuke Iwasawa}.} \bibinfo{year}{2022}\natexlab{}.
\newblock \showarticletitle{Large Language Models are Zero-Shot Reasoners}. In \bibinfo{booktitle}{\emph{Advances in Neural Information Processing Systems 35: Annual Conference on Neural Information Processing Systems 2022, NeurIPS 2022, New Orleans, LA, USA, November 28 - December 9, 2022}}, \bibfield{editor}{\bibinfo{person}{Sanmi Koyejo}, \bibinfo{person}{S.~Mohamed}, \bibinfo{person}{A.~Agarwal}, \bibinfo{person}{Danielle Belgrave}, \bibinfo{person}{K.~Cho}, {and} \bibinfo{person}{A.~Oh}} (Eds.).
\newblock
\urldef\tempurl%
\url{http://papers.nips.cc/paper\_files/paper/2022/hash/8bb0d291acd4acf06ef112099c16f326-Abstract-Conference.html}
\showURL{%
\tempurl}


\bibitem[Kong et~al\mbox{.}(2024)]%
        {DSMR2024}
\bibfield{author}{\bibinfo{person}{Juri Kong}, \bibinfo{person}{Hong Liang}, \bibinfo{person}{Yuan Zhang}, \bibinfo{person}{Hongxiang Li}, \bibinfo{person}{Pengcheng Shen}, {and} \bibinfo{person}{Fang Lu}.} \bibinfo{year}{2024}\natexlab{}.
\newblock \showarticletitle{Dynamic semantic memory retention in large language models: An exploration of spontaneous retrieval mechanisms}.
\newblock  (\bibinfo{year}{2024}).
\newblock


\bibitem[Kwon et~al\mbox{.}(2023)]%
        {vLLM}
\bibfield{author}{\bibinfo{person}{Woosuk Kwon}, \bibinfo{person}{Zhuohan Li}, \bibinfo{person}{Siyuan Zhuang}, \bibinfo{person}{Ying Sheng}, \bibinfo{person}{Lianmin Zheng}, \bibinfo{person}{Cody~Hao Yu}, \bibinfo{person}{Joseph Gonzalez}, \bibinfo{person}{Hao Zhang}, {and} \bibinfo{person}{Ion Stoica}.} \bibinfo{year}{2023}\natexlab{}.
\newblock \showarticletitle{Efficient Memory Management for Large Language Model Serving with PagedAttention}. In \bibinfo{booktitle}{\emph{Proceedings of the 29th Symposium on Operating Systems Principles, {SOSP} 2023, Koblenz, Germany, October 23-26, 2023}}, \bibfield{editor}{\bibinfo{person}{Jason Flinn}, \bibinfo{person}{Margo~I. Seltzer}, \bibinfo{person}{Peter Druschel}, \bibinfo{person}{Antoine Kaufmann}, {and} \bibinfo{person}{Jonathan Mace}} (Eds.). \bibinfo{publisher}{{ACM}}, \bibinfo{pages}{611--626}.
\newblock
\urldef\tempurl%
\url{https://doi.org/10.1145/3600006.3613165}
\showDOI{\tempurl}


\bibitem[Lang et~al\mbox{.}(2024)]%
        {modelquantizesurvey2024}
\bibfield{author}{\bibinfo{person}{Jiedong Lang}, \bibinfo{person}{Zhehao Guo}, {and} \bibinfo{person}{Shuyu Huang}.} \bibinfo{year}{2024}\natexlab{}.
\newblock \showarticletitle{A Comprehensive Study on Quantization Techniques for Large Language Models}.
\newblock \bibinfo{journal}{\emph{arXiv preprint arXiv:2411.02530}} (\bibinfo{year}{2024}).
\newblock


\bibitem[Lee et~al\mbox{.}(2024c)]%
        {CXLdatabasekernels}
\bibfield{author}{\bibinfo{person}{Sangjin Lee}, \bibinfo{person}{Alberto Lerner}, \bibinfo{person}{Philippe Bonnet}, {and} \bibinfo{person}{Philippe Cudr{\'{e}}{-}Mauroux}.} \bibinfo{year}{2024}\natexlab{c}.
\newblock \showarticletitle{Database Kernels: Seamless Integration of Database Systems and Fast Storage via {CXL}}. In \bibinfo{booktitle}{\emph{14th Conference on Innovative Data Systems Research, {CIDR} 2024, Chaminade, HI, USA, January 14-17, 2024}}. \bibinfo{publisher}{www.cidrdb.org}.
\newblock
\urldef\tempurl%
\url{https://www.cidrdb.org/cidr2024/papers/p43-lee.pdf}
\showURL{%
\tempurl}


\bibitem[Lee et~al\mbox{.}(2024b)]%
        {InfiniGen}
\bibfield{author}{\bibinfo{person}{Wonbeom Lee}, \bibinfo{person}{Jungi Lee}, \bibinfo{person}{Junghwan Seo}, {and} \bibinfo{person}{Jaewoong Sim}.} \bibinfo{year}{2024}\natexlab{b}.
\newblock \showarticletitle{InfiniGen: Efficient Generative Inference of Large Language Models with Dynamic {KV} Cache Management}. In \bibinfo{booktitle}{\emph{18th {USENIX} Symposium on Operating Systems Design and Implementation, {OSDI} 2024, Santa Clara, CA, USA, July 10-12, 2024}}, \bibfield{editor}{\bibinfo{person}{Ada Gavrilovska} {and} \bibinfo{person}{Douglas~B. Terry}} (Eds.). \bibinfo{publisher}{{USENIX} Association}, \bibinfo{pages}{155--172}.
\newblock
\urldef\tempurl%
\url{https://www.usenix.org/conference/osdi24/presentation/lee}
\showURL{%
\tempurl}


\bibitem[Lee et~al\mbox{.}(2024a)]%
        {tablepromptinglearningtoreduce}
\bibfield{author}{\bibinfo{person}{Younghun Lee}, \bibinfo{person}{Sungchul Kim}, \bibinfo{person}{Tong Yu}, \bibinfo{person}{Ryan~A. Rossi}, {and} \bibinfo{person}{Xiang Chen}.} \bibinfo{year}{2024}\natexlab{a}.
\newblock \showarticletitle{Learning to Reduce: Optimal Representations of Structured Data in Prompting Large Language Models}.
\newblock \bibinfo{journal}{\emph{CoRR}}  \bibinfo{volume}{abs/2402.14195} (\bibinfo{year}{2024}).
\newblock
\urldef\tempurl%
\url{https://doi.org/10.48550/ARXIV.2402.14195}
\showDOI{\tempurl}
\showeprint[arXiv]{2402.14195}


\bibitem[Lei et~al\mbox{.}(2024)]%
        {spider2}
\bibfield{author}{\bibinfo{person}{Fangyu Lei}, \bibinfo{person}{Jixuan Chen}, \bibinfo{person}{Yuxiao Ye}, \bibinfo{person}{Ruisheng Cao}, \bibinfo{person}{Dongchan Shin}, \bibinfo{person}{Hongjin Su}, \bibinfo{person}{Zhaoqing Suo}, \bibinfo{person}{Hongcheng Gao}, \bibinfo{person}{Wenjing Hu}, \bibinfo{person}{Pengcheng Yin}, {et~al\mbox{.}}} \bibinfo{year}{2024}\natexlab{}.
\newblock \showarticletitle{Spider 2.0: Evaluating language models on real-world enterprise text-to-sql workflows}.
\newblock \bibinfo{journal}{\emph{arXiv preprint arXiv:2411.07763}} (\bibinfo{year}{2024}).
\newblock


\bibitem[Lester et~al\mbox{.}(2021)]%
        {prompttuning2021}
\bibfield{author}{\bibinfo{person}{Brian Lester}, \bibinfo{person}{Rami Al{-}Rfou}, {and} \bibinfo{person}{Noah Constant}.} \bibinfo{year}{2021}\natexlab{}.
\newblock \showarticletitle{The Power of Scale for Parameter-Efficient Prompt Tuning}. In \bibinfo{booktitle}{\emph{Proceedings of the 2021 Conference on Empirical Methods in Natural Language Processing, {EMNLP} 2021, Virtual Event / Punta Cana, Dominican Republic, 7-11 November, 2021}}, \bibfield{editor}{\bibinfo{person}{Marie{-}Francine Moens}, \bibinfo{person}{Xuanjing Huang}, \bibinfo{person}{Lucia Specia}, {and} \bibinfo{person}{Scott~Wen{-}tau Yih}} (Eds.). \bibinfo{publisher}{Association for Computational Linguistics}, \bibinfo{pages}{3045--3059}.
\newblock
\urldef\tempurl%
\url{https://doi.org/10.18653/V1/2021.EMNLP-MAIN.243}
\showDOI{\tempurl}


\bibitem[Leviathan et~al\mbox{.}(2023)]%
        {SpeculativeDecoding}
\bibfield{author}{\bibinfo{person}{Yaniv Leviathan}, \bibinfo{person}{Matan Kalman}, {and} \bibinfo{person}{Yossi Matias}.} \bibinfo{year}{2023}\natexlab{}.
\newblock \showarticletitle{Fast Inference from Transformers via Speculative Decoding}. In \bibinfo{booktitle}{\emph{International Conference on Machine Learning, {ICML} 2023, 23-29 July 2023, Honolulu, Hawaii, {USA}}} \emph{(\bibinfo{series}{Proceedings of Machine Learning Research}, Vol.~\bibinfo{volume}{202})}, \bibfield{editor}{\bibinfo{person}{Andreas Krause}, \bibinfo{person}{Emma Brunskill}, \bibinfo{person}{Kyunghyun Cho}, \bibinfo{person}{Barbara Engelhardt}, \bibinfo{person}{Sivan Sabato}, {and} \bibinfo{person}{Jonathan Scarlett}} (Eds.). \bibinfo{publisher}{{PMLR}}, \bibinfo{pages}{19274--19286}.
\newblock
\urldef\tempurl%
\url{https://proceedings.mlr.press/v202/leviathan23a.html}
\showURL{%
\tempurl}


\bibitem[Lewis et~al\mbox{.}(2020)]%
        {RAG}
\bibfield{author}{\bibinfo{person}{Patrick S.~H. Lewis}, \bibinfo{person}{Ethan Perez}, \bibinfo{person}{Aleksandra Piktus}, \bibinfo{person}{Fabio Petroni}, \bibinfo{person}{Vladimir Karpukhin}, \bibinfo{person}{Naman Goyal}, \bibinfo{person}{Heinrich K{\"{u}}ttler}, \bibinfo{person}{Mike Lewis}, \bibinfo{person}{Wen{-}tau Yih}, \bibinfo{person}{Tim Rockt{\"{a}}schel}, \bibinfo{person}{Sebastian Riedel}, {and} \bibinfo{person}{Douwe Kiela}.} \bibinfo{year}{2020}\natexlab{}.
\newblock \showarticletitle{Retrieval-Augmented Generation for Knowledge-Intensive {NLP} Tasks}. In \bibinfo{booktitle}{\emph{Advances in Neural Information Processing Systems 33: Annual Conference on Neural Information Processing Systems 2020, NeurIPS 2020, December 6-12, 2020, virtual}}, \bibfield{editor}{\bibinfo{person}{Hugo Larochelle}, \bibinfo{person}{Marc'Aurelio Ranzato}, \bibinfo{person}{Raia Hadsell}, \bibinfo{person}{Maria{-}Florina Balcan}, {and} \bibinfo{person}{Hsuan{-}Tien Lin}} (Eds.).
\newblock
\urldef\tempurl%
\url{https://proceedings.neurips.cc/paper/2020/hash/6b493230205f780e1bc26945df7481e5-Abstract.html}
\showURL{%
\tempurl}


\bibitem[Li et~al\mbox{.}(2024f)]%
        {loramixrora2024}
\bibfield{author}{\bibinfo{person}{Dengchun Li}, \bibinfo{person}{Yingzi Ma}, \bibinfo{person}{Naizheng Wang}, \bibinfo{person}{Zhiyuan Cheng}, \bibinfo{person}{Lei Duan}, \bibinfo{person}{Jie Zuo}, \bibinfo{person}{Cal Yang}, {and} \bibinfo{person}{Mingjie Tang}.} \bibinfo{year}{2024}\natexlab{f}.
\newblock \showarticletitle{MixLoRA: Enhancing Large Language Models Fine-Tuning with LoRA based Mixture of Experts}.
\newblock \bibinfo{journal}{\emph{CoRR}}  \bibinfo{volume}{abs/2404.15159} (\bibinfo{year}{2024}).
\newblock
\urldef\tempurl%
\url{https://doi.org/10.48550/ARXIV.2404.15159}
\showDOI{\tempurl}
\showeprint[arXiv]{2404.15159}


\bibitem[Li et~al\mbox{.}(2021c)]%
        {AItutorial1}
\bibfield{author}{\bibinfo{person}{Guoliang Li}, \bibinfo{person}{Xuanhe Zhou}, {and} \bibinfo{person}{Lei Cao}.} \bibinfo{year}{2021}\natexlab{c}.
\newblock \showarticletitle{{AI} Meets Database: {AI4DB} and {DB4AI}}. In \bibinfo{booktitle}{\emph{{SIGMOD} '21: International Conference on Management of Data, Virtual Event, China, June 20-25, 2021}}, \bibfield{editor}{\bibinfo{person}{Guoliang Li}, \bibinfo{person}{Zhanhuai Li}, \bibinfo{person}{Stratos Idreos}, {and} \bibinfo{person}{Divesh Srivastava}} (Eds.). \bibinfo{publisher}{{ACM}}, \bibinfo{pages}{2859--2866}.
\newblock
\urldef\tempurl%
\url{https://doi.org/10.1145/3448016.3457542}
\showDOI{\tempurl}


\bibitem[Li et~al\mbox{.}(2024b)]%
        {nextgenerationdata2024}
\bibfield{author}{\bibinfo{person}{Jeffrey Li}, \bibinfo{person}{Alex Fang}, \bibinfo{person}{Georgios Smyrnis}, \bibinfo{person}{Maor Ivgi}, \bibinfo{person}{Matt Jordan}, \bibinfo{person}{Samir~Yitzhak Gadre}, \bibinfo{person}{Hritik Bansal}, \bibinfo{person}{Etash~Kumar Guha}, \bibinfo{person}{Sedrick Keh}, \bibinfo{person}{Kushal Arora}, \bibinfo{person}{Saurabh Garg}, \bibinfo{person}{Rui Xin}, \bibinfo{person}{Niklas Muennighoff}, \bibinfo{person}{Reinhard Heckel}, \bibinfo{person}{Jean Mercat}, \bibinfo{person}{Mayee Chen}, \bibinfo{person}{Suchin Gururangan}, \bibinfo{person}{Mitchell Wortsman}, \bibinfo{person}{Alon Albalak}, \bibinfo{person}{Yonatan Bitton}, \bibinfo{person}{Marianna Nezhurina}, \bibinfo{person}{Amro Abbas}, \bibinfo{person}{Cheng{-}Yu Hsieh}, \bibinfo{person}{Dhruba Ghosh}, \bibinfo{person}{Josh Gardner}, \bibinfo{person}{Maciej Kilian}, \bibinfo{person}{Hanlin Zhang}, \bibinfo{person}{Rulin Shao}, \bibinfo{person}{Sarah~M. Pratt}, \bibinfo{person}{Sunny Sanyal},
  \bibinfo{person}{Gabriel Ilharco}, \bibinfo{person}{Giannis Daras}, \bibinfo{person}{Kalyani Marathe}, \bibinfo{person}{Aaron Gokaslan}, \bibinfo{person}{Jieyu Zhang}, \bibinfo{person}{Khyathi~Raghavi Chandu}, \bibinfo{person}{Thao Nguyen}, \bibinfo{person}{Igor Vasiljevic}, \bibinfo{person}{Sham~M. Kakade}, \bibinfo{person}{Shuran Song}, \bibinfo{person}{Sujay Sanghavi}, \bibinfo{person}{Fartash Faghri}, \bibinfo{person}{Sewoong Oh}, \bibinfo{person}{Luke Zettlemoyer}, \bibinfo{person}{Kyle Lo}, \bibinfo{person}{Alaaeldin El{-}Nouby}, \bibinfo{person}{Hadi Pouransari}, \bibinfo{person}{Alexander Toshev}, \bibinfo{person}{Stephanie Wang}, \bibinfo{person}{Dirk Groeneveld}, \bibinfo{person}{Luca Soldaini}, \bibinfo{person}{Pang~Wei Koh}, \bibinfo{person}{Jenia Jitsev}, \bibinfo{person}{Thomas Kollar}, \bibinfo{person}{Alexandros~G. Dimakis}, \bibinfo{person}{Yair Carmon}, \bibinfo{person}{Achal Dave}, \bibinfo{person}{Ludwig Schmidt}, {and} \bibinfo{person}{Vaishaal Shankar}.}
  \bibinfo{year}{2024}\natexlab{b}.
\newblock \showarticletitle{DataComp-LM: In search of the next generation of training sets for language models}.
\newblock \bibinfo{journal}{\emph{CoRR}}  \bibinfo{volume}{abs/2406.11794} (\bibinfo{year}{2024}).
\newblock
\urldef\tempurl%
\url{https://doi.org/10.48550/ARXIV.2406.11794}
\showDOI{\tempurl}
\showeprint[arXiv]{2406.11794}


\bibitem[Li et~al\mbox{.}(2023a)]%
        {tableRAGtext2sqlbenchmark2023}
\bibfield{author}{\bibinfo{person}{Jinyang Li}, \bibinfo{person}{Binyuan Hui}, \bibinfo{person}{Ge Qu}, \bibinfo{person}{Jiaxi Yang}, \bibinfo{person}{Binhua Li}, \bibinfo{person}{Bowen Li}, \bibinfo{person}{Bailin Wang}, \bibinfo{person}{Bowen Qin}, \bibinfo{person}{Ruiying Geng}, \bibinfo{person}{Nan Huo}, \bibinfo{person}{Xuanhe Zhou}, \bibinfo{person}{Chenhao Ma}, \bibinfo{person}{Guoliang Li}, \bibinfo{person}{Kevin~Chen{-}Chuan Chang}, \bibinfo{person}{Fei Huang}, \bibinfo{person}{Reynold Cheng}, {and} \bibinfo{person}{Yongbin Li}.} \bibinfo{year}{2023}\natexlab{a}.
\newblock \showarticletitle{Can {LLM} Already Serve as {A} Database Interface? {A} BIg Bench for Large-Scale Database Grounded Text-to-SQLs}. In \bibinfo{booktitle}{\emph{Advances in Neural Information Processing Systems 36: Annual Conference on Neural Information Processing Systems 2023, NeurIPS 2023, New Orleans, LA, USA, December 10 - 16, 2023}}, \bibfield{editor}{\bibinfo{person}{Alice Oh}, \bibinfo{person}{Tristan Naumann}, \bibinfo{person}{Amir Globerson}, \bibinfo{person}{Kate Saenko}, \bibinfo{person}{Moritz Hardt}, {and} \bibinfo{person}{Sergey Levine}} (Eds.).
\newblock
\urldef\tempurl%
\url{http://papers.nips.cc/paper\_files/paper/2023/hash/83fc8fab1710363050bbd1d4b8cc0021-Abstract-Datasets\_and\_Benchmarks.html}
\showURL{%
\tempurl}


\bibitem[Li et~al\mbox{.}(2024g)]%
        {reflectionbench2024}
\bibfield{author}{\bibinfo{person}{Lingyu Li}, \bibinfo{person}{Yixu Wang}, \bibinfo{person}{Haiquan Zhao}, \bibinfo{person}{Shuqi Kong}, \bibinfo{person}{Yan Teng}, \bibinfo{person}{Chunbo Li}, {and} \bibinfo{person}{Yingchun Wang}.} \bibinfo{year}{2024}\natexlab{g}.
\newblock \showarticletitle{Reflection-Bench: probing AI intelligence with reflection}.
\newblock \bibinfo{journal}{\emph{arXiv preprint arXiv:2410.16270}} (\bibinfo{year}{2024}).
\newblock


\bibitem[Li et~al\mbox{.}(2024c)]%
        {TableGPTfindtunedGPTfortabletasks2024}
\bibfield{author}{\bibinfo{person}{Peng Li}, \bibinfo{person}{Yeye He}, \bibinfo{person}{Dror Yashar}, \bibinfo{person}{Weiwei Cui}, \bibinfo{person}{Song Ge}, \bibinfo{person}{Haidong Zhang}, \bibinfo{person}{Danielle~Rifinski Fainman}, \bibinfo{person}{Dongmei Zhang}, {and} \bibinfo{person}{Surajit Chaudhuri}.} \bibinfo{year}{2024}\natexlab{c}.
\newblock \showarticletitle{Table-GPT: Table Fine-tuned {GPT} for Diverse Table Tasks}.
\newblock \bibinfo{journal}{\emph{Proc. {ACM} Manag. Data}} \bibinfo{volume}{2}, \bibinfo{number}{3} (\bibinfo{year}{2024}), \bibinfo{pages}{176}.
\newblock
\urldef\tempurl%
\url{https://doi.org/10.1145/3654979}
\showDOI{\tempurl}


\bibitem[Li et~al\mbox{.}(2024d)]%
        {FlexNN}
\bibfield{author}{\bibinfo{person}{Xiangyu Li}, \bibinfo{person}{Yuanchun Li}, \bibinfo{person}{Yuanzhe Li}, \bibinfo{person}{Ting Cao}, {and} \bibinfo{person}{Yunxin Liu}.} \bibinfo{year}{2024}\natexlab{d}.
\newblock \showarticletitle{FlexNN: Efficient and Adaptive {DNN} Inference on Memory-Constrained Edge Devices}. In \bibinfo{booktitle}{\emph{Proceedings of the 30th Annual International Conference on Mobile Computing and Networking, {ACM} MobiCom 2024, Washington D.C., DC, USA, November 18-22, 2024}}, \bibfield{editor}{\bibinfo{person}{Weisong Shi}, \bibinfo{person}{Deepak Ganesan}, {and} \bibinfo{person}{Nicholas~D. Lane}} (Eds.). \bibinfo{publisher}{{ACM}}, \bibinfo{pages}{709--723}.
\newblock
\urldef\tempurl%
\url{https://doi.org/10.1145/3636534.3649391}
\showDOI{\tempurl}


\bibitem[Li and Liang(2021)]%
        {prefixtuning2021}
\bibfield{author}{\bibinfo{person}{Xiang~Lisa Li} {and} \bibinfo{person}{Percy Liang}.} \bibinfo{year}{2021}\natexlab{}.
\newblock \showarticletitle{Prefix-Tuning: Optimizing Continuous Prompts for Generation}. In \bibinfo{booktitle}{\emph{Proceedings of the 59th Annual Meeting of the Association for Computational Linguistics and the 11th International Joint Conference on Natural Language Processing, {ACL/IJCNLP} 2021, (Volume 1: Long Papers), Virtual Event, August 1-6, 2021}}, \bibfield{editor}{\bibinfo{person}{Chengqing Zong}, \bibinfo{person}{Fei Xia}, \bibinfo{person}{Wenjie Li}, {and} \bibinfo{person}{Roberto Navigli}} (Eds.). \bibinfo{publisher}{Association for Computational Linguistics}, \bibinfo{pages}{4582--4597}.
\newblock
\urldef\tempurl%
\url{https://doi.org/10.18653/V1/2021.ACL-LONG.353}
\showDOI{\tempurl}


\bibitem[Li et~al\mbox{.}(2024a)]%
        {llmfordataagumentation2024}
\bibfield{author}{\bibinfo{person}{Yichuan Li}, \bibinfo{person}{Kaize Ding}, \bibinfo{person}{Jianling Wang}, {and} \bibinfo{person}{Kyumin Lee}.} \bibinfo{year}{2024}\natexlab{a}.
\newblock \showarticletitle{Empowering Large Language Models for Textual Data Augmentation}. In \bibinfo{booktitle}{\emph{Findings of the Association for Computational Linguistics, {ACL} 2024, Bangkok, Thailand and virtual meeting, August 11-16, 2024}}, \bibfield{editor}{\bibinfo{person}{Lun{-}Wei Ku}, \bibinfo{person}{Andre Martins}, {and} \bibinfo{person}{Vivek Srikumar}} (Eds.). \bibinfo{publisher}{Association for Computational Linguistics}, \bibinfo{pages}{12734--12751}.
\newblock
\urldef\tempurl%
\url{https://doi.org/10.18653/V1/2024.FINDINGS-ACL.756}
\showDOI{\tempurl}


\bibitem[Li et~al\mbox{.}(2021a)]%
        {modelquantizebenchmark2021}
\bibfield{author}{\bibinfo{person}{Yuhang Li}, \bibinfo{person}{Mingzhu Shen}, \bibinfo{person}{Jian Ma}, \bibinfo{person}{Yan Ren}, \bibinfo{person}{Mingxin Zhao}, \bibinfo{person}{Qi Zhang}, \bibinfo{person}{Ruihao Gong}, \bibinfo{person}{Fengwei Yu}, {and} \bibinfo{person}{Junjie Yan}.} \bibinfo{year}{2021}\natexlab{a}.
\newblock \showarticletitle{MQBench: Towards Reproducible and Deployable Model Quantization Benchmark}. In \bibinfo{booktitle}{\emph{Proceedings of the Neural Information Processing Systems Track on Datasets and Benchmarks 1, NeurIPS Datasets and Benchmarks 2021, December 2021, virtual}}, \bibfield{editor}{\bibinfo{person}{Joaquin Vanschoren} {and} \bibinfo{person}{Sai{-}Kit Yeung}} (Eds.).
\newblock
\urldef\tempurl%
\url{https://datasets-benchmarks-proceedings.neurips.cc/paper/2021/hash/c20ad4d76fe97759aa27a0c99bff6710-Abstract-round1.html}
\showURL{%
\tempurl}


\bibitem[Li et~al\mbox{.}(2021b)]%
        {AItutorial2}
\bibfield{author}{\bibinfo{person}{Yuliang Li}, \bibinfo{person}{Xiaolan Wang}, \bibinfo{person}{Zhengjie Miao}, {and} \bibinfo{person}{Wang{-}Chiew Tan}.} \bibinfo{year}{2021}\natexlab{b}.
\newblock \showarticletitle{Data Augmentation for ML-driven Data Preparation and Integration}.
\newblock \bibinfo{journal}{\emph{Proc. {VLDB} Endow.}} \bibinfo{volume}{14}, \bibinfo{number}{12} (\bibinfo{year}{2021}), \bibinfo{pages}{3182--3185}.
\newblock
\urldef\tempurl%
\url{https://doi.org/10.14778/3476311.3476403}
\showDOI{\tempurl}


\bibitem[Li et~al\mbox{.}(2024e)]%
        {RAGorlongcontext2024}
\bibfield{author}{\bibinfo{person}{Zhuowan Li}, \bibinfo{person}{Cheng Li}, \bibinfo{person}{Mingyang Zhang}, \bibinfo{person}{Qiaozhu Mei}, {and} \bibinfo{person}{Michael Bendersky}.} \bibinfo{year}{2024}\natexlab{e}.
\newblock \showarticletitle{Retrieval Augmented Generation or Long-Context LLMs? {A} Comprehensive Study and Hybrid Approach}. In \bibinfo{booktitle}{\emph{Proceedings of the 2024 Conference on Empirical Methods in Natural Language Processing: {EMNLP} 2024 - Industry Track, Miami, Florida, USA, November 12-16, 2024}}, \bibfield{editor}{\bibinfo{person}{Franck Dernoncourt}, \bibinfo{person}{Daniel Preotiuc{-}Pietro}, {and} \bibinfo{person}{Anastasia Shimorina}} (Eds.). \bibinfo{publisher}{Association for Computational Linguistics}, \bibinfo{pages}{881--893}.
\newblock
\urldef\tempurl%
\url{https://aclanthology.org/2024.emnlp-industry.66}
\showURL{%
\tempurl}


\bibitem[Li et~al\mbox{.}(2023b)]%
        {offlineRLHF2023}
\bibfield{author}{\bibinfo{person}{Zihao Li}, \bibinfo{person}{Zhuoran Yang}, {and} \bibinfo{person}{Mengdi Wang}.} \bibinfo{year}{2023}\natexlab{b}.
\newblock \showarticletitle{Reinforcement Learning with Human Feedback: Learning Dynamic Choices via Pessimism}.
\newblock \bibinfo{journal}{\emph{CoRR}}  \bibinfo{volume}{abs/2305.18438} (\bibinfo{year}{2023}).
\newblock
\urldef\tempurl%
\url{https://doi.org/10.48550/ARXIV.2305.18438}
\showDOI{\tempurl}
\showeprint[arXiv]{2305.18438}


\bibitem[Li et~al\mbox{.}(2024h)]%
        {LLMQueryOptLLMR2}
\bibfield{author}{\bibinfo{person}{Zhaodonghui Li}, \bibinfo{person}{Haitao Yuan}, \bibinfo{person}{Huiming Wang}, \bibinfo{person}{Gao Cong}, {and} \bibinfo{person}{Lidong Bing}.} \bibinfo{year}{2024}\natexlab{h}.
\newblock \showarticletitle{{LLM-R2:} {A} Large Language Model Enhanced Rule-based Rewrite System for Boosting Query Efficiency}.
\newblock \bibinfo{journal}{\emph{CoRR}}  \bibinfo{volume}{abs/2404.12872} (\bibinfo{year}{2024}).
\newblock
\urldef\tempurl%
\url{https://doi.org/10.48550/ARXIV.2404.12872}
\showDOI{\tempurl}
\showeprint[arXiv]{2404.12872}


\bibitem[Liang et~al\mbox{.}(2024)]%
        {graphRAGsurvey2024}
\bibfield{author}{\bibinfo{person}{Ke Liang}, \bibinfo{person}{Lingyuan Meng}, \bibinfo{person}{Meng Liu}, \bibinfo{person}{Yue Liu}, \bibinfo{person}{Wenxuan Tu}, \bibinfo{person}{Siwei Wang}, \bibinfo{person}{Sihang Zhou}, \bibinfo{person}{Xinwang Liu}, \bibinfo{person}{Fuchun Sun}, {and} \bibinfo{person}{Kunlun He}.} \bibinfo{year}{2024}\natexlab{}.
\newblock \showarticletitle{A Survey of Knowledge Graph Reasoning on Graph Types: Static, Dynamic, and Multi-Modal}.
\newblock \bibinfo{journal}{\emph{{IEEE} Trans. Pattern Anal. Mach. Intell.}} \bibinfo{volume}{46}, \bibinfo{number}{12} (\bibinfo{year}{2024}), \bibinfo{pages}{9456--9478}.
\newblock
\urldef\tempurl%
\url{https://doi.org/10.1109/TPAMI.2024.3417451}
\showDOI{\tempurl}


\bibitem[Lie(2023)]%
        {cerebrasHWSWcodesign2023}
\bibfield{author}{\bibinfo{person}{Sean Lie}.} \bibinfo{year}{2023}\natexlab{}.
\newblock \showarticletitle{Cerebras Architecture Deep Dive: First Look Inside the Hardware/Software Co-Design for Deep Learning}.
\newblock \bibinfo{journal}{\emph{{IEEE} Micro}} \bibinfo{volume}{43}, \bibinfo{number}{3} (\bibinfo{year}{2023}), \bibinfo{pages}{18--30}.
\newblock
\urldef\tempurl%
\url{https://doi.org/10.1109/MM.2023.3256384}
\showDOI{\tempurl}


\bibitem[Lieber et~al\mbox{.}(2024)]%
        {SSMjamba2024}
\bibfield{author}{\bibinfo{person}{Opher Lieber}, \bibinfo{person}{Barak Lenz}, \bibinfo{person}{Hofit Bata}, \bibinfo{person}{Gal Cohen}, \bibinfo{person}{Jhonathan Osin}, \bibinfo{person}{Itay Dalmedigos}, \bibinfo{person}{Erez Safahi}, \bibinfo{person}{Shaked Meirom}, \bibinfo{person}{Yonatan Belinkov}, \bibinfo{person}{Shai Shalev{-}Shwartz}, \bibinfo{person}{Omri Abend}, \bibinfo{person}{Raz Alon}, \bibinfo{person}{Tomer Asida}, \bibinfo{person}{Amir Bergman}, \bibinfo{person}{Roman Glozman}, \bibinfo{person}{Michael Gokhman}, \bibinfo{person}{Avashalom Manevich}, \bibinfo{person}{Nir Ratner}, \bibinfo{person}{Noam Rozen}, \bibinfo{person}{Erez Shwartz}, \bibinfo{person}{Mor Zusman}, {and} \bibinfo{person}{Yoav Shoham}.} \bibinfo{year}{2024}\natexlab{}.
\newblock \showarticletitle{Jamba: {A} Hybrid Transformer-Mamba Language Model}.
\newblock \bibinfo{journal}{\emph{CoRR}}  \bibinfo{volume}{abs/2403.19887} (\bibinfo{year}{2024}).
\newblock
\urldef\tempurl%
\url{https://doi.org/10.48550/ARXIV.2403.19887}
\showDOI{\tempurl}
\showeprint[arXiv]{2403.19887}


\bibitem[Light et~al\mbox{.}(2024)]%
        {multiagentLLMstrategist2024}
\bibfield{author}{\bibinfo{person}{Jonathan Light}, \bibinfo{person}{Min Cai}, \bibinfo{person}{Weiqin Chen}, \bibinfo{person}{Guanzhi Wang}, \bibinfo{person}{Xiusi Chen}, \bibinfo{person}{Wei Cheng}, \bibinfo{person}{Yisong Yue}, {and} \bibinfo{person}{Ziniu Hu}.} \bibinfo{year}{2024}\natexlab{}.
\newblock \showarticletitle{Strategist: Learning Strategic Skills by LLMs via Bi-Level Tree Search}.
\newblock \bibinfo{journal}{\emph{CoRR}}  \bibinfo{volume}{abs/2408.10635} (\bibinfo{year}{2024}).
\newblock
\urldef\tempurl%
\url{https://doi.org/10.48550/ARXIV.2408.10635}
\showDOI{\tempurl}
\showeprint[arXiv]{2408.10635}


\bibitem[Lin et~al\mbox{.}(2024a)]%
        {Parrot}
\bibfield{author}{\bibinfo{person}{Chaofan Lin}, \bibinfo{person}{Zhenhua Han}, \bibinfo{person}{Chengruidong Zhang}, \bibinfo{person}{Yuqing Yang}, \bibinfo{person}{Fan Yang}, \bibinfo{person}{Chen Chen}, {and} \bibinfo{person}{Lili Qiu}.} \bibinfo{year}{2024}\natexlab{a}.
\newblock \showarticletitle{Parrot: Efficient Serving of LLM-based Applications with Semantic Variable}. In \bibinfo{booktitle}{\emph{18th {USENIX} Symposium on Operating Systems Design and Implementation, {OSDI} 2024, Santa Clara, CA, USA, July 10-12, 2024}}, \bibfield{editor}{\bibinfo{person}{Ada Gavrilovska} {and} \bibinfo{person}{Douglas~B. Terry}} (Eds.). \bibinfo{publisher}{{USENIX} Association}, \bibinfo{pages}{929--945}.
\newblock
\urldef\tempurl%
\url{https://www.usenix.org/conference/osdi24/presentation/lin-chaofan}
\showURL{%
\tempurl}


\bibitem[Lin and Mehrotra(2024)]%
        {PLAQUE}
\bibfield{author}{\bibinfo{person}{Yiming Lin} {and} \bibinfo{person}{Sharad Mehrotra}.} \bibinfo{year}{2024}\natexlab{}.
\newblock \showarticletitle{{PLAQUE:} Automated Predicate Learning at Query Time}.
\newblock \bibinfo{journal}{\emph{Proc. {ACM} Manag. Data}} \bibinfo{volume}{2}, \bibinfo{number}{1} (\bibinfo{year}{2024}), \bibinfo{pages}{46:1--46:25}.
\newblock
\urldef\tempurl%
\url{https://doi.org/10.1145/3639301}
\showDOI{\tempurl}


\bibitem[Lin et~al\mbox{.}(2024b)]%
        {criticaltokensmatterDPO2024}
\bibfield{author}{\bibinfo{person}{Zicheng Lin}, \bibinfo{person}{Tian Liang}, \bibinfo{person}{Jiahao Xu}, \bibinfo{person}{Xing Wang}, \bibinfo{person}{Ruilin Luo}, \bibinfo{person}{Chufan Shi}, \bibinfo{person}{Siheng Li}, \bibinfo{person}{Yujiu Yang}, {and} \bibinfo{person}{Zhaopeng Tu}.} \bibinfo{year}{2024}\natexlab{b}.
\newblock \showarticletitle{Critical Tokens Matter: Token-Level Contrastive Estimation Enhence LLM's Reasoning Capability}.
\newblock \bibinfo{journal}{\emph{arXiv preprint arXiv:2411.19943}} (\bibinfo{year}{2024}).
\newblock


\bibitem[Liu et~al\mbox{.}(2024f)]%
        {KVcompressminicache2024}
\bibfield{author}{\bibinfo{person}{Akide Liu}, \bibinfo{person}{Jing Liu}, \bibinfo{person}{Zizheng Pan}, \bibinfo{person}{Yefei He}, \bibinfo{person}{Gholamreza Haffari}, {and} \bibinfo{person}{Bohan Zhuang}.} \bibinfo{year}{2024}\natexlab{f}.
\newblock \showarticletitle{MiniCache: {KV} Cache Compression in Depth Dimension for Large Language Models}.
\newblock \bibinfo{journal}{\emph{CoRR}}  \bibinfo{volume}{abs/2405.14366} (\bibinfo{year}{2024}).
\newblock
\urldef\tempurl%
\url{https://doi.org/10.48550/ARXIV.2405.14366}
\showDOI{\tempurl}
\showeprint[arXiv]{2405.14366}


\bibitem[Liu et~al\mbox{.}(2024h)]%
        {palimpzest2024}
\bibfield{author}{\bibinfo{person}{Chunwei Liu}, \bibinfo{person}{Matthew Russo}, \bibinfo{person}{Michael~J. Cafarella}, \bibinfo{person}{Lei Cao}, \bibinfo{person}{Peter~Baile Chen}, \bibinfo{person}{Zui Chen}, \bibinfo{person}{Michael~J. Franklin}, \bibinfo{person}{Tim Kraska}, \bibinfo{person}{Samuel Madden}, {and} \bibinfo{person}{Gerardo Vitagliano}.} \bibinfo{year}{2024}\natexlab{h}.
\newblock \showarticletitle{A Declarative System for Optimizing {AI} Workloads}.
\newblock \bibinfo{journal}{\emph{CoRR}}  \bibinfo{volume}{abs/2405.14696} (\bibinfo{year}{2024}).
\newblock
\urldef\tempurl%
\url{https://doi.org/10.48550/ARXIV.2405.14696}
\showDOI{\tempurl}
\showeprint[arXiv]{2405.14696}


\bibitem[Liu et~al\mbox{.}(2024b)]%
        {retrievalattention2024}
\bibfield{author}{\bibinfo{person}{Di Liu}, \bibinfo{person}{Meng Chen}, \bibinfo{person}{Baotong Lu}, \bibinfo{person}{Huiqiang Jiang}, \bibinfo{person}{Zhenhua Han}, \bibinfo{person}{Qianxi Zhang}, \bibinfo{person}{Qi Chen}, \bibinfo{person}{Chengruidong Zhang}, \bibinfo{person}{Bailu Ding}, \bibinfo{person}{Kai Zhang}, \bibinfo{person}{Chen Chen}, \bibinfo{person}{Fan Yang}, \bibinfo{person}{Yuqing Yang}, {and} \bibinfo{person}{Lili Qiu}.} \bibinfo{year}{2024}\natexlab{b}.
\newblock \showarticletitle{RetrievalAttention: Accelerating Long-Context {LLM} Inference via Vector Retrieval}.
\newblock \bibinfo{journal}{\emph{CoRR}}  \bibinfo{volume}{abs/2409.10516} (\bibinfo{year}{2024}).
\newblock
\urldef\tempurl%
\url{https://doi.org/10.48550/ARXIV.2409.10516}
\showDOI{\tempurl}
\showeprint[arXiv]{2409.10516}


\bibitem[Liu and Abbeel(2023)]%
        {agenticLLMagentictransformer2023}
\bibfield{author}{\bibinfo{person}{Hao Liu} {and} \bibinfo{person}{Pieter Abbeel}.} \bibinfo{year}{2023}\natexlab{}.
\newblock \showarticletitle{Emergent Agentic Transformer from Chain of Hindsight Experience}. In \bibinfo{booktitle}{\emph{International Conference on Machine Learning, {ICML} 2023, 23-29 July 2023, Honolulu, Hawaii, {USA}}} \emph{(\bibinfo{series}{Proceedings of Machine Learning Research}, Vol.~\bibinfo{volume}{202})}, \bibfield{editor}{\bibinfo{person}{Andreas Krause}, \bibinfo{person}{Emma Brunskill}, \bibinfo{person}{Kyunghyun Cho}, \bibinfo{person}{Barbara Engelhardt}, \bibinfo{person}{Sivan Sabato}, {and} \bibinfo{person}{Jonathan Scarlett}} (Eds.). \bibinfo{publisher}{{PMLR}}, \bibinfo{pages}{21362--21374}.
\newblock
\urldef\tempurl%
\url{https://proceedings.mlr.press/v202/liu23a.html}
\showURL{%
\tempurl}


\bibitem[Liu et~al\mbox{.}(2024c)]%
        {CXLdissectingmemoryperformance}
\bibfield{author}{\bibinfo{person}{Jinshu Liu}, \bibinfo{person}{Hamid Hadian}, \bibinfo{person}{Hanchen Xu}, \bibinfo{person}{Daniel~S. Berger}, {and} \bibinfo{person}{Huaicheng Li}.} \bibinfo{year}{2024}\natexlab{c}.
\newblock \showarticletitle{Dissecting {CXL} Memory Performance at Scale: Analysis, Modeling, and Optimization}.
\newblock \bibinfo{journal}{\emph{CoRR}}  \bibinfo{volume}{abs/2409.14317} (\bibinfo{year}{2024}).
\newblock
\urldef\tempurl%
\url{https://doi.org/10.48550/ARXIV.2409.14317}
\showDOI{\tempurl}
\showeprint[arXiv]{2409.14317}


\bibitem[Liu et~al\mbox{.}(2024e)]%
        {lostinthemiddle}
\bibfield{author}{\bibinfo{person}{Nelson~F. Liu}, \bibinfo{person}{Kevin Lin}, \bibinfo{person}{John Hewitt}, \bibinfo{person}{Ashwin Paranjape}, \bibinfo{person}{Michele Bevilacqua}, \bibinfo{person}{Fabio Petroni}, {and} \bibinfo{person}{Percy Liang}.} \bibinfo{year}{2024}\natexlab{e}.
\newblock \showarticletitle{Lost in the Middle: How Language Models Use Long Contexts}.
\newblock \bibinfo{journal}{\emph{Trans. Assoc. Comput. Linguistics}}  \bibinfo{volume}{12} (\bibinfo{year}{2024}), \bibinfo{pages}{157--173}.
\newblock
\urldef\tempurl%
\url{https://doi.org/10.1162/TACL\_A\_00638}
\showDOI{\tempurl}


\bibitem[Liu et~al\mbox{.}(2024a)]%
        {optimizingrelationalLLMqueries}
\bibfield{author}{\bibinfo{person}{Shu Liu}, \bibinfo{person}{Asim Biswal}, \bibinfo{person}{Audrey Cheng}, \bibinfo{person}{Xiangxi Mo}, \bibinfo{person}{Shiyi Cao}, \bibinfo{person}{Joseph~E. Gonzalez}, \bibinfo{person}{Ion Stoica}, {and} \bibinfo{person}{Matei Zaharia}.} \bibinfo{year}{2024}\natexlab{a}.
\newblock \showarticletitle{Optimizing {LLM} Queries in Relational Workloads}.
\newblock \bibinfo{journal}{\emph{CoRR}}  \bibinfo{volume}{abs/2403.05821} (\bibinfo{year}{2024}).
\newblock
\urldef\tempurl%
\url{https://doi.org/10.48550/ARXIV.2403.05821}
\showDOI{\tempurl}
\showeprint[arXiv]{2403.05821}


\bibitem[Liu et~al\mbox{.}(2024i)]%
        {memoryRAGmemlong2024}
\bibfield{author}{\bibinfo{person}{Weijie Liu}, \bibinfo{person}{Zecheng Tang}, \bibinfo{person}{Juntao Li}, \bibinfo{person}{Kehai Chen}, {and} \bibinfo{person}{Min Zhang}.} \bibinfo{year}{2024}\natexlab{i}.
\newblock \showarticletitle{MemLong: Memory-Augmented Retrieval for Long Text Modeling}.
\newblock \bibinfo{journal}{\emph{CoRR}}  \bibinfo{volume}{abs/2408.16967} (\bibinfo{year}{2024}).
\newblock
\urldef\tempurl%
\url{https://doi.org/10.48550/ARXIV.2408.16967}
\showDOI{\tempurl}
\showeprint[arXiv]{2408.16967}


\bibitem[Liu et~al\mbox{.}(2024d)]%
        {CacheGenKVcompress}
\bibfield{author}{\bibinfo{person}{Yuhan Liu}, \bibinfo{person}{Hanchen Li}, \bibinfo{person}{Yihua Cheng}, \bibinfo{person}{Siddhant Ray}, \bibinfo{person}{Yuyang Huang}, \bibinfo{person}{Qizheng Zhang}, \bibinfo{person}{Kuntai Du}, \bibinfo{person}{Jiayi Yao}, \bibinfo{person}{Shan Lu}, \bibinfo{person}{Ganesh Ananthanarayanan}, \bibinfo{person}{Michael Maire}, \bibinfo{person}{Henry Hoffmann}, \bibinfo{person}{Ari Holtzman}, {and} \bibinfo{person}{Junchen Jiang}.} \bibinfo{year}{2024}\natexlab{d}.
\newblock \showarticletitle{CacheGen: {KV} Cache Compression and Streaming for Fast Large Language Model Serving}. In \bibinfo{booktitle}{\emph{Proceedings of the {ACM} {SIGCOMM} 2024 Conference, {ACM} {SIGCOMM} 2024, Sydney, NSW, Australia, August 4-8, 2024}}. \bibinfo{publisher}{{ACM}}, \bibinfo{pages}{38--56}.
\newblock
\urldef\tempurl%
\url{https://doi.org/10.1145/3651890.3672274}
\showDOI{\tempurl}


\bibitem[Liu et~al\mbox{.}(2024g)]%
        {RAGiterativeselffeedback2024}
\bibfield{author}{\bibinfo{person}{Yanming Liu}, \bibinfo{person}{Xinyue Peng}, \bibinfo{person}{Xuhong Zhang}, \bibinfo{person}{Weihao Liu}, \bibinfo{person}{Jianwei Yin}, \bibinfo{person}{Jiannan Cao}, {and} \bibinfo{person}{Tianyu Du}.} \bibinfo{year}{2024}\natexlab{g}.
\newblock \showarticletitle{{RA-ISF:} Learning to Answer and Understand from Retrieval Augmentation via Iterative Self-Feedback}. In \bibinfo{booktitle}{\emph{Findings of the Association for Computational Linguistics, {ACL} 2024, Bangkok, Thailand and virtual meeting, August 11-16, 2024}}, \bibfield{editor}{\bibinfo{person}{Lun{-}Wei Ku}, \bibinfo{person}{Andre Martins}, {and} \bibinfo{person}{Vivek Srikumar}} (Eds.). \bibinfo{publisher}{Association for Computational Linguistics}, \bibinfo{pages}{4730--4749}.
\newblock
\urldef\tempurl%
\url{https://doi.org/10.18653/V1/2024.FINDINGS-ACL.281}
\showDOI{\tempurl}


\bibitem[Lou et~al\mbox{.}(2024)]%
        {sparseattention2024}
\bibfield{author}{\bibinfo{person}{Chao Lou}, \bibinfo{person}{Zixia Jia}, \bibinfo{person}{Zilong Zheng}, {and} \bibinfo{person}{Kewei Tu}.} \bibinfo{year}{2024}\natexlab{}.
\newblock \showarticletitle{Sparser is Faster and Less is More: Efficient Sparse Attention for Long-Range Transformers}.
\newblock \bibinfo{journal}{\emph{CoRR}}  \bibinfo{volume}{abs/2406.16747} (\bibinfo{year}{2024}).
\newblock
\urldef\tempurl%
\url{https://doi.org/10.48550/ARXIV.2406.16747}
\showDOI{\tempurl}
\showeprint[arXiv]{2406.16747}


\bibitem[Lu et~al\mbox{.}(2024c)]%
        {AIscientist2024}
\bibfield{author}{\bibinfo{person}{Chris Lu}, \bibinfo{person}{Cong Lu}, \bibinfo{person}{Robert~Tjarko Lange}, \bibinfo{person}{Jakob Foerster}, \bibinfo{person}{Jeff Clune}, {and} \bibinfo{person}{David Ha}.} \bibinfo{year}{2024}\natexlab{c}.
\newblock \showarticletitle{The {AI} Scientist: Towards Fully Automated Open-Ended Scientific Discovery}.
\newblock \bibinfo{journal}{\emph{CoRR}}  \bibinfo{volume}{abs/2408.06292} (\bibinfo{year}{2024}).
\newblock
\urldef\tempurl%
\url{https://doi.org/10.48550/ARXIV.2408.06292}
\showDOI{\tempurl}
\showeprint[arXiv]{2408.06292}


\bibitem[Lu et~al\mbox{.}(2024a)]%
        {toolsandbox2024}
\bibfield{author}{\bibinfo{person}{Jiarui Lu}, \bibinfo{person}{Thomas Holleis}, \bibinfo{person}{Yizhe Zhang}, \bibinfo{person}{Bernhard Aumayer}, \bibinfo{person}{Feng Nan}, \bibinfo{person}{Felix Bai}, \bibinfo{person}{Shuang Ma}, \bibinfo{person}{Shen Ma}, \bibinfo{person}{Mengyu Li}, \bibinfo{person}{Guoli Yin}, \bibinfo{person}{Zirui Wang}, {and} \bibinfo{person}{Ruoming Pang}.} \bibinfo{year}{2024}\natexlab{a}.
\newblock \showarticletitle{ToolSandbox: {A} Stateful, Conversational, Interactive Evaluation Benchmark for {LLM} Tool Use Capabilities}.
\newblock \bibinfo{journal}{\emph{CoRR}}  \bibinfo{volume}{abs/2408.04682} (\bibinfo{year}{2024}).
\newblock
\urldef\tempurl%
\url{https://doi.org/10.48550/ARXIV.2408.04682}
\showDOI{\tempurl}
\showeprint[arXiv]{2408.04682}


\bibitem[Lu et~al\mbox{.}(2024b)]%
        {SLMsurvey2024}
\bibfield{author}{\bibinfo{person}{Zhenyan Lu}, \bibinfo{person}{Xiang Li}, \bibinfo{person}{Dongqi Cai}, \bibinfo{person}{Rongjie Yi}, \bibinfo{person}{Fangming Liu}, \bibinfo{person}{Xiwen Zhang}, \bibinfo{person}{Nicholas~D. Lane}, {and} \bibinfo{person}{Mengwei Xu}.} \bibinfo{year}{2024}\natexlab{b}.
\newblock \showarticletitle{Small Language Models: Survey, Measurements, and Insights}.
\newblock \bibinfo{journal}{\emph{CoRR}}  \bibinfo{volume}{abs/2409.15790} (\bibinfo{year}{2024}).
\newblock
\urldef\tempurl%
\url{https://doi.org/10.48550/ARXIV.2409.15790}
\showDOI{\tempurl}
\showeprint[arXiv]{2409.15790}


\bibitem[Ma et~al\mbox{.}(2022)]%
        {tableRAGchainofreasoning2022}
\bibfield{author}{\bibinfo{person}{Kaixin Ma}, \bibinfo{person}{Hao Cheng}, \bibinfo{person}{Xiaodong Liu}, \bibinfo{person}{Eric Nyberg}, {and} \bibinfo{person}{Jianfeng Gao}.} \bibinfo{year}{2022}\natexlab{}.
\newblock \showarticletitle{Open-domain Question Answering via Chain of Reasoning over Heterogeneous Knowledge}. In \bibinfo{booktitle}{\emph{Findings of the Association for Computational Linguistics: {EMNLP} 2022, Abu Dhabi, United Arab Emirates, December 7-11, 2022}}, \bibfield{editor}{\bibinfo{person}{Yoav Goldberg}, \bibinfo{person}{Zornitsa Kozareva}, {and} \bibinfo{person}{Yue Zhang}} (Eds.). \bibinfo{publisher}{Association for Computational Linguistics}, \bibinfo{pages}{5360--5374}.
\newblock
\urldef\tempurl%
\url{https://doi.org/10.18653/V1/2022.FINDINGS-EMNLP.392}
\showDOI{\tempurl}


\bibitem[Mao et~al\mbox{.}(2024)]%
        {skyservedatalocality2024}
\bibfield{author}{\bibinfo{person}{Ziming Mao}, \bibinfo{person}{Tian Xia}, \bibinfo{person}{Zhanghao Wu}, \bibinfo{person}{Wei{-}Lin Chiang}, \bibinfo{person}{Tyler Griggs}, \bibinfo{person}{Romil Bhardwaj}, \bibinfo{person}{Zongheng Yang}, \bibinfo{person}{Scott Shenker}, {and} \bibinfo{person}{Ion Stoica}.} \bibinfo{year}{2024}\natexlab{}.
\newblock \showarticletitle{SkyServe: Serving {AI} Models across Regions and Clouds with Spot Instances}.
\newblock \bibinfo{journal}{\emph{CoRR}}  \bibinfo{volume}{abs/2411.01438} (\bibinfo{year}{2024}).
\newblock
\urldef\tempurl%
\url{https://doi.org/10.48550/ARXIV.2411.01438}
\showDOI{\tempurl}
\showeprint[arXiv]{2411.01438}


\bibitem[McKinney(2010)]%
        {pandas}
\bibfield{author}{\bibinfo{person}{Wes McKinney}.} \bibinfo{year}{2010}\natexlab{}.
\newblock \showarticletitle{Data Structures for Statistical Computing in Python}. In \bibinfo{booktitle}{\emph{Proceedings of the 9th Python in Science Conference}}, \bibfield{editor}{\bibinfo{person}{Stéfan van~der Walt} {and} \bibinfo{person}{Jarrod Millman}} (Eds.). \bibinfo{pages}{56--61}.
\newblock
\urldef\tempurl%
\url{https://doi.org/10.25080/Majora-92bf1922-00a}
\showDOI{\tempurl}


\bibitem[Meng et~al\mbox{.}(2023)]%
        {memoryretrievalmasseditingmemory2023}
\bibfield{author}{\bibinfo{person}{Kevin Meng}, \bibinfo{person}{Arnab~Sen Sharma}, \bibinfo{person}{Alex~J. Andonian}, \bibinfo{person}{Yonatan Belinkov}, {and} \bibinfo{person}{David Bau}.} \bibinfo{year}{2023}\natexlab{}.
\newblock \showarticletitle{Mass-Editing Memory in a Transformer}. In \bibinfo{booktitle}{\emph{The Eleventh International Conference on Learning Representations, {ICLR} 2023, Kigali, Rwanda, May 1-5, 2023}}. \bibinfo{publisher}{OpenReview.net}.
\newblock
\urldef\tempurl%
\url{https://openreview.net/forum?id=MkbcAHIYgyS}
\showURL{%
\tempurl}


\bibitem[Miao et~al\mbox{.}(2024)]%
        {SIGMOD2024tutorial}
\bibfield{author}{\bibinfo{person}{Xupeng Miao}, \bibinfo{person}{Zhihao Jia}, {and} \bibinfo{person}{Bin Cui}.} \bibinfo{year}{2024}\natexlab{}.
\newblock \showarticletitle{Demystifying Data Management for Large Language Models}. In \bibinfo{booktitle}{\emph{Companion of the 2024 International Conference on Management of Data, {SIGMOD/PODS} 2024, Santiago AA, Chile, June 9-15, 2024}}, \bibfield{editor}{\bibinfo{person}{Pablo Barcel{\'{o}}}, \bibinfo{person}{Nayat S{\'{a}}nchez{-}Pi}, \bibinfo{person}{Alexandra Meliou}, {and} \bibinfo{person}{S.~Sudarshan}} (Eds.). \bibinfo{publisher}{{ACM}}, \bibinfo{pages}{547--555}.
\newblock
\urldef\tempurl%
\url{https://doi.org/10.1145/3626246.3654683}
\showDOI{\tempurl}


\bibitem[Mishra et~al\mbox{.}(2024)]%
        {detectedithallucination2024}
\bibfield{author}{\bibinfo{person}{Abhika Mishra}, \bibinfo{person}{Akari Asai}, \bibinfo{person}{Vidhisha Balachandran}, \bibinfo{person}{Yizhong Wang}, \bibinfo{person}{Graham Neubig}, \bibinfo{person}{Yulia Tsvetkov}, {and} \bibinfo{person}{Hannaneh Hajishirzi}.} \bibinfo{year}{2024}\natexlab{}.
\newblock \showarticletitle{Fine-grained Hallucination Detection and Editing for Language Models}.
\newblock \bibinfo{journal}{\emph{CoRR}}  \bibinfo{volume}{abs/2401.06855} (\bibinfo{year}{2024}).
\newblock
\urldef\tempurl%
\url{https://doi.org/10.48550/ARXIV.2401.06855}
\showDOI{\tempurl}
\showeprint[arXiv]{2401.06855}


\bibitem[Mojarradi et~al\mbox{.}(2024)]%
        {SLMincontextlearning2024}
\bibfield{author}{\bibinfo{person}{M.~Mehdi Mojarradi}, \bibinfo{person}{Lingyi Yang}, \bibinfo{person}{Robert McCraith}, {and} \bibinfo{person}{Adam Mahdi}.} \bibinfo{year}{2024}\natexlab{}.
\newblock \showarticletitle{Improving In-Context Learning with Small Language Model Ensembles}.
\newblock \bibinfo{journal}{\emph{CoRR}}  \bibinfo{volume}{abs/2410.21868} (\bibinfo{year}{2024}).
\newblock
\urldef\tempurl%
\url{https://doi.org/10.48550/ARXIV.2410.21868}
\showDOI{\tempurl}
\showeprint[arXiv]{2410.21868}


\bibitem[Mombaerts et~al\mbox{.}(2024)]%
        {toolRAGmetaknowledge2024}
\bibfield{author}{\bibinfo{person}{Laurent Mombaerts}, \bibinfo{person}{Terry Ding}, \bibinfo{person}{Adi Banerjee}, \bibinfo{person}{Florian Felice}, \bibinfo{person}{Jonathan Taws}, {and} \bibinfo{person}{Tarik Borogovac}.} \bibinfo{year}{2024}\natexlab{}.
\newblock \showarticletitle{Meta Knowledge for Retrieval Augmented Large Language Models}.
\newblock \bibinfo{journal}{\emph{CoRR}}  \bibinfo{volume}{abs/2408.09017} (\bibinfo{year}{2024}).
\newblock
\urldef\tempurl%
\url{https://doi.org/10.48550/ARXIV.2408.09017}
\showDOI{\tempurl}
\showeprint[arXiv]{2408.09017}


\bibitem[Muennighoff et~al\mbox{.}(2024a)]%
        {instructiontuningcode}
\bibfield{author}{\bibinfo{person}{Niklas Muennighoff}, \bibinfo{person}{Qian Liu}, \bibinfo{person}{Armel~Randy Zebaze}, \bibinfo{person}{Qinkai Zheng}, \bibinfo{person}{Binyuan Hui}, \bibinfo{person}{Terry~Yue Zhuo}, \bibinfo{person}{Swayam Singh}, \bibinfo{person}{Xiangru Tang}, \bibinfo{person}{Leandro von Werra}, {and} \bibinfo{person}{Shayne Longpre}.} \bibinfo{year}{2024}\natexlab{a}.
\newblock \showarticletitle{OctoPack: Instruction Tuning Code Large Language Models}. In \bibinfo{booktitle}{\emph{The Twelfth International Conference on Learning Representations, {ICLR} 2024, Vienna, Austria, May 7-11, 2024}}. \bibinfo{publisher}{OpenReview.net}.
\newblock
\urldef\tempurl%
\url{https://openreview.net/forum?id=mw1PWNSWZP}
\showURL{%
\tempurl}


\bibitem[Muennighoff et~al\mbox{.}(2024b)]%
        {GRITinstructiontuning2024}
\bibfield{author}{\bibinfo{person}{Niklas Muennighoff}, \bibinfo{person}{Hongjin Su}, \bibinfo{person}{Liang Wang}, \bibinfo{person}{Nan Yang}, \bibinfo{person}{Furu Wei}, \bibinfo{person}{Tao Yu}, \bibinfo{person}{Amanpreet Singh}, {and} \bibinfo{person}{Douwe Kiela}.} \bibinfo{year}{2024}\natexlab{b}.
\newblock \showarticletitle{Generative Representational Instruction Tuning}.
\newblock \bibinfo{journal}{\emph{CoRR}}  \bibinfo{volume}{abs/2402.09906} (\bibinfo{year}{2024}).
\newblock
\urldef\tempurl%
\url{https://doi.org/10.48550/ARXIV.2402.09906}
\showDOI{\tempurl}
\showeprint[arXiv]{2402.09906}


\bibitem[Nagel et~al\mbox{.}(2021)]%
        {modelquantizewhitepaper2021}
\bibfield{author}{\bibinfo{person}{Markus Nagel}, \bibinfo{person}{Marios Fournarakis}, \bibinfo{person}{Rana~Ali Amjad}, \bibinfo{person}{Yelysei Bondarenko}, \bibinfo{person}{Mart van Baalen}, {and} \bibinfo{person}{Tijmen Blankevoort}.} \bibinfo{year}{2021}\natexlab{}.
\newblock \showarticletitle{A White Paper on Neural Network Quantization}.
\newblock \bibinfo{journal}{\emph{CoRR}}  \bibinfo{volume}{abs/2106.08295} (\bibinfo{year}{2021}).
\newblock
\showeprint[arXiv]{2106.08295}
\urldef\tempurl%
\url{https://arxiv.org/abs/2106.08295}
\showURL{%
\tempurl}


\bibitem[Nargesian et~al\mbox{.}(2022)]%
        {AItutorial4}
\bibfield{author}{\bibinfo{person}{Fatemeh Nargesian}, \bibinfo{person}{Abolfazl Asudeh}, {and} \bibinfo{person}{H.~V. Jagadish}.} \bibinfo{year}{2022}\natexlab{}.
\newblock \showarticletitle{Responsible Data Integration: Next-generation Challenges}. In \bibinfo{booktitle}{\emph{{SIGMOD} '22: International Conference on Management of Data, Philadelphia, PA, USA, June 12 - 17, 2022}}, \bibfield{editor}{\bibinfo{person}{Zachary~G. Ives}, \bibinfo{person}{Angela Bonifati}, {and} \bibinfo{person}{Amr~El Abbadi}} (Eds.). \bibinfo{publisher}{{ACM}}, \bibinfo{pages}{2458--2464}.
\newblock
\urldef\tempurl%
\url{https://doi.org/10.1145/3514221.3522567}
\showDOI{\tempurl}


\bibitem[Nguyen et~al\mbox{.}(2024)]%
        {SLMsurvey2024.2}
\bibfield{author}{\bibinfo{person}{Chien~Van Nguyen}, \bibinfo{person}{Xuan Shen}, \bibinfo{person}{Ryan Aponte}, \bibinfo{person}{Yu Xia}, \bibinfo{person}{Samyadeep Basu}, \bibinfo{person}{Zhengmian Hu}, \bibinfo{person}{Jian Chen}, \bibinfo{person}{Mihir Parmar}, \bibinfo{person}{Sasidhar Kunapuli}, \bibinfo{person}{Joe Barrow}, \bibinfo{person}{Junda Wu}, \bibinfo{person}{Ashish Singh}, \bibinfo{person}{Yu Wang}, \bibinfo{person}{Jiuxiang Gu}, \bibinfo{person}{Franck Dernoncourt}, \bibinfo{person}{Nesreen~K. Ahmed}, \bibinfo{person}{Nedim Lipka}, \bibinfo{person}{Ruiyi Zhang}, \bibinfo{person}{Xiang Chen}, \bibinfo{person}{Tong Yu}, \bibinfo{person}{Sungchul Kim}, \bibinfo{person}{Hanieh Deilamsalehy}, \bibinfo{person}{Namyong Park}, \bibinfo{person}{Mike Rimer}, \bibinfo{person}{Zhehao Zhang}, \bibinfo{person}{Huanrui Yang}, \bibinfo{person}{Ryan~A. Rossi}, {and} \bibinfo{person}{Thien~Huu Nguyen}.} \bibinfo{year}{2024}\natexlab{}.
\newblock \showarticletitle{A Survey of Small Language Models}.
\newblock \bibinfo{journal}{\emph{CoRR}}  \bibinfo{volume}{abs/2410.20011} (\bibinfo{year}{2024}).
\newblock
\urldef\tempurl%
\url{https://doi.org/10.48550/ARXIV.2410.20011}
\showDOI{\tempurl}
\showeprint[arXiv]{2410.20011}


\bibitem[Ni et~al\mbox{.}(2022)]%
        {largedualencoder2024}
\bibfield{author}{\bibinfo{person}{Jianmo Ni}, \bibinfo{person}{Chen Qu}, \bibinfo{person}{Jing Lu}, \bibinfo{person}{Zhuyun Dai}, \bibinfo{person}{Gustavo~Hern{\'{a}}ndez {\'{A}}brego}, \bibinfo{person}{Ji Ma}, \bibinfo{person}{Vincent~Y. Zhao}, \bibinfo{person}{Yi Luan}, \bibinfo{person}{Keith~B. Hall}, \bibinfo{person}{Ming{-}Wei Chang}, {and} \bibinfo{person}{Yinfei Yang}.} \bibinfo{year}{2022}\natexlab{}.
\newblock \showarticletitle{Large Dual Encoders Are Generalizable Retrievers}. In \bibinfo{booktitle}{\emph{Proceedings of the 2022 Conference on Empirical Methods in Natural Language Processing, {EMNLP} 2022, Abu Dhabi, United Arab Emirates, December 7-11, 2022}}, \bibfield{editor}{\bibinfo{person}{Yoav Goldberg}, \bibinfo{person}{Zornitsa Kozareva}, {and} \bibinfo{person}{Yue Zhang}} (Eds.). \bibinfo{publisher}{Association for Computational Linguistics}, \bibinfo{pages}{9844--9855}.
\newblock
\urldef\tempurl%
\url{https://doi.org/10.18653/V1/2022.EMNLP-MAIN.669}
\showDOI{\tempurl}


\bibitem[Nonkes et~al\mbox{.}(2024)]%
        {detecthallucinationgraph2024}
\bibfield{author}{\bibinfo{person}{Noa Nonkes}, \bibinfo{person}{Sergei Agaronian}, \bibinfo{person}{Evangelos Kanoulas}, {and} \bibinfo{person}{Roxana Petcu}.} \bibinfo{year}{2024}\natexlab{}.
\newblock \showarticletitle{Leveraging Graph Structures to Detect Hallucinations in Large Language Models}.
\newblock \bibinfo{journal}{\emph{CoRR}}  \bibinfo{volume}{abs/2407.04485} (\bibinfo{year}{2024}).
\newblock
\urldef\tempurl%
\url{https://doi.org/10.48550/ARXIV.2407.04485}
\showDOI{\tempurl}
\showeprint[arXiv]{2407.04485}


\bibitem[OpenAI(2024)]%
        {openaio1}
\bibfield{author}{\bibinfo{person}{OpenAI}.} \bibinfo{year}{2024}\natexlab{}.
\newblock \bibinfo{title}{Introducing OpenAI o1}.
\newblock \bibinfo{howpublished}{\url{https://openai.com/o1}}.
\newblock
\newblock
\shownote{Accessed: 2024-12-15}.


\bibitem[Orr et~al\mbox{.}(2019)]%
        {diPs}
\bibfield{author}{\bibinfo{person}{Laurel~J. Orr}, \bibinfo{person}{Srikanth Kandula}, {and} \bibinfo{person}{Surajit Chaudhuri}.} \bibinfo{year}{2019}\natexlab{}.
\newblock \showarticletitle{Pushing Data-Induced Predicates Through Joins in Big-Data Clusters}.
\newblock \bibinfo{journal}{\emph{Proc. {VLDB} Endow.}} \bibinfo{volume}{13}, \bibinfo{number}{3} (\bibinfo{year}{2019}), \bibinfo{pages}{252--265}.
\newblock
\urldef\tempurl%
\url{https://doi.org/10.14778/3368289.3368292}
\showDOI{\tempurl}


\bibitem[Orr et~al\mbox{.}(2021)]%
        {AItutorial7}
\bibfield{author}{\bibinfo{person}{Laurel~J. Orr}, \bibinfo{person}{Atindriyo Sanyal}, \bibinfo{person}{Xiao Ling}, \bibinfo{person}{Karan Goel}, {and} \bibinfo{person}{Megan Leszczynski}.} \bibinfo{year}{2021}\natexlab{}.
\newblock \showarticletitle{Managing {ML} Pipelines: Feature Stores and the Coming Wave of Embedding Ecosystems}.
\newblock \bibinfo{journal}{\emph{Proc. {VLDB} Endow.}} \bibinfo{volume}{14}, \bibinfo{number}{12} (\bibinfo{year}{2021}), \bibinfo{pages}{3178--3181}.
\newblock
\urldef\tempurl%
\url{https://doi.org/10.14778/3476311.3476402}
\showDOI{\tempurl}


\bibitem[Ouyang et~al\mbox{.}(2022)]%
        {RLHF2022}
\bibfield{author}{\bibinfo{person}{Long Ouyang}, \bibinfo{person}{Jeffrey Wu}, \bibinfo{person}{Xu Jiang}, \bibinfo{person}{Diogo Almeida}, \bibinfo{person}{Carroll~L. Wainwright}, \bibinfo{person}{Pamela Mishkin}, \bibinfo{person}{Chong Zhang}, \bibinfo{person}{Sandhini Agarwal}, \bibinfo{person}{Katarina Slama}, \bibinfo{person}{Alex Ray}, \bibinfo{person}{John Schulman}, \bibinfo{person}{Jacob Hilton}, \bibinfo{person}{Fraser Kelton}, \bibinfo{person}{Luke Miller}, \bibinfo{person}{Maddie Simens}, \bibinfo{person}{Amanda Askell}, \bibinfo{person}{Peter Welinder}, \bibinfo{person}{Paul~F. Christiano}, \bibinfo{person}{Jan Leike}, {and} \bibinfo{person}{Ryan Lowe}.} \bibinfo{year}{2022}\natexlab{}.
\newblock \showarticletitle{Training language models to follow instructions with human feedback}. In \bibinfo{booktitle}{\emph{Advances in Neural Information Processing Systems 35: Annual Conference on Neural Information Processing Systems 2022, NeurIPS 2022, New Orleans, LA, USA, November 28 - December 9, 2022}}, \bibfield{editor}{\bibinfo{person}{Sanmi Koyejo}, \bibinfo{person}{S.~Mohamed}, \bibinfo{person}{A.~Agarwal}, \bibinfo{person}{Danielle Belgrave}, \bibinfo{person}{K.~Cho}, {and} \bibinfo{person}{A.~Oh}} (Eds.).
\newblock
\urldef\tempurl%
\url{http://papers.nips.cc/paper\_files/paper/2022/hash/b1efde53be364a73914f58805a001731-Abstract-Conference.html}
\showURL{%
\tempurl}


\bibitem[Pagnoni et~al\mbox{.}(2024)]%
        {bytelatenttransformer2024}
\bibfield{author}{\bibinfo{person}{Artidoro Pagnoni}, \bibinfo{person}{Ram Pasunuru}, \bibinfo{person}{Pedro Rodriguez}, \bibinfo{person}{John Nguyen}, \bibinfo{person}{Benjamin Muller}, \bibinfo{person}{Margaret Li}, \bibinfo{person}{Chunting Zhou}, \bibinfo{person}{Lili Yu}, \bibinfo{person}{Jason Weston}, \bibinfo{person}{Luke Zettlemoyer}, \bibinfo{person}{Gargi Ghosh}, \bibinfo{person}{Mike Lewis}, \bibinfo{person}{Ari Holtzman}, {and} \bibinfo{person}{Srini Iyer}.} \bibinfo{year}{2024}\natexlab{}.
\newblock \showarticletitle{Byte Latent Transformer: Patches Scale Better Than Tokens}.
\newblock  (\bibinfo{year}{2024}).
\newblock
\urldef\tempurl%
\url{https://ai.meta.com/research/publications/byte-latent-transformer-patches-scale-better-than-tokens/}
\showURL{%
\tempurl}


\bibitem[Pan et~al\mbox{.}(2024d)]%
        {searchvectordatabasesurvey2024}
\bibfield{author}{\bibinfo{person}{James~Jie Pan}, \bibinfo{person}{Jianguo Wang}, {and} \bibinfo{person}{Guoliang Li}.} \bibinfo{year}{2024}\natexlab{d}.
\newblock \showarticletitle{Survey of vector database management systems}.
\newblock \bibinfo{journal}{\emph{{VLDB} J.}} \bibinfo{volume}{33}, \bibinfo{number}{5} (\bibinfo{year}{2024}), \bibinfo{pages}{1591--1615}.
\newblock
\urldef\tempurl%
\url{https://doi.org/10.1007/S00778-024-00864-X}
\showDOI{\tempurl}


\bibitem[Pan et~al\mbox{.}(2024c)]%
        {graphRAGroadmap2024}
\bibfield{author}{\bibinfo{person}{Shirui Pan}, \bibinfo{person}{Linhao Luo}, \bibinfo{person}{Yufei Wang}, \bibinfo{person}{Chen Chen}, \bibinfo{person}{Jiapu Wang}, {and} \bibinfo{person}{Xindong Wu}.} \bibinfo{year}{2024}\natexlab{c}.
\newblock \showarticletitle{Unifying large language models and knowledge graphs: A roadmap}.
\newblock \bibinfo{journal}{\emph{IEEE Transactions on Knowledge and Data Engineering}} (\bibinfo{year}{2024}).
\newblock


\bibitem[Pan et~al\mbox{.}(2024a)]%
        {multiagentLLMverylargescale2024}
\bibfield{author}{\bibinfo{person}{Xuchen Pan}, \bibinfo{person}{Dawei Gao}, \bibinfo{person}{Yuexiang Xie}, \bibinfo{person}{Zhewei Wei}, \bibinfo{person}{Yaliang Li}, \bibinfo{person}{Bolin Ding}, \bibinfo{person}{Ji{-}Rong Wen}, {and} \bibinfo{person}{Jingren Zhou}.} \bibinfo{year}{2024}\natexlab{a}.
\newblock \showarticletitle{Very Large-Scale Multi-Agent Simulation in AgentScope}.
\newblock \bibinfo{journal}{\emph{CoRR}}  \bibinfo{volume}{abs/2407.17789} (\bibinfo{year}{2024}).
\newblock
\urldef\tempurl%
\url{https://doi.org/10.48550/ARXIV.2407.17789}
\showDOI{\tempurl}
\showeprint[arXiv]{2407.17789}


\bibitem[Pan et~al\mbox{.}(2024b)]%
        {InstInfer}
\bibfield{author}{\bibinfo{person}{Xiurui Pan}, \bibinfo{person}{Endian Li}, \bibinfo{person}{Qiao Li}, \bibinfo{person}{Shengwen Liang}, \bibinfo{person}{Yizhou Shan}, \bibinfo{person}{Ke Zhou}, \bibinfo{person}{Yingwei Luo}, \bibinfo{person}{Xiaolin Wang}, {and} \bibinfo{person}{Jie Zhang}.} \bibinfo{year}{2024}\natexlab{b}.
\newblock \showarticletitle{InstInfer: In-Storage Attention Offloading for Cost-Effective Long-Context {LLM} Inference}.
\newblock \bibinfo{journal}{\emph{CoRR}}  \bibinfo{volume}{abs/2409.04992} (\bibinfo{year}{2024}).
\newblock
\urldef\tempurl%
\url{https://doi.org/10.48550/ARXIV.2409.04992}
\showDOI{\tempurl}
\showeprint[arXiv]{2409.04992}


\bibitem[Park et~al\mbox{.}(2023)]%
        {multiagentLLMgame2023}
\bibfield{author}{\bibinfo{person}{Joon~Sung Park}, \bibinfo{person}{Joseph~C. O'Brien}, \bibinfo{person}{Carrie~Jun Cai}, \bibinfo{person}{Meredith~Ringel Morris}, \bibinfo{person}{Percy Liang}, {and} \bibinfo{person}{Michael~S. Bernstein}.} \bibinfo{year}{2023}\natexlab{}.
\newblock \showarticletitle{Generative Agents: Interactive Simulacra of Human Behavior}. In \bibinfo{booktitle}{\emph{Proceedings of the 36th Annual {ACM} Symposium on User Interface Software and Technology, {UIST} 2023, San Francisco, CA, USA, 29 October 2023- 1 November 2023}}, \bibfield{editor}{\bibinfo{person}{Sean Follmer}, \bibinfo{person}{Jeff Han}, \bibinfo{person}{J{\"{u}}rgen Steimle}, {and} \bibinfo{person}{Nathalie~Henry Riche}} (Eds.). \bibinfo{publisher}{{ACM}}, \bibinfo{pages}{2:1--2:22}.
\newblock
\urldef\tempurl%
\url{https://doi.org/10.1145/3586183.3606763}
\showDOI{\tempurl}


\bibitem[Patel and Ahmad(2023)]%
        {chatgptcost}
\bibfield{author}{\bibinfo{person}{Dylan Patel} {and} \bibinfo{person}{Afzal Ahmad}.} \bibinfo{year}{2023}\natexlab{}.
\newblock \bibinfo{title}{The Inference Cost of Search Disruption – Large Language Model Cost Analysis}.
\newblock \bibinfo{howpublished}{\url{https://www.semianalysis.com/p/the-inference-cost-of-search-disruption}}.
\newblock
\newblock
\shownote{Accessed: 2024-12-15}.


\bibitem[Patel et~al\mbox{.}(2024b)]%
        {LOTUS}
\bibfield{author}{\bibinfo{person}{Liana Patel}, \bibinfo{person}{Siddharth Jha}, \bibinfo{person}{Carlos Guestrin}, {and} \bibinfo{person}{Matei Zaharia}.} \bibinfo{year}{2024}\natexlab{b}.
\newblock \showarticletitle{{LOTUS:} Enabling Semantic Queries with LLMs Over Tables of Unstructured and Structured Data}.
\newblock \bibinfo{journal}{\emph{CoRR}}  \bibinfo{volume}{abs/2407.11418} (\bibinfo{year}{2024}).
\newblock
\urldef\tempurl%
\url{https://doi.org/10.48550/ARXIV.2407.11418}
\showDOI{\tempurl}
\showeprint[arXiv]{2407.11418}


\bibitem[Patel et~al\mbox{.}(2024a)]%
        {Splitwise}
\bibfield{author}{\bibinfo{person}{Pratyush Patel}, \bibinfo{person}{Esha Choukse}, \bibinfo{person}{Chaojie Zhang}, \bibinfo{person}{Aashaka Shah}, \bibinfo{person}{{\'{I}}{\~{n}}igo Goiri}, \bibinfo{person}{Saeed Maleki}, {and} \bibinfo{person}{Ricardo Bianchini}.} \bibinfo{year}{2024}\natexlab{a}.
\newblock \showarticletitle{Splitwise: Efficient Generative {LLM} Inference Using Phase Splitting}. In \bibinfo{booktitle}{\emph{51st {ACM/IEEE} Annual International Symposium on Computer Architecture, {ISCA} 2024, Buenos Aires, Argentina, June 29 - July 3, 2024}}. \bibinfo{publisher}{{IEEE}}, \bibinfo{pages}{118--132}.
\newblock
\urldef\tempurl%
\url{https://doi.org/10.1109/ISCA59077.2024.00019}
\showDOI{\tempurl}


\bibitem[Patil et~al\mbox{.}(2023)]%
        {toolRAGgorilla2023}
\bibfield{author}{\bibinfo{person}{Shishir~G. Patil}, \bibinfo{person}{Tianjun Zhang}, \bibinfo{person}{Xin Wang}, {and} \bibinfo{person}{Joseph~E. Gonzalez}.} \bibinfo{year}{2023}\natexlab{}.
\newblock \showarticletitle{Gorilla: Large Language Model Connected with Massive APIs}.
\newblock \bibinfo{journal}{\emph{CoRR}}  \bibinfo{volume}{abs/2305.15334} (\bibinfo{year}{2023}).
\newblock
\urldef\tempurl%
\url{https://doi.org/10.48550/ARXIV.2305.15334}
\showDOI{\tempurl}
\showeprint[arXiv]{2305.15334}


\bibitem[Paul et~al\mbox{.}(2020)]%
        {Pyper}
\bibfield{author}{\bibinfo{person}{Johns Paul}, \bibinfo{person}{Bingsheng He}, \bibinfo{person}{Shengliang Lu}, {and} \bibinfo{person}{Chiew~Tong Lau}.} \bibinfo{year}{2020}\natexlab{}.
\newblock \showarticletitle{Improving execution efficiency of just-in-time compilation based query processing on GPUs}.
\newblock \bibinfo{journal}{\emph{Proceedings of the VLDB Endowment}} \bibinfo{volume}{14}, \bibinfo{number}{2} (\bibinfo{year}{2020}), \bibinfo{pages}{202--214}.
\newblock


\bibitem[Paul et~al\mbox{.}(2021)]%
        {databaseonGPU}
\bibfield{author}{\bibinfo{person}{Johns Paul}, \bibinfo{person}{Shengliang Lu}, {and} \bibinfo{person}{Bingsheng He}.} \bibinfo{year}{2021}\natexlab{}.
\newblock \showarticletitle{Database Systems on GPUs}.
\newblock \bibinfo{journal}{\emph{Found. Trends Databases}} \bibinfo{volume}{11}, \bibinfo{number}{1} (\bibinfo{year}{2021}), \bibinfo{pages}{1--108}.
\newblock
\urldef\tempurl%
\url{https://doi.org/10.1561/1900000076}
\showDOI{\tempurl}


\bibitem[Pearce and Song(2024)]%
        {reconcilingchinchilla2024}
\bibfield{author}{\bibinfo{person}{Tim Pearce} {and} \bibinfo{person}{Jinyeop Song}.} \bibinfo{year}{2024}\natexlab{}.
\newblock \showarticletitle{Reconciling Kaplan and Chinchilla Scaling Laws}.
\newblock \bibinfo{journal}{\emph{CoRR}}  \bibinfo{volume}{abs/2406.12907} (\bibinfo{year}{2024}).
\newblock
\urldef\tempurl%
\url{https://doi.org/10.48550/ARXIV.2406.12907}
\showDOI{\tempurl}
\showeprint[arXiv]{2406.12907}


\bibitem[Pfeiffer et~al\mbox{.}(2021)]%
        {adapterfusion2021}
\bibfield{author}{\bibinfo{person}{Jonas Pfeiffer}, \bibinfo{person}{Aishwarya Kamath}, \bibinfo{person}{Andreas R{\"{u}}ckl{\'{e}}}, \bibinfo{person}{Kyunghyun Cho}, {and} \bibinfo{person}{Iryna Gurevych}.} \bibinfo{year}{2021}\natexlab{}.
\newblock \showarticletitle{AdapterFusion: Non-Destructive Task Composition for Transfer Learning}. In \bibinfo{booktitle}{\emph{Proceedings of the 16th Conference of the European Chapter of the Association for Computational Linguistics: Main Volume, {EACL} 2021, Online, April 19 - 23, 2021}}, \bibfield{editor}{\bibinfo{person}{Paola Merlo}, \bibinfo{person}{J{\"{o}}rg Tiedemann}, {and} \bibinfo{person}{Reut Tsarfaty}} (Eds.). \bibinfo{publisher}{Association for Computational Linguistics}, \bibinfo{pages}{487--503}.
\newblock
\urldef\tempurl%
\url{https://doi.org/10.18653/V1/2021.EACL-MAIN.39}
\showDOI{\tempurl}


\bibitem[Pilault et~al\mbox{.}(2023)]%
        {SSMblockstatetransformer2023}
\bibfield{author}{\bibinfo{person}{Jonathan Pilault}, \bibinfo{person}{Mahan Fathi}, \bibinfo{person}{Orhan Firat}, \bibinfo{person}{Chris Pal}, \bibinfo{person}{Pierre{-}Luc Bacon}, {and} \bibinfo{person}{Ross Goroshin}.} \bibinfo{year}{2023}\natexlab{}.
\newblock \showarticletitle{Block-State Transformers}. In \bibinfo{booktitle}{\emph{Advances in Neural Information Processing Systems 36: Annual Conference on Neural Information Processing Systems 2023, NeurIPS 2023, New Orleans, LA, USA, December 10 - 16, 2023}}, \bibfield{editor}{\bibinfo{person}{Alice Oh}, \bibinfo{person}{Tristan Naumann}, \bibinfo{person}{Amir Globerson}, \bibinfo{person}{Kate Saenko}, \bibinfo{person}{Moritz Hardt}, {and} \bibinfo{person}{Sergey Levine}} (Eds.).
\newblock
\urldef\tempurl%
\url{http://papers.nips.cc/paper\_files/paper/2023/hash/16ccd203e9e3696a7ab0dcf568316379-Abstract-Conference.html}
\showURL{%
\tempurl}


\bibitem[Pourreza et~al\mbox{.}(2024)]%
        {multipathchasesql2024}
\bibfield{author}{\bibinfo{person}{Mohammadreza Pourreza}, \bibinfo{person}{Hailong Li}, \bibinfo{person}{Ruoxi Sun}, \bibinfo{person}{Yeounoh Chung}, \bibinfo{person}{Shayan Talaei}, \bibinfo{person}{Gaurav~Tarlok Kakkar}, \bibinfo{person}{Yu Gan}, \bibinfo{person}{Amin Saberi}, \bibinfo{person}{Fatma Ozcan}, {and} \bibinfo{person}{Sercan~{\"{O}}. Arik}.} \bibinfo{year}{2024}\natexlab{}.
\newblock \showarticletitle{{CHASE-SQL:} Multi-Path Reasoning and Preference Optimized Candidate Selection in Text-to-SQL}.
\newblock \bibinfo{journal}{\emph{CoRR}}  \bibinfo{volume}{abs/2410.01943} (\bibinfo{year}{2024}).
\newblock
\urldef\tempurl%
\url{https://doi.org/10.48550/ARXIV.2410.01943}
\showDOI{\tempurl}
\showeprint[arXiv]{2410.01943}


\bibitem[Prudhvith et~al\mbox{.}(2024)]%
        {graphRAG2024}
\bibfield{author}{\bibinfo{person}{Tavva Prudhvith}, \bibinfo{person}{Chakrabarty Swattik}, {and} \bibinfo{person}{Selvakumar Prakash}.} \bibinfo{year}{2024}\natexlab{}.
\newblock \showarticletitle{Enhancing Retrieval Augmented Generation Systems with Knowledge Graphs}. In \bibinfo{booktitle}{\emph{2024 International Conference on Electrical, Computer and Energy Technologies (ICECET}}. IEEE, \bibinfo{pages}{1--8}.
\newblock


\bibitem[Qi et~al\mbox{.}(2024)]%
        {rstarmultipath2024}
\bibfield{author}{\bibinfo{person}{Zhenting Qi}, \bibinfo{person}{Mingyuan Ma}, \bibinfo{person}{Jiahang Xu}, \bibinfo{person}{Li~Lyna Zhang}, \bibinfo{person}{Fan Yang}, {and} \bibinfo{person}{Mao Yang}.} \bibinfo{year}{2024}\natexlab{}.
\newblock \showarticletitle{Mutual Reasoning Makes Smaller LLMs Stronger Problem-Solvers}.
\newblock \bibinfo{journal}{\emph{CoRR}}  \bibinfo{volume}{abs/2408.06195} (\bibinfo{year}{2024}).
\newblock
\urldef\tempurl%
\url{https://doi.org/10.48550/ARXIV.2408.06195}
\showDOI{\tempurl}
\showeprint[arXiv]{2408.06195}


\bibitem[Qian et~al\mbox{.}(2024b)]%
        {multiagentscaling2024}
\bibfield{author}{\bibinfo{person}{Chen Qian}, \bibinfo{person}{Zihao Xie}, \bibinfo{person}{Yifei Wang}, \bibinfo{person}{Wei Liu}, \bibinfo{person}{Yufan Dang}, \bibinfo{person}{Zhuoyun Du}, \bibinfo{person}{Weize Chen}, \bibinfo{person}{Cheng Yang}, \bibinfo{person}{Zhiyuan Liu}, {and} \bibinfo{person}{Maosong Sun}.} \bibinfo{year}{2024}\natexlab{b}.
\newblock \showarticletitle{Scaling Large-Language-Model-based Multi-Agent Collaboration}.
\newblock \bibinfo{journal}{\emph{CoRR}}  \bibinfo{volume}{abs/2406.07155} (\bibinfo{year}{2024}).
\newblock
\urldef\tempurl%
\url{https://doi.org/10.48550/ARXIV.2406.07155}
\showDOI{\tempurl}
\showeprint[arXiv]{2406.07155}


\bibitem[Qian et~al\mbox{.}(2024a)]%
        {efficientMoEpipeline2024}
\bibfield{author}{\bibinfo{person}{Yulei Qian}, \bibinfo{person}{Fengcun Li}, \bibinfo{person}{Xiangyang Ji}, \bibinfo{person}{Xiaoyu Zhao}, \bibinfo{person}{Jianchao Tan}, \bibinfo{person}{Kefeng Zhang}, {and} \bibinfo{person}{Xunliang Cai}.} \bibinfo{year}{2024}\natexlab{a}.
\newblock \showarticletitle{EPS-MoE: Expert Pipeline Scheduler for Cost-Efficient MoE Inference}.
\newblock \bibinfo{journal}{\emph{CoRR}}  \bibinfo{volume}{abs/2410.12247} (\bibinfo{year}{2024}).
\newblock
\urldef\tempurl%
\url{https://doi.org/10.48550/ARXIV.2410.12247}
\showDOI{\tempurl}
\showeprint[arXiv]{2410.12247}


\bibitem[Qin et~al\mbox{.}(2024)]%
        {toolRAGtoolLLM16000}
\bibfield{author}{\bibinfo{person}{Yujia Qin}, \bibinfo{person}{Shihao Liang}, \bibinfo{person}{Yining Ye}, \bibinfo{person}{Kunlun Zhu}, \bibinfo{person}{Lan Yan}, \bibinfo{person}{Yaxi Lu}, \bibinfo{person}{Yankai Lin}, \bibinfo{person}{Xin Cong}, \bibinfo{person}{Xiangru Tang}, \bibinfo{person}{Bill Qian}, \bibinfo{person}{Sihan Zhao}, \bibinfo{person}{Lauren Hong}, \bibinfo{person}{Runchu Tian}, \bibinfo{person}{Ruobing Xie}, \bibinfo{person}{Jie Zhou}, \bibinfo{person}{Mark Gerstein}, \bibinfo{person}{Dahai Li}, \bibinfo{person}{Zhiyuan Liu}, {and} \bibinfo{person}{Maosong Sun}.} \bibinfo{year}{2024}\natexlab{}.
\newblock \showarticletitle{ToolLLM: Facilitating Large Language Models to Master 16000+ Real-world APIs}. In \bibinfo{booktitle}{\emph{The Twelfth International Conference on Learning Representations, {ICLR} 2024, Vienna, Austria, May 7-11, 2024}}. \bibinfo{publisher}{OpenReview.net}.
\newblock
\urldef\tempurl%
\url{https://openreview.net/forum?id=dHng2O0Jjr}
\showURL{%
\tempurl}


\bibitem[Qiu et~al\mbox{.}(2024a)]%
        {longcontexthallucinationbenchmark2024}
\bibfield{author}{\bibinfo{person}{Han Qiu}, \bibinfo{person}{Jiaxing Huang}, \bibinfo{person}{Peng Gao}, \bibinfo{person}{Qin Qi}, \bibinfo{person}{Xiaoqin Zhang}, \bibinfo{person}{Ling Shao}, {and} \bibinfo{person}{Shijian Lu}.} \bibinfo{year}{2024}\natexlab{a}.
\newblock \showarticletitle{LongHalQA: Long-Context Hallucination Evaluation for MultiModal Large Language Models}.
\newblock \bibinfo{journal}{\emph{CoRR}}  \bibinfo{volume}{abs/2410.09962} (\bibinfo{year}{2024}).
\newblock
\urldef\tempurl%
\url{https://doi.org/10.48550/ARXIV.2410.09962}
\showDOI{\tempurl}
\showeprint[arXiv]{2410.09962}


\bibitem[Qiu et~al\mbox{.}(2024b)]%
        {OPrediction}
\bibfield{author}{\bibinfo{person}{Haoran Qiu}, \bibinfo{person}{Weichao Mao}, \bibinfo{person}{Archit Patke}, \bibinfo{person}{Shengkun Cui}, \bibinfo{person}{Saurabh Jha}, \bibinfo{person}{Chen Wang}, \bibinfo{person}{Hubertus Franke}, \bibinfo{person}{Zbigniew~T. Kalbarczyk}, \bibinfo{person}{Tamer Basar}, {and} \bibinfo{person}{Ravishankar~K. Iyer}.} \bibinfo{year}{2024}\natexlab{b}.
\newblock \showarticletitle{Efficient Interactive {LLM} Serving with Proxy Model-based Sequence Length Prediction}.
\newblock \bibinfo{journal}{\emph{CoRR}}  \bibinfo{volume}{abs/2404.08509} (\bibinfo{year}{2024}).
\newblock
\urldef\tempurl%
\url{https://doi.org/10.48550/ARXIV.2404.08509}
\showDOI{\tempurl}
\showeprint[arXiv]{2404.08509}


\bibitem[Qu et~al\mbox{.}(2024)]%
        {toolRAGcolt2024}
\bibfield{author}{\bibinfo{person}{Changle Qu}, \bibinfo{person}{Sunhao Dai}, \bibinfo{person}{Xiaochi Wei}, \bibinfo{person}{Hengyi Cai}, \bibinfo{person}{Shuaiqiang Wang}, \bibinfo{person}{Dawei Yin}, \bibinfo{person}{Jun Xu}, {and} \bibinfo{person}{Ji{-}Rong Wen}.} \bibinfo{year}{2024}\natexlab{}.
\newblock \showarticletitle{Towards Completeness-Oriented Tool Retrieval for Large Language Models}. In \bibinfo{booktitle}{\emph{Proceedings of the 33rd {ACM} International Conference on Information and Knowledge Management, {CIKM} 2024, Boise, ID, USA, October 21-25, 2024}}, \bibfield{editor}{\bibinfo{person}{Edoardo Serra} {and} \bibinfo{person}{Francesca Spezzano}} (Eds.). \bibinfo{publisher}{{ACM}}, \bibinfo{pages}{1930--1940}.
\newblock
\urldef\tempurl%
\url{https://doi.org/10.1145/3627673.3679847}
\showDOI{\tempurl}


\bibitem[Quevedo et~al\mbox{.}(2024)]%
        {detecthallucinationtokenprobability2024}
\bibfield{author}{\bibinfo{person}{Ernesto Quevedo}, \bibinfo{person}{Jorge Yero}, \bibinfo{person}{Rachel Koerner}, \bibinfo{person}{Pablo Rivas}, {and} \bibinfo{person}{Tom{\'{a}}s Cern{\'{y}}}.} \bibinfo{year}{2024}\natexlab{}.
\newblock \showarticletitle{Detecting Hallucinations in Large Language Model Generation: {A} Token Probability Approach}.
\newblock \bibinfo{journal}{\emph{CoRR}}  \bibinfo{volume}{abs/2405.19648} (\bibinfo{year}{2024}).
\newblock
\urldef\tempurl%
\url{https://doi.org/10.48550/ARXIV.2405.19648}
\showDOI{\tempurl}
\showeprint[arXiv]{2405.19648}


\bibitem[Rafailov et~al\mbox{.}(2023)]%
        {DPO2023}
\bibfield{author}{\bibinfo{person}{Rafael Rafailov}, \bibinfo{person}{Archit Sharma}, \bibinfo{person}{Eric Mitchell}, \bibinfo{person}{Christopher~D. Manning}, \bibinfo{person}{Stefano Ermon}, {and} \bibinfo{person}{Chelsea Finn}.} \bibinfo{year}{2023}\natexlab{}.
\newblock \showarticletitle{Direct Preference Optimization: Your Language Model is Secretly a Reward Model}. In \bibinfo{booktitle}{\emph{Advances in Neural Information Processing Systems 36: Annual Conference on Neural Information Processing Systems 2023, NeurIPS 2023, New Orleans, LA, USA, December 10 - 16, 2023}}, \bibfield{editor}{\bibinfo{person}{Alice Oh}, \bibinfo{person}{Tristan Naumann}, \bibinfo{person}{Amir Globerson}, \bibinfo{person}{Kate Saenko}, \bibinfo{person}{Moritz Hardt}, {and} \bibinfo{person}{Sergey Levine}} (Eds.).
\newblock
\urldef\tempurl%
\url{http://papers.nips.cc/paper\_files/paper/2023/hash/a85b405ed65c6477a4fe8302b5e06ce7-Abstract-Conference.html}
\showURL{%
\tempurl}


\bibitem[Raffel et~al\mbox{.}(2020)]%
        {texttotextdata2020}
\bibfield{author}{\bibinfo{person}{Colin Raffel}, \bibinfo{person}{Noam Shazeer}, \bibinfo{person}{Adam Roberts}, \bibinfo{person}{Katherine Lee}, \bibinfo{person}{Sharan Narang}, \bibinfo{person}{Michael Matena}, \bibinfo{person}{Yanqi Zhou}, \bibinfo{person}{Wei Li}, {and} \bibinfo{person}{Peter~J. Liu}.} \bibinfo{year}{2020}\natexlab{}.
\newblock \showarticletitle{Exploring the Limits of Transfer Learning with a Unified Text-to-Text Transformer}.
\newblock \bibinfo{journal}{\emph{J. Mach. Learn. Res.}}  \bibinfo{volume}{21} (\bibinfo{year}{2020}), \bibinfo{pages}{140:1--140:67}.
\newblock
\urldef\tempurl%
\url{https://jmlr.org/papers/v21/20-074.html}
\showURL{%
\tempurl}


\bibitem[Rehg(2024)]%
        {KVcompressKVcompress2024}
\bibfield{author}{\bibinfo{person}{Isaac Rehg}.} \bibinfo{year}{2024}\natexlab{}.
\newblock \showarticletitle{KV-Compress: Paged KV-Cache Compression with Variable Compression Rates per Attention Head}.
\newblock \bibinfo{journal}{\emph{CoRR}}  \bibinfo{volume}{abs/2410.00161} (\bibinfo{year}{2024}).
\newblock
\urldef\tempurl%
\url{https://doi.org/10.48550/ARXIV.2410.00161}
\showDOI{\tempurl}
\showeprint[arXiv]{2410.00161}


\bibitem[Ren et~al\mbox{.}(2024)]%
        {SSMsamba}
\bibfield{author}{\bibinfo{person}{Liliang Ren}, \bibinfo{person}{Yang Liu}, \bibinfo{person}{Yadong Lu}, \bibinfo{person}{Yelong Shen}, \bibinfo{person}{Chen Liang}, {and} \bibinfo{person}{Weizhu Chen}.} \bibinfo{year}{2024}\natexlab{}.
\newblock \showarticletitle{Samba: Simple Hybrid State Space Models for Efficient Unlimited Context Language Modeling}.
\newblock \bibinfo{journal}{\emph{arXiv preprint arXiv:2406.07522}} (\bibinfo{year}{2024}).
\newblock


\bibitem[Rhu et~al\mbox{.}(2016)]%
        {vDNN}
\bibfield{author}{\bibinfo{person}{Minsoo Rhu}, \bibinfo{person}{Natalia Gimelshein}, \bibinfo{person}{Jason Clemons}, \bibinfo{person}{Arslan Zulfiqar}, {and} \bibinfo{person}{Stephen~W. Keckler}.} \bibinfo{year}{2016}\natexlab{}.
\newblock \showarticletitle{vDNN: Virtualized deep neural networks for scalable, memory-efficient neural network design}. In \bibinfo{booktitle}{\emph{49th Annual {IEEE/ACM} International Symposium on Microarchitecture, {MICRO} 2016, Taipei, Taiwan, October 15-19, 2016}}. \bibinfo{publisher}{{IEEE} Computer Society}, \bibinfo{pages}{18:1--18:13}.
\newblock
\urldef\tempurl%
\url{https://doi.org/10.1109/MICRO.2016.7783721}
\showDOI{\tempurl}


\bibitem[Rosa et~al\mbox{.}(2022)]%
        {searchcrossencoder2022}
\bibfield{author}{\bibinfo{person}{Guilherme Rosa}, \bibinfo{person}{Luiz Bonifacio}, \bibinfo{person}{Vitor Jeronymo}, \bibinfo{person}{Hugo Abonizio}, \bibinfo{person}{Marzieh Fadaee}, \bibinfo{person}{Roberto Lotufo}, {and} \bibinfo{person}{Rodrigo Nogueira}.} \bibinfo{year}{2022}\natexlab{}.
\newblock \showarticletitle{In defense of cross-encoders for zero-shot retrieval}.
\newblock \bibinfo{journal}{\emph{arXiv preprint arXiv:2212.06121}} (\bibinfo{year}{2022}).
\newblock


\bibitem[Sahoo et~al\mbox{.}(2024)]%
        {systematicsurveyprompt2024}
\bibfield{author}{\bibinfo{person}{Pranab Sahoo}, \bibinfo{person}{Ayush~Kumar Singh}, \bibinfo{person}{Sriparna Saha}, \bibinfo{person}{Vinija Jain}, \bibinfo{person}{Samrat Mondal}, {and} \bibinfo{person}{Aman Chadha}.} \bibinfo{year}{2024}\natexlab{}.
\newblock \showarticletitle{A Systematic Survey of Prompt Engineering in Large Language Models: Techniques and Applications}.
\newblock \bibinfo{journal}{\emph{CoRR}}  \bibinfo{volume}{abs/2402.07927} (\bibinfo{year}{2024}).
\newblock
\urldef\tempurl%
\url{https://doi.org/10.48550/ARXIV.2402.07927}
\showDOI{\tempurl}
\showeprint[arXiv]{2402.07927}


\bibitem[Sahu et~al\mbox{.}(2022)]%
        {dataaugmentforintentclassificationwithllm2022}
\bibfield{author}{\bibinfo{person}{Gaurav Sahu}, \bibinfo{person}{Pau Rodr{\'{\i}}guez}, \bibinfo{person}{Issam~H. Laradji}, \bibinfo{person}{Parmida Atighehchian}, \bibinfo{person}{David V{\'{a}}zquez}, {and} \bibinfo{person}{Dzmitry Bahdanau}.} \bibinfo{year}{2022}\natexlab{}.
\newblock \showarticletitle{Data Augmentation for Intent Classification with Off-the-shelf Large Language Models}. In \bibinfo{booktitle}{\emph{Proceedings of the 4th Workshop on {NLP} for Conversational AI, ConvAI@ACL 2022, Dublin, Ireland, May 27, 2022}}, \bibfield{editor}{\bibinfo{person}{Bing Liu}, \bibinfo{person}{Alexandros Papangelis}, \bibinfo{person}{Stefan Ultes}, \bibinfo{person}{Abhinav Rastogi}, \bibinfo{person}{Yun{-}Nung Chen}, \bibinfo{person}{Georgios Spithourakis}, \bibinfo{person}{Elnaz Nouri}, {and} \bibinfo{person}{Weiyan Shi}} (Eds.). \bibinfo{publisher}{Association for Computational Linguistics}, \bibinfo{pages}{47--57}.
\newblock
\urldef\tempurl%
\url{https://doi.org/10.18653/V1/2022.NLP4CONVAI-1.5}
\showDOI{\tempurl}


\bibitem[Sanca and Ailamaki(2024)]%
        {relationalvector}
\bibfield{author}{\bibinfo{person}{Viktor Sanca} {and} \bibinfo{person}{Anastasia Ailamaki}.} \bibinfo{year}{2024}\natexlab{}.
\newblock \showarticletitle{Efficient Data Access Paths for Mixed Vector-Relational Search}. In \bibinfo{booktitle}{\emph{Proceedings of the 20th International Workshop on Data Management on New Hardware, DaMoN 2024, Santiago, Chile, 10 June 2024}}, \bibfield{editor}{\bibinfo{person}{Carsten Binnig} {and} \bibinfo{person}{Nesime Tatbul}} (Eds.). \bibinfo{publisher}{{ACM}}, \bibinfo{pages}{6:1--6:9}.
\newblock
\urldef\tempurl%
\url{https://doi.org/10.1145/3662010.3663448}
\showDOI{\tempurl}


\bibitem[Sanh et~al\mbox{.}(2022)]%
        {multitaskinstruction2022}
\bibfield{author}{\bibinfo{person}{Victor Sanh}, \bibinfo{person}{Albert Webson}, \bibinfo{person}{Colin Raffel}, \bibinfo{person}{Stephen~H. Bach}, \bibinfo{person}{Lintang Sutawika}, \bibinfo{person}{Zaid Alyafeai}, \bibinfo{person}{Antoine Chaffin}, \bibinfo{person}{Arnaud Stiegler}, \bibinfo{person}{Arun Raja}, \bibinfo{person}{Manan Dey}, \bibinfo{person}{M~Saiful Bari}, \bibinfo{person}{Canwen Xu}, \bibinfo{person}{Urmish Thakker}, \bibinfo{person}{Shanya~Sharma Sharma}, \bibinfo{person}{Eliza Szczechla}, \bibinfo{person}{Taewoon Kim}, \bibinfo{person}{Gunjan Chhablani}, \bibinfo{person}{Nihal~V. Nayak}, \bibinfo{person}{Debajyoti Datta}, \bibinfo{person}{Jonathan Chang}, \bibinfo{person}{Mike~Tian{-}Jian Jiang}, \bibinfo{person}{Han Wang}, \bibinfo{person}{Matteo Manica}, \bibinfo{person}{Sheng Shen}, \bibinfo{person}{Zheng~Xin Yong}, \bibinfo{person}{Harshit Pandey}, \bibinfo{person}{Rachel Bawden}, \bibinfo{person}{Thomas Wang}, \bibinfo{person}{Trishala Neeraj}, \bibinfo{person}{Jos Rozen},
  \bibinfo{person}{Abheesht Sharma}, \bibinfo{person}{Andrea Santilli}, \bibinfo{person}{Thibault F{\'{e}}vry}, \bibinfo{person}{Jason~Alan Fries}, \bibinfo{person}{Ryan Teehan}, \bibinfo{person}{Teven~Le Scao}, \bibinfo{person}{Stella Biderman}, \bibinfo{person}{Leo Gao}, \bibinfo{person}{Thomas Wolf}, {and} \bibinfo{person}{Alexander~M. Rush}.} \bibinfo{year}{2022}\natexlab{}.
\newblock \showarticletitle{Multitask Prompted Training Enables Zero-Shot Task Generalization}. In \bibinfo{booktitle}{\emph{The Tenth International Conference on Learning Representations, {ICLR} 2022, Virtual Event, April 25-29, 2022}}. \bibinfo{publisher}{OpenReview.net}.
\newblock
\urldef\tempurl%
\url{https://openreview.net/forum?id=9Vrb9D0WI4}
\showURL{%
\tempurl}


\bibitem[Santhanam et~al\mbox{.}(2024)]%
        {ALTOcompoundAI}
\bibfield{author}{\bibinfo{person}{Keshav Santhanam}, \bibinfo{person}{Deepti Raghavan}, \bibinfo{person}{Muhammad~Shahir Rahman}, \bibinfo{person}{Thejas Venkatesh}, \bibinfo{person}{Neha Kunjal}, \bibinfo{person}{Pratiksha Thaker}, \bibinfo{person}{Philip~Alexander Levis}, {and} \bibinfo{person}{Matei Zaharia}.} \bibinfo{year}{2024}\natexlab{}.
\newblock \showarticletitle{{ALTO:} An Efficient Network Orchestrator for Compound {AI} Systems}. In \bibinfo{booktitle}{\emph{Proceedings of the 4th Workshop on Machine Learning and Systems, EuroMLSys 2024, Athens, Greece, 22 April 2024}}. \bibinfo{publisher}{{ACM}}, \bibinfo{pages}{117--125}.
\newblock
\urldef\tempurl%
\url{https://doi.org/10.1145/3642970.3655844}
\showDOI{\tempurl}


\bibitem[Sarmah et~al\mbox{.}(2024)]%
        {graphRAGhybrid2024}
\bibfield{author}{\bibinfo{person}{Bhaskarjit Sarmah}, \bibinfo{person}{Dhagash Mehta}, \bibinfo{person}{Benika Hall}, \bibinfo{person}{Rohan Rao}, \bibinfo{person}{Sunil Patel}, {and} \bibinfo{person}{Stefano Pasquali}.} \bibinfo{year}{2024}\natexlab{}.
\newblock \showarticletitle{HybridRAG: Integrating Knowledge Graphs and Vector Retrieval Augmented Generation for Efficient Information Extraction}. In \bibinfo{booktitle}{\emph{Proceedings of the 5th ACM International Conference on AI in Finance}}. \bibinfo{pages}{608--616}.
\newblock


\bibitem[Schick et~al\mbox{.}(2023)]%
        {toolRAGtoolformer2023}
\bibfield{author}{\bibinfo{person}{Timo Schick}, \bibinfo{person}{Jane Dwivedi{-}Yu}, \bibinfo{person}{Roberto Dess{\`{\i}}}, \bibinfo{person}{Roberta Raileanu}, \bibinfo{person}{Maria Lomeli}, \bibinfo{person}{Eric Hambro}, \bibinfo{person}{Luke Zettlemoyer}, \bibinfo{person}{Nicola Cancedda}, {and} \bibinfo{person}{Thomas Scialom}.} \bibinfo{year}{2023}\natexlab{}.
\newblock \showarticletitle{Toolformer: Language Models Can Teach Themselves to Use Tools}. In \bibinfo{booktitle}{\emph{Advances in Neural Information Processing Systems 36: Annual Conference on Neural Information Processing Systems 2023, NeurIPS 2023, New Orleans, LA, USA, December 10 - 16, 2023}}, \bibfield{editor}{\bibinfo{person}{Alice Oh}, \bibinfo{person}{Tristan Naumann}, \bibinfo{person}{Amir Globerson}, \bibinfo{person}{Kate Saenko}, \bibinfo{person}{Moritz Hardt}, {and} \bibinfo{person}{Sergey Levine}} (Eds.).
\newblock
\urldef\tempurl%
\url{http://papers.nips.cc/paper\_files/paper/2023/hash/d842425e4bf79ba039352da0f658a906-Abstract-Conference.html}
\showURL{%
\tempurl}


\bibitem[Schulhoff et~al\mbox{.}(2024)]%
        {promptreportsurvey2024}
\bibfield{author}{\bibinfo{person}{Sander Schulhoff}, \bibinfo{person}{Michael Ilie}, \bibinfo{person}{Nishant Balepur}, \bibinfo{person}{Konstantine Kahadze}, \bibinfo{person}{Amanda Liu}, \bibinfo{person}{Chenglei Si}, \bibinfo{person}{Yinheng Li}, \bibinfo{person}{Aayush Gupta}, \bibinfo{person}{HyoJung Han}, \bibinfo{person}{Sevien Schulhoff}, \bibinfo{person}{Pranav~Sandeep Dulepet}, \bibinfo{person}{Saurav Vidyadhara}, \bibinfo{person}{Dayeon Ki}, \bibinfo{person}{Sweta Agrawal}, \bibinfo{person}{Chau Pham}, \bibinfo{person}{Gerson~C. Kroiz}, \bibinfo{person}{Feileen Li}, \bibinfo{person}{Hudson Tao}, \bibinfo{person}{Ashay Srivastava}, \bibinfo{person}{Hevander~Da Costa}, \bibinfo{person}{Saloni Gupta}, \bibinfo{person}{Megan~L. Rogers}, \bibinfo{person}{Inna Goncearenco}, \bibinfo{person}{Giuseppe Sarli}, \bibinfo{person}{Igor Galynker}, \bibinfo{person}{Denis Peskoff}, \bibinfo{person}{Marine Carpuat}, \bibinfo{person}{Jules White}, \bibinfo{person}{Shyamal Anadkat},
  \bibinfo{person}{Alexander~Miserlis Hoyle}, {and} \bibinfo{person}{Philip Resnik}.} \bibinfo{year}{2024}\natexlab{}.
\newblock \showarticletitle{The Prompt Report: {A} Systematic Survey of Prompting Techniques}.
\newblock \bibinfo{journal}{\emph{CoRR}}  \bibinfo{volume}{abs/2406.06608} (\bibinfo{year}{2024}).
\newblock
\urldef\tempurl%
\url{https://doi.org/10.48550/ARXIV.2406.06608}
\showDOI{\tempurl}
\showeprint[arXiv]{2406.06608}


\bibitem[Schulze et~al\mbox{.}(2024)]%
        {ClickHouse}
\bibfield{author}{\bibinfo{person}{Robert Schulze}, \bibinfo{person}{Tom Schreiber}, \bibinfo{person}{Ilya Yatsishin}, \bibinfo{person}{Ryadh Dahimene}, {and} \bibinfo{person}{Alexey Milovidov}.} \bibinfo{year}{2024}\natexlab{}.
\newblock \showarticletitle{ClickHouse-Lightning Fast Analytics for Everyone}.
\newblock \bibinfo{journal}{\emph{Proceedings of the VLDB Endowment}} \bibinfo{volume}{17}, \bibinfo{number}{12} (\bibinfo{year}{2024}), \bibinfo{pages}{3731--3744}.
\newblock


\bibitem[Shah et~al\mbox{.}(2024)]%
        {FlashAttention_3}
\bibfield{author}{\bibinfo{person}{Jay Shah}, \bibinfo{person}{Ganesh Bikshandi}, \bibinfo{person}{Ying Zhang}, \bibinfo{person}{Vijay Thakkar}, \bibinfo{person}{Pradeep Ramani}, {and} \bibinfo{person}{Tri Dao}.} \bibinfo{year}{2024}\natexlab{}.
\newblock \showarticletitle{FlashAttention-3: Fast and Accurate Attention with Asynchrony and Low-precision}.
\newblock \bibinfo{journal}{\emph{CoRR}}  \bibinfo{volume}{abs/2407.08608} (\bibinfo{year}{2024}).
\newblock
\urldef\tempurl%
\url{https://doi.org/10.48550/ARXIV.2407.08608}
\showDOI{\tempurl}
\showeprint[arXiv]{2407.08608}


\bibitem[Shani et~al\mbox{.}(2024)]%
        {multiturnRLHF2024}
\bibfield{author}{\bibinfo{person}{Lior Shani}, \bibinfo{person}{Aviv Rosenberg}, \bibinfo{person}{Asaf~B. Cassel}, \bibinfo{person}{Oran Lang}, \bibinfo{person}{Daniele Calandriello}, \bibinfo{person}{Avital Zipori}, \bibinfo{person}{Hila Noga}, \bibinfo{person}{Orgad Keller}, \bibinfo{person}{Bilal Piot}, \bibinfo{person}{Idan Szpektor}, \bibinfo{person}{Avinatan Hassidim}, \bibinfo{person}{Yossi Matias}, {and} \bibinfo{person}{R{\'{e}}mi Munos}.} \bibinfo{year}{2024}\natexlab{}.
\newblock \showarticletitle{Multi-turn Reinforcement Learning from Preference Human Feedback}.
\newblock \bibinfo{journal}{\emph{CoRR}}  \bibinfo{volume}{abs/2405.14655} (\bibinfo{year}{2024}).
\newblock
\urldef\tempurl%
\url{https://doi.org/10.48550/ARXIV.2405.14655}
\showDOI{\tempurl}
\showeprint[arXiv]{2405.14655}


\bibitem[Shao et~al\mbox{.}(2024)]%
        {modelquantizeomniquant2024}
\bibfield{author}{\bibinfo{person}{Wenqi Shao}, \bibinfo{person}{Mengzhao Chen}, \bibinfo{person}{Zhaoyang Zhang}, \bibinfo{person}{Peng Xu}, \bibinfo{person}{Lirui Zhao}, \bibinfo{person}{Zhiqian Li}, \bibinfo{person}{Kaipeng Zhang}, \bibinfo{person}{Peng Gao}, \bibinfo{person}{Yu Qiao}, {and} \bibinfo{person}{Ping Luo}.} \bibinfo{year}{2024}\natexlab{}.
\newblock \showarticletitle{OmniQuant: Omnidirectionally Calibrated Quantization for Large Language Models}. In \bibinfo{booktitle}{\emph{The Twelfth International Conference on Learning Representations, {ICLR} 2024, Vienna, Austria, May 7-11, 2024}}. \bibinfo{publisher}{OpenReview.net}.
\newblock
\urldef\tempurl%
\url{https://openreview.net/forum?id=8Wuvhh0LYW}
\showURL{%
\tempurl}


\bibitem[Sheng et~al\mbox{.}(2024)]%
        {VTCvirtualtokencounter2024}
\bibfield{author}{\bibinfo{person}{Ying Sheng}, \bibinfo{person}{Shiyi Cao}, \bibinfo{person}{Dacheng Li}, \bibinfo{person}{Banghua Zhu}, \bibinfo{person}{Zhuohan Li}, \bibinfo{person}{Danyang Zhuo}, \bibinfo{person}{Joseph~E. Gonzalez}, {and} \bibinfo{person}{Ion Stoica}.} \bibinfo{year}{2024}\natexlab{}.
\newblock \showarticletitle{Fairness in Serving Large Language Models}. In \bibinfo{booktitle}{\emph{18th {USENIX} Symposium on Operating Systems Design and Implementation, {OSDI} 2024, Santa Clara, CA, USA, July 10-12, 2024}}, \bibfield{editor}{\bibinfo{person}{Ada Gavrilovska} {and} \bibinfo{person}{Douglas~B. Terry}} (Eds.). \bibinfo{publisher}{{USENIX} Association}, \bibinfo{pages}{965--988}.
\newblock
\urldef\tempurl%
\url{https://www.usenix.org/conference/osdi24/presentation/sheng}
\showURL{%
\tempurl}


\bibitem[Sheng et~al\mbox{.}(2023)]%
        {FlexGen}
\bibfield{author}{\bibinfo{person}{Ying Sheng}, \bibinfo{person}{Lianmin Zheng}, \bibinfo{person}{Binhang Yuan}, \bibinfo{person}{Zhuohan Li}, \bibinfo{person}{Max Ryabinin}, \bibinfo{person}{Beidi Chen}, \bibinfo{person}{Percy Liang}, \bibinfo{person}{Christopher R{\'{e}}}, \bibinfo{person}{Ion Stoica}, {and} \bibinfo{person}{Ce Zhang}.} \bibinfo{year}{2023}\natexlab{}.
\newblock \showarticletitle{FlexGen: High-Throughput Generative Inference of Large Language Models with a Single {GPU}}. In \bibinfo{booktitle}{\emph{International Conference on Machine Learning, {ICML} 2023, 23-29 July 2023, Honolulu, Hawaii, {USA}}} \emph{(\bibinfo{series}{Proceedings of Machine Learning Research}, Vol.~\bibinfo{volume}{202})}, \bibfield{editor}{\bibinfo{person}{Andreas Krause}, \bibinfo{person}{Emma Brunskill}, \bibinfo{person}{Kyunghyun Cho}, \bibinfo{person}{Barbara Engelhardt}, \bibinfo{person}{Sivan Sabato}, {and} \bibinfo{person}{Jonathan Scarlett}} (Eds.). \bibinfo{publisher}{{PMLR}}, \bibinfo{pages}{31094--31116}.
\newblock
\urldef\tempurl%
\url{https://proceedings.mlr.press/v202/sheng23a.html}
\showURL{%
\tempurl}


\bibitem[Sheoran et~al\mbox{.}(2023)]%
        {DeepOnlineAgg2023}
\bibfield{author}{\bibinfo{person}{Nikhil Sheoran}, \bibinfo{person}{Supawit Chockchowwat}, \bibinfo{person}{Arav Chheda}, \bibinfo{person}{Suwen Wang}, \bibinfo{person}{Riya Verma}, {and} \bibinfo{person}{Yongjoo Park}.} \bibinfo{year}{2023}\natexlab{}.
\newblock \showarticletitle{A Step Toward Deep Online Aggregation}.
\newblock \bibinfo{journal}{\emph{Proc. {ACM} Manag. Data}} \bibinfo{volume}{1}, \bibinfo{number}{2} (\bibinfo{year}{2023}), \bibinfo{pages}{124:1--124:28}.
\newblock
\urldef\tempurl%
\url{https://doi.org/10.1145/3589269}
\showDOI{\tempurl}


\bibitem[Shin et~al\mbox{.}(2020)]%
        {autoprompt2020}
\bibfield{author}{\bibinfo{person}{Taylor Shin}, \bibinfo{person}{Yasaman Razeghi}, \bibinfo{person}{Robert L.~Logan IV}, \bibinfo{person}{Eric Wallace}, {and} \bibinfo{person}{Sameer Singh}.} \bibinfo{year}{2020}\natexlab{}.
\newblock \showarticletitle{AutoPrompt: Eliciting Knowledge from Language Models with Automatically Generated Prompts}. In \bibinfo{booktitle}{\emph{Proceedings of the 2020 Conference on Empirical Methods in Natural Language Processing, {EMNLP} 2020, Online, November 16-20, 2020}}, \bibfield{editor}{\bibinfo{person}{Bonnie Webber}, \bibinfo{person}{Trevor Cohn}, \bibinfo{person}{Yulan He}, {and} \bibinfo{person}{Yang Liu}} (Eds.). \bibinfo{publisher}{Association for Computational Linguistics}, \bibinfo{pages}{4222--4235}.
\newblock
\urldef\tempurl%
\url{https://doi.org/10.18653/V1/2020.EMNLP-MAIN.346}
\showDOI{\tempurl}


\bibitem[Shinn et~al\mbox{.}(2023)]%
        {agenticLLMreflexion2023}
\bibfield{author}{\bibinfo{person}{Noah Shinn}, \bibinfo{person}{Federico Cassano}, \bibinfo{person}{Ashwin Gopinath}, \bibinfo{person}{Karthik Narasimhan}, {and} \bibinfo{person}{Shunyu Yao}.} \bibinfo{year}{2023}\natexlab{}.
\newblock \showarticletitle{Reflexion: language agents with verbal reinforcement learning}. In \bibinfo{booktitle}{\emph{Advances in Neural Information Processing Systems 36: Annual Conference on Neural Information Processing Systems 2023, NeurIPS 2023, New Orleans, LA, USA, December 10 - 16, 2023}}, \bibfield{editor}{\bibinfo{person}{Alice Oh}, \bibinfo{person}{Tristan Naumann}, \bibinfo{person}{Amir Globerson}, \bibinfo{person}{Kate Saenko}, \bibinfo{person}{Moritz Hardt}, {and} \bibinfo{person}{Sergey Levine}} (Eds.).
\newblock
\urldef\tempurl%
\url{http://papers.nips.cc/paper\_files/paper/2023/hash/1b44b878bb782e6954cd888628510e90-Abstract-Conference.html}
\showURL{%
\tempurl}


\bibitem[Shoeybi et~al\mbox{.}(2019)]%
        {megatronmodelparallelism2019}
\bibfield{author}{\bibinfo{person}{Mohammad Shoeybi}, \bibinfo{person}{Mostofa Patwary}, \bibinfo{person}{Raul Puri}, \bibinfo{person}{Patrick LeGresley}, \bibinfo{person}{Jared Casper}, {and} \bibinfo{person}{Bryan Catanzaro}.} \bibinfo{year}{2019}\natexlab{}.
\newblock \showarticletitle{Megatron-LM: Training Multi-Billion Parameter Language Models Using Model Parallelism}.
\newblock \bibinfo{journal}{\emph{CoRR}}  \bibinfo{volume}{abs/1909.08053} (\bibinfo{year}{2019}).
\newblock
\showeprint[arXiv]{1909.08053}
\urldef\tempurl%
\url{http://arxiv.org/abs/1909.08053}
\showURL{%
\tempurl}


\bibitem[Si et~al\mbox{.}(2023)]%
        {reliablepromptgpt32023}
\bibfield{author}{\bibinfo{person}{Chenglei Si}, \bibinfo{person}{Zhe Gan}, \bibinfo{person}{Zhengyuan Yang}, \bibinfo{person}{Shuohang Wang}, \bibinfo{person}{Jianfeng Wang}, \bibinfo{person}{Jordan~L. Boyd{-}Graber}, {and} \bibinfo{person}{Lijuan Wang}.} \bibinfo{year}{2023}\natexlab{}.
\newblock \showarticletitle{Prompting {GPT-3} To Be Reliable}. In \bibinfo{booktitle}{\emph{The Eleventh International Conference on Learning Representations, {ICLR} 2023, Kigali, Rwanda, May 1-5, 2023}}. \bibinfo{publisher}{OpenReview.net}.
\newblock
\urldef\tempurl%
\url{https://openreview.net/forum?id=98p5x51L5af}
\showURL{%
\tempurl}


\bibitem[Simon(2024)]%
        {AIscientistsofscientist2024}
\bibfield{author}{\bibinfo{person}{Tomer Simon}.} \bibinfo{year}{2024}\natexlab{}.
\newblock \showarticletitle{The scientist of the scientist}.
\newblock \bibinfo{journal}{\emph{{AI} Soc.}} \bibinfo{volume}{39}, \bibinfo{number}{2} (\bibinfo{year}{2024}), \bibinfo{pages}{803--804}.
\newblock
\urldef\tempurl%
\url{https://doi.org/10.1007/S00146-022-01544-6}
\showDOI{\tempurl}


\bibitem[Sioulas et~al\mbox{.}(2021)]%
        {speculation}
\bibfield{author}{\bibinfo{person}{Panagiotis Sioulas}, \bibinfo{person}{Viktor Sanca}, \bibinfo{person}{Ioannis Mytilinis}, {and} \bibinfo{person}{Anastasia Ailamaki}.} \bibinfo{year}{2021}\natexlab{}.
\newblock \showarticletitle{Accelerating Complex Analytics using Speculation}. In \bibinfo{booktitle}{\emph{11th Conference on Innovative Data Systems Research, {CIDR} 2021, Virtual Event, January 11-15, 2021, Online Proceedings}}. \bibinfo{publisher}{www.cidrdb.org}.
\newblock
\urldef\tempurl%
\url{http://cidrdb.org/cidr2021/papers/cidr2021\_paper03.pdf}
\showURL{%
\tempurl}


\bibitem[Song et~al\mbox{.}(2024)]%
        {PowerInfer}
\bibfield{author}{\bibinfo{person}{Yixin Song}, \bibinfo{person}{Zeyu Mi}, \bibinfo{person}{Haotong Xie}, {and} \bibinfo{person}{Haibo Chen}.} \bibinfo{year}{2024}\natexlab{}.
\newblock \showarticletitle{PowerInfer: Fast Large Language Model Serving with a Consumer-grade {GPU}}. In \bibinfo{booktitle}{\emph{Proceedings of the {ACM} {SIGOPS} 30th Symposium on Operating Systems Principles, {SOSP} 2024, Austin, TX, USA, November 4-6, 2024}}, \bibfield{editor}{\bibinfo{person}{Emmett Witchel}, \bibinfo{person}{Christopher~J. Rossbach}, \bibinfo{person}{Andrea~C. Arpaci{-}Dusseau}, {and} \bibinfo{person}{Kimberly Keeton}} (Eds.). \bibinfo{publisher}{{ACM}}, \bibinfo{pages}{590--606}.
\newblock
\urldef\tempurl%
\url{https://doi.org/10.1145/3694715.3695964}
\showDOI{\tempurl}


\bibitem[Stiennon et~al\mbox{.}(2020)]%
        {RLHF2020}
\bibfield{author}{\bibinfo{person}{Nisan Stiennon}, \bibinfo{person}{Long Ouyang}, \bibinfo{person}{Jeffrey Wu}, \bibinfo{person}{Daniel~M. Ziegler}, \bibinfo{person}{Ryan Lowe}, \bibinfo{person}{Chelsea Voss}, \bibinfo{person}{Alec Radford}, \bibinfo{person}{Dario Amodei}, {and} \bibinfo{person}{Paul~F. Christiano}.} \bibinfo{year}{2020}\natexlab{}.
\newblock \showarticletitle{Learning to summarize with human feedback}. In \bibinfo{booktitle}{\emph{Advances in Neural Information Processing Systems 33: Annual Conference on Neural Information Processing Systems 2020, NeurIPS 2020, December 6-12, 2020, virtual}}, \bibfield{editor}{\bibinfo{person}{Hugo Larochelle}, \bibinfo{person}{Marc'Aurelio Ranzato}, \bibinfo{person}{Raia Hadsell}, \bibinfo{person}{Maria{-}Florina Balcan}, {and} \bibinfo{person}{Hsuan{-}Tien Lin}} (Eds.).
\newblock
\urldef\tempurl%
\url{https://proceedings.neurips.cc/paper/2020/hash/1f89885d556929e98d3ef9b86448f951-Abstract.html}
\showURL{%
\tempurl}


\bibitem[Strati et~al\mbox{.}(2024)]%
        {Dejavu}
\bibfield{author}{\bibinfo{person}{Foteini Strati}, \bibinfo{person}{Sara McAllister}, \bibinfo{person}{Amar Phanishayee}, \bibinfo{person}{Jakub Tarnawski}, {and} \bibinfo{person}{Ana Klimovic}.} \bibinfo{year}{2024}\natexlab{}.
\newblock \showarticletitle{D{\'{e}}j{\`{a}}Vu: KV-cache Streaming for Fast, Fault-tolerant Generative {LLM} Serving}. In \bibinfo{booktitle}{\emph{Forty-first International Conference on Machine Learning, {ICML} 2024, Vienna, Austria, July 21-27, 2024}}. \bibinfo{publisher}{OpenReview.net}.
\newblock
\urldef\tempurl%
\url{https://openreview.net/forum?id=AbGbGZFYOD}
\showURL{%
\tempurl}


\bibitem[Su et~al\mbox{.}(2024)]%
        {realtimehallucinationdetect2024}
\bibfield{author}{\bibinfo{person}{Weihang Su}, \bibinfo{person}{Changyue Wang}, \bibinfo{person}{Qingyao Ai}, \bibinfo{person}{Yiran Hu}, \bibinfo{person}{Zhijing Wu}, \bibinfo{person}{Yujia Zhou}, {and} \bibinfo{person}{Yiqun Liu}.} \bibinfo{year}{2024}\natexlab{}.
\newblock \showarticletitle{Unsupervised Real-Time Hallucination Detection based on the Internal States of Large Language Models}. In \bibinfo{booktitle}{\emph{Findings of the Association for Computational Linguistics, {ACL} 2024, Bangkok, Thailand and virtual meeting, August 11-16, 2024}}, \bibfield{editor}{\bibinfo{person}{Lun{-}Wei Ku}, \bibinfo{person}{Andre Martins}, {and} \bibinfo{person}{Vivek Srikumar}} (Eds.). \bibinfo{publisher}{Association for Computational Linguistics}, \bibinfo{pages}{14379--14391}.
\newblock
\urldef\tempurl%
\url{https://doi.org/10.18653/V1/2024.FINDINGS-ACL.854}
\showDOI{\tempurl}


\bibitem[Sui et~al\mbox{.}(2024)]%
        {tableRAGbenchmark2024}
\bibfield{author}{\bibinfo{person}{Yuan Sui}, \bibinfo{person}{Mengyu Zhou}, \bibinfo{person}{Mingjie Zhou}, \bibinfo{person}{Shi Han}, {and} \bibinfo{person}{Dongmei Zhang}.} \bibinfo{year}{2024}\natexlab{}.
\newblock \showarticletitle{Table meets llm: Can large language models understand structured table data? a benchmark and empirical study}. In \bibinfo{booktitle}{\emph{Proceedings of the 17th ACM International Conference on Web Search and Data Mining}}. \bibinfo{pages}{645--654}.
\newblock


\bibitem[Sun et~al\mbox{.}(2023)]%
        {graphRAGthinkongraph2023}
\bibfield{author}{\bibinfo{person}{Jiashuo Sun}, \bibinfo{person}{Chengjin Xu}, \bibinfo{person}{Lumingyuan Tang}, \bibinfo{person}{Saizhuo Wang}, \bibinfo{person}{Chen Lin}, \bibinfo{person}{Yeyun Gong}, \bibinfo{person}{Heung{-}Yeung Shum}, {and} \bibinfo{person}{Jian Guo}.} \bibinfo{year}{2023}\natexlab{}.
\newblock \showarticletitle{Think-on-Graph: Deep and Responsible Reasoning of Large Language Model with Knowledge Graph}.
\newblock \bibinfo{journal}{\emph{CoRR}}  \bibinfo{volume}{abs/2307.07697} (\bibinfo{year}{2023}).
\newblock
\urldef\tempurl%
\url{https://doi.org/10.48550/ARXIV.2307.07697}
\showDOI{\tempurl}
\showeprint[arXiv]{2307.07697}


\bibitem[Tan et~al\mbox{.}(2024)]%
        {teolacompoundAI}
\bibfield{author}{\bibinfo{person}{Xin Tan}, \bibinfo{person}{Yimin Jiang}, \bibinfo{person}{Yitao Yang}, {and} \bibinfo{person}{Hong Xu}.} \bibinfo{year}{2024}\natexlab{}.
\newblock \showarticletitle{Teola: Towards End-to-End Optimization of LLM-based Applications}.
\newblock \bibinfo{journal}{\emph{CoRR}}  \bibinfo{volume}{abs/2407.00326} (\bibinfo{year}{2024}).
\newblock
\urldef\tempurl%
\url{https://doi.org/10.48550/ARXIV.2407.00326}
\showDOI{\tempurl}
\showeprint[arXiv]{2407.00326}


\bibitem[Thorne et~al\mbox{.}(2021)]%
        {neuraldatabases2021}
\bibfield{author}{\bibinfo{person}{James Thorne}, \bibinfo{person}{Majid Yazdani}, \bibinfo{person}{Marzieh Saeidi}, \bibinfo{person}{Fabrizio Silvestri}, \bibinfo{person}{Sebastian Riedel}, {and} \bibinfo{person}{Alon~Y. Levy}.} \bibinfo{year}{2021}\natexlab{}.
\newblock \showarticletitle{From Natural Language Processing to Neural Databases}.
\newblock \bibinfo{journal}{\emph{Proc. {VLDB} Endow.}} \bibinfo{volume}{14}, \bibinfo{number}{6} (\bibinfo{year}{2021}), \bibinfo{pages}{1033--1039}.
\newblock
\urldef\tempurl%
\url{https://doi.org/10.14778/3447689.3447706}
\showDOI{\tempurl}


\bibitem[Tian et~al\mbox{.}(2024)]%
        {ANNS2024}
\bibfield{author}{\bibinfo{person}{Bing Tian}, \bibinfo{person}{Haikun Liu}, \bibinfo{person}{Yuhang Tang}, \bibinfo{person}{Shihai Xiao}, \bibinfo{person}{Zhuohui Duan}, \bibinfo{person}{Xiaofei Liao}, \bibinfo{person}{Xuecang Zhang}, \bibinfo{person}{Junhua Zhu}, {and} \bibinfo{person}{Yu Zhang}.} \bibinfo{year}{2024}\natexlab{}.
\newblock \showarticletitle{FusionANNS: An Efficient {CPU/GPU} Cooperative Processing Architecture for Billion-scale Approximate Nearest Neighbor Search}.
\newblock \bibinfo{journal}{\emph{CoRR}}  \bibinfo{volume}{abs/2409.16576} (\bibinfo{year}{2024}).
\newblock
\urldef\tempurl%
\url{https://doi.org/10.48550/ARXIV.2409.16576}
\showDOI{\tempurl}
\showeprint[arXiv]{2409.16576}


\bibitem[Tonmoy et~al\mbox{.}(2024)]%
        {surveyhallucinationmitigation2024}
\bibfield{author}{\bibinfo{person}{S.~M. Towhidul~Islam Tonmoy}, \bibinfo{person}{S.~M.~Mehedi Zaman}, \bibinfo{person}{Vinija Jain}, \bibinfo{person}{Anku Rani}, \bibinfo{person}{Vipula Rawte}, \bibinfo{person}{Aman Chadha}, {and} \bibinfo{person}{Amitava Das}.} \bibinfo{year}{2024}\natexlab{}.
\newblock \showarticletitle{A Comprehensive Survey of Hallucination Mitigation Techniques in Large Language Models}.
\newblock \bibinfo{journal}{\emph{CoRR}}  \bibinfo{volume}{abs/2401.01313} (\bibinfo{year}{2024}).
\newblock
\urldef\tempurl%
\url{https://doi.org/10.48550/ARXIV.2401.01313}
\showDOI{\tempurl}
\showeprint[arXiv]{2401.01313}


\bibitem[Trummer(2022)]%
        {LLMDatabaseTuningManualDBBERT}
\bibfield{author}{\bibinfo{person}{Immanuel Trummer}.} \bibinfo{year}{2022}\natexlab{}.
\newblock \showarticletitle{{DB-BERT:} {A} Database Tuning Tool that "Reads the Manual"}. In \bibinfo{booktitle}{\emph{{SIGMOD} '22: International Conference on Management of Data, Philadelphia, PA, USA, June 12 - 17, 2022}}, \bibfield{editor}{\bibinfo{person}{Zachary~G. Ives}, \bibinfo{person}{Angela Bonifati}, {and} \bibinfo{person}{Amr~El Abbadi}} (Eds.). \bibinfo{publisher}{{ACM}}, \bibinfo{pages}{190--203}.
\newblock
\urldef\tempurl%
\url{https://doi.org/10.1145/3514221.3517843}
\showDOI{\tempurl}


\bibitem[Trummer(2023)]%
        {VLDB2023tutorial}
\bibfield{author}{\bibinfo{person}{Immanuel Trummer}.} \bibinfo{year}{2023}\natexlab{}.
\newblock \showarticletitle{From {BERT} to {GPT-3} Codex: Harnessing the Potential of Very Large Language Models for Data Management}.
\newblock \bibinfo{journal}{\emph{CoRR}}  \bibinfo{volume}{abs/2306.09339} (\bibinfo{year}{2023}).
\newblock
\urldef\tempurl%
\url{https://doi.org/10.48550/ARXIV.2306.09339}
\showDOI{\tempurl}
\showeprint[arXiv]{2306.09339}


\bibitem[Urban and Binnig(2024)]%
        {CAESURA}
\bibfield{author}{\bibinfo{person}{Matthias Urban} {and} \bibinfo{person}{Carsten Binnig}.} \bibinfo{year}{2024}\natexlab{}.
\newblock \showarticletitle{{CAESURA:} Language Models as Multi-Modal Query Planners}. In \bibinfo{booktitle}{\emph{14th Conference on Innovative Data Systems Research, {CIDR} 2024, Chaminade, HI, USA, January 14-17, 2024}}. \bibinfo{publisher}{www.cidrdb.org}.
\newblock
\urldef\tempurl%
\url{https://www.cidrdb.org/cidr2024/papers/p14-urban.pdf}
\showURL{%
\tempurl}


\bibitem[Vaswani et~al\mbox{.}(2017)]%
        {attention}
\bibfield{author}{\bibinfo{person}{Ashish Vaswani}, \bibinfo{person}{Noam Shazeer}, \bibinfo{person}{Niki Parmar}, \bibinfo{person}{Jakob Uszkoreit}, \bibinfo{person}{Llion Jones}, \bibinfo{person}{Aidan~N. Gomez}, \bibinfo{person}{Lukasz Kaiser}, {and} \bibinfo{person}{Illia Polosukhin}.} \bibinfo{year}{2017}\natexlab{}.
\newblock \showarticletitle{Attention is All you Need}. In \bibinfo{booktitle}{\emph{Advances in Neural Information Processing Systems 30: Annual Conference on Neural Information Processing Systems 2017, December 4-9, 2017, Long Beach, CA, {USA}}}, \bibfield{editor}{\bibinfo{person}{Isabelle Guyon}, \bibinfo{person}{Ulrike von Luxburg}, \bibinfo{person}{Samy Bengio}, \bibinfo{person}{Hanna~M. Wallach}, \bibinfo{person}{Rob Fergus}, \bibinfo{person}{S.~V.~N. Vishwanathan}, {and} \bibinfo{person}{Roman Garnett}} (Eds.). \bibinfo{pages}{5998--6008}.
\newblock
\urldef\tempurl%
\url{https://proceedings.neurips.cc/paper/2017/hash/3f5ee243547dee91fbd053c1c4a845aa-Abstract.html}
\showURL{%
\tempurl}


\bibitem[Vatsal and Dubey(2024)]%
        {surveyofpromptingnlp2024}
\bibfield{author}{\bibinfo{person}{Shubham Vatsal} {and} \bibinfo{person}{Harsh Dubey}.} \bibinfo{year}{2024}\natexlab{}.
\newblock \showarticletitle{A Survey of Prompt Engineering Methods in Large Language Models for Different {NLP} Tasks}.
\newblock \bibinfo{journal}{\emph{CoRR}}  \bibinfo{volume}{abs/2407.12994} (\bibinfo{year}{2024}).
\newblock
\urldef\tempurl%
\url{https://doi.org/10.48550/ARXIV.2407.12994}
\showDOI{\tempurl}
\showeprint[arXiv]{2407.12994}


\bibitem[Waldendorf et~al\mbox{.}(2024)]%
        {contrastivedecoding2024}
\bibfield{author}{\bibinfo{person}{Jonas Waldendorf}, \bibinfo{person}{Barry Haddow}, {and} \bibinfo{person}{Alexandra Birch}.} \bibinfo{year}{2024}\natexlab{}.
\newblock \showarticletitle{Contrastive Decoding Reduces Hallucinations in Large Multilingual Machine Translation Models}. In \bibinfo{booktitle}{\emph{Proceedings of the 18th Conference of the European Chapter of the Association for Computational Linguistics, {EACL} 2024 - Volume 1: Long Papers, St. Julian's, Malta, March 17-22, 2024}}, \bibfield{editor}{\bibinfo{person}{Yvette Graham} {and} \bibinfo{person}{Matthew Purver}} (Eds.). \bibinfo{publisher}{Association for Computational Linguistics}, \bibinfo{pages}{2526--2539}.
\newblock
\urldef\tempurl%
\url{https://aclanthology.org/2024.eacl-long.155}
\showURL{%
\tempurl}


\bibitem[Wang et~al\mbox{.}(2024e)]%
        {astuteRAGimperfect}
\bibfield{author}{\bibinfo{person}{Fei Wang}, \bibinfo{person}{Xingchen Wan}, \bibinfo{person}{Ruoxi Sun}, \bibinfo{person}{Jiefeng Chen}, {and} \bibinfo{person}{Sercan~{\"{O}}. Arik}.} \bibinfo{year}{2024}\natexlab{e}.
\newblock \showarticletitle{Astute {RAG:} Overcoming Imperfect Retrieval Augmentation and Knowledge Conflicts for Large Language Models}.
\newblock \bibinfo{journal}{\emph{CoRR}}  \bibinfo{volume}{abs/2410.07176} (\bibinfo{year}{2024}).
\newblock
\urldef\tempurl%
\url{https://doi.org/10.48550/ARXIV.2410.07176}
\showDOI{\tempurl}
\showeprint[arXiv]{2410.07176}


\bibitem[Wang et~al\mbox{.}(2023c)]%
        {multihopQAselfpromptedcot2023}
\bibfield{author}{\bibinfo{person}{Jinyuan Wang}, \bibinfo{person}{Junlong Li}, {and} \bibinfo{person}{Hai Zhao}.} \bibinfo{year}{2023}\natexlab{c}.
\newblock \showarticletitle{Self-prompted chain-of-thought on large language models for open-domain multi-hop reasoning}.
\newblock \bibinfo{journal}{\emph{arXiv preprint arXiv:2310.13552}} (\bibinfo{year}{2023}).
\newblock


\bibitem[Wang et~al\mbox{.}(2024f)]%
        {surveydatasynthesisaugmentation2024}
\bibfield{author}{\bibinfo{person}{Ke Wang}, \bibinfo{person}{Jiahui Zhu}, \bibinfo{person}{Minjie Ren}, \bibinfo{person}{Zeming Liu}, \bibinfo{person}{Shiwei Li}, \bibinfo{person}{Zongye Zhang}, \bibinfo{person}{Chenkai Zhang}, \bibinfo{person}{Xiaoyu Wu}, \bibinfo{person}{Qiqi Zhan}, \bibinfo{person}{Qingjie Liu}, {and} \bibinfo{person}{Yunhong Wang}.} \bibinfo{year}{2024}\natexlab{f}.
\newblock \showarticletitle{A Survey on Data Synthesis and Augmentation for Large Language Models}.
\newblock \bibinfo{journal}{\emph{CoRR}}  \bibinfo{volume}{abs/2410.12896} (\bibinfo{year}{2024}).
\newblock
\urldef\tempurl%
\url{https://doi.org/10.48550/ARXIV.2410.12896}
\showDOI{\tempurl}
\showeprint[arXiv]{2410.12896}


\bibitem[Wang et~al\mbox{.}(2024a)]%
        {semanticbackpropagationagenticsystems2024}
\bibfield{author}{\bibinfo{person}{Wenyi Wang}, \bibinfo{person}{Hisham~A Alyahya}, \bibinfo{person}{Dylan~R Ashley}, \bibinfo{person}{Oleg Serikov}, \bibinfo{person}{Dmitrii Khizbullin}, \bibinfo{person}{Francesco Faccio}, {and} \bibinfo{person}{J{\"u}rgen Schmidhuber}.} \bibinfo{year}{2024}\natexlab{a}.
\newblock \showarticletitle{How to Correctly do Semantic Backpropagation on Language-based Agentic Systems}.
\newblock \bibinfo{journal}{\emph{arXiv preprint arXiv:2412.03624}} (\bibinfo{year}{2024}).
\newblock


\bibitem[Wang et~al\mbox{.}(2023a)]%
        {memoryRAGaugmentmemory2023}
\bibfield{author}{\bibinfo{person}{Weizhi Wang}, \bibinfo{person}{Li Dong}, \bibinfo{person}{Hao Cheng}, \bibinfo{person}{Xiaodong Liu}, \bibinfo{person}{Xifeng Yan}, \bibinfo{person}{Jianfeng Gao}, {and} \bibinfo{person}{Furu Wei}.} \bibinfo{year}{2023}\natexlab{a}.
\newblock \showarticletitle{Augmenting Language Models with Long-Term Memory}. In \bibinfo{booktitle}{\emph{Advances in Neural Information Processing Systems 36: Annual Conference on Neural Information Processing Systems 2023, NeurIPS 2023, New Orleans, LA, USA, December 10 - 16, 2023}}, \bibfield{editor}{\bibinfo{person}{Alice Oh}, \bibinfo{person}{Tristan Naumann}, \bibinfo{person}{Amir Globerson}, \bibinfo{person}{Kate Saenko}, \bibinfo{person}{Moritz Hardt}, {and} \bibinfo{person}{Sergey Levine}} (Eds.).
\newblock
\urldef\tempurl%
\url{http://papers.nips.cc/paper\_files/paper/2023/hash/ebd82705f44793b6f9ade5a669d0f0bf-Abstract-Conference.html}
\showURL{%
\tempurl}


\bibitem[Wang et~al\mbox{.}(2024c)]%
        {multihopQAselfguidingzeroshotprompt2024}
\bibfield{author}{\bibinfo{person}{Xiaochen Wang}, \bibinfo{person}{Junqing He}, \bibinfo{person}{Liang Chen}, \bibinfo{person}{Reza Haf~Zhe Yang}, \bibinfo{person}{Yiru Wang}, \bibinfo{person}{Xiangdi Meng}, \bibinfo{person}{Kunhao Pan}, {and} \bibinfo{person}{Zhifang Sui}.} \bibinfo{year}{2024}\natexlab{c}.
\newblock \showarticletitle{SG-FSM: A Self-Guiding Zero-Shot Prompting Paradigm for Multi-Hop Question Answering Based on Finite State Machine}.
\newblock \bibinfo{journal}{\emph{arXiv preprint arXiv:2410.17021}} (\bibinfo{year}{2024}).
\newblock


\bibitem[Wang et~al\mbox{.}(2023d)]%
        {selfconsistency2023}
\bibfield{author}{\bibinfo{person}{Xuezhi Wang}, \bibinfo{person}{Jason Wei}, \bibinfo{person}{Dale Schuurmans}, \bibinfo{person}{Quoc~V. Le}, \bibinfo{person}{Ed~H. Chi}, \bibinfo{person}{Sharan Narang}, \bibinfo{person}{Aakanksha Chowdhery}, {and} \bibinfo{person}{Denny Zhou}.} \bibinfo{year}{2023}\natexlab{d}.
\newblock \showarticletitle{Self-Consistency Improves Chain of Thought Reasoning in Language Models}. In \bibinfo{booktitle}{\emph{The Eleventh International Conference on Learning Representations, {ICLR} 2023, Kigali, Rwanda, May 1-5, 2023}}. \bibinfo{publisher}{OpenReview.net}.
\newblock
\urldef\tempurl%
\url{https://openreview.net/forum?id=1PL1NIMMrw}
\showURL{%
\tempurl}


\bibitem[Wang and Zhou(2024)]%
        {CoTwithoutprompting2024}
\bibfield{author}{\bibinfo{person}{Xuezhi Wang} {and} \bibinfo{person}{Denny Zhou}.} \bibinfo{year}{2024}\natexlab{}.
\newblock \showarticletitle{Chain-of-Thought Reasoning Without Prompting}.
\newblock \bibinfo{journal}{\emph{CoRR}}  \bibinfo{volume}{abs/2402.10200} (\bibinfo{year}{2024}).
\newblock
\urldef\tempurl%
\url{https://doi.org/10.48550/ARXIV.2402.10200}
\showDOI{\tempurl}
\showeprint[arXiv]{2402.10200}


\bibitem[Wang et~al\mbox{.}(2023b)]%
        {selfinstructtuning2023}
\bibfield{author}{\bibinfo{person}{Yizhong Wang}, \bibinfo{person}{Yeganeh Kordi}, \bibinfo{person}{Swaroop Mishra}, \bibinfo{person}{Alisa Liu}, \bibinfo{person}{Noah~A. Smith}, \bibinfo{person}{Daniel Khashabi}, {and} \bibinfo{person}{Hannaneh Hajishirzi}.} \bibinfo{year}{2023}\natexlab{b}.
\newblock \showarticletitle{Self-Instruct: Aligning Language Models with Self-Generated Instructions}. In \bibinfo{booktitle}{\emph{Proceedings of the 61st Annual Meeting of the Association for Computational Linguistics (Volume 1: Long Papers), {ACL} 2023, Toronto, Canada, July 9-14, 2023}}, \bibfield{editor}{\bibinfo{person}{Anna Rogers}, \bibinfo{person}{Jordan~L. Boyd{-}Graber}, {and} \bibinfo{person}{Naoaki Okazaki}} (Eds.). \bibinfo{publisher}{Association for Computational Linguistics}, \bibinfo{pages}{13484--13508}.
\newblock
\urldef\tempurl%
\url{https://doi.org/10.18653/V1/2023.ACL-LONG.754}
\showDOI{\tempurl}


\bibitem[Wang et~al\mbox{.}(2024b)]%
        {toolRAGwhataretoolsanyway}
\bibfield{author}{\bibinfo{person}{Zhiruo Wang}, \bibinfo{person}{Zhoujun Cheng}, \bibinfo{person}{Hao Zhu}, \bibinfo{person}{Daniel Fried}, {and} \bibinfo{person}{Graham Neubig}.} \bibinfo{year}{2024}\natexlab{b}.
\newblock \showarticletitle{What Are Tools Anyway? {A} Survey from the Language Model Perspective}.
\newblock \bibinfo{journal}{\emph{CoRR}}  \bibinfo{volume}{abs/2403.15452} (\bibinfo{year}{2024}).
\newblock
\urldef\tempurl%
\url{https://doi.org/10.48550/ARXIV.2403.15452}
\showDOI{\tempurl}
\showeprint[arXiv]{2403.15452}


\bibitem[Wang et~al\mbox{.}(2024d)]%
        {partitionedmemoryretrieval2024}
\bibfield{author}{\bibinfo{person}{Zheng Wang}, \bibinfo{person}{Shu~Xian Teo}, \bibinfo{person}{Jieer Ouyang}, \bibinfo{person}{Yongjun Xu}, {and} \bibinfo{person}{Wei Shi}.} \bibinfo{year}{2024}\natexlab{d}.
\newblock \showarticletitle{{M-RAG:} Reinforcing Large Language Model Performance through Retrieval-Augmented Generation with Multiple Partitions}. In \bibinfo{booktitle}{\emph{Proceedings of the 62nd Annual Meeting of the Association for Computational Linguistics (Volume 1: Long Papers), {ACL} 2024, Bangkok, Thailand, August 11-16, 2024}}, \bibfield{editor}{\bibinfo{person}{Lun{-}Wei Ku}, \bibinfo{person}{Andre Martins}, {and} \bibinfo{person}{Vivek Srikumar}} (Eds.). \bibinfo{publisher}{Association for Computational Linguistics}, \bibinfo{pages}{1966--1978}.
\newblock
\urldef\tempurl%
\url{https://doi.org/10.18653/V1/2024.ACL-LONG.108}
\showDOI{\tempurl}


\bibitem[Wei et~al\mbox{.}(2022a)]%
        {instructiontuning2021}
\bibfield{author}{\bibinfo{person}{Jason Wei}, \bibinfo{person}{Maarten Bosma}, \bibinfo{person}{Vincent~Y. Zhao}, \bibinfo{person}{Kelvin Guu}, \bibinfo{person}{Adams~Wei Yu}, \bibinfo{person}{Brian Lester}, \bibinfo{person}{Nan Du}, \bibinfo{person}{Andrew~M. Dai}, {and} \bibinfo{person}{Quoc~V. Le}.} \bibinfo{year}{2022}\natexlab{a}.
\newblock \showarticletitle{Finetuned Language Models are Zero-Shot Learners}. In \bibinfo{booktitle}{\emph{The Tenth International Conference on Learning Representations, {ICLR} 2022, Virtual Event, April 25-29, 2022}}. \bibinfo{publisher}{OpenReview.net}.
\newblock
\urldef\tempurl%
\url{https://openreview.net/forum?id=gEZrGCozdqR}
\showURL{%
\tempurl}


\bibitem[Wei et~al\mbox{.}(2022b)]%
        {cot2022}
\bibfield{author}{\bibinfo{person}{Jason Wei}, \bibinfo{person}{Xuezhi Wang}, \bibinfo{person}{Dale Schuurmans}, \bibinfo{person}{Maarten Bosma}, \bibinfo{person}{Brian Ichter}, \bibinfo{person}{Fei Xia}, \bibinfo{person}{Ed~H. Chi}, \bibinfo{person}{Quoc~V. Le}, {and} \bibinfo{person}{Denny Zhou}.} \bibinfo{year}{2022}\natexlab{b}.
\newblock \showarticletitle{Chain-of-Thought Prompting Elicits Reasoning in Large Language Models}. In \bibinfo{booktitle}{\emph{Advances in Neural Information Processing Systems 35: Annual Conference on Neural Information Processing Systems 2022, NeurIPS 2022, New Orleans, LA, USA, November 28 - December 9, 2022}}, \bibfield{editor}{\bibinfo{person}{Sanmi Koyejo}, \bibinfo{person}{S.~Mohamed}, \bibinfo{person}{A.~Agarwal}, \bibinfo{person}{Danielle Belgrave}, \bibinfo{person}{K.~Cho}, {and} \bibinfo{person}{A.~Oh}} (Eds.).
\newblock
\urldef\tempurl%
\url{http://papers.nips.cc/paper\_files/paper/2022/hash/9d5609613524ecf4f15af0f7b31abca4-Abstract-Conference.html}
\showURL{%
\tempurl}


\bibitem[Whang and Lee(2020)]%
        {AItutorial3}
\bibfield{author}{\bibinfo{person}{Steven Whang} {and} \bibinfo{person}{Jae{-}Gil Lee}.} \bibinfo{year}{2020}\natexlab{}.
\newblock \showarticletitle{Data Collection and Quality Challenges for Deep Learning}.
\newblock \bibinfo{journal}{\emph{Proc. {VLDB} Endow.}} \bibinfo{volume}{13}, \bibinfo{number}{12} (\bibinfo{year}{2020}), \bibinfo{pages}{3429--3432}.
\newblock
\urldef\tempurl%
\url{https://doi.org/10.14778/3415478.3415562}
\showDOI{\tempurl}


\bibitem[Wu et~al\mbox{.}(2023)]%
        {FastServemultilevelfeedbackqueue2023}
\bibfield{author}{\bibinfo{person}{Bingyang Wu}, \bibinfo{person}{Yinmin Zhong}, \bibinfo{person}{Zili Zhang}, \bibinfo{person}{Gang Huang}, \bibinfo{person}{Xuanzhe Liu}, {and} \bibinfo{person}{Xin Jin}.} \bibinfo{year}{2023}\natexlab{}.
\newblock \showarticletitle{Fast Distributed Inference Serving for Large Language Models}.
\newblock \bibinfo{journal}{\emph{CoRR}}  \bibinfo{volume}{abs/2305.05920} (\bibinfo{year}{2023}).
\newblock
\urldef\tempurl%
\url{https://doi.org/10.48550/ARXIV.2305.05920}
\showDOI{\tempurl}
\showeprint[arXiv]{2305.05920}


\bibitem[Wu et~al\mbox{.}(2024)]%
        {moelora2024}
\bibfield{author}{\bibinfo{person}{Xun Wu}, \bibinfo{person}{Shaohan Huang}, {and} \bibinfo{person}{Furu Wei}.} \bibinfo{year}{2024}\natexlab{}.
\newblock \showarticletitle{Mixture of LoRA Experts}. In \bibinfo{booktitle}{\emph{The Twelfth International Conference on Learning Representations, {ICLR} 2024, Vienna, Austria, May 7-11, 2024}}. \bibinfo{publisher}{OpenReview.net}.
\newblock
\urldef\tempurl%
\url{https://openreview.net/forum?id=uWvKBCYh4S}
\showURL{%
\tempurl}


\bibitem[Wu et~al\mbox{.}(2016)]%
        {healing}
\bibfield{author}{\bibinfo{person}{Yingjun Wu}, \bibinfo{person}{Chee~Yong Chan}, {and} \bibinfo{person}{Kian{-}Lee Tan}.} \bibinfo{year}{2016}\natexlab{}.
\newblock \showarticletitle{Transaction Healing: Scaling Optimistic Concurrency Control on Multicores}. In \bibinfo{booktitle}{\emph{Proceedings of the 2016 International Conference on Management of Data, {SIGMOD} Conference 2016, San Francisco, CA, USA, June 26 - July 01, 2016}}, \bibfield{editor}{\bibinfo{person}{Fatma {\"{O}}zcan}, \bibinfo{person}{Georgia Koutrika}, {and} \bibinfo{person}{Sam Madden}} (Eds.). \bibinfo{publisher}{{ACM}}, \bibinfo{pages}{1689--1704}.
\newblock
\urldef\tempurl%
\url{https://doi.org/10.1145/2882903.2915202}
\showDOI{\tempurl}


\bibitem[Wu et~al\mbox{.}(2022)]%
        {memoryRAGmemorizingtransformer2022}
\bibfield{author}{\bibinfo{person}{Yuhuai Wu}, \bibinfo{person}{Markus~Norman Rabe}, \bibinfo{person}{DeLesley Hutchins}, {and} \bibinfo{person}{Christian Szegedy}.} \bibinfo{year}{2022}\natexlab{}.
\newblock \showarticletitle{Memorizing Transformers}. In \bibinfo{booktitle}{\emph{The Tenth International Conference on Learning Representations, {ICLR} 2022, Virtual Event, April 25-29, 2022}}. \bibinfo{publisher}{OpenReview.net}.
\newblock
\urldef\tempurl%
\url{https://openreview.net/forum?id=TrjbxzRcnf-}
\showURL{%
\tempurl}


\bibitem[Xiao et~al\mbox{.}(2023)]%
        {modelquantizesmoothquant2023}
\bibfield{author}{\bibinfo{person}{Guangxuan Xiao}, \bibinfo{person}{Ji Lin}, \bibinfo{person}{Micka{\"{e}}l Seznec}, \bibinfo{person}{Hao Wu}, \bibinfo{person}{Julien Demouth}, {and} \bibinfo{person}{Song Han}.} \bibinfo{year}{2023}\natexlab{}.
\newblock \showarticletitle{SmoothQuant: Accurate and Efficient Post-Training Quantization for Large Language Models}. In \bibinfo{booktitle}{\emph{International Conference on Machine Learning, {ICML} 2023, 23-29 July 2023, Honolulu, Hawaii, {USA}}} \emph{(\bibinfo{series}{Proceedings of Machine Learning Research}, Vol.~\bibinfo{volume}{202})}, \bibfield{editor}{\bibinfo{person}{Andreas Krause}, \bibinfo{person}{Emma Brunskill}, \bibinfo{person}{Kyunghyun Cho}, \bibinfo{person}{Barbara Engelhardt}, \bibinfo{person}{Sivan Sabato}, {and} \bibinfo{person}{Jonathan Scarlett}} (Eds.). \bibinfo{publisher}{{PMLR}}, \bibinfo{pages}{38087--38099}.
\newblock
\urldef\tempurl%
\url{https://proceedings.mlr.press/v202/xiao23c.html}
\showURL{%
\tempurl}


\bibitem[Xing et~al\mbox{.}(2024)]%
        {LLMforDBtableLLMspecialist2024}
\bibfield{author}{\bibinfo{person}{Junjie Xing}, \bibinfo{person}{Yeye He}, \bibinfo{person}{Mengyu Zhou}, \bibinfo{person}{Haoyu Dong}, \bibinfo{person}{Shi Han}, \bibinfo{person}{Dongmei Zhang}, {and} \bibinfo{person}{Surajit Chaudhuri}.} \bibinfo{year}{2024}\natexlab{}.
\newblock \showarticletitle{Table-LLM-Specialist: Language Model Specialists for Tables using Iterative Generator-Validator Fine-tuning}.
\newblock \bibinfo{journal}{\emph{CoRR}}  \bibinfo{volume}{abs/2410.12164} (\bibinfo{year}{2024}).
\newblock
\urldef\tempurl%
\url{https://doi.org/10.48550/ARXIV.2410.12164}
\showDOI{\tempurl}
\showeprint[arXiv]{2410.12164}


\bibitem[Xiong et~al\mbox{.}(2024)]%
        {temporalscalinglaws2024}
\bibfield{author}{\bibinfo{person}{Yizhe Xiong}, \bibinfo{person}{Xiansheng Chen}, \bibinfo{person}{Xin Ye}, \bibinfo{person}{Hui Chen}, \bibinfo{person}{Zijia Lin}, \bibinfo{person}{Haoran Lian}, \bibinfo{person}{Jianwei Niu}, {and} \bibinfo{person}{Guiguang Ding}.} \bibinfo{year}{2024}\natexlab{}.
\newblock \showarticletitle{Temporal Scaling Law for Large Language Models}.
\newblock \bibinfo{journal}{\emph{CoRR}}  \bibinfo{volume}{abs/2404.17785} (\bibinfo{year}{2024}).
\newblock
\urldef\tempurl%
\url{https://doi.org/10.48550/ARXIV.2404.17785}
\showDOI{\tempurl}
\showeprint[arXiv]{2404.17785}


\bibitem[Xu et~al\mbox{.}(2024h)]%
        {vTensor}
\bibfield{author}{\bibinfo{person}{Jiale Xu}, \bibinfo{person}{Rui Zhang}, \bibinfo{person}{Cong Guo}, \bibinfo{person}{Weiming Hu}, \bibinfo{person}{Zihan Liu}, \bibinfo{person}{Feiyang Wu}, \bibinfo{person}{Yu Feng}, \bibinfo{person}{Shixuan Sun}, \bibinfo{person}{Changxu Shao}, \bibinfo{person}{Yuhong Guo}, \bibinfo{person}{Junping Zhao}, \bibinfo{person}{Ke Zhang}, \bibinfo{person}{Minyi Guo}, {and} \bibinfo{person}{Jingwen Leng}.} \bibinfo{year}{2024}\natexlab{h}.
\newblock \showarticletitle{vTensor: Flexible Virtual Tensor Management for Efficient {LLM} Serving}.
\newblock \bibinfo{journal}{\emph{CoRR}}  \bibinfo{volume}{abs/2407.15309} (\bibinfo{year}{2024}).
\newblock
\urldef\tempurl%
\url{https://doi.org/10.48550/ARXIV.2407.15309}
\showDOI{\tempurl}
\showeprint[arXiv]{2407.15309}


\bibitem[Xu et~al\mbox{.}(2024c)]%
        {multiagentLLMmagic2024}
\bibfield{author}{\bibinfo{person}{Lin Xu}, \bibinfo{person}{Zhiyuan Hu}, \bibinfo{person}{Daquan Zhou}, \bibinfo{person}{Hongyu Ren}, \bibinfo{person}{Zhen Dong}, \bibinfo{person}{Kurt Keutzer}, \bibinfo{person}{See{-}Kiong Ng}, {and} \bibinfo{person}{Jiashi Feng}.} \bibinfo{year}{2024}\natexlab{c}.
\newblock \showarticletitle{MAgIC: Investigation of Large Language Model Powered Multi-Agent in Cognition, Adaptability, Rationality and Collaboration}. In \bibinfo{booktitle}{\emph{Proceedings of the 2024 Conference on Empirical Methods in Natural Language Processing, {EMNLP} 2024, Miami, FL, USA, November 12-16, 2024}}, \bibfield{editor}{\bibinfo{person}{Yaser Al{-}Onaizan}, \bibinfo{person}{Mohit Bansal}, {and} \bibinfo{person}{Yun{-}Nung Chen}} (Eds.). \bibinfo{publisher}{Association for Computational Linguistics}, \bibinfo{pages}{7315--7332}.
\newblock
\urldef\tempurl%
\url{https://aclanthology.org/2024.emnlp-main.416}
\showURL{%
\tempurl}


\bibitem[Xu et~al\mbox{.}(2024f)]%
        {iterativetoolretrieval2024}
\bibfield{author}{\bibinfo{person}{Qiancheng Xu}, \bibinfo{person}{Yongqi Li}, \bibinfo{person}{Heming Xia}, {and} \bibinfo{person}{Wenjie Li}.} \bibinfo{year}{2024}\natexlab{f}.
\newblock \showarticletitle{Enhancing Tool Retrieval with Iterative Feedback from Large Language Models}. In \bibinfo{booktitle}{\emph{Findings of the Association for Computational Linguistics: {EMNLP} 2024, Miami, Florida, USA, November 12-16, 2024}}, \bibfield{editor}{\bibinfo{person}{Yaser Al{-}Onaizan}, \bibinfo{person}{Mohit Bansal}, {and} \bibinfo{person}{Yun{-}Nung Chen}} (Eds.). \bibinfo{publisher}{Association for Computational Linguistics}, \bibinfo{pages}{9609--9619}.
\newblock
\urldef\tempurl%
\url{https://aclanthology.org/2024.findings-emnlp.561}
\showURL{%
\tempurl}


\bibitem[Xu et~al\mbox{.}(2024b)]%
        {graphRAGdualpathway2024}
\bibfield{author}{\bibinfo{person}{Sheng Xu}, \bibinfo{person}{Mike Chen}, {and} \bibinfo{person}{Shuwen Chen}.} \bibinfo{year}{2024}\natexlab{b}.
\newblock \showarticletitle{Enhancing Retrieval-Augmented Generation Models with Knowledge Graphs: Innovative Practices Through a Dual-Pathway Approach}. In \bibinfo{booktitle}{\emph{International Conference on Intelligent Computing}}. Springer, \bibinfo{pages}{398--409}.
\newblock


\bibitem[Xu et~al\mbox{.}(2024a)]%
        {repromptingcot2024}
\bibfield{author}{\bibinfo{person}{Weijia Xu}, \bibinfo{person}{Andrzej Banburski}, {and} \bibinfo{person}{Nebojsa Jojic}.} \bibinfo{year}{2024}\natexlab{a}.
\newblock \showarticletitle{Reprompting: Automated Chain-of-Thought Prompt Inference Through Gibbs Sampling}. In \bibinfo{booktitle}{\emph{Forty-first International Conference on Machine Learning, {ICML} 2024, Vienna, Austria, July 21-27, 2024}}. \bibinfo{publisher}{OpenReview.net}.
\newblock
\urldef\tempurl%
\url{https://openreview.net/forum?id=D8zn1DnTuj}
\showURL{%
\tempurl}


\bibitem[Xu et~al\mbox{.}(2023)]%
        {SPFresh}
\bibfield{author}{\bibinfo{person}{Yuming Xu}, \bibinfo{person}{Hengyu Liang}, \bibinfo{person}{Jin Li}, \bibinfo{person}{Shuotao Xu}, \bibinfo{person}{Qi Chen}, \bibinfo{person}{Qianxi Zhang}, \bibinfo{person}{Cheng Li}, \bibinfo{person}{Ziyue Yang}, \bibinfo{person}{Fan Yang}, \bibinfo{person}{Yuqing Yang}, \bibinfo{person}{Peng Cheng}, {and} \bibinfo{person}{Mao Yang}.} \bibinfo{year}{2023}\natexlab{}.
\newblock \showarticletitle{SPFresh: Incremental In-Place Update for Billion-Scale Vector Search}. In \bibinfo{booktitle}{\emph{Proceedings of the 29th Symposium on Operating Systems Principles, {SOSP} 2023, Koblenz, Germany, October 23-26, 2023}}, \bibfield{editor}{\bibinfo{person}{Jason Flinn}, \bibinfo{person}{Margo~I. Seltzer}, \bibinfo{person}{Peter Druschel}, \bibinfo{person}{Antoine Kaufmann}, {and} \bibinfo{person}{Jonathan Mace}} (Eds.). \bibinfo{publisher}{{ACM}}, \bibinfo{pages}{545--561}.
\newblock
\urldef\tempurl%
\url{https://doi.org/10.1145/3600006.3613166}
\showDOI{\tempurl}


\bibitem[Xu et~al\mbox{.}(2024g)]%
        {PIEpoolingcpu2024}
\bibfield{author}{\bibinfo{person}{Yi Xu}, \bibinfo{person}{Ziming Mao}, \bibinfo{person}{Xiangxi Mo}, \bibinfo{person}{Shu Liu}, {and} \bibinfo{person}{Ion Stoica}.} \bibinfo{year}{2024}\natexlab{g}.
\newblock \showarticletitle{Pie: Pooling CPU Memory for LLM Inference}.
\newblock \bibinfo{journal}{\emph{arXiv preprint arXiv:2411.09317}} (\bibinfo{year}{2024}).
\newblock


\bibitem[Xu et~al\mbox{.}(2024d)]%
        {hallucinationisinevitable}
\bibfield{author}{\bibinfo{person}{Ziwei Xu}, \bibinfo{person}{Sanjay Jain}, {and} \bibinfo{person}{Mohan~S. Kankanhalli}.} \bibinfo{year}{2024}\natexlab{d}.
\newblock \showarticletitle{Hallucination is Inevitable: An Innate Limitation of Large Language Models}.
\newblock \bibinfo{journal}{\emph{CoRR}}  \bibinfo{volume}{abs/2401.11817} (\bibinfo{year}{2024}).
\newblock
\urldef\tempurl%
\url{https://doi.org/10.48550/ARXIV.2401.11817}
\showDOI{\tempurl}
\showeprint[arXiv]{2401.11817}


\bibitem[Xu et~al\mbox{.}(2024e)]%
        {quantizedscalinglaws2024}
\bibfield{author}{\bibinfo{person}{Zifei Xu}, \bibinfo{person}{Alexander Lan}, \bibinfo{person}{Wanzin Yazar}, \bibinfo{person}{Tristan Webb}, \bibinfo{person}{Sayeh Sharify}, {and} \bibinfo{person}{Xin Wang}.} \bibinfo{year}{2024}\natexlab{e}.
\newblock \showarticletitle{Scaling laws for post-training quantized large language models}.
\newblock \bibinfo{journal}{\emph{CoRR}}  \bibinfo{volume}{abs/2410.12119} (\bibinfo{year}{2024}).
\newblock
\urldef\tempurl%
\url{https://doi.org/10.48550/ARXIV.2410.12119}
\showDOI{\tempurl}
\showeprint[arXiv]{2410.12119}


\bibitem[Xue et~al\mbox{.}(2024)]%
        {DatabricksAdaptiveQE}
\bibfield{author}{\bibinfo{person}{Maryann Xue}, \bibinfo{person}{Yingyi Bu}, \bibinfo{person}{Abhishek Somani}, \bibinfo{person}{Wenchen Fan}, \bibinfo{person}{Ziqi Liu}, \bibinfo{person}{Steven Chen}, \bibinfo{person}{Herman~Van Hovell}, \bibinfo{person}{Bart Samwel}, \bibinfo{person}{Mostafa Mokhtar}, \bibinfo{person}{Rk Korlapati}, \bibinfo{person}{Andy Lam}, \bibinfo{person}{Yunxiao Ma}, \bibinfo{person}{Vuk Ercegovac}, \bibinfo{person}{Jiexing Li}, \bibinfo{person}{Alexander Behm}, \bibinfo{person}{Yuanjian Li}, \bibinfo{person}{Xiao Li}, \bibinfo{person}{Sriram Krishnamurthy}, \bibinfo{person}{Amit Shukla}, \bibinfo{person}{Michalis Petropoulos}, \bibinfo{person}{Sameer Paranjpye}, \bibinfo{person}{Reynold Xin}, {and} \bibinfo{person}{Matei Zaharia}.} \bibinfo{year}{2024}\natexlab{}.
\newblock \showarticletitle{Adaptive and Robust Query Execution for Lakehouses At Scale}.
\newblock \bibinfo{journal}{\emph{Proc. {VLDB} Endow.}} \bibinfo{volume}{17}, \bibinfo{number}{12} (\bibinfo{year}{2024}), \bibinfo{pages}{3947--3959}.
\newblock
\urldef\tempurl%
\url{https://www.vldb.org/pvldb/vol17/p3947-bu.pdf}
\showURL{%
\tempurl}


\bibitem[Yak et~al\mbox{.}(2023)]%
        {ingestablestabularfoundationmodel2023}
\bibfield{author}{\bibinfo{person}{Scott Yak}, \bibinfo{person}{Yihe Dong}, \bibinfo{person}{Javier Gonzalvo}, {and} \bibinfo{person}{Sercan Arik}.} \bibinfo{year}{2023}\natexlab{}.
\newblock \showarticletitle{IngesTables: Scalable and Efficient Training of LLM-Enabled Tabular Foundation Models}. In \bibinfo{booktitle}{\emph{NeurIPS 2023 Second Table Representation Learning Workshop}}.
\newblock


\bibitem[Yang et~al\mbox{.}(2024a)]%
        {CRAG}
\bibfield{author}{\bibinfo{person}{Xiao Yang}, \bibinfo{person}{Kai Sun}, \bibinfo{person}{Hao Xin}, \bibinfo{person}{Yushi Sun}, \bibinfo{person}{Nikita Bhalla}, \bibinfo{person}{Xiangsen Chen}, \bibinfo{person}{Sajal Choudhary}, \bibinfo{person}{Rongze~Daniel Gui}, \bibinfo{person}{Ziran~Will Jiang}, \bibinfo{person}{Ziyu Jiang}, \bibinfo{person}{Lingkun Kong}, \bibinfo{person}{Brian Moran}, \bibinfo{person}{Jiaqi Wang}, \bibinfo{person}{Yifan~Ethan Xu}, \bibinfo{person}{An Yan}, \bibinfo{person}{Chenyu Yang}, \bibinfo{person}{Eting Yuan}, \bibinfo{person}{Hanwen Zha}, \bibinfo{person}{Nan Tang}, \bibinfo{person}{Lei Chen}, \bibinfo{person}{Nicolas Scheffer}, \bibinfo{person}{Yue Liu}, \bibinfo{person}{Nirav Shah}, \bibinfo{person}{Rakesh Wanga}, \bibinfo{person}{Anuj Kumar}, \bibinfo{person}{Wen{-}tau Yih}, {and} \bibinfo{person}{Xin~Luna Dong}.} \bibinfo{year}{2024}\natexlab{a}.
\newblock \showarticletitle{{CRAG} - Comprehensive {RAG} Benchmark}.
\newblock \bibinfo{journal}{\emph{CoRR}}  \bibinfo{volume}{abs/2406.04744} (\bibinfo{year}{2024}).
\newblock
\urldef\tempurl%
\url{https://doi.org/10.48550/ARXIV.2406.04744}
\showDOI{\tempurl}
\showeprint[arXiv]{2406.04744}


\bibitem[Yang et~al\mbox{.}(2024b)]%
        {FlexpushdownDB}
\bibfield{author}{\bibinfo{person}{Yifei Yang}, \bibinfo{person}{Xiangyao Yu}, \bibinfo{person}{Marco Serafini}, \bibinfo{person}{Ashraf Aboulnaga}, {and} \bibinfo{person}{Michael Stonebraker}.} \bibinfo{year}{2024}\natexlab{b}.
\newblock \showarticletitle{FlexpushdownDB: rethinking computation pushdown for cloud {OLAP} DBMSs}.
\newblock \bibinfo{journal}{\emph{{VLDB} J.}} \bibinfo{volume}{33}, \bibinfo{number}{5} (\bibinfo{year}{2024}), \bibinfo{pages}{1643--1670}.
\newblock
\urldef\tempurl%
\url{https://doi.org/10.1007/S00778-024-00867-8}
\showDOI{\tempurl}


\bibitem[Yao et~al\mbox{.}(2024)]%
        {CacheBlend}
\bibfield{author}{\bibinfo{person}{Jiayi Yao}, \bibinfo{person}{Hanchen Li}, \bibinfo{person}{Yuhan Liu}, \bibinfo{person}{Siddhant Ray}, \bibinfo{person}{Yihua Cheng}, \bibinfo{person}{Qizheng Zhang}, \bibinfo{person}{Kuntai Du}, \bibinfo{person}{Shan Lu}, {and} \bibinfo{person}{Junchen Jiang}.} \bibinfo{year}{2024}\natexlab{}.
\newblock \showarticletitle{CacheBlend: Fast Large Language Model Serving for {RAG} with Cached Knowledge Fusion}.
\newblock \bibinfo{journal}{\emph{CoRR}}  \bibinfo{volume}{abs/2405.16444} (\bibinfo{year}{2024}).
\newblock
\urldef\tempurl%
\url{https://doi.org/10.48550/ARXIV.2405.16444}
\showDOI{\tempurl}
\showeprint[arXiv]{2405.16444}


\bibitem[Yao et~al\mbox{.}(2023a)]%
        {tot2023}
\bibfield{author}{\bibinfo{person}{Shunyu Yao}, \bibinfo{person}{Dian Yu}, \bibinfo{person}{Jeffrey Zhao}, \bibinfo{person}{Izhak Shafran}, \bibinfo{person}{Tom Griffiths}, \bibinfo{person}{Yuan Cao}, {and} \bibinfo{person}{Karthik Narasimhan}.} \bibinfo{year}{2023}\natexlab{a}.
\newblock \showarticletitle{Tree of Thoughts: Deliberate Problem Solving with Large Language Models}. In \bibinfo{booktitle}{\emph{Advances in Neural Information Processing Systems 36: Annual Conference on Neural Information Processing Systems 2023, NeurIPS 2023, New Orleans, LA, USA, December 10 - 16, 2023}}, \bibfield{editor}{\bibinfo{person}{Alice Oh}, \bibinfo{person}{Tristan Naumann}, \bibinfo{person}{Amir Globerson}, \bibinfo{person}{Kate Saenko}, \bibinfo{person}{Moritz Hardt}, {and} \bibinfo{person}{Sergey Levine}} (Eds.).
\newblock
\urldef\tempurl%
\url{http://papers.nips.cc/paper\_files/paper/2023/hash/271db9922b8d1f4dd7aaef84ed5ac703-Abstract-Conference.html}
\showURL{%
\tempurl}


\bibitem[Yao et~al\mbox{.}(2023b)]%
        {agenticLLMreact2023}
\bibfield{author}{\bibinfo{person}{Shunyu Yao}, \bibinfo{person}{Jeffrey Zhao}, \bibinfo{person}{Dian Yu}, \bibinfo{person}{Nan Du}, \bibinfo{person}{Izhak Shafran}, \bibinfo{person}{Karthik~R. Narasimhan}, {and} \bibinfo{person}{Yuan Cao}.} \bibinfo{year}{2023}\natexlab{b}.
\newblock \showarticletitle{ReAct: Synergizing Reasoning and Acting in Language Models}. In \bibinfo{booktitle}{\emph{The Eleventh International Conference on Learning Representations, {ICLR} 2023, Kigali, Rwanda, May 1-5, 2023}}. \bibinfo{publisher}{OpenReview.net}.
\newblock
\urldef\tempurl%
\url{https://openreview.net/forum?id=WE\_vluYUL-X}
\showURL{%
\tempurl}


\bibitem[Yasunaga et~al\mbox{.}(2023)]%
        {imageRAG2023}
\bibfield{author}{\bibinfo{person}{Michihiro Yasunaga}, \bibinfo{person}{Armen Aghajanyan}, \bibinfo{person}{Weijia Shi}, \bibinfo{person}{Richard James}, \bibinfo{person}{Jure Leskovec}, \bibinfo{person}{Percy Liang}, \bibinfo{person}{Mike Lewis}, \bibinfo{person}{Luke Zettlemoyer}, {and} \bibinfo{person}{Wen{-}Tau Yih}.} \bibinfo{year}{2023}\natexlab{}.
\newblock \showarticletitle{Retrieval-Augmented Multimodal Language Modeling}. In \bibinfo{booktitle}{\emph{International Conference on Machine Learning, {ICML} 2023, 23-29 July 2023, Honolulu, Hawaii, {USA}}} \emph{(\bibinfo{series}{Proceedings of Machine Learning Research}, Vol.~\bibinfo{volume}{202})}, \bibfield{editor}{\bibinfo{person}{Andreas Krause}, \bibinfo{person}{Emma Brunskill}, \bibinfo{person}{Kyunghyun Cho}, \bibinfo{person}{Barbara Engelhardt}, \bibinfo{person}{Sivan Sabato}, {and} \bibinfo{person}{Jonathan Scarlett}} (Eds.). \bibinfo{publisher}{{PMLR}}, \bibinfo{pages}{39755--39769}.
\newblock
\urldef\tempurl%
\url{https://proceedings.mlr.press/v202/yasunaga23a.html}
\showURL{%
\tempurl}


\bibitem[Ye et~al\mbox{.}(2024a)]%
        {differentialtransformer2024}
\bibfield{author}{\bibinfo{person}{Tianzhu Ye}, \bibinfo{person}{Li Dong}, \bibinfo{person}{Yuqing Xia}, \bibinfo{person}{Yutao Sun}, \bibinfo{person}{Yi Zhu}, \bibinfo{person}{Gao Huang}, {and} \bibinfo{person}{Furu Wei}.} \bibinfo{year}{2024}\natexlab{a}.
\newblock \showarticletitle{Differential transformer}.
\newblock \bibinfo{journal}{\emph{arXiv preprint arXiv:2410.05258}} (\bibinfo{year}{2024}).
\newblock


\bibitem[Ye et~al\mbox{.}(2024b)]%
        {casecadesharedprefixbatchdecode2024}
\bibfield{author}{\bibinfo{person}{Zihao Ye}, \bibinfo{person}{Ruihang Lai}, \bibinfo{person}{Bo-Ru Lu}, \bibinfo{person}{Chien-Yu Lin}, \bibinfo{person}{Size Zheng}, \bibinfo{person}{Lequn Chen}, \bibinfo{person}{Tianqi Chen}, {and} \bibinfo{person}{Luis Ceze}.} \bibinfo{year}{2024}\natexlab{b}.
\newblock \bibinfo{title}{Cascade inference: Memory bandwidth efficient shared prefix batch decoding}.
\newblock
\newblock


\bibitem[Yu et~al\mbox{.}(2024)]%
        {TwinPilots}
\bibfield{author}{\bibinfo{person}{Chengye Yu}, \bibinfo{person}{Tianyu Wang}, \bibinfo{person}{Zili Shao}, \bibinfo{person}{Linjie Zhu}, \bibinfo{person}{Xu Zhou}, {and} \bibinfo{person}{Song Jiang}.} \bibinfo{year}{2024}\natexlab{}.
\newblock \showarticletitle{Twinpilots: A new computing paradigm for gpu-cpu parallel llm inference}. In \bibinfo{booktitle}{\emph{Proceedings of the 17th ACM International Systems and Storage Conference}}. \bibinfo{pages}{91--103}.
\newblock


\bibitem[Yu et~al\mbox{.}(2022)]%
        {Orca}
\bibfield{author}{\bibinfo{person}{Gyeong{-}In Yu}, \bibinfo{person}{Joo~Seong Jeong}, \bibinfo{person}{Geon{-}Woo Kim}, \bibinfo{person}{Soojeong Kim}, {and} \bibinfo{person}{Byung{-}Gon Chun}.} \bibinfo{year}{2022}\natexlab{}.
\newblock \showarticletitle{Orca: {A} Distributed Serving System for Transformer-Based Generative Models}. In \bibinfo{booktitle}{\emph{16th {USENIX} Symposium on Operating Systems Design and Implementation, {OSDI} 2022, Carlsbad, CA, USA, July 11-13, 2022}}, \bibfield{editor}{\bibinfo{person}{Marcos~K. Aguilera} {and} \bibinfo{person}{Hakim Weatherspoon}} (Eds.). \bibinfo{publisher}{{USENIX} Association}, \bibinfo{pages}{521--538}.
\newblock
\urldef\tempurl%
\url{https://www.usenix.org/conference/osdi22/presentation/yu}
\showURL{%
\tempurl}


\bibitem[Yu et~al\mbox{.}(2023)]%
        {bettercotpromptsurvey2023}
\bibfield{author}{\bibinfo{person}{Zihan Yu}, \bibinfo{person}{Liang He}, \bibinfo{person}{Zhen Wu}, \bibinfo{person}{Xinyu Dai}, {and} \bibinfo{person}{Jiajun Chen}.} \bibinfo{year}{2023}\natexlab{}.
\newblock \showarticletitle{Towards Better Chain-of-Thought Prompting Strategies: {A} Survey}.
\newblock \bibinfo{journal}{\emph{CoRR}}  \bibinfo{volume}{abs/2310.04959} (\bibinfo{year}{2023}).
\newblock
\urldef\tempurl%
\url{https://doi.org/10.48550/ARXIV.2310.04959}
\showDOI{\tempurl}
\showeprint[arXiv]{2310.04959}


\bibitem[Yuan et~al\mbox{.}(2024a)]%
        {tableRAGmultigranularity2024}
\bibfield{author}{\bibinfo{person}{Ruize Yuan}, \bibinfo{person}{Xiang Ao}, \bibinfo{person}{Li Zeng}, {and} \bibinfo{person}{Qing He}.} \bibinfo{year}{2024}\natexlab{a}.
\newblock \showarticletitle{{DRAMA:} Dynamic Multi-Granularity Graph Estimate Retrieval over Tabular and Textual Question Answering}. In \bibinfo{booktitle}{\emph{Proceedings of the 2024 Joint International Conference on Computational Linguistics, Language Resources and Evaluation, {LREC/COLING} 2024, 20-25 May, 2024, Torino, Italy}}, \bibfield{editor}{\bibinfo{person}{Nicoletta Calzolari}, \bibinfo{person}{Min{-}Yen Kan}, \bibinfo{person}{V{\'{e}}ronique Hoste}, \bibinfo{person}{Alessandro Lenci}, \bibinfo{person}{Sakriani Sakti}, {and} \bibinfo{person}{Nianwen Xue}} (Eds.). \bibinfo{publisher}{{ELRA} and {ICCL}}, \bibinfo{pages}{5365--5375}.
\newblock
\urldef\tempurl%
\url{https://aclanthology.org/2024.lrec-main.477}
\showURL{%
\tempurl}


\bibitem[Yuan et~al\mbox{.}(2024c)]%
        {nsDB}
\bibfield{author}{\bibinfo{person}{Ye Yuan}, \bibinfo{person}{Bo Tang}, \bibinfo{person}{Tianfei Zhou}, \bibinfo{person}{Zhiwei Zhang}, {and} \bibinfo{person}{Jianbin Qin}.} \bibinfo{year}{2024}\natexlab{c}.
\newblock \showarticletitle{nsDB: Architecting the Next Generation Database by Integrating Neural and Symbolic Systems (Vision)}.
\newblock \bibinfo{journal}{\emph{Proc. {VLDB} Endow.}} \bibinfo{volume}{17}, \bibinfo{number}{11} (\bibinfo{year}{2024}), \bibinfo{pages}{3283--3289}.
\newblock
\urldef\tempurl%
\url{https://www.vldb.org/pvldb/vol17/p3283-tang.pdf}
\showURL{%
\tempurl}


\bibitem[Yuan et~al\mbox{.}(2024b)]%
        {LLMViewer}
\bibfield{author}{\bibinfo{person}{Zhihang Yuan}, \bibinfo{person}{Yuzhang Shang}, \bibinfo{person}{Yang Zhou}, \bibinfo{person}{Zhen Dong}, \bibinfo{person}{Zhe Zhou}, \bibinfo{person}{Chenhao Xue}, \bibinfo{person}{Bingzhe Wu}, \bibinfo{person}{Zhikai Li}, \bibinfo{person}{Qingyi Gu}, \bibinfo{person}{Yong~Jae Lee}, \bibinfo{person}{Yan Yan}, \bibinfo{person}{Beidi Chen}, \bibinfo{person}{Guangyu Sun}, {and} \bibinfo{person}{Kurt Keutzer}.} \bibinfo{year}{2024}\natexlab{b}.
\newblock \showarticletitle{{LLM} Inference Unveiled: Survey and Roofline Model Insights}.
\newblock \bibinfo{journal}{\emph{CoRR}}  \bibinfo{volume}{abs/2402.16363} (\bibinfo{year}{2024}).
\newblock
\urldef\tempurl%
\url{https://doi.org/10.48550/ARXIV.2402.16363}
\showDOI{\tempurl}
\showeprint[arXiv]{2402.16363}


\bibitem[Zaharia et~al\mbox{.}(2024)]%
        {compoundAI}
\bibfield{author}{\bibinfo{person}{Matei Zaharia}, \bibinfo{person}{Omar Khattab}, \bibinfo{person}{Lingjiao Chen}, \bibinfo{person}{Jared~Quincy Davis}, \bibinfo{person}{Heather Miller}, \bibinfo{person}{Chris Potts}, \bibinfo{person}{James Zou}, \bibinfo{person}{Michael Carbin}, \bibinfo{person}{Jonathan Frankle}, \bibinfo{person}{Naveen Rao}, {and} \bibinfo{person}{Ali Ghodsi}.} \bibinfo{year}{2024}\natexlab{}.
\newblock \bibinfo{title}{The Shift from Models to Compound AI Systems}.
\newblock \bibinfo{howpublished}{\url{https://bair.berkeley.edu/blog/2024/02/18/compound-ai-systems/}}.
\newblock


\bibitem[Zhang et~al\mbox{.}(2024f)]%
        {multipathllamaberrypairwiseoptimizationolympiad2024}
\bibfield{author}{\bibinfo{person}{Di Zhang}, \bibinfo{person}{Jianbo Wu}, \bibinfo{person}{Jingdi Lei}, \bibinfo{person}{Tong Che}, \bibinfo{person}{Jiatong Li}, \bibinfo{person}{Tong Xie}, \bibinfo{person}{Xiaoshui Huang}, \bibinfo{person}{Shufei Zhang}, \bibinfo{person}{Marco Pavone}, \bibinfo{person}{Yuqiang Li}, \bibinfo{person}{Wanli Ouyang}, {and} \bibinfo{person}{Dongzhan Zhou}.} \bibinfo{year}{2024}\natexlab{f}.
\newblock \showarticletitle{LLaMA-Berry: Pairwise Optimization for O1-like Olympiad-Level Mathematical Reasoning}.
\newblock \bibinfo{journal}{\emph{CoRR}}  \bibinfo{volume}{abs/2410.02884} (\bibinfo{year}{2024}).
\newblock
\urldef\tempurl%
\url{https://doi.org/10.48550/ARXIV.2410.02884}
\showDOI{\tempurl}
\showeprint[arXiv]{2410.02884}


\bibitem[Zhang et~al\mbox{.}(2024c)]%
        {cascademultiobjective2024}
\bibfield{author}{\bibinfo{person}{Kai Zhang}, \bibinfo{person}{Liqian Peng}, \bibinfo{person}{Congchao Wang}, \bibinfo{person}{Alec Go}, {and} \bibinfo{person}{Xiaozhong Liu}.} \bibinfo{year}{2024}\natexlab{c}.
\newblock \showarticletitle{{LLM} Cascade with Multi-Objective Optimal Consideration}.
\newblock \bibinfo{journal}{\emph{CoRR}}  \bibinfo{volume}{abs/2410.08014} (\bibinfo{year}{2024}).
\newblock
\urldef\tempurl%
\url{https://doi.org/10.48550/ARXIV.2410.08014}
\showDOI{\tempurl}
\showeprint[arXiv]{2410.08014}


\bibitem[Zhang et~al\mbox{.}(2023a)]%
        {RAGanything2023}
\bibfield{author}{\bibinfo{person}{Peitian Zhang}, \bibinfo{person}{Shitao Xiao}, \bibinfo{person}{Zheng Liu}, \bibinfo{person}{Zhicheng Dou}, {and} \bibinfo{person}{Jian{-}Yun Nie}.} \bibinfo{year}{2023}\natexlab{a}.
\newblock \showarticletitle{Retrieve Anything To Augment Large Language Models}.
\newblock \bibinfo{journal}{\emph{CoRR}}  \bibinfo{volume}{abs/2310.07554} (\bibinfo{year}{2023}).
\newblock
\urldef\tempurl%
\url{https://doi.org/10.48550/ARXIV.2310.07554}
\showDOI{\tempurl}
\showeprint[arXiv]{2310.07554}


\bibitem[Zhang et~al\mbox{.}(2023b)]%
        {relationalvectorvbase2023}
\bibfield{author}{\bibinfo{person}{Qianxi Zhang}, \bibinfo{person}{Shuotao Xu}, \bibinfo{person}{Qi Chen}, \bibinfo{person}{Guoxin Sui}, \bibinfo{person}{Jiadong Xie}, \bibinfo{person}{Zhizhen Cai}, \bibinfo{person}{Yaoqi Chen}, \bibinfo{person}{Yinxuan He}, \bibinfo{person}{Yuqing Yang}, \bibinfo{person}{Fan Yang}, \bibinfo{person}{Mao Yang}, {and} \bibinfo{person}{Lidong Zhou}.} \bibinfo{year}{2023}\natexlab{b}.
\newblock \showarticletitle{{VBASE:} Unifying Online Vector Similarity Search and Relational Queries via Relaxed Monotonicity}. In \bibinfo{booktitle}{\emph{17th {USENIX} Symposium on Operating Systems Design and Implementation, {OSDI} 2023, Boston, MA, USA, July 10-12, 2023}}, \bibfield{editor}{\bibinfo{person}{Roxana Geambasu} {and} \bibinfo{person}{Ed~Nightingale}} (Eds.). \bibinfo{publisher}{{USENIX} Association}, \bibinfo{pages}{377--395}.
\newblock
\urldef\tempurl%
\url{https://www.usenix.org/conference/osdi23/presentation/zhang-qianxi}
\showURL{%
\tempurl}


\bibitem[Zhang et~al\mbox{.}(2024b)]%
        {datacleaningusingllm2024}
\bibfield{author}{\bibinfo{person}{Shuo Zhang}, \bibinfo{person}{Zezhou Huang}, {and} \bibinfo{person}{Eugene Wu}.} \bibinfo{year}{2024}\natexlab{b}.
\newblock \showarticletitle{Data Cleaning Using Large Language Models}.
\newblock \bibinfo{journal}{\emph{arXiv preprint arXiv:2410.15547}} (\bibinfo{year}{2024}).
\newblock


\bibitem[Zhang et~al\mbox{.}(2024e)]%
        {multihopQAhierarchicalrethink2024}
\bibfield{author}{\bibinfo{person}{Xiaoming Zhang}, \bibinfo{person}{Ming Wang}, \bibinfo{person}{Xiaocui Yang}, \bibinfo{person}{Daling Wang}, \bibinfo{person}{Shi Feng}, {and} \bibinfo{person}{Yifei Zhang}.} \bibinfo{year}{2024}\natexlab{e}.
\newblock \showarticletitle{Hierarchical Retrieval-Augmented Generation Model with Rethink for Multi-hop Question Answering}.
\newblock \bibinfo{journal}{\emph{arXiv preprint arXiv:2408.11875}} (\bibinfo{year}{2024}).
\newblock


\bibitem[Zhang et~al\mbox{.}(2022)]%
        {promptgen2024}
\bibfield{author}{\bibinfo{person}{Yue Zhang}, \bibinfo{person}{Hongliang Fei}, \bibinfo{person}{Dingcheng Li}, {and} \bibinfo{person}{Ping Li}.} \bibinfo{year}{2022}\natexlab{}.
\newblock \showarticletitle{PromptGen: Automatically Generate Prompts using Generative Models}. In \bibinfo{booktitle}{\emph{Findings of the Association for Computational Linguistics: {NAACL} 2022, Seattle, WA, United States, July 10-15, 2022}}, \bibfield{editor}{\bibinfo{person}{Marine Carpuat}, \bibinfo{person}{Marie{-}Catherine de~Marneffe}, {and} \bibinfo{person}{Iv{\'{a}}n Vladimir~Meza Ru{\'{\i}}z}} (Eds.). \bibinfo{publisher}{Association for Computational Linguistics}, \bibinfo{pages}{30--37}.
\newblock
\urldef\tempurl%
\url{https://doi.org/10.18653/V1/2022.FINDINGS-NAACL.3}
\showDOI{\tempurl}


\bibitem[Zhang et~al\mbox{.}(2024a)]%
        {adaptiveRAGshortformopendomainQA2024}
\bibfield{author}{\bibinfo{person}{Zihan Zhang}, \bibinfo{person}{Meng Fang}, {and} \bibinfo{person}{Ling Chen}.} \bibinfo{year}{2024}\natexlab{a}.
\newblock \showarticletitle{RetrievalQA: Assessing Adaptive Retrieval-Augmented Generation for Short-form Open-Domain Question Answering}. In \bibinfo{booktitle}{\emph{Findings of the Association for Computational Linguistics, {ACL} 2024, Bangkok, Thailand and virtual meeting, August 11-16, 2024}}, \bibfield{editor}{\bibinfo{person}{Lun{-}Wei Ku}, \bibinfo{person}{Andre Martins}, {and} \bibinfo{person}{Vivek Srikumar}} (Eds.). \bibinfo{publisher}{Association for Computational Linguistics}, \bibinfo{pages}{6963--6975}.
\newblock
\urldef\tempurl%
\url{https://doi.org/10.18653/V1/2024.FINDINGS-ACL.415}
\showDOI{\tempurl}


\bibitem[Zhang et~al\mbox{.}(2024d)]%
        {multipathnashcot2024}
\bibfield{author}{\bibinfo{person}{Ziqi Zhang}, \bibinfo{person}{Cunxiang Wang}, \bibinfo{person}{Xiao Xiong}, \bibinfo{person}{Yue Zhang}, {and} \bibinfo{person}{Donglin Wang}.} \bibinfo{year}{2024}\natexlab{d}.
\newblock \showarticletitle{Nash CoT: Multi-Path Inference with Preference Equilibrium}. In \bibinfo{booktitle}{\emph{Proceedings of the 2024 Conference on Empirical Methods in Natural Language Processing, {EMNLP} 2024, Miami, FL, USA, November 12-16, 2024}}, \bibfield{editor}{\bibinfo{person}{Yaser Al{-}Onaizan}, \bibinfo{person}{Mohit Bansal}, {and} \bibinfo{person}{Yun{-}Nung Chen}} (Eds.). \bibinfo{publisher}{Association for Computational Linguistics}, \bibinfo{pages}{14572--14587}.
\newblock
\urldef\tempurl%
\url{https://aclanthology.org/2024.emnlp-main.807}
\showURL{%
\tempurl}


\bibitem[Zhao et~al\mbox{.}(2024a)]%
        {HeteGen}
\bibfield{author}{\bibinfo{person}{Xuanlei Zhao}, \bibinfo{person}{Bin Jia}, \bibinfo{person}{Haotian Zhou}, \bibinfo{person}{Ziming Liu}, \bibinfo{person}{Shenggan Cheng}, {and} \bibinfo{person}{Yang You}.} \bibinfo{year}{2024}\natexlab{a}.
\newblock \showarticletitle{HeteGen: Efficient Heterogeneous Parallel Inference for Large Language Models on Resource-Constrained Devices}. In \bibinfo{booktitle}{\emph{Proceedings of the Seventh Annual Conference on Machine Learning and Systems, MLSys 2024, Santa Clara, CA, USA, May 13-16, 2024}}, \bibfield{editor}{\bibinfo{person}{Phillip~B. Gibbons}, \bibinfo{person}{Gennady Pekhimenko}, {and} \bibinfo{person}{Christopher~De Sa}} (Eds.). \bibinfo{publisher}{mlsys.org}.
\newblock
\urldef\tempurl%
\url{https://proceedings.mlsys.org/paper\_files/paper/2024/hash/5431dca75a8d2abc1fb51e89e8324f10-Abstract-Conference.html}
\showURL{%
\tempurl}


\bibitem[Zhao et~al\mbox{.}(2024b)]%
        {BlendServeprefixsharing2024}
\bibfield{author}{\bibinfo{person}{Yilong Zhao}, \bibinfo{person}{Shuo Yang}, \bibinfo{person}{Kan Zhu}, \bibinfo{person}{Lianmin Zheng}, \bibinfo{person}{Baris Kasikci}, \bibinfo{person}{Yang Zhou}, \bibinfo{person}{Jiarong Xing}, {and} \bibinfo{person}{Ion Stoica}.} \bibinfo{year}{2024}\natexlab{b}.
\newblock \showarticletitle{BlendServe: Optimizing Offline Inference for Auto-regressive Large Models with Resource-aware Batching}.
\newblock \bibinfo{journal}{\emph{arXiv preprint arXiv:2411.16102}} (\bibinfo{year}{2024}).
\newblock


\bibitem[Zheng et~al\mbox{.}(2023b)]%
        {SGLang}
\bibfield{author}{\bibinfo{person}{Lianmin Zheng}, \bibinfo{person}{Liangsheng Yin}, \bibinfo{person}{Zhiqiang Xie}, \bibinfo{person}{Jeff Huang}, \bibinfo{person}{Chuyue Sun}, \bibinfo{person}{Cody~Hao Yu}, \bibinfo{person}{Shiyi Cao}, \bibinfo{person}{Christos Kozyrakis}, \bibinfo{person}{Ion Stoica}, \bibinfo{person}{Joseph~E. Gonzalez}, \bibinfo{person}{Clark~W. Barrett}, {and} \bibinfo{person}{Ying Sheng}.} \bibinfo{year}{2023}\natexlab{b}.
\newblock \showarticletitle{Efficiently Programming Large Language Models using SGLang}.
\newblock \bibinfo{journal}{\emph{CoRR}}  \bibinfo{volume}{abs/2312.07104} (\bibinfo{year}{2023}).
\newblock
\urldef\tempurl%
\url{https://doi.org/10.48550/ARXIV.2312.07104}
\showDOI{\tempurl}
\showeprint[arXiv]{2312.07104}


\bibitem[Zheng et~al\mbox{.}(2024)]%
        {CITERmodelcascadestokenlevelrouting2024}
\bibfield{author}{\bibinfo{person}{Wenhao Zheng}, \bibinfo{person}{Yixiao Chen}, \bibinfo{person}{Weitong Zhang}, \bibinfo{person}{Souvik Kundu}, \bibinfo{person}{Yun Li}, \bibinfo{person}{Zhengzhong Liu}, \bibinfo{person}{Eric~P Xing}, \bibinfo{person}{Hongyi Wang}, {and} \bibinfo{person}{Huaxiu Yao}.} \bibinfo{year}{2024}\natexlab{}.
\newblock \showarticletitle{CITER: Collaborative Inference for Efficient Large Language Model Decoding with Token-Level Routing}. In \bibinfo{booktitle}{\emph{Adaptive Foundation Models: Evolving AI for Personalized and Efficient Learning}}.
\newblock


\bibitem[Zheng et~al\mbox{.}(2023a)]%
        {OPrediction2}
\bibfield{author}{\bibinfo{person}{Zangwei Zheng}, \bibinfo{person}{Xiaozhe Ren}, \bibinfo{person}{Fuzhao Xue}, \bibinfo{person}{Yang Luo}, \bibinfo{person}{Xin Jiang}, {and} \bibinfo{person}{Yang You}.} \bibinfo{year}{2023}\natexlab{a}.
\newblock \showarticletitle{Response Length Perception and Sequence Scheduling: An LLM-Empowered {LLM} Inference Pipeline}. In \bibinfo{booktitle}{\emph{Advances in Neural Information Processing Systems 36: Annual Conference on Neural Information Processing Systems 2023, NeurIPS 2023, New Orleans, LA, USA, December 10 - 16, 2023}}, \bibfield{editor}{\bibinfo{person}{Alice Oh}, \bibinfo{person}{Tristan Naumann}, \bibinfo{person}{Amir Globerson}, \bibinfo{person}{Kate Saenko}, \bibinfo{person}{Moritz Hardt}, {and} \bibinfo{person}{Sergey Levine}} (Eds.).
\newblock
\urldef\tempurl%
\url{http://papers.nips.cc/paper\_files/paper/2023/hash/ce7ff3405c782f761fac7f849b41ae9a-Abstract-Conference.html}
\showURL{%
\tempurl}


\bibitem[Zhong et~al\mbox{.}(2024a)]%
        {RTODPOPPO2024}
\bibfield{author}{\bibinfo{person}{Han Zhong}, \bibinfo{person}{Guhao Feng}, \bibinfo{person}{Wei Xiong}, \bibinfo{person}{Li Zhao}, \bibinfo{person}{Di He}, \bibinfo{person}{Jiang Bian}, {and} \bibinfo{person}{Liwei Wang}.} \bibinfo{year}{2024}\natexlab{a}.
\newblock \showarticletitle{{DPO} Meets {PPO:} Reinforced Token Optimization for {RLHF}}.
\newblock \bibinfo{journal}{\emph{CoRR}}  \bibinfo{volume}{abs/2404.18922} (\bibinfo{year}{2024}).
\newblock
\urldef\tempurl%
\url{https://doi.org/10.48550/ARXIV.2404.18922}
\showDOI{\tempurl}
\showeprint[arXiv]{2404.18922}


\bibitem[Zhong et~al\mbox{.}(2024b)]%
        {DistServe}
\bibfield{author}{\bibinfo{person}{Yinmin Zhong}, \bibinfo{person}{Shengyu Liu}, \bibinfo{person}{Junda Chen}, \bibinfo{person}{Jianbo Hu}, \bibinfo{person}{Yibo Zhu}, \bibinfo{person}{Xuanzhe Liu}, \bibinfo{person}{Xin Jin}, {and} \bibinfo{person}{Hao Zhang}.} \bibinfo{year}{2024}\natexlab{b}.
\newblock \showarticletitle{DistServe: Disaggregating Prefill and Decoding for Goodput-optimized Large Language Model Serving}. In \bibinfo{booktitle}{\emph{18th {USENIX} Symposium on Operating Systems Design and Implementation, {OSDI} 2024, Santa Clara, CA, USA, July 10-12, 2024}}, \bibfield{editor}{\bibinfo{person}{Ada Gavrilovska} {and} \bibinfo{person}{Douglas~B. Terry}} (Eds.). \bibinfo{publisher}{{USENIX} Association}, \bibinfo{pages}{193--210}.
\newblock
\urldef\tempurl%
\url{https://www.usenix.org/conference/osdi24/presentation/zhong-yinmin}
\showURL{%
\tempurl}


\bibitem[Zhou et~al\mbox{.}(2023)]%
        {LLMasDBA}
\bibfield{author}{\bibinfo{person}{Xuanhe Zhou}, \bibinfo{person}{Guoliang Li}, {and} \bibinfo{person}{Zhiyuan Liu}.} \bibinfo{year}{2023}\natexlab{}.
\newblock \showarticletitle{{LLM} As {DBA}}.
\newblock \bibinfo{journal}{\emph{CoRR}}  \bibinfo{volume}{abs/2308.05481} (\bibinfo{year}{2023}).
\newblock
\urldef\tempurl%
\url{https://doi.org/10.48550/ARXIV.2308.05481}
\showDOI{\tempurl}
\showeprint[arXiv]{2308.05481}


\bibitem[Zhou et~al\mbox{.}(2024a)]%
        {surveydataaugment2024}
\bibfield{author}{\bibinfo{person}{Yue Zhou}, \bibinfo{person}{Chenlu Guo}, \bibinfo{person}{Xu Wang}, \bibinfo{person}{Yi Chang}, {and} \bibinfo{person}{Yuan Wu}.} \bibinfo{year}{2024}\natexlab{a}.
\newblock \showarticletitle{A Survey on Data Augmentation in Large Model Era}.
\newblock \bibinfo{journal}{\emph{CoRR}}  \bibinfo{volume}{abs/2401.15422} (\bibinfo{year}{2024}).
\newblock
\urldef\tempurl%
\url{https://doi.org/10.48550/ARXIV.2401.15422}
\showDOI{\tempurl}
\showeprint[arXiv]{2401.15422}


\bibitem[Zhou et~al\mbox{.}(2024b)]%
        {LLMinferenceefficiencysurvey2024}
\bibfield{author}{\bibinfo{person}{Zixuan Zhou}, \bibinfo{person}{Xuefei Ning}, \bibinfo{person}{Ke Hong}, \bibinfo{person}{Tianyu Fu}, \bibinfo{person}{Jiaming Xu}, \bibinfo{person}{Shiyao Li}, \bibinfo{person}{Yuming Lou}, \bibinfo{person}{Luning Wang}, \bibinfo{person}{Zhihang Yuan}, \bibinfo{person}{Xiuhong Li}, \bibinfo{person}{Shengen Yan}, \bibinfo{person}{Guohao Dai}, \bibinfo{person}{Xiao{-}Ping Zhang}, \bibinfo{person}{Yuhan Dong}, {and} \bibinfo{person}{Yu Wang}.} \bibinfo{year}{2024}\natexlab{b}.
\newblock \showarticletitle{A Survey on Efficient Inference for Large Language Models}.
\newblock \bibinfo{journal}{\emph{CoRR}}  \bibinfo{volume}{abs/2404.14294} (\bibinfo{year}{2024}).
\newblock
\urldef\tempurl%
\url{https://doi.org/10.48550/ARXIV.2404.14294}
\showDOI{\tempurl}
\showeprint[arXiv]{2404.14294}


\bibitem[Zhu et~al\mbox{.}(2024)]%
        {NanoFlow}
\bibfield{author}{\bibinfo{person}{Kan Zhu}, \bibinfo{person}{Yilong Zhao}, \bibinfo{person}{Liangyu Zhao}, \bibinfo{person}{Gefei Zuo}, \bibinfo{person}{Yile Gu}, \bibinfo{person}{Dedong Xie}, \bibinfo{person}{Yufei Gao}, \bibinfo{person}{Qinyu Xu}, \bibinfo{person}{Tian Tang}, \bibinfo{person}{Zihao Ye}, {et~al\mbox{.}}} \bibinfo{year}{2024}\natexlab{}.
\newblock \showarticletitle{NanoFlow: Towards Optimal Large Language Model Serving Throughput}.
\newblock \bibinfo{journal}{\emph{arXiv preprint arXiv:2408.12757}} (\bibinfo{year}{2024}).
\newblock


\end{thebibliography}

\end{document}